\pdfoutput=1
\documentclass[11pt]{article}
\usepackage{amsfonts, amsmath, amssymb, amsthm}
\usepackage[english]{babel}
\usepackage{booktabs}
\usepackage{bm}
\usepackage{eucal}
\usepackage{enumerate}
\usepackage{float}
\usepackage{dsfont}
\usepackage{epsfig}
\usepackage{fontenc}
\usepackage[hmargin=1.6cm,vmargin=3cm]{geometry}
\usepackage{lscape}
\usepackage{mathrsfs}
\usepackage{natbib}
\usepackage{rotating}
\usepackage{verbatim}
\usepackage[table]{xcolor}
\usepackage{dirtytalk}
\usepackage{multirow}
\usepackage{subfigure}
\usepackage{mleftright}
\usepackage{enumitem}
\usepackage{mleftright}
\usepackage{alltt}
\usepackage[colorlinks,bookmarksopen,bookmarksnumbered,citecolor=blue,urlcolor=green,hypertexnames=false]{hyperref}

\usepackage{framed}
\usepackage{titling}
\usepackage[]{color}
\usepackage{bibunits}

\date{}

\renewcommand{\thefootnote}{$\dagger$}

\begin{document}
\begin{bibunit}[hapalike]
\title{Bayesian nonparametric inference for the covariate-adjusted ROC curve}
\author{\textsc{Vanda~In\'acio de Carvalho} and \textsc{Mar\'ia Xos\'e Rodr\'iguez-\'Alvarez}}
\date{}
\maketitle 

\begin{abstract}
\footnotesize{Accurate diagnosis of disease is of fundamental importance in clinical practice and medical research. Before a medical diagnostic test is routinely used in practice, its ability to distinguish between diseased and nondiseased states must be rigorously assessed through statistical analysis. The receiver operating characteristic (ROC) curve is the most popular used tool for evaluating the discriminatory ability of continuous-outcome diagnostic tests. It has been acknowledged that several factors (e.g., subject-specific characteristics, such as age and/or gender) can affect the test's accuracy beyond disease status. Recently, the covariate-adjusted ROC curve has been proposed and successfully applied as a global summary measure of diagnostic accuracy that takes covariate information into account. We motivate the use of the covariate-adjusted ROC curve and develop a highly robust model based on a combination of B-splines dependent Dirichlet process mixture models and the Bayesian bootstrap. Multiple simulation studies demonstrate the ability of our model to successfully recover the true covariate-adjusted ROC curve and to produce valid inferences in a variety of complex scenarios. Our methods are motivated by and applied to an endocrine study where the main goal is to assess the accuracy of the body mass index, adjusted for age and gender, for predicting clusters of cardiovascular disease risk factors. The \texttt{R}-package \texttt{AROC}, implementing our proposed methods, is provided.}
\end{abstract}

\let\thefootnote\relax\footnotetext{Vanda In\'acio de Carvalho, School of Mathematics, University of Edinburgh, Scotland, UK (\textit{Vanda.Inacio@ed.ac.uk}). Mar\'ia Xos\'e Rodr\'iguez-\'Alvarez, BCAM-Basque Center for Applied Mathematics \& IKERBASQUE, Basque Foundation for Science, Bilbao, Basque Country, Spain (\textit{mxrodriguez@bcamath.org}).}

\textsc{key words:} Bayesian bootstrap; Covariate-adjustment; Diagnostic test; Dependent Dirichlet process mixtures; Receiver operating characteristic curve

\section{\large{\textsf{INTRODUCTION}}}
Evaluating the performance of medical tests for screening and diagnosing disease is of great importance in clinical practice and medical research. Prior to approving the use of a diagnostic test in routine practice, its ability to distinguish diseased from nondiseased individuals must be rigorously vetted through statistical analysis. Throughout this article we use the terms `medical test' and `diagnostic test' to broadly include any continuous classifier (e.g., a single biological marker or a composite univariate score obtained from a combination of multiple biomarkers) for a well-defined condition (termed `disease', with `nondiseased' used to indicate the absence of the condition). \\
\indent The receiver operating characteristic (ROC) curve is the most popular graphical tool to assess the performance of continuous outcome diagnostic tests in discriminating between diseased and nondiseased states. It is a plot of pairs of true positive fraction (or sensitivity) against the false positive fraction (or 1-specificity) for all possible threshold values that can be used to convert continuous outcomes into dichotomous ones. Specifically, let $Y_{D}$ and $Y_{\bar{D}}$ be continuous random variables denoting the diagnostic test outcomes in the diseased and nondiseased groups, respectively, with cumulative distribution functions given by $F_{D}$ and $F_{\bar{D}}$. Further, let $c \in\mathbb{R}$ be a threshold value for defining a positive test result. Without loss of generality, we assume that a subject is classified as diseased (nondiseased) if the test outcome is equal or greater (less) than $c$. Hence, each threshold value $c$ chosen will give rise to a true positive fraction, $\text{TPF}(c)=\Pr(Y_D\geq c)=1-F_{D}(c)$ (the probability that a diseased subject tests positive), and a false positive fraction, $\text{FPF}(c)=\Pr(Y_{\bar{D}}\geq c)=1-F_{\bar{D}}(c)$ (the probability that a nondiseased subject tests positive). The ROC curve is defined as the the set of all $\text{FPF}-\text{TPF}$ pairs that can be obtained on the threshold value $c$ varying, i.e., $\{\text{FPF}(c), \text{TPF}(c): c\in\mathbb{R}\}$. If we let the FPF at threshold $c$ to be $t$, i.e., $t = \text{FPF}(c) = 1-F_{\bar{D}}(c)$, the ROC curve can be equivalently written as $\left\{\left(t, \text{ROC}(t)\right): t \in [0,1] \right\}$, where
\begin{equation*}
\text{ROC}(t)=\Pr\{Y_{D}>F_{\bar{D}}^{-1}(1-t)\}=1-F_{D}\{F_{\bar{D}}^{-1}(1-t)\},\qquad 0\leq t\leq 1.
\end{equation*}
When the distributions of test outcomes in the diseased and nondiseased groups completely overlap, the ROC curve is the diagonal line of the unit square (that is, $\text{TPF}(c)=\text{FPF}(c)$ for all c), thus indicating a useless test. On the other hand, the more separated the distributions the closer the ROC curve is to the point $(0,1)$ in the unit square. A curve that reaches the point $(0,1)$ has $\text{TPF}(c)=1$ and $\text{FPF}(c)=0$ for some threshold c, and hence corresponds to a perfect test. Related to the ROC curve, several indexes can be used as summaries of the accuracy of a diagnostic test. The  area under the ROC curve (AUC), $\int_0^1 \text{ROC}(t)\text{d}t$, is the one most widely used. For a test that perfectly discriminates diseased individuals from nondiseased ones, $\text{AUC}=1$, while for a useless test $\text{AUC}=0.5$. \\
\indent Recently, it has been acknowledged that other factors can affect test's accuracy beyond disease status; examples of such factors include different test settings and subject-specific characteristics (e.g., age and gender). Whenever available, covariate information must be incorporated in ROC analysis as failure to do so might result in oversimplified and biased conclusions about a test's accuracy \citep[see, e.g., ][Section 2]{Pardo14}, whereas stratifying by covariates might be impractical for continuous ones. Denoting as $\mathbf{X}_{D}$ and $\mathbf{X}_{\bar{D}}$ the diseased and nondiseased vector of covariates we are interested in, the covariate-specific ROC curve, given a covariate value $\mathbf{x}$, is
\begin{align*}
\text{ROC}(t\mid\mathbf{x})&=\Pr\{Y_{D}>F_{\bar{D}}^{-1}(1-t\mid\mathbf{X}_{\bar{D}}=\mathbf{x})\mid\mathbf{X}_{D}=\mathbf{x}\}\\
&=1-F_{D}\{F_{\bar{D}}^{-1}(1-t\mid\mathbf{X}_{\bar{D}}=\mathbf{x})\mid\mathbf{X}_{D}=\mathbf{x}\},
\end{align*}
where $F_{D}(y\mid\mathbf{X}_{D}=\mathbf{x})=\Pr(Y_{D}\leq y\mid\mathbf{X}_{D}=\mathbf{x})$ is the conditional cumulative distribution function for $Y_D$ conditional on $\mathbf{X}_{D}=\mathbf{x}$ and with $F_{\bar{D}}(y\mid\mathbf{X}_{\bar{D}}=\mathbf{x})$ being similarly defined. Similarly to the unconditional case, the covariate-specific AUC is $\text{AUC}(\mathbf{x})=\int_{0}^{1}\text{ROC}(t\mid\mathbf{x})\text{d}t$. Note that for each possible value $\mathbf{x}$ we might obtain a different ROC curve/AUC and therefore a possibly different diagnostic accuracy. The covariate-specific ROC curve is, therefore, an important tool that helps to understand the optimal and suboptimal populations where to perform the tests on.\\
\indent Although the covariate-specific ROC curve and corresponding AUC depict the accuracy of the test for every possible covariate value $\mathbf{x}$ that we might be interested in, it would be undoubtedly useful to have a global summary measure that also takes covariate information into account. Such summary measure was developed by \cite{Janes09a}, who proposed the covariate-adjusted ROC (AROC) curve, defined as
\begin{equation*}
\text{AROC}(t)=\int \text{ROC}(t\mid\mathbf{x})\text{d}H_{D}(\mathbf{x}),
\end{equation*}
where $H_{D}(\mathbf{x})=\Pr(\mathbf{X}_{D}\leq \mathbf{x})$ is the cumulative distribution function of $\mathbf{X}_{D}$. That is, the AROC curve is an average of covariate-specific ROC curves, weighted according to the distribution of the covariates in the diseased group. The covariate-adjusted AUC (AAUC) is thus expressed as
\begin{equation*}
\text{AAUC}=\int_{0}^{1}\text{AROC}(t)\text{d}t=\int_{0}^{1}\int \text{ROC}(t\mid\mathbf{x})\text{d}H_{D}(\mathbf{x}) \text{d}t=\int \text{AUC}(\mathbf{x})\text{d}H_{D}(\mathbf{x}).
\end{equation*}
Note that the AAUC is also a weighted average of covariate-specific AUCs. \cite{Janes09a} shown that the AROC can be equivalently represented as
\begin{align}\label{aroc}
\text{AROC}(t)&=\Pr\{Y_{D}>F_{\bar{D}}^{-1}(1-t\mid \mathbf{X}_{D})\} \nonumber \\
&=\Pr\{1-F_{\bar{D}}(Y_D\mid\mathbf{X}_{D})\leq t\}.
\end{align}
This representation makes clear that the AROC at a FPF of $t$ is the overall true positive fraction when the thresholds used for defining a positive test are covariate-specific and chosen to ensure that the FPF is $t$ in each subpopulation defined by the covariate values. The AROC thus summarises the covariate-specific performance of the test and is particularly useful when sample sizes are reduced and covariate-specific ROC curves cannot be estimated precisely.\\
\indent In this paper, besides providing insight about the AROC curve and its relationship to the pooled ROC curve and the covariate-specific ROC curve, we develop a Bayesian nonparametric approach to obtain data-driven inference for the AROC curve and the AAUC. Specifically, our method combines a B-splines dependent Dirichlet process mixture model to estimate the conditional cumulative distribution function of test outcomes in the nondiseased population $F_{\bar{D}}(\cdot\mid \mathbf{X}_{\bar{D}} = \mathbf{x})$ and the Bayesian bootstrap to estimate the outside probability in \eqref{aroc}. This results in a highly flexible estimator that can adapt to intricate distributional features, such as nonlinearities, multimodality, skewness, and/or extreme variability, without the need to know them in advance. Hence, the Bayesian nonparametric model we propose is a widely applicable approach to inference for the AROC/AAUC that can be used for many populations and for a large number of diseases and continuous diagnostic measures. To the best of our knowledge, the only literature on estimating the AROC involves semiparametric linear models \citep{Janes09a} and kernel-type of estimators \citep{MX11a}. We also mention the work of \cite{Guan2012} where semiparametric techniques for estimating the AROC are adopted, but the method is only devised for categorical covariates thus limiting its applicability in practice. Compared to these approaches, our method has a number of advantages, namely: (i) in contrast to \cite{Janes09a}, by using B-splines regression we overcome the linearity assumption which, more often than not, is violated in practice, and (ii) by opposition to the kernel type of approach, we allow the entire distribution of test outcomes to smoothly change as a function of the covariates, and not only the mean and variance. Besides, and even more importantly, we can easily handle multiple covariates, either continuous or categorical. Further, point estimates and credible intervals for the AROC curve and its corresponding AAUC are obtained in a single integrated framework. Software is provided in the \texttt{R} package \citep{R18} \texttt{AROC} available on \texttt{https://CRAN.R-project.org/package=AROC}. Recent developments of Bayesian flexible models that have been successfully applied in medical diagnostic testing research abound \citep{Erkanli2006,Branscum2008,Branscum2015,Gu2008,Hanson2008,Inacio2011,Inacio13,Inacio17,Inacio18,Rodriguez2013,Hwang2015,Zhao2016,Inacio2018}.\\
\indent The remainder of this paper is organised as follows. In the next section, we provide deep reasoning about the AROC curve. In Section \ref{BModel} we introduce our new approach to Bayesian nonparametric inference for the AROC curve and the AAUC. The performance of our methods is validated in Section \ref{simulation} using multiple simulation studies. Section \ref{application} applies our methods to assess the accuracy of the body mass index, adjusted for age and gender, for predicting clusters of cardiovascular disease risk factors. Concluding remarks are provided in Section \ref{remarks}. Some technical details and extra simulations are available in (Web) Appendices, where we also describe the \texttt{R}-package \texttt{AROC} that accompanies this paper.

\section{\large{\textsf{MOTIVATIONAL EXAMPLES}}}\label{examples}
We shed light on the differences and similarities among the pooled or marginal ROC curve (obtained by pooling the test outcomes in each group altogether regardless of the covariate values), the covariate-specific ROC curve and the AROC curve introduced in the previous section. It may be surprising that the pooled/marginal ROC curve is not an average of covariate-specific ROC curves and, thus, it does not always coincide with the AROC.

We start by re-expressing the different ROC curves (pooled/marginal, covariate-specific and AROC) in such a way that the differences and similarities become clearer. Associated with each threshold value $c\in\mathbb{R}$, we may define the marginal ($t$) and covariate-specific ($t_{\mathbf{x}}$) FPF as follows
\begin{align*}
t & = \Pr(Y_{\bar{D}} > c),\\
t_{\mathbf{x}} & = \Pr(Y_{\bar{D}} > c \mid \mathbf{X}_{\bar{D}} = \mathbf{x}).
\end{align*}
We note that \citep[][Chapter 6]{Pepe03},
\begin{align}
t = \Pr(Y_{\bar{D}} > c) = \int \Pr(Y_{\bar{D}} > c \mid \mathbf{X}_{\bar{D}} = \mathbf{x})\mbox{d}H_{\bar{D}}(\mathbf{x}) = \int t_{\mathbf{x}}\mbox{d}H_{\bar{D}}(\mathbf{x}),
\label{marg_FPF}
\end{align}
where $H_{\bar{D}}(\mathbf{x}) = \Pr\left(\mathbf{X}_{\bar{D}} \leq \mathbf{x}\right)$. Thus, for a particular threshold value $c$, the marginal FPF, $t$, is a weighted average of covariate-specific FPFs, $t_{\mathbf{x}}$, when the same threshold value $c$ is used for defining a positive test result in each of the subpopulations defined by the covariates. On the other hand, by specifying the FPF, $t$, we are interested in, it is possible to define the covariate-specific threshold values $c_{\mathbf{x}}$ which give rise to a FPF of $t$ in each of the subpopulations defined by the covariates, i.e.,
\[
c_{\mathbf{x}} = F_{\bar{D}}^{-1}\left(1-t \mid \mathbf{X}_{\bar{D}} = \mathbf{x} \right).
\]
Bearing this in mind, the covariate-specific, pooled/marginal and AROC curves can be expressed, respectively, as follows
\begin{align}
\mbox{ROC}(t \mid \mathbf{x}) & = \Pr(Y_D > c_{\mathbf{x}} \mid \mathbf{X}_{D} = \mathbf{x}), \nonumber \\
\mbox{ROC}(t) & = \Pr(Y_D > c) = \int \Pr(Y_D > c \mid \mathbf{X}_{D} = \mathbf{x})\mbox{d}H_{D}(\mathbf{x}) = \int \mbox{ROC}(t_{\mathbf{x}}\mid \mathbf{x})\mbox{d}H_{D}( \mathbf{x}),\label{marg_TPF}\\
\mbox{AROC}(t) & = \int \mbox{ROC}(t\mid \mathbf{x})\mbox{d}H_{D}( \mathbf{x}) = \int \Pr(Y_D > c_{\mathbf{x}} \mid \mathbf{X}_{D} = \mathbf{x})\mbox{d}H_{D}(\mathbf{x}).\label{AROC_TPF}
\end{align}
Thus, the marginal/pooled ROC curve, $\mbox{ROC}(\cdot)$, and the AROC, $\mbox{AROC}(\cdot)$, can both be expressed in terms of covariate-specific ROC curves, $\mbox{ROC}(\cdot \mid \mathbf{x})$. However, there are important differences. An operating point (i.e., a FPF-TPF pair), in the marginal/pooled ROC curve is obtained by using a common threshold value for defining a positive test result in each of the subpopulations defined by the covariates. The marginal FPF at threshold $c$ is then the average of covariate-specific FPFs (using $c$ as threshold), weighted according to the distribution of the covariates in the nondiseased group (see (\ref{marg_FPF})). The same applies to the marginal TPF (i.e., the ROC curve), but here the weights correspond to the distribution of the covariates in the diseased group (see (\ref{marg_TPF})). On the other hand, an operating point in the AROC is obtained by using covariate-specific threshold values, $c_{\mathbf{x}}$, which ensure the same (covariate-specific) FPF across the subpopulations. The TPF, in this case, is the average of covariate-specific TPFs calculated at the covariate-specific thresholds (see (\ref{AROC_TPF})).

In Figure~\ref{mot_example} we offer deeper insight into the practical implications of these differences. In all cases, we consider a single continuous covariate (although results generalise to multiple covariates, both continuous and categorical) that might or might not be associated with the diagnostic test outcome, and that might or might not impact its diagnostic accuracy. Figure~\ref{no_association} depicts the pooled/marginal and AROC curves in a situation in which the covariate is not associated with the diagnostic test outcome ($Y_{\bar{D}} \sim N(0.5, 0.3^2)$ and $Y_{D} \sim N(1, 0.3^2)$). In this case, it is easy to show that the covariate-specific ROC curve is the same for all covariate values, and it coincides with both the pooled/marginal ROC curve and the AROC. In turn, in Figures~\ref{no_effect}~and~\ref{effect} we show several examples where the covariate is associated with the diagnostic test outcome. For the examples shown in Figure~\ref{no_effect}, the covariate does not impact the accuracy of the diagnostic test ($Y_{\bar{D}} \sim N(0.5 + X_{\bar{D}}, 0.3^2)$ and $Y_{D} \sim N(0.75 + X_{D}, 0.3^2)$). Thus, the covariate-specific ROC curve is the same for all covariate values, and it coincides with the AROC. On the other hand, for the examples shown in Figure~\ref{effect}, the covariate does have an impact on diagnostic accuracy, and thus we have a different covariate-specific ROC curve for each possible covariate value ($Y_{\bar{D}} \sim N(0.25 + 0.5X_{\bar{D}}, 0.3^2)$ and $Y_{D} \sim N(0.75 + X_{D}, 0.3^2)$). Here, the larger the covariate value, the better the diagnostic accuracy. In both Figures~\ref{no_effect}~and~\ref{effect}, the left-hand side plots show the pooled/marginal ROC curve and the AROC when the covariate has the same distribution in the nondiseased and diseased groups ($X_{\bar{D}} \sim N(0, 0.15^2)$ and $X_{D} \sim N(0, 0.15^2)$). Note that this implies that the covariate is not associated with disease status. Under this assumption, and assuming that the covariate-specific ROC curves are concave, it can be shown that the pooled ROC curve lies below (i.e., is attenuated with respect to) the AROC (see Appendix \ref{app1}). The amount of separation will depend on the strength of association between the covariate and the diagnostic test outcome \citep{Janes2008b}. Even when the discriminatory capacity of the test is not affected by the covariate (Figure~\ref{no_effect}), the pooled ROC curve does not coincide with the common covariate-specific ROC curve or AROC. The reason behind is that the threshold that gives rise to a particular operating point on the common covariate-specific ROC curve may not be the same in each subpopulation defined by the covariate \cite[see, e.g.,][]{Janes2008a, Janes2008b, Pardo14}. However, the pooled ROC curve is based on averaging covariate-specific FPFs and TPFs at a common threshold. Perhaps even more surprising are the results shown in the middle- and right-hand side of Figures~\ref{no_effect}~and~\ref{effect}, where differences between the pooled ROC curve and the AROC are more marked and in opposite directions. These plots show the situation where the covariate is associated with disease status. In this case, the pooled ROC curve will ``incorporate'' the portion of diagnostic accuracy attributable to the covariate. The middle-hand plots show the situation where large values of the covariate are associated with the disease ($X_{\bar{D}} \sim N(0, 0.15^2)$ and $X_{D} \sim N(0.4, 0.15^2)$), and thus, for this example, the pooled ROC curve lies above the AROC (optimistic result). On the other hand, the right-hand plots show the situation where the larger the covariate, the more likely the individual will be nondiseased ($X_{\bar{D}} \sim N(0.4, 0.15^2)$ and $X_{D} \sim N(0, 0.15^2)$). In this case and for this example, the pooled ROC curve lies below the AROC (pessimistic result). We would like to mention that the above examples represent particular cases. In general, the `relation' between the AROC curve and the pooled ROC curve might be more complex (with, e.g., the curves crossing), as illustrated in the simulation study reported in Section \ref{simulation}. 

Summarising, in those situations where the diagnostic test outcome is affected by covariates (though not necessarily its discriminatory capacity), covariate information must be incorporated into the ROC analysis. If a classification rule is used that relies on the same threshold value for all populations (by pooling the data regardless of the values of the covariate), this will result in the test having a diagnostic accuracy that is biased vis-\`a-vis its `true potential' diagnostic accuracy. As a result, optimistic or pessimistic results can be obtained, and by extension, misleading conclusions. In such circumstances it is, therefore, advisable to report, when possible, the covariate-specific ROC curves, and, as a global summary of the covariate-specific performance of the test, the AROC.
\section{\large{\textsf{BAYESIAN NONPARAMETRIC MODEL FOR THE COVARIATE-ADJUSTED ROC CURVE}}\label{BModel}}
In this section we detail our nonparametric regression model for conducting inference about the covariate-adjusted ROC curve. We start by noting that
\begin{equation*}
\text{AROC}(t)=\Pr\{Y_D>F_{\bar{D}}^{-1}(1-t\mid \mathbf{X}_{D})\}=\Pr\{1-F_{\bar{D}}(Y_D\mid \mathbf{X}_D)\leq t\} = \Pr(U_{D} \leq t), \qquad 0\leq t \leq 1,
\end{equation*}
where the random variable $U_{D}=1-F_{\bar{D}}(Y_D\mid \mathbf{X}_D)$ is the so-called placement value of the test outcome in the diseased population, i.e., the standardisation of $Y_{D}$ to the conditional distribution of $Y_{\bar{D}}$. Thus, the AROC can be interpreted as the cumulative distribution function of $U_D$. Our modelling procedure involves three steps: 1) modelling the conditional distribution of test outcomes in the nondiseased group, $F_{\bar{D}}(\cdot \mid \mathbf{X}_{\bar{D}} = \mathbf{x})$, which we do by using a B-splines dependent Dirichlet process mixture of normals model, 2) computing the placement value $U_D$, and 3) modelling the cumulative distribution function of $U_D$, which we do also nonparametrically through the Bayesian bootstrap. In what follows, we assume that $\{(\mathbf{x}_{\bar{D}i},y_{\bar{D}i})\}_{i=1}^{n_{\bar{D}}}$ and $\{(\mathbf{x}_{Dj},y_{Dj})\}_{j=1}^{n_{D}}$ are random samples of size $n_{\bar{D}}$ and $n_{D}$ from the nondiseased and diseased populations, respectively. Further, for all $i=1,\ldots,n_{\bar{D}}$ and $j=1,\ldots,n_{D}$, let $\mathbf{x}_{\bar{D}i}=(x_{\bar{D}i,1},\ldots,x_{\bar{D}i,p})^{\prime}$ and $\mathbf{x}_{Dj}=(x_{Dj,1},\ldots,x_{Dj,p})^{\prime}$ be $p-$dimensional vectors of covariates. To avoid notational overhead, we assume that all covariates are continuous. However, our modelling approach can easily deal with categorical covariates (through, as usual, dummy variables), as well as, with the interaction between continuous and categorical covariates.

\subsection{Three step modelling approach}
\noindent \textbf{\large{Step 1. Modelling $F_{\bar{D}}(\cdot \mid \mathbf{X}_{\bar{D}} = \mathbf{x})$}}\\
In a single-weights dependent Dirichlet process mixture of normals model \citep{Iorio2009}, the conditional distribution $F_{\bar{D}}(y_{\bar{D}i}\mid \mathbf{x}_{\bar{D}i})$ is modelled as
\begin{equation}\label{lddp}
F_{\bar{D}}(y_{\bar{D}i}\mid \mathbf{X}_{\bar{D}} = \mathbf{x}_{\bar{D}i})=\int \Phi(y_{\bar{D}i}\mid\mu(\mathbf{x}_{\bar{D}i};\boldsymbol{\beta}),\sigma^2)\text{d}G(\boldsymbol{\beta},\sigma^2),
\end{equation}
where $\Phi(y\mid\mu,\sigma^2)$ is the cumulative function of the normal distribution with mean $\mu$ and variance $\sigma^2$, evaluated at $y$, and $G$ follows the Dirichlet process (DP) prior \citep{Ferguson1973}, with base measure $E(G)=G_0(\boldsymbol{\beta},\sigma^2)$ and precision parameter $\alpha>0$, denoted by $G\sim\text{DP}(\alpha,G_0)$. The parameter $\alpha$ determines, among other properties, the variation of $G$ around $G_0$, with smaller (larger) values of $\alpha$ implying higher (lower) uncertainty. For ease of modelling, we express the DP prior $G$ in the stick-breaking form \citep{Sethuraman1994} as
\begin{equation}\label{sethu}
G(\cdot)=\sum_{l=1}^{\infty}\omega_l\delta_{(\boldsymbol{\beta}_l,\sigma_l^2)}(\cdot),\qquad \omega_l=v_{l}\prod_{r<l}(1-v_r),
\end{equation}
where $\delta_\textbf{a}$ denotes a point mass at $\textbf{a}$, $v_l\overset{\text{iid}}\sim\text{Beta}(1,\alpha)$ and $(\boldsymbol{\beta}_l,\sigma_l^2)\overset{\text{iid}}\sim G_0$ are mutually independent, for $l\geq 1$. This stick-breaking representation of the DP allows us to rewrite \eqref{lddp} as
\begin{equation*}
F_{\bar{D}}(y_{\bar{D}i}\mid \mathbf{X}_{\bar{D}} = \mathbf{x}_{\bar{D}i})=\sum_{l=1}^{\infty}\omega_l\Phi(y_{\bar{D}i}\mid \mu(\mathbf{x}_{\bar{D}i};\boldsymbol{\beta}_l),\sigma_l^2).
\end{equation*}
Regarding the specification of $\mu(\mathbf{x}_{\bar{D}i};\boldsymbol{\beta}_l)$, the usual, but somewhat rigid, choice is to assume a linear combination of the covariates in each component, i.e, 
\[
\mu\left(\mathbf{x}_{\bar{D}i}; \boldsymbol{\beta}_l\right) = \mu_l(\mathbf{x}_{\bar{D}i})  = \beta_{l0} + \beta_{l1}x_{\bar{D}i,1} + \cdots + \beta_{lp}x_{\bar{D}i,p} = \mathbf{x}_{\bar{D}i}^{*\prime}\boldsymbol{\beta}_l,
\]
where $\mathbf{x}_{\bar{D}i}^{*\prime}=(1,\mathbf{x}_{\bar{D}i}^{\prime})$. However, this formulation implies that the expected test outcome changes linearly with the covariates, $E(y_{\bar{D}i}\mid  \mathbf{x}_{\bar{D}i})=\sum_{l=1}^{\infty}\omega_l\mathbf{x}_{\bar{D}i}^{*\prime}\boldsymbol{\beta}_l=\mathbf{x}_{\bar{D}i}^{*\prime}\bar{\boldsymbol{\beta}}$, with the $h$th component of the vector $\bar{\boldsymbol{\beta}}$ being $\sum_{l=1}^{\infty}\omega_l\beta_{lh}$, for $h=0,\ldots,p$. To allow for greater flexibility and since nonlinear relationships between test results and continuous covariates often occur in practice, in this paper we assume a flexible model for the mean of each component
\begin{equation}
\mu_l(\mathbf{x}_{\bar{D}i}) = \beta_{l0} + f_{l1}(x_{\bar{D}i,1}) + \cdots + f_{lp}(x_{\bar{D}i,p}),
\label{DDP_mu_flex}
\end{equation}
where $f_{lh}(\cdot)$ are smooth and unknown functions, $l\geq 1$ and $h=1,\ldots, p$. In particular, we propose to approximate each smooth function $f_{lh}(\cdot)$ by a linear combination of cubic B-splines basis functions defined over a sequence of knots $\xi_{h0}<\xi_{h1}<\cdots<\xi_{hK_{h}}<\xi_{h,K_{h}+1}$, $h=1,\ldots, p$. The knots $\xi_{h0}$ and $\xi_{hK_{h}+1}$ are boundary knots, while the remaining ones are interior knots. Thus, model (\ref{DDP_mu_flex}) is then expressed as
\begin{equation*}
\mu_{l}(\mathbf{x}_{\bar{D}i})=\beta_{l0} + \underbrace{\mathbf{B}^{\prime}_{\mathbf{\xi}_1}(x_{\bar{D}i,1})\boldsymbol{\beta}_{l1}}_{
f_{l1}(x_{\bar{D}i,1})} + \cdots + \underbrace{\mathbf{B}^{\prime}_{\mathbf{\xi}_p}(x_{\bar{D}i,p})\boldsymbol{\beta}_{lp}}_{f_{lp}(x_{\bar{D}i,p})} = \mathbf{z}_{\bar{D}i}^{\prime}\boldsymbol{\beta}_l,
\end{equation*}
where $\mathbf{B}_{\mathbf{\xi}_h}(x_{\bar{D}i,h}) = (B_{h1}(x_{\bar{D}i,h}),\ldots,B_{h,K_h+3}(x_{\bar{D}i,h}))^{\prime}$, with $B_{hk}(x)$ denoting the $k$th cubic B-spline basis function (evaluated at $x$) defined by the vector of knots $\mathbf{\xi}_h$ \citep[][Chapter 9]{deBoor1978}, and $\boldsymbol{\beta}_{lh}=(\beta_{lh1},\ldots,\beta_{lh,K_h+3})^{\prime}$, for $l\geq 1$ and $h = 1, \ldots,p$. Finally, $\mathbf{z}_{\bar{D}i}^{\prime} = (1, \mathbf{B}^{\prime}_{\mathbf{\xi}_1}(x_{\bar{D}i,1}), \ldots, \mathbf{B}^{\prime}_{\mathbf{\xi}_p}(x_{\bar{D}i,p}))$ and $\boldsymbol{\beta}_l = (\beta_{l0}, \boldsymbol{\beta}_{l1}^{\prime}, \ldots, \boldsymbol{\beta}_{lp}^{\prime})^{\prime}$. In general, both the number and location of the knots that characterise the B-splines basis functions have the potential to influence inferences, more so for the former rather than the latter. As noted by \cite{Durrleman1989}, in practice, only a few number of knots are needed to adequately describe most of the phenomena likely to be observed in medical studies. According to the authors, a maximum of three or four knots will often be enough. Here, the selection of the number of knots is assisted by a model selection criterion, for example, the log pseudo marginal likelihood (LPML) \citep{Geisser1979} or the widely applicable information criterion (WAIC) \citep{Gelman2014}. With regard to the location of the $K_h$ interior knots, we use the quantiles of $\mathbf{x}_{\bar{D},h}=(x_{\bar{D}1,h},\ldots,x_{\bar{D}n_{\bar{D}},h})^{\prime}$, $h=1,\ldots, p$. Particularly, following \cite{Rosenberg1995}, $\xi_{hk}$ is set equal to the $k/(K_h+1)$ quantile of $\mathbf{x}_{\bar{D},h}$, for $k=1,\ldots, K_h$. This assures an approximate equal number of observations at each interval defined by the knots. The boundary knots $\xi_{h0}$ and $\xi_{h,K_{h}+1}$ are set equal to the minimum and maximum of $\mathbf{x}_{\bar{D},h}$, respectively.\\
\indent To complete our model we specify a conditionally conjugate base measure, which greatly simplifies computations, namely
\begin{equation*}
(\boldsymbol{\beta}_l,\sigma_l^2)\overset{\text{iid}}\sim N_{Q}(\boldsymbol{\beta}_l\mid \mathbf{m},\mathbf{S})\Gamma(\sigma^{-2}_l\mid a, b),
\end{equation*}
with conjugate hyperprior $\mathbf{m}\sim\text{N}_{Q}(\mathbf{m}_0,\mathbf{S}_0)$ and $\mathbf{S}^{-1}\sim\text{W}(\nu,(\nu\Psi)^{-1})$, where $Q$ denotes the dimension of vector $\mathbf{z}_{\bar{D}}$. Here $\Gamma(\cdot\mid a,b)$ denotes a gamma distribution with shape $a$ and rate $b$ and $\text{W}(\nu,(\nu\Psi)^{-1})$ denotes a Wishart distribution with $\nu$ degrees of freedom and expectation $\Psi^{-1}$. For ease of posterior of simulation and because it provides a highly accurate approximation we use the blocked Gibbs sampler of \cite{Ishwaran2001}, which relies on truncating the stick-breaking representation in \eqref{sethu} to a finite number of components, say $L$. Therefore, our B-splines dependent Dirichlet process mixture of normals model, hereafter shortly referred to as B-splines DDP, for the conditional distribution of test outcomes in the nondiseased group is expressed as
\begin{equation*}
F_{\bar{D}}(y_{\bar{D}i}\mid \mathbf{X}_{\bar{D}} = \mathbf{x}_{\bar{D}i})=\sum_{l=1}^{L}\omega_l\Phi(y_{\bar{D}i}\mid\mathbf{z}_{\bar{D}i}^{\prime}\boldsymbol{\beta}_l,\sigma_l^2),
\end{equation*}
where $L$ is pre-specified, the $\omega_l$'s result from a truncated version of the stick-breaking construction: $\omega_1=v_1$, $\omega_l=v_l\prod_{r<l}(1-v_r)$, for $l=2,\ldots,L$, $v_1,\ldots,v_{L-1}\overset{\text{iid}}\sim\text{Beta}(1,\alpha)$, and $v_L=1$. Note that $L$ is not the exact number of components we expect to observe but instead an upper bound on the number of components. An appropriate value of $L$ can be determined by considering properties of the high-order $\omega_l$ values in the infinite sum representation \eqref{sethu}. Specifically, \cite{Ishwaran2000} shown that $E(\sum_{l=L+1}^{\infty}\omega_l\mid\alpha)=\alpha^{L}(1+\alpha)^{-L}$. For example, setting $\alpha=1$ and $L=10$, as in our simulation studies and application, leads to $E(\sum_{l=L+1}^{\infty}\omega_l)\approx 0.00098$, so the truncation error is about $0.1\%$. The conditional distribution is then estimated by a finite mixture of normal regression models with the mixing weights automatically determined by the data. The full conditional distributions have the conjugate forms detailed in \ref{gibbs_sampler} of the Supplementary Materials.\\

\noindent\textbf{\large{Step 2. Modelling the placement value $U_D$}}\\
Once the Step 1 has been done, it is rather straightforward to model the placement value $U_D$. Given a posterior sample from the parameters of interest $\left(\omega_1^{(s)},\ldots,\omega_L^{(s)},\boldsymbol{\beta}_1^{(s)},\ldots,\boldsymbol{\beta}_L^{(s)},(\sigma_{1}^{(s)})^2,\ldots,(\sigma_{L}^{(s)})^2\right)$, we compute the corresponding realisation of the placement value of a diseased subject in the nondiseased population as
\begin{equation*}
U_{Dj}^{(s)}=1-F_{\bar{D}}^{(s)}(y_{Dj}\mid\mathbf{x}_{Dj})=1-\sum_{l=1}^{L}\omega_{l}^{(s)}\Phi\left(y_{Dj}\mid\mathbf{z}_{Dj}^{\prime}\boldsymbol{\beta}_l^{(s)},(\sigma_l^{(s)})^2\right),\quad j=1,\ldots,n_{D},\quad s=1,\ldots,S,
\end{equation*}
where $S$ denotes the number of posterior samples after burn-in and $\mathbf{z}_{Dj}^{\prime} = (1, \mathbf{B}^{\prime}_{\mathbf{\xi}_1}(x_{Dj,1}), \ldots, \mathbf{B}^{\prime}_{\mathbf{\xi}_p}(x_{Dj,p}))$.\\

\noindent\textbf{\large{Step 3. Modelling the cumulative distribution function of $U_D$}}\\
The last step in our modelling strategy to estimate the AROC involves the need of estimating the cumulative distribution function of $U_D$. We propose to it do through the Bayesian bootstrap (BB) proposed by \cite{Rubin1981}. In Efron's frequentist nonparametric bootstrap \citep{Efron1979}, for a bootstrap sample of size $n$, the weights associated to each observation belong to the discrete set $\{0,1/n,\ldots,n/n\}$, corresponding to the number of times each observation appears in such bootstrap sample. By opposition, in the Bayesian bootstrap, the weights are considered unknown, and their posterior distribution is derived. \cite{Rubin1981} showed that, by considering a non-informative prior for the weights, samples from the BB correspond to discrete distributions supported at the observed data points with weights distributed according to a $\text{Dirichlet}(n;1,\ldots,1)$ distribution. Thus, the weights in the BB are smoother than those from the classical bootstrap. Note that in the BB data should be regarded as fixed, so that we do not resample from it. Interestingly, the BB has also connections with the DP; specifically, by letting the precision parameter $\alpha$ converging to zero, the BB can be regarded as a noninformative version of the DP \citep[][Theorem 2]{Gasparini1995}.\\
\indent According to the BB, given a posterior realisation from $U_{Dj}^{(s)}$, $j=1,\ldots,n_D$ (available from Step 2), a posterior realisation for the cumulative distribution function of $U_D$, i.e., for the AROC curve, can be computed as
\begin{equation*}
\text{AROC}^{(s)}(t)=\sum_{j=1}^{n_D}q_{j}^{(s)}I(U_{Dj}^{(s)}\leq t),\quad (q_1^{(s)},\ldots,q_{n_D}^{(s)})\sim\text{Dirichlet}(n_D;1,\ldots,1), \quad s=1,\ldots,S, \quad 0\leq t\leq 1.
\end{equation*}
Thus, the $S$ posterior samples give rise to an ensemble of AROC curves $\{\text{AROC}^{(1)}(t),\ldots,\text{AROC}^{(S)}(t)\}$, from which the posterior mean (or median) can be computed
\begin{equation*} 
\widehat{\text{AROC}}(t)=\frac{1}{S}\sum_{s=1}^{S}\text{AROC}^{(s)}(t),\qquad 0\leq t\leq 1,
\end{equation*}
with pointwise credible bands derived from the percentiles of the ensemble. In addition, our method also leads to closed form expressions for the area and partial area under the AROC curve. It is easy to show that
\begin{align*}
\text{AAUC}^{(s)}&=\int_0^1\text{AROC}^{(s)}(t)\text{dt}=1-\sum_{j=1}^{n_D}q_{j}^{(s)}U_{Dj}^{(s)},\\
\text{pAAUC}^{(s)}(t_0)&=\int_{0}^{t_0}\text{AROC}^{(s)}(t)\text{dt}=t_0-\sum_{j=1}^{n_D}q_{j}^{(s)}\min\left\{t_0,U_{Dj}^{(s)}\right\},
\end{align*}
where $\text{pAAUC}(t_0)$ stands for the partial area under the AROC curve for a pre-specified maximum false positive fraction of $t_0$. Point and interval estimates for the AAUC and pAAUC are found similarly as for the AROC curve.

\subsection{Model comparison criteria}
Since it is ill-advised to compare models based on a single criterion, both the log pseudo marginal likelihood (LPML) \citep{Geisser1979} and the widely applicable information criterion (WAIC) \citep{Gelman2014,Watanabe2010} are employed with a two-fold purpose: i) assist the selection of the number of knots necessary for adequately modelling $F_{\bar{D}}(\cdot \mid \mathbf{X}_{\bar{D}} = \mathbf{x})$ and (ii) compare competing models for $F_{\bar{D}}(\cdot \mid \mathbf{X}_{\bar{D}} = \mathbf{x})$ (e.g., our B-splines DDP mixture model against a normal linear model).

\subsubsection{Log pseudo marginal likelihood (LPML)}
\indent The main ingredient to calculate the LPML is the conditional predictive ordinate (CPO). For the $i$th nondiseased observation, the $\text{CPO}_{\bar{D}i}$ is defined as
\begin{equation}\label{cpo}
\text{CPO}_{\bar{D}i}=f\left(y_{\bar{D}i}\mid \mathbf{y}_{\bar{D}}^{(-i)}\right)=\int_{\boldsymbol{\Theta}}f\left(y_{\bar{D}i}\mid\mathbf{x}_{\bar{D}i},\boldsymbol{\theta}\right)p\left(\boldsymbol{\theta}\mid  \mathbf{y}_{\bar{D}}^{(-i)}\right)\text{d}\boldsymbol{\theta},
\end{equation}
where $\boldsymbol{\theta}=(\omega_{1},\ldots,\omega_{L},\boldsymbol{\beta}_1,\ldots,\boldsymbol{\beta}_L,\sigma_1^2,\ldots,\sigma_L^2)$, $\mathbf{y}_{\bar{D}}^{(-i)}$ is $\mathbf{y}_{\bar{D}}=(y_{\bar{D}1},\ldots,y_{\bar{D}n_{\bar{D}}})$ with the $i$th observation omitted, and $p\left(\boldsymbol{\theta}\mid \mathbf{y}_{\bar{D}}^{(-i)}\right)$ is the posterior density of $\boldsymbol{\theta}$ based on the data $\mathbf{y}_{\bar{D}}^{(-i)}$. Under the model specification in Section \ref{BModel} (Step 1), we have
\begin{equation*}
f(y_{\bar{D}i}\mid\mathbf{x}_{\bar{D}i},\boldsymbol{\theta})=\sum_{l=1}^{L}\omega_l\phi\left(y_{\bar{D}i}\mid\mathbf{z}_{\bar{D}i}^{\prime}\boldsymbol{\beta}_l,\sigma_{l}^{2}\right),
\end{equation*}
where $\phi(y\mid\mu,\sigma^2)$ stands for the density function of the normal distribution, evaluated at $y$, with mean $\mu$ and variance $\sigma^2$. Thus $\text{CPO}_{\bar{D}i}$ is the marginal posterior predictive density of $y_{\bar{D}i}$ given $\mathbf{y}_{\bar{D}}^{(-i)}$ and can be interpreted as the height of this marginal density at $y_{\bar{D}i}$. A higher value of $\text{CPO}_{\bar{D}i}$ under one model for the nondiseased group implies a better fit of that model for the $i$th observation, $i=1,\ldots,n_{\bar{D}}$. \cite{Gelfand1994} showed that $\text{CPO}_{\bar{D}i}$ is easily estimated from a posterior sample via
\begin{equation*}
\text{CPO}_{\bar{D}i}	\approx\left\{\frac{1}{S}\sum_{s=1}^{S}\frac{1}{f\left(y_{\bar{D}i}\mid\mathbf{x}_{\bar{D}i},\boldsymbol{\theta}^{(s)}\right)}\right\}^{-1},
\end{equation*}
where the result is based on showing that $\text{CPO}_{\bar{D}i}$ as defined in \eqref{cpo} can be rewritten as
\begin{equation*}
\text{CPO}_{\bar{D}i}=\left\{\int_{\boldsymbol{\Theta}}\frac{1}{f(y_{\bar{D}i}\mid\mathbf{x}_{\bar{D}i},\boldsymbol{\theta)}}p(\boldsymbol{\theta}\mid \mathbf{y}_{\bar{D}})\text{d}\boldsymbol{\theta}\right\}^{-1}=\left\{E_{\boldsymbol{\theta}\mid\mathbf{y}_{\bar{D}}}\left[\frac{1}{f(y_{\bar{D}i}\mid\mathbf{x}_{\bar{D}i},\boldsymbol{\theta})}\right]\right\}^{-1}.
\end{equation*}
The LPML defined as $\text{LPML}=\sum_{i=1}^{n_{\bar{D}}}\log (\text{CPO}_{\bar{D}i})$ gives an aggregate summary measure of a model's predictive ability and the larger the LPML is, the better the fit of the model under consideration.

\subsubsection{Widely applicable information criterion (WAIC)}
The widely applicable information criterion is decomposed as terms for fit and complexity. The idea is to compute the log pointwise posterior predictive density (lppd) given by
\begin{equation*}
\text{lppd}=\sum_{i=1}^{n_{\bar{D}}}\log E_{\boldsymbol{\theta}\mid \mathbf{y}_{\bar{D}}}[f(y_{\bar{D}i}\mid\mathbf{x}_{\bar{D}i},\boldsymbol{\theta})]\approx \sum_{i=1}^{n_{\bar{D}}}\log\left(\frac{1}{S}\sum_{s=1}^{S}f\left(y_{\bar{D}i}\mid \mathbf{x}_{\bar{D}i},\boldsymbol{\theta}^{(s)}\right)\right),
\end{equation*}
and then, to adjust for overfitting, add a term to correct for effective number of parameters
\begin{align*}
\rho_{\text{WAIC}}&=\sum_{i=1}^{n_{\bar{D}}}\text{Var}_{\boldsymbol{\theta}\mid \mathbf{y}_{\bar{D}}}[\log f(y_{\bar{D}i}\mid\mathbf{x}_{\bar{D}i},\boldsymbol{\theta})]\\
&\approx \sum_{i=1}^{n_{\bar{D}}}\left\{\frac{1}{S-1}\sum_{s=1}^{S}\left(\log f\left(y_{\bar{D}i}\mid \mathbf{x}_{\bar{D}i},\boldsymbol{\theta}^{(s)}\right)-\frac{1}{S}\sum_{s=1}^{S} \log f\left(y_{\bar{D}i}\mid \mathbf{x}_{\bar{D}i},\boldsymbol{\theta}^{(s)}\right)\right)^2\right\}.
\end{align*}
To set the WAIC in the deviance scale, \cite{Gelman2014} suggested to use
\begin{equation*}
\text{WAIC}=-2(\text{lppd}-\rho_{\text{WAIC}}),
\end{equation*}
so that the model that provides the best fit is the one with the smallest WAIC.
\section{\large{\textsf{SIMULATION STUDY}}}\label{simulation}
This section reports the results of a simulation study conducted to study the empirical performance of our nonparametric method for conducting inference about the AROC curve and the AAUC. Extra simulation results can be found in the Supplementary Materials. All simulations were done in the \texttt{R} environment \citep[version 3.4.4,][]{R18}, using the \texttt{R}-package \texttt{AROC} that accompanies this paper.

\subsection{Simulation Scenarios}\label{scenarios}
We consider a wide range of simulation scenarios, covering cases where covariates do not affect test outcomes at all, where the performance of the test is affected by covariates but not its discriminatory ability, and cases where covariates do affect the discriminatory capacity of the test. For the latter case, we consider linear and nonlinear simple regression models for test outcomes, as well as, multiple regression models. The scenarios we consider are as follows. 
\begin{itemize}
\item Scenario I

In this scenario independent data are generated as
\begin{equation*}
y_{\bar{D}i}\sim\text{N}(0.5,0.5^2),\quad y_{Dj}\sim\text{N}(1,1^2),\quad i=1,\ldots,n_{\bar{D}},\quad j=1,\ldots,n_D.
\end{equation*}
The primary purpose of including this scenario is to investigate the loss of efficiency of using our AROC estimator, instead of the pooled ROC curve, when covariates have no effect on test outcomes. 

\item Scenario II 

This scenario involves linear mean regression models, namely
\begin{equation*}
y_{\bar{D}i}\sim\text{N}(0.5+(2x_{\bar{D}i,1}-10)/23,0.5^2),\quad y_{Dj}\sim\text{N}(1+(2x_{\bar{D}j,1}-10)/23,1^2).
\end{equation*}
In this scenario, although covariates do affect test outcomes, they do not affect the discriminatory ability of the test. As already mentioned, even in this case, the pooled ROC curve can lead to erroneous inferences, and the AROC curve should be instead reported. 

\item Scenario III 

We consider again linear regression models but, in contrast to Scenario II, now covariates do affect the discriminatory capacity of the test
\begin{equation*}
y_{\bar{D}i}\sim\text{N}(0.25+0.5(2x_{\bar{D}i,1}-10)/23,0.5^2),\quad y_{Dj}\sim\text{N}(0.75+(2x_{\bar{D}j,1}-10)/23,1^2).
\end{equation*}
\end{itemize}
Scenarios II and III allow us to assess a possible increase in posterior uncertainty of our nonparametric AROC estimator when a standard normal linear model is the true data generating mechanism. Such popular model is violated in Scenarios IV--VI. 
\begin{itemize}
\item Scenario IV

Data for this scenario are governed by a nonlinear regression model for the nondiseased group and a linear one for the diseased subjects
\begin{align*}
y_{\bar{D}i}&\sim\text{N}(5 + 3((x_{\bar{D}i,1}+8)/23)^2 -25\left(((x_{\bar{D}i,1}+8)/23)-0.2\right)_{+}^{3} + 250\left(((x_{\bar{D}i,1}+8)/23)-0.65\right)_{+}^{3}, 0.5^2),\\
y_{Dj}&\sim\text{N}(-3 - 0.6((x_{Dj,1}+8)/23), 1^2).
\end{align*}

\item Scenario V

Here we consider the case where two continuous covariates affect, nonlinearly, the test outcomes in the nondiseased group, while in the diseased group we specify a simple nonlinear regression model
\begin{align*}
y_{\bar{D}i}&\sim\text{N}(0.5\exp\{(2x_{\bar{D}i,1} - 10)/10) - 2((2x_{\bar{D}i,2}^2-10)/10), 0.5^2),\\
y_{Dj}&\sim\text{N}(0.5\sin(\pi((2x_{Dj,1} -10)/10 + 1)) + 0.5\exp((2x_{Dj,1}-10)/10), 1^2).
\end{align*}

\item Scenario VI

A nonlinear regression model with an interaction between a continuous and a binary covariate is considered for the nondiseased group, whereas in the diseased group, as in the previous scenario, we consider a simple nonlinear regression model
\begin{align*}
y_{\bar{D}i}&\sim\text{N}(-\sin(0.7\pi((2x_{\bar{D}i,1}-10)/10 + 30))x_{\bar{D}i,3} + (((2x_{\bar{D}i,1}-10)/10)^2)(1-x_{\bar{D}i,3}), 0.5^2),\\
y_{Dj}&\sim\text{N}(0.5 + (((2x_{Dj,1}-10)/10)^2), 1^2).
\end{align*}
\end{itemize}
For all scenarios, the continuous covariates $x_1$ and $x_2$, are independently generated from a skew normal distribution, namely
\begin{align*}
x_{\bar{D}i,1},x_{\bar{D}i,2}\sim\text{SN}(0,5^2,2),\qquad x_{Dj,1},x_{Dj,2}\sim\text{SN}(3,4^2,1),
\end{align*}
where $\text{SN}(\mu,\sigma^2,\lambda)$ denotes a skew normal distribution with mean $\mu$, variance $\sigma^2$, and skewness parameter $\lambda$. This choice is due to the fact that the covariate distribution in our data application is also skewed (see Web Figure \ref{covariates} of the Supplementary Materials). Further, the binary covariate $x_3$ is generated from a Bernoulli distribution, $x_{\bar{D}i,3}\sim\text{Bern}(0.5)$.

\subsection{Models}
For each simulated dataset, the B-splines DDP mixture model for the conditional distribution of test outcomes in the nondiseased group was fitted considering four interior knots ($K = 4$) for each continuous covariate (a further inspection to this choice is discussed in the next subsection). Following the rule discussed in Section \ref{BModel} (Step 1), the knots are located at the $0.2$, $0.4$, $0.6$, and $0.8$ quantiles of the covariates. Regarding prior information, $\alpha$ was set equal to one, which according to \cite{Hanson2006} is the default value in the absence of prior information on the number of components needed to adequately describe $F_{\bar{D}}(\cdot \mid \mathbf{X}_{\bar{D}} = \mathbf{x})$. Using results from \cite{Antoniak1974} and \cite{Escobar1994}, this choice leads to a prior expected number of components of approximately of $5$ when $n_{\bar{D}}=50$ and $70$, and approximately of $6$ when $n_{\bar{D}}=200$ and $300$. For the total number of mixture components, we set $L=10$, and allow the shrinkage induced by the stick-breaking prior to adaptively delete redundant components not necessary to characterise the data in the nondiseased group. To facilitate prior specification of the hyperparameters associated with the centring distribution, responses were scaled by dividing by the standard deviation when fitting the model. We transformed back to the original scale when presenting the results. Thus, for the normal-gamma prior, we used $\mathbf{m}_0=\mathbf{0}_Q$, $\mathbf{S}_0=100I_{Q}$, $\nu=Q+2$, $\Psi^{-1}=I_{Q}$, where $I_{Q}$ denotes the $Q\times Q$ identity matrix and we set $a=2$ and $b=0.5$. The normal prior for $\boldsymbol{\beta}_l$ is relatively diffuse since the variances in $\mathbf{S}_0$ are large, and the degrees of freedom in the Wishart distribution are small. Note that $a=2$ leads to a prior with infinite variance (hence, in some sense, vague) that is centred around a finite mean ($b=0.5$). The prior on $\sigma^2$, therefore, favours variances less than one; note that the scaled data have a marginal variance of one, so the within-component variance $\sigma_l^2$ is expected to be smaller than the marginal variance.\\
\indent We compared our method to results from a Bayesian semiparametric approach where the conditional distribution for the nondiseased subjects is modelled using a normal linear regression model, that is
\begin{equation*}
F_{\bar{D}}(y_{\bar{D}i}\mid \mathbf{x}_{\bar{D}i})=\Phi(y_{\bar{D}i}\mid \mathbf{x}_{\bar{D}i}^{*\prime}\boldsymbol{\beta}^{*},\sigma^{2*}),\quad i=1,\ldots,n_{\bar{D}},
\end{equation*}
with $\mathbf{x}_{\bar{D}i}^{*\prime}=(1,\mathbf{x}_{\bar{D}i}^{\prime})$ and $\boldsymbol{\beta}^{*}=(\beta_0^{*},\beta_{1}^{*},\ldots,\beta_{p}^{*})^{\prime}$. The cumulative distribution function of $U_D$ is modelled, as in our approach, using the Bayesian bootstrap, so that the two approaches only differ in Step 1. We use the following priors that align with those from our nonparametric analysis,
\begin{equation*}
\boldsymbol{\beta}^{*}\sim\text{N}_{p+1}(\mathbf{m}^{*},\mathbf{S}^{*}),\quad \sigma^{-2*}\sim\Gamma(a^{*},b^{*}),\quad \mathbf{m}^{*}\sim\text{N}_{p+1}(\mathbf{m}_0^{*},\mathbf{S}_0^{*}),\quad \mathbf{S}^{*-1}\sim\text{W}_{p+1}(\nu^{*},(\nu^{*}\Psi^{*})^{-1}),
\end{equation*}
with 
\begin{equation*}
\mathbf{m}_0^{*}=\mathbf{0}_{p+1},\quad \mathbf{S}_0^{*}=100I_{p+1},\quad \nu^{*}=p+3,\quad (\Psi^{*})^{-1}=I_{p+1},\quad a^{*}=2,\quad b^{*}=0.5.
\end{equation*}
This model can be regarded as the Bayesian counterpart of the frequentist model of \cite{Janes09a}. For both Bayesian approaches, 8000 samples were kept after a burn-in period of 2000 iterations of the Gibbs sampler. We also compared our model to the nonparametric kernel-based method of \cite{MX11a} (details appear in \ref{kernel_est} of the Supplementary Materials). Estimation in this case relies on the \texttt{R}-package \texttt{np} \citep{nppackage}, and the bandwidth parameters involved were selected using least-squares cross-validation. Since the kernel approach, as it stands now, can only deal with one continuous covariate, comparisons with this method were only performed for Scenarios I--IV. As a reinforcement of what has been exposed in Section \ref{examples}, we have also computed the pooled ROC curve for all scenarios to assess how ignoring covariate effect may impact the inferences. We have chosen to estimate the pooled ROC using the Bayesian bootstrap approach of \cite{Gu2008} which is an extremely robust and computationally simple one. 

\subsection{Results}
For each of the six scenarios described in Section \ref{scenarios}, 100 datasets were generated using both balanced and unbalanced sample sizes, namely $(n_{\bar{D}},n_{D})=(50,50)$, $(n_{\bar{D}},n_{D})=(200,70)$, $(n_{\bar{D}},n_{D})=(200,200)$, and $(n_{\bar{D}},n_{D})=(300,100)$. The discrepancy between the estimated and true AROC curve was assessed using the empirical root mean squared error
\begin{equation*}
\text{ERMSE}=\sqrt{\frac{1}{n_{T}}\sum_{r=1}^{n_T}\left\{\widehat{\text{AROC}}(t_r)-\text{AROC}(t_r)\right\}^2}\approx \sqrt{\int_{0}^{1}\left\{\widehat{\text{AROC}}(t)-\text{AROC}(t)\right\}^2\text{d}t},
\end{equation*}
where $n_T=101$ and the $t_r$'s are evenly spaced over $[0,1]$. As far as the AAUC is concerned, the behaviour was evaluated in terms of the bias. Table \ref{MSE_K_4} and Web Table \ref{AUC_K_4} in the Supplementary Materials summarise, respectively, the ERMSE and bias for each scenario, sample size, and approach considered. The estimated AROC curves along with the 2.5 and 97.5 simulation quantiles for all scenarios and sample size $(n_{\bar{D}},n_D)=(200,200)$ are depicted in Figure \ref{aROC_sim}. The plots for the remaining sample sizes are shown in Web Figures \ref{sim_ndx_5_I}--\ref{sim_ndx_5_VI} in the Supplementary Materials. Our estimator has the ability to successfully recover the form of the true AROC for all scenarios and samples sizes considered. Even in scenarios where the standard normal linear regression model holds (Scenarios II and III) our estimator still has an excellent performance, indeed presenting a very competitive performance with the Bayesian semiparametric estimator where the conditional distribution of test results for nondiseased subjects is modelled using a normal linear regression model. Further, our Bayesian nonparametric method presents a superior performance in almost all scenarios and sample sizes, as measured by the ERMSE, when compared to its nonparametric competitor, the kernel-based approach. As expected, posterior uncertainty associated with our estimator decreases as sample size increases. By opposition, the Bayesian semiparametric estimator is unsuitable for Scenarios IV--VI and its performance fails to improve as the sample size increases. It is worth commenting that for Scenario I, where the performance of the test is not affected by covariates at all, our Bayesian nonparametric estimator achieves a smaller ERMSE than that of the pooled ROC curve, which might seem counterintuitive. However, the fact that we are estimating the pooled curve by the Bayesian bootstrap where absolutely no structure is imposed, while our estimator in Step 1 still involves a mixture model (and thus, to a certain extent, a little more structure) might explain the difference. On the other hand, using the pooled ROC curve when covariates affect test outcomes/discriminatory capacity of the test, as attested by the results of Scenarios II--VI, generally leads to erroneous inferences. \\
\indent We have also investigated, in Scenarios I--IV, how the three different approaches estimate the corresponding covariate-specific threshold values used for defining a positive test result, for FPFs of $0.1$ and $0.3$. The results are shown in Web Figures \ref{thresholds_sim_ndx_5_I}--\ref{thresholds_sim_ndx_5_IV} of the Supplementary Materials. For Scenarios I, II, and III and for the two FPFs considered, the three approaches were able to recover the true covariate-specific threshold curve. Whereas our estimates are unbiased but present some larger variability at the boundaries, the kernel estimates lead to slightly biased estimates. In turn, in Scenario IV, the one involving a nonlinear trend in the nondiseased group, the Bayesian semiparametric estimates fail to recover the true conditional threshold curve. \ref{threshold_est} of the Supplementary Materials describes how the covariate-specific thresholds were computed.\\
\indent The frequentist coverage of the $95\%$ credible intervals for $\text{AROC}(t)$ (averaged over the 101 FPFs considered), and for the corresponding AAUC are presented in Table \ref{covAROC_K_4} and Web Table \ref{covAAROC_K_4} of the Supplementary Materials, respectively. For our Bayesian nonparametric estimator, we found the coverages to be between $92-99\%$, showing the validity of our inferences. In turn, for the Bayesian semiparametric approach and for the scenarios involving nonlinear mean functions, the respective coverage probabilities are, obviously, much lower. \\
\indent We rely on the LPML and WAIC discussed in Section 3.2 to assist in: (1) the selection of the number of knots needed to adequately describe the conditional distribution of test results in the nondiseased group; and, (2) the comparison between competing models for this distribution, namely our Bayesian B-splines DDP mixture model against a Bayesian normal linear regression model. Thus, we also investigated the behaviour of these two criteria in performing the aforementioned two tasks. Specifically, over the $100$ simulated datasets, for each scenario considered and for the different sample sizes in the nondiseased group, $n_{\bar{D}}\in\{50,200,300\}$, we computed the number of times the LPML and WAIC favour the model with four interior knots ($K = 4$) over a model with no interior knots ($K = 0$). Results are displayed in Web Table \ref{WAIC_NPML_ndx_5_ndx_1} of the Supplementary Materials and show that, most of the time, both criteria favour the model with no interior knots for Scenarios I, II, III, and V, while for Scenarios IV and VI the model with four interior knots is, in general, preferred. Despite this, our method seems to be robust to the misspecification of the number of knots. This can be observed in Web Figures \ref{sim_ndx_1_I}--\ref{sim_ndx_1_VI} and Web Tables \ref{MSE_K_0}--\ref{covAAROC_K_0} of the Supplementary Materials. In these tables and figures we show, for the Bayesian nonparametric approach, the comparisons of the simulation results for no interior knots ($K = 0$) and four interior knots ($K = 4$). All in all, the results remain very similar. The major differences are observed for the estimates of the corresponding covariate-specific threshold values used for defining a positive test result (Web Figures \ref{thresholds_sim_ndx_1_I}--\ref{thresholds_sim_ndx_1_IV}). Note that for Scenarios I, II and III, the estimates using no interior knots present lower variability than those using four interior knots. This is specially remarkable at the boundaries and for small sample sizes. On the other hand, for Scenario VI (where both criteria favour the model with four interior knots), the estimates using no interior knots exhibit a large bias at the boundaries. Finally, following the same reasoning we also checked the percentage of time the LPML and WAIC criteria favour the B-splines DDP for modelling test outcomes in the nondiseased group over the less flexible Bayesian normal linear regression model. As our intuition would dictate, with the exception of Scenarios I--III, both the LPML and the WAIC chose the B-splines DDP mixture model the majority of the time (see Web Tables \ref{WAIC_LPML_ndx5} and \ref{WAIC_LPML_ndx1} for, respectively, $K = 4$ and $K = 0$). Unlike the results on the number of knots, the misspecification of the model, as seen before, would lead to erroneous inferences. 

\section{\large{\textsf{APPLICATION}}}\label{application}
According to the World Health Organization (WHO), cardiovascular diseases (CVDs) are the leading cause of death worldwide. Approximately 17.7 million people died from CVD in 2015, with this number representing $31\%$ of worldwide deaths. As can be read at the WHO website (\texttt{www.who.int/mediacentre/factsheets/fs317/}) \emph{``People with cardiovascular disease or who are at high cardiovascular risk (...) need early detection (...)''}. In this work, we are interested in studying the possible modifying effect of age and gender on the discriminatory capacity of the body mass index (BMI) for detecting the presence of CVD risk factors.\\
\indent Our data comes from a cross-sectional study carried out by the Galician Endocrinology and Nutrition Foundation (FENGA). A stratified sampling design was used to obtain a representative sample of the non-institutionalised civilian population of Galicia (northwest of Spain), consisting of $2840$ individuals with an age range of $18$-$85$ years \cite[for further details, see][]{Tome09}. A diseased subject was defined as any person presenting with two or more CVD risk factors (raised triglycerides, reduced HDL-cholesterol, raised blood pressure and raised fasting plasma glucose). Based on this definition, out of the total of $2840$ individuals, $691$ were classified as diseased ($273$ women and $418$ men) and $2149$ as nondiseased ($1250 $ women and $899$ men). As it has been extensively discussed in the literature, anthropometric measures (as the BMI) perform differently according to both age and gender. This can be observed in Web Table \ref{EndoDataDescriptive} of the Supplementary Materials, where some summary statistics of the data are presented. As a result, it is expected that the discriminatory capacity of the BMI varies with age and gender. This is confirmed by Figure~\ref{cROC_aROC_gender_a}, which shows, separately for men and women, the estimated age-specific AUCs obtained using the Bayesian nonparametric approach of \cite{Inacio13}. Similar results have been reported in \cite{MX11a} and \cite{MX11b}. Note that whereas for men the discriminatory capacity of the BMI remains more or less constant along age, for women, age displays a marked effect, with accuracy decreasing until an age of about 70 years old, and then slightly increasing until the age of 80. Nonetheless, even for men, the estimated AUC obtained by pooling the data (separately in men and women) without regard to age is, in general, larger than the age-specific AUC (see Figure~\ref{cROC_aROC_gender_a} and Table~\ref{AUCDescriptive}). As a consequence, analyses based upon the pooled ROC curve would lead to optimistic conclusions about the discriminatory capacity of the BMI when predicting the presence of CVD factors in both men and women. As discussed in Section \ref{examples}, this optimism (bias) is because, for the pooled ROC curve, a common threshold value is used for all individuals regardless of their covariate values. In particular, in our study, CVD risk factors are more prevalent in old individuals, where the BMI is also higher. It implies that the pooled ROC curve for the BMI in both men and women `incorporates' the capacity of distinguishing between diseased and nondiseased individuals attributable to age.\\
\indent In line with the above-discussed results, we started by estimating the age-adjusted ROC curve (AROC) separately in men and women using our proposed Bayesian nonparametric estimator. Results were also compared to the ones obtained by using the Bayesian semiparametric and kernel methods described in Section \ref{simulation}. For our proposal, we modelled age effects using a B-spline basis of dimension $3$, thus corresponding to no interior knots; the selection of the basis dimension was done by the LPML and WAIC (results shown in Web Table~\ref{WAIC_LPML_knots_appl} of the Supplementary Materials). For both Bayesian approaches, posterior inference was based on $45 000$ Gibbs sampler iterates after a burn-in of the first 5000 realisations was discarded and the same prior specification described in Section \ref{simulation} was applied. Inspection of traceplots and Geweke criterion suggest convergence of the chains. The estimated AROC curves are shown in Figure~\ref{cROC_aROC_gender_b}, where we also depict the pooled ROC curves (estimated as described in Section~\ref{simulation} and separately for men and women). As expected, all approaches provided an AROC curve that lies below the corresponding pooled ROC curve, with this being especially remarkable in women. These results highlight the need for including covariate information into the ROC analysis and the danger that focusing on the pooled ROC curve may cause. Although the estimated AROC curves under our Bayesian nonparametric model and the kernel-based approach are very similar, the differences to the results obtained under the Bayesian semiparametric method are quite pronounced in women. It is worth remembering that the only difference between our Bayesian nonparametric model and the semiparametric one is how the conditional cumulative distribution function of test outcomes in the nondiseased group is modelled. Specifically, under our proposal, such distribution is modelled through a B-splines DDP mixture model, whereas under the semiparametric approach it is done via a Bayesian normal linear model (as detailed in Section \ref{simulation}). Both the LPML and WAIC criteria support the use of our flexible B-splines DDP for modelling the distribution of test results in the nondiseased group over the more restrictive Bayesian normal linear model (results not shown). Recall that, as the AROC is an average of covariate-specific ROC curves, the AAUC is also an average of covariate-specific AUCs. This is patent in Figure~\ref{cROC_aROC_gender_a} where we present, separately for men and women, the age-specific AUCs along with the pooled AUC and the area under the age-adjusted ROC curve (AAUC) estimated using our approach.\\
\indent Thus far, AROC estimates were obtained separately for men and women. We now jointly analyse the whole sample, with interest lying in estimating a single AROC curve by adjusting for both age and gender. This will provide us with a global summary of the age- and gender-specific accuracy of the BMI to detect clusters of CVD. Estimation was carried out using the proposed Bayesian nonparametric estimator and the Bayesian semiparametric one. The results shown above suggest the presence of an interaction between age and gender, which was included in both models. The kernel-based approach, as it stands now, can only deal with one continuous covariate, and therefore cannot be applied in this setting. AROC estimates are shown in Figure~\ref{aROC_a}. The AAUC under our Bayesian nonparametric approach is $0.667$ $(0.639,0.695)$, thus revealing a reasonable overall age- and gender-specific accuracy of the BMI to predict CVD. Again, the estimated AROC curve lies below the pooled ROC curve (estimated with the whole sample), with the pooled AUC being $0.766$ $(0.747,0.783)$. Here, the pooled ROC curve `incorporates' the capacity of distinguishing between diseased and nondiseased individuals attributable to both age and gender. For this joint analyses setting, as it was also the case for the separate analyses in men and women, both the LPML and WAIC criteria favour the use of our B-splines DDP for modelling the test outcomes in the nondiseased group over the normal linear model. Particularly, the LPML is $-5868$ for our approach and $-6058$ for the normal linear competitor. A deeper inspection was done via a plot of the log CPO ratio of the B-splines DDP mixture model over the normal linear model against observation numbers. Thus, a point larger than $0$ supports our approach. Results are displayed in Figure~\ref{predchecks_c} and it shows that approximately $66\%$ of the observations support our model. The superiority of our method was further validated by the WAIC, which was $11736$ under our approach and $12117$ under the normal linear model. Since the LPML and WAIC are relative criteria, posterior predictive checks were also implemented. We generated replicate datasets from the posterior predictive distribution in the nondiseased group and compared to the data (BMI values) in such group using specific test quantities. For the choice of the test quantities, we follow \cite{Gabry2017}, who suggest choosing statistics that are orthogonal to the model parameters. Since we are using a location-scale mixture, we decided to investigate how well the posterior predictive distribution captures the skewness and kurtosis. From Figures~\ref{predchecks_a}~and~\ref{predchecks_b} we see that our model does an excellent job capturing both quantities, while the normal linear model fails. Also shown in Figure~\ref{predchecks} are the kernel density estimates of $500$ randomly selected datasets drawn from the posterior predictive distribution (nondiseased group) compared to the kernel density estimate of the BMI in the nondiseased group. It is evident that our model, as opposed to the normal linear model, is able to simulate data that is very much similar to the observed BMI values. Further, although the difference between the estimated AROCs by our approach and the Bayesian semiparametric method is not large, the estimated threshold values used to classify an individual as testing positive, shown in Figure~\ref{aROC_b} for a FPF of $0.3$, differ by a large extent. This is also in line with what has been discussed in the simulation study presented in Section~\ref{simulation}.\\
\indent Finally, a sensitivity analysis to our prior specification, for the model fitted in the nondiseased group, was conducted. The three prior configurations considered are catalogued in Web Table \ref{sens_analysis_app} of the Supplementary Materials. The first and second prior configurations changed $\alpha$, the parameter that controls the number of mixture components, and $L$, the upper bound on the number of mixture components. Specifically, the first configuration decreased $\alpha$ from $1$ to $0.1$ and increased $L$ from $10$ to $30$, whereas for the second configuration, $\alpha$ was increased from $1$ to $10$ and $L$ was again set to 30. In turn, the third configuration, changed the within-component variance parameters from $a=2$ and $b=0.5$ to $a=2$ and $b=3$; this new configuration leads to larger within-component variances. The AAUC results for each of the three configurations are, respectively, $0.666$ $(0.638,0.693)$, $0.663$ $(0.636, 0.690)$, and $0.667$ $(0.638,0.696)$, which show that inferences are essentially unchanged, partially due, possibly, to the large sample size ($2149$ nondiseased individuals).

\section{\large{\textsf{CONCLUDING REMARKS}}\label{remarks}}
We developed a Bayesian nonparametric regression model to estimate the covariate-adjusted ROC curve. The flexibility of our model arises from using B-splines dependent Dirichlet process mixtures and the Bayesian bootstrap. Under our model framework, and by opposition to the existing nonparametric kernel-based method, continuous and categorical covariates, as well as, interactions, can be easily handled. The model developed has thus the potential to be applicable to many populations, as well as, for a large number of diseases and (continuous) medical tests. Our simulation study illustrated the ability of the model to respond to complex data distributions in a variety of scenarios, with little price to be paid in terms of decreased posterior precision for the extra generality of our nonparametric estimator when compared with parametric estimates (even when the parametric model holds). Further, when compared to the kernel method, the general superiority of our method, as measured by the empirical root mean squared error between estimated and true covariate-adjusted ROC curves, was also evident in the simulation study. Additionally, the frequentist validity of our inferences, in terms of coverage probability, was also demonstrated. Our data application revealed a reasonable overall age- and gender-specific accuracy of the BMI to predict CVD and also how erroneous conclusions can be if focus is placed on the pooled ROC curve instead of on covariate-adjusted measures. The B-splines dependent Dirichlet process mixture model used for the conditional distribution of test outcomes in the nondiseased group assumes a known number of knots and model selection criteria (LPML and WAIC) were used to choose between candidate models with different number of knots. Although a prior can be placed on the number of knots and their location \citep{Dimatteo2001}, this could be computationally difficult to implement efficiently in practice (e.g., involving reversible jump Markov chain Monte Carlo). A viable alternative is to use penalised splines \citep{Eilers1996}, thus leading to a penalised splines dependent Dirichlet process mixture model that is, in essence, a dependent Dirichlet process mixture of generalised additive models. Another interesting avenue for future research is the development of AROC estimators based on direct ROC regression methodology \citep{Pepe2000}. 

\section*{\large{\textsf{ACKNOWLEDGEMENTS}}}
The authors would like to thank the Galician Endocrinology and Nutrition Foundation (\textit{Fundaci\'{o}n de Endocrinolox\'{\i}a e Nutrici\'{o}n Galega - FENGA}) and Carmen Cadarso Su\'arez for having supplied the database used in this study. M.X. Rodr\'{\i}guez-\'{A}lvarez expresses her gratitude for the support received by the Basque Government through the BERC 2018-2021 program, by the Spanish Ministry of Economy and Competitiveness MINECO through BCAM Severo Ochoa excellence accreditation SEV-2013-0323 and through project MTM2017-82379-R funded by (AEI/FEDER, UE) and acronym ``AFTERAM''. V. In\'acio de Carvalho acknowledges support from FCT--Funda\c c\~ao para a Ci\^encia e a Tecnologia, Portugal, through the project \linebreak UID/MAT/00006/2013. We are grateful to Miguel de Carvalho for the detailed reading of the paper and his many suggestions.
\appendix
\section{\large{\textsf{APPENDIX}}}\label{app1}
\begin{Large}\textbf{Result}\end{Large}\\
Suppose that, for all $\mathbf{x}$, the covariate-specific ROC curve, $\mbox{ROC}(\cdot \mid \mathbf{x})$, is concave and the distribution of the covariates is the same in the healthy and diseased groups, i.e, $H_{\bar{D}}(\mathbf{x}) = H_{D}(\mathbf{x})$. Then,
\[
\mbox{AROC}(t) \geq \mbox{ROC}(t)\;\;\mbox{for all}\;\;t\in [0,1].
\]
A similar result has been proved in \citet[][Chapter 6]{Pepe03} and \cite{Janes2008b}, but under the more restrictive assumption that the covariate-specific ROC curve is the same across all covariate values. This assumption is relaxed here.
\proof
Using the notation of Section~\ref{examples}, we have
\begin{align*}
\mbox{AROC}(t) & = \int \mbox{ROC}(t \mid \mathbf{x})\mbox{d}H_{D}(\mathbf{x})\\
& = \int \mbox{ROC}\left(\int t_{\mathbf{x}}\mbox{d}H_{\bar{D}}(\mathbf{x}) \mathrel{\bigg|} \mathbf{x}\right)\mbox{d}H_{D}( \mathbf{x})\\
& \geq \int \int \mbox{ROC}(t_{\mathbf{x}}\mid \mathbf{x})\mbox{d}H_{\bar{D}}(\mathbf{x})\mbox{d}H_{D}( \mathbf{x})\\
& = \int \int \mbox{ROC}(t_{\mathbf{x}}\mid\mathbf{x})\mbox{d}H_{D}(\mathbf{x})\mbox{d}H_{\bar{D}}(\mathbf{x})\\
& = \int \mbox{ROC}(t)\mbox{d}H_{\bar{D}}(\mathbf{x})\\
& = \mbox{ROC}(t),
\end{align*}
where the inequality follows from the concavity of the covariate-specific ROC curves (using Jensen's inequality).

\putbib[References_AROC]
\end{bibunit}
\newpage

\begin{figure}[H]
    \begin{center}
    	\subfigure[No association between the covariate and the diagnostic test outcome]{
		\includegraphics[width =4cm]{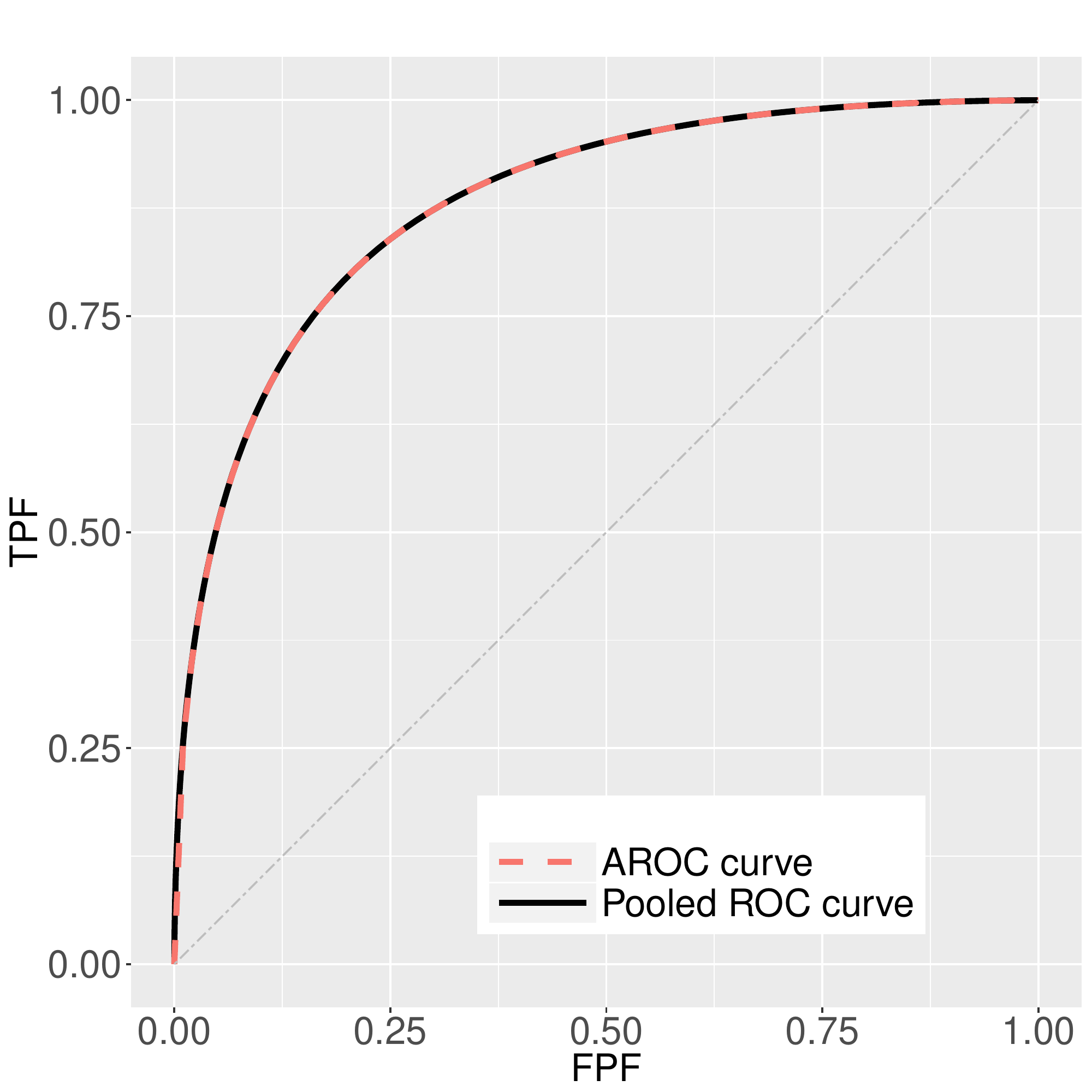}
		\label{no_association}
		}
		\subfigure[No effect of the covariate on diagnostic accuracy]{
		\includegraphics[height=4cm]{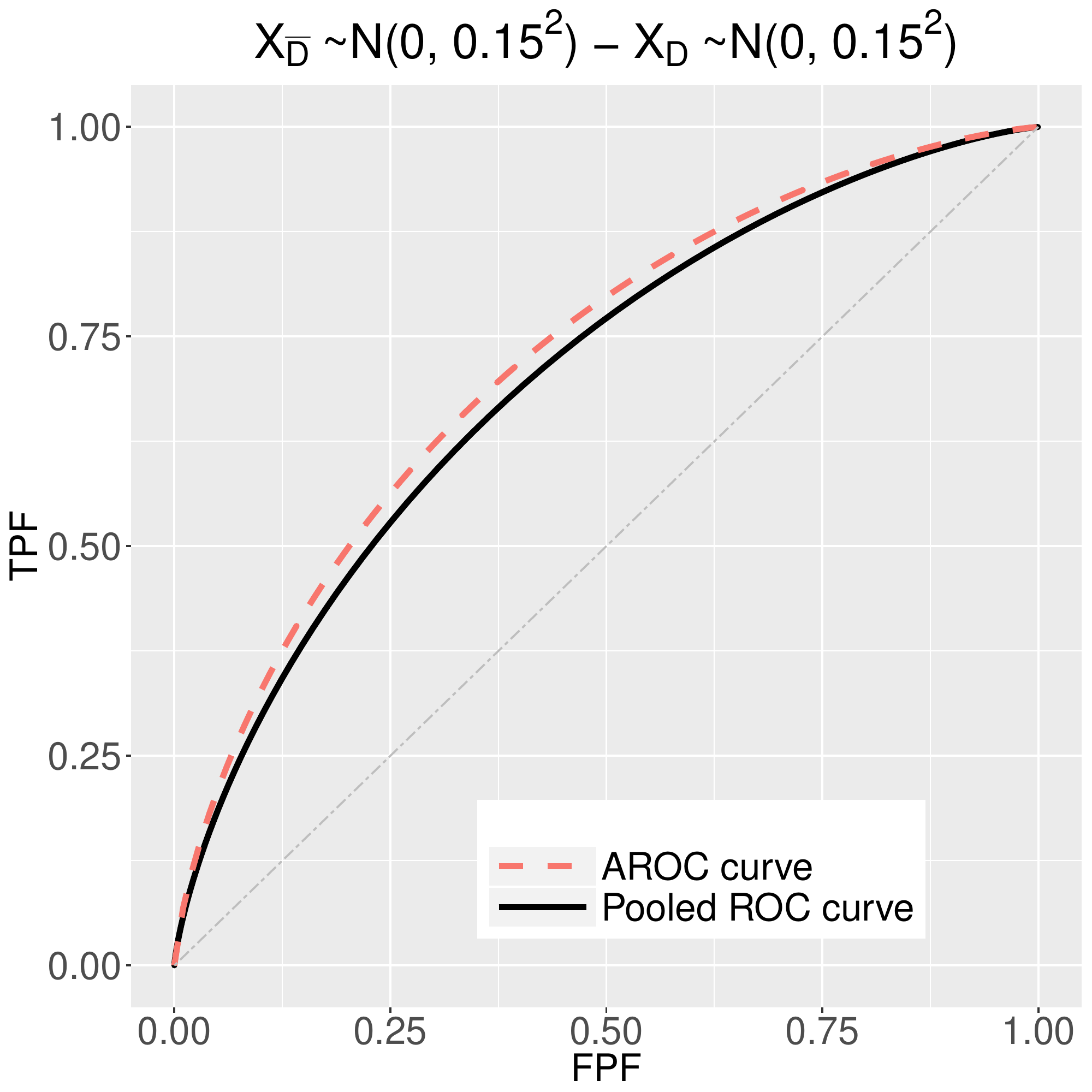}
		\includegraphics[height=4cm]{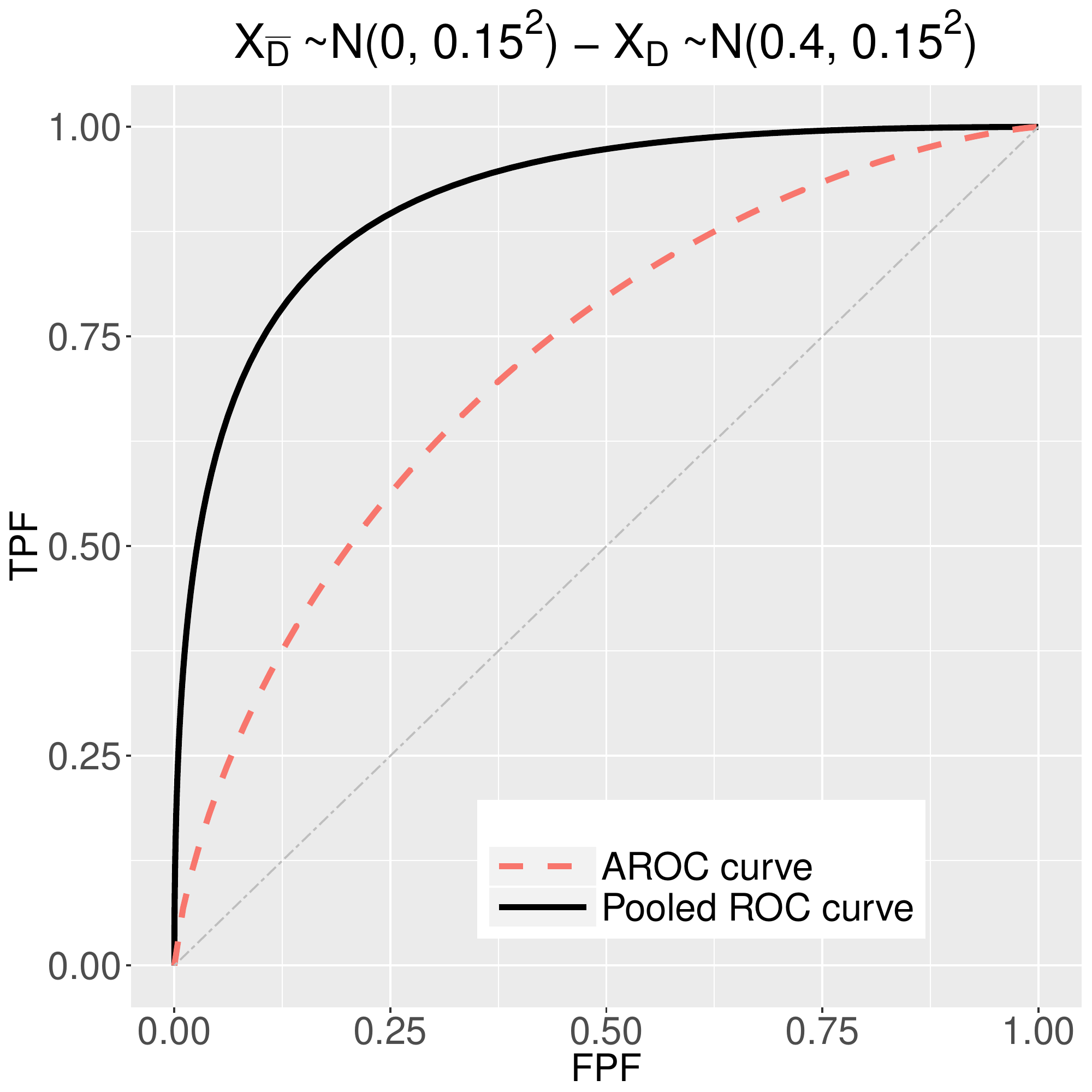}
		\includegraphics[height=4cm]{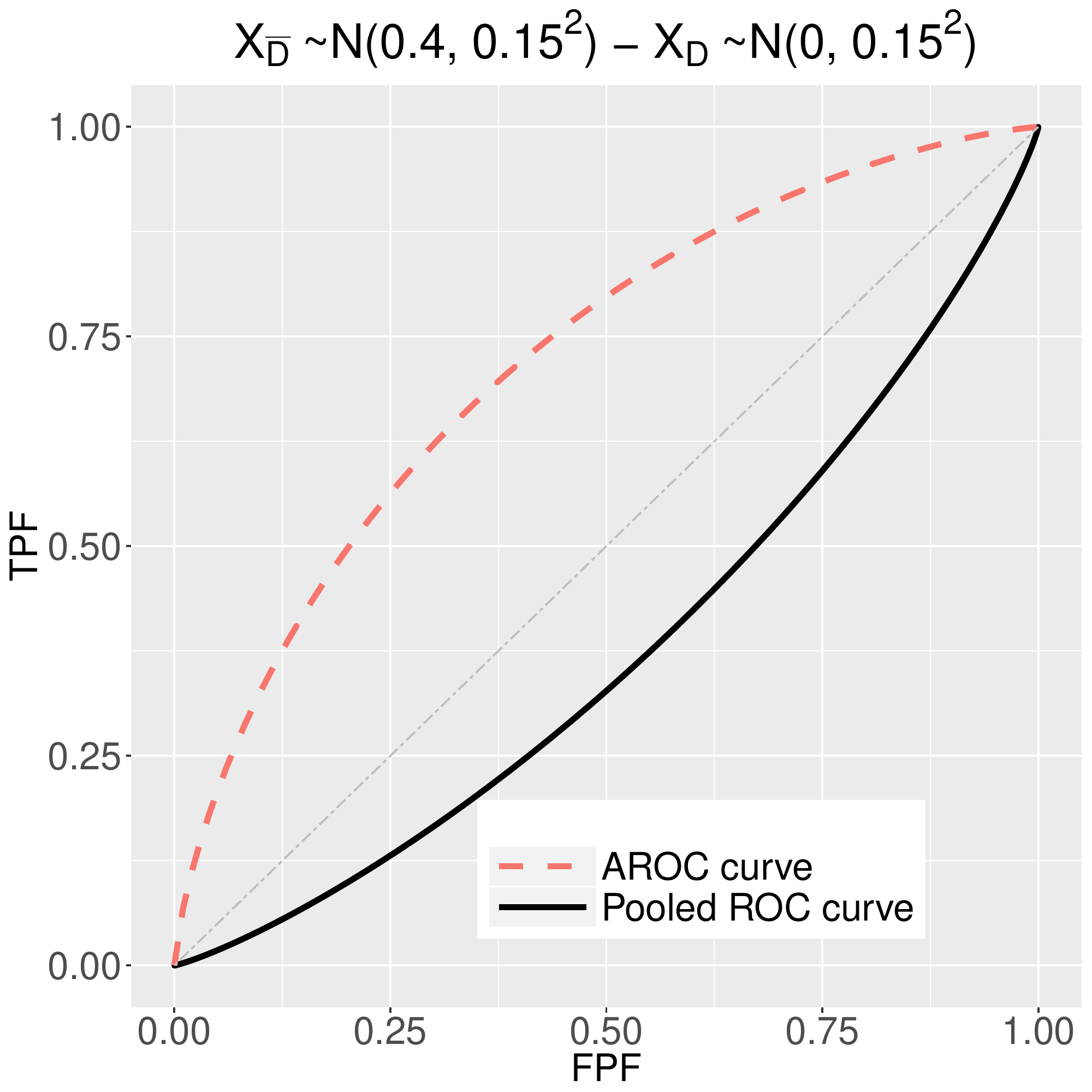}
		\label{no_effect}}
		\subfigure[Effect of the covariate on diagnostic accuracy]{
		\includegraphics[height=4cm]{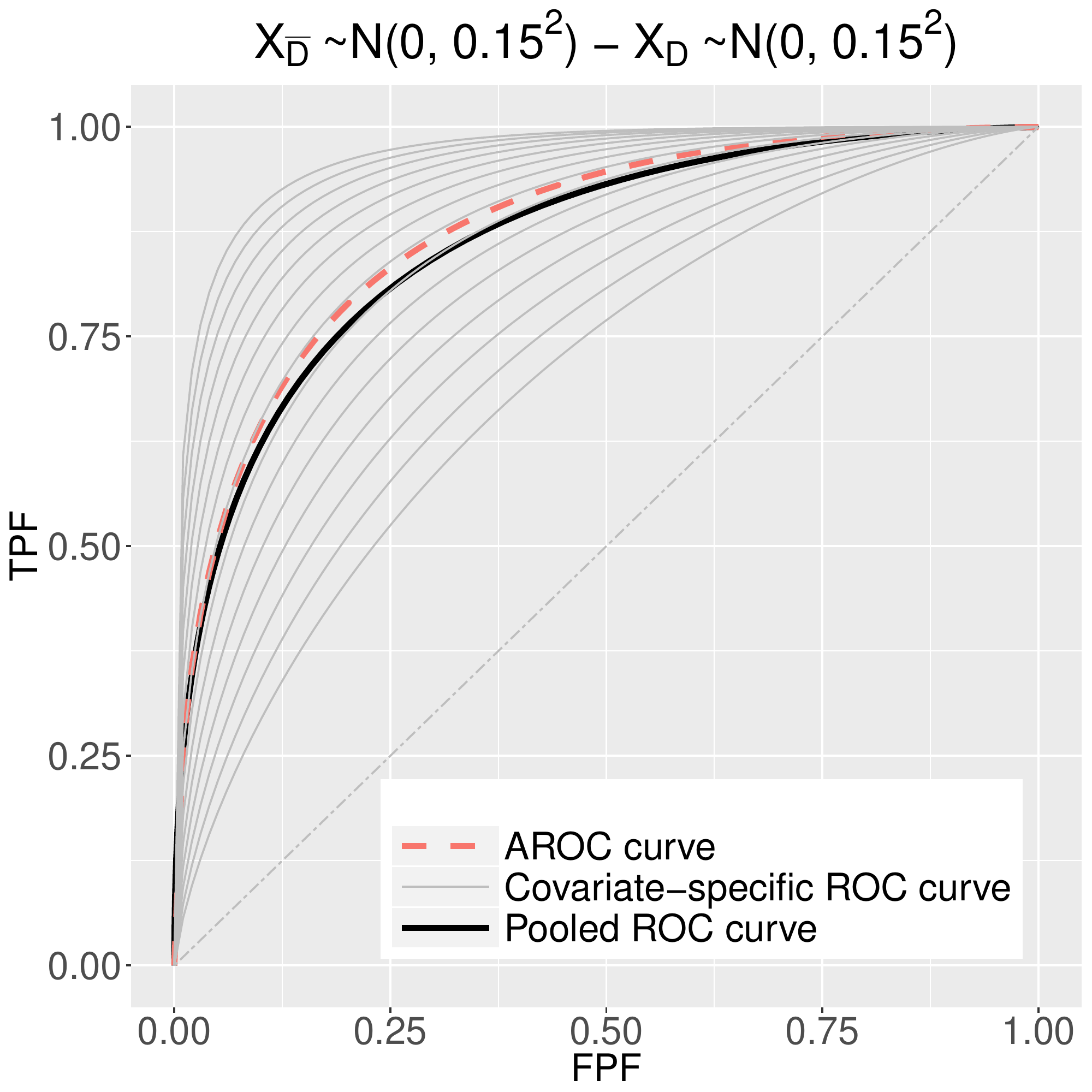}
		\includegraphics[height=4cm]{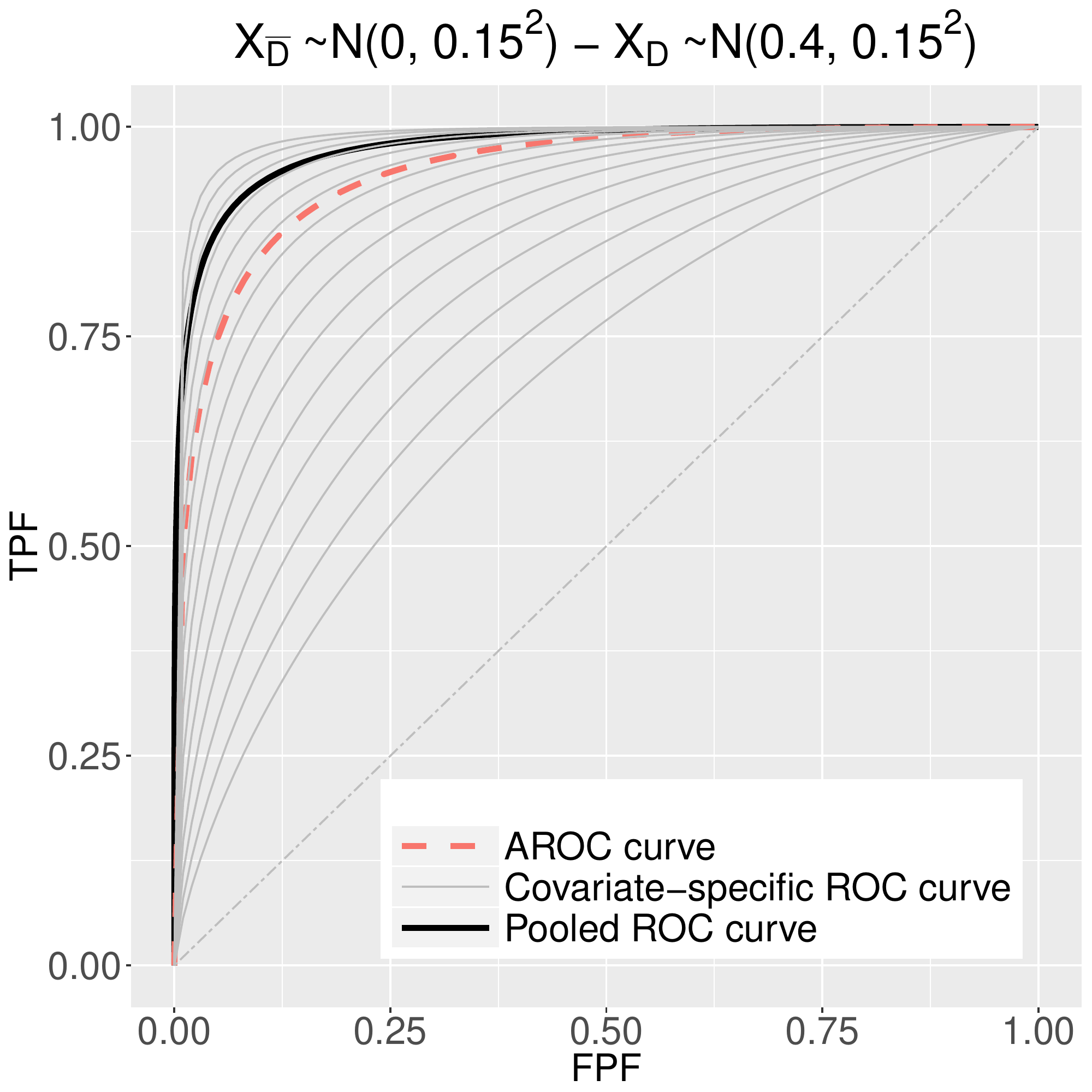}
		\includegraphics[height=4cm]{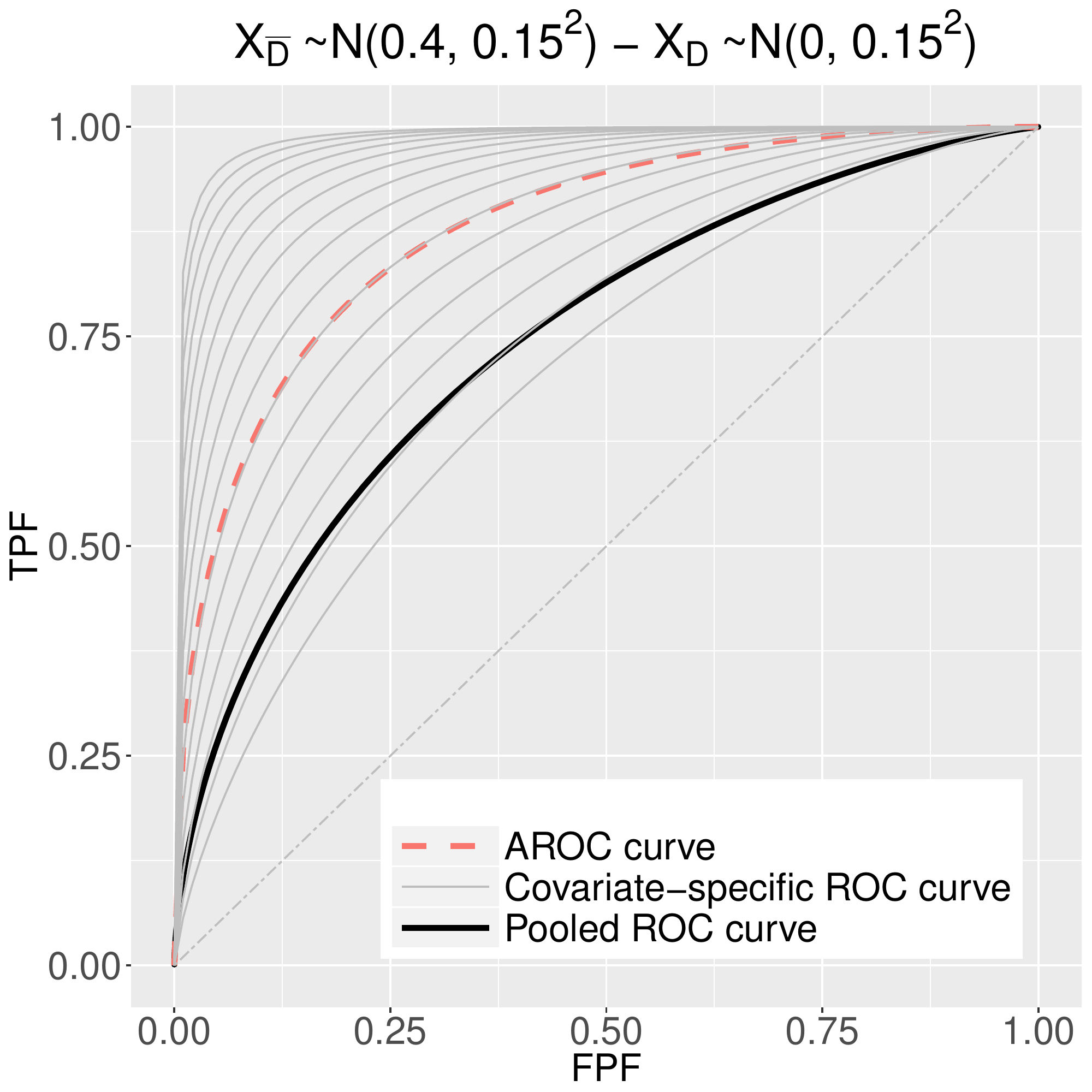}
		\label{effect}}
	\end{center}
		 \caption{For a hypothetical diagnostic test $Y$, comparisons among the pooled ROC curve (black solid line), the AROC curve (red dashed line) and the covariate-specific ROC curve (grey solid lines) in the presence of a continuous covariate $X$. (a) The covariate is not associated with the diagnostic test outcome ($Y_{\bar{D}} \sim N(0.5, 0.3^2)$ and $Y_{D} \sim N(1, 0.3^2)$). (b) The covariate is associated with the diagnostic test outcome, but it has no impact on the diagnostic accuracy of the test ($Y_{\bar{D}} \sim N(0.5 + X_{\bar{D}}, 0.3^2)$ and $Y_{D} \sim N(0.75 + X_{D}, 0.3^2)$). (c) The same as in (b) but the covariate has an impact on the diagnostic accuracy of the test ($Y_{\bar{D}} \sim N(0.25 + 0.5X_{\bar{D}}, 0.3^2)$ and $Y_{D} \sim N(0.75 + X_{D}, 0.3^2)$). The left-hand side plots in (b) and (c) show the situation where the distribution of the covariate is the same in the healthy and diseased group ($X_{\bar{D}} \sim N(0, 0.15^2)$ and $X_{D} \sim N(0, 0.15^2)$). The middle- and right-hand side plots in (b) and (c) depict the situation where the distribution of the covariate is different in the healthy and diseased group. Specifically, in the middle-hand side plots $X_{\bar{D}} \sim N(0, 0.15^2)$ and $X_{D} \sim N(0.4, 0.15^2)$; and in the right-hand side plots $X_{\bar{D}} \sim N(0.4, 0.15^2)$ and $X_{D} \sim N(0, 0.15^2)$.}
		\label{mot_example}
\end{figure}

\begin{table}[H]
\caption{Average (standard deviation) ($\times$ 100), across simulations, of the empirical root mean squared error for the different approaches under consideration and for $K=4$. The results are presented for each of the simulated scenarios and sample sizes.}\label{MSE_K_4}
 \begin{center}
 \footnotesize	
	\begin{tabular}{cccccc}
	& & \multicolumn{4}{c}{Sample size}\\
	& & \multicolumn{4}{c}{$(n_{\bar{D}},n_{D})$}\\\hline
	Scenario & Approach & $(50,50)$ & $(200,70)$ & $(200,200)$ &  $(300,100)$ \\\hline
	\multirow{4}{*}{I} &	
Bayesian nonparametric & 6.013 (3.494) & 4.438 (2.296) & 3.174 (1.644) & 4.070 (2.121)\\
 & Bayesian semiparametric & 6.345 (3.483) & 4.465 (2.260) & 3.163 (1.548) & 4.195 (2.122)\\
 & Kernel & 8.210 (2.964) & 5.363 (2.057) & 4.216 (1.305) & 4.937 (1.897)\\
 & Pooled & 7.233 (2.928) & 5.123 (2.108) & 4.006 (1.283) & 4.752 (1.875)\\ \hline
	\multirow{4}{*}{II} &	
Bayesian nonparametric & 5.990 (3.498) & 4.434 (2.293) & 3.164 (1.630) & 4.071 (2.124)\\
 & Bayesian semiparametric & 6.328 (3.484) & 4.449 (2.257) & 3.167 (1.545) & 4.194 (2.114)\\
 & Kernel & 8.363 (3.107) & 5.448 (2.151) & 4.425 (1.459) & 5.015 (2.054)\\
 & Pooled & 7.740 (3.208) & 5.722 (2.552) & 5.240 (1.891) & 5.807 (2.551)\\ \hline
	\multirow{4}{*}{III} &	
Bayesian nonparametric & 6.021 (3.454) & 4.405 (2.324) & 3.196 (1.592) & 4.118 (2.133)\\
 & Bayesian semiparametric & 6.177 (3.431) & 4.433 (2.274) & 3.206 (1.498) & 4.206 (2.149)\\
 & Kernel & 8.127 (2.954) & 5.398 (2.179) & 4.408 (1.354) & 5.023 (2.052)\\
 & Pooled & 7.216 (2.949) & 5.251 (2.344) & 4.395 (1.630) & 5.083 (2.284)\\ \hline
	\multirow{4}{*}{IV} &	
Bayesian nonparametric & 6.060 (3.249) & 3.973 (2.288) & 2.914 (1.539) & 3.888 (2.085)\\
 & Bayesian semiparametric & 8.424 (4.373) & 8.271 (4.228) & 7.817 (4.042) & 8.837 (4.035)\\
 & Kernel & 8.475 (2.482) & 6.548 (1.891) & 5.794 (1.351) & 6.198 (1.684)\\
 & Pooled & 16.571 (2.100) & 16.450 (1.723) & 16.497 (1.194) & 16.654 (1.259)\\ \hline
	\multirow{4}{*}{V} &	
Bayesian nonparametric & 6.361 (2.832) & 5.553 (1.944) & 4.394 (1.316) & 4.646 (1.396)\\
 & Bayesian semiparametric& 15.745 (3.610) & 17.277 (2.731) & 17.058 (2.270) & 17.311 (2.073)\\
 & Kernel & - & - & - & -\\
 & Pooled & 11.203 (2.281) & 10.468 (1.448) & 9.889 (1.183) & 9.840 (1.228)\\ \hline
	\multirow{4}{*}{VI} &	
Bayesian nonparametric & 5.673 (3.056) & 4.304 (2.180) & 3.063 (1.748) & 3.781 (2.120)\\
 & Bayesian semiparametric & 8.674 (4.397) & 8.948 (2.880) & 8.104 (2.471) & 8.442 (2.805)\\
 & Kernel & - & - & - & -\\
 & Pooled & 12.063 (3.627) & 11.749 (3.125) & 11.086 (2.240) & 11.143 (2.485)\\ \hline
	\end{tabular}
 \end{center}
\end{table}

\begin{figure}[H]
    \begin{center}
		\includegraphics[width =5.95cm]{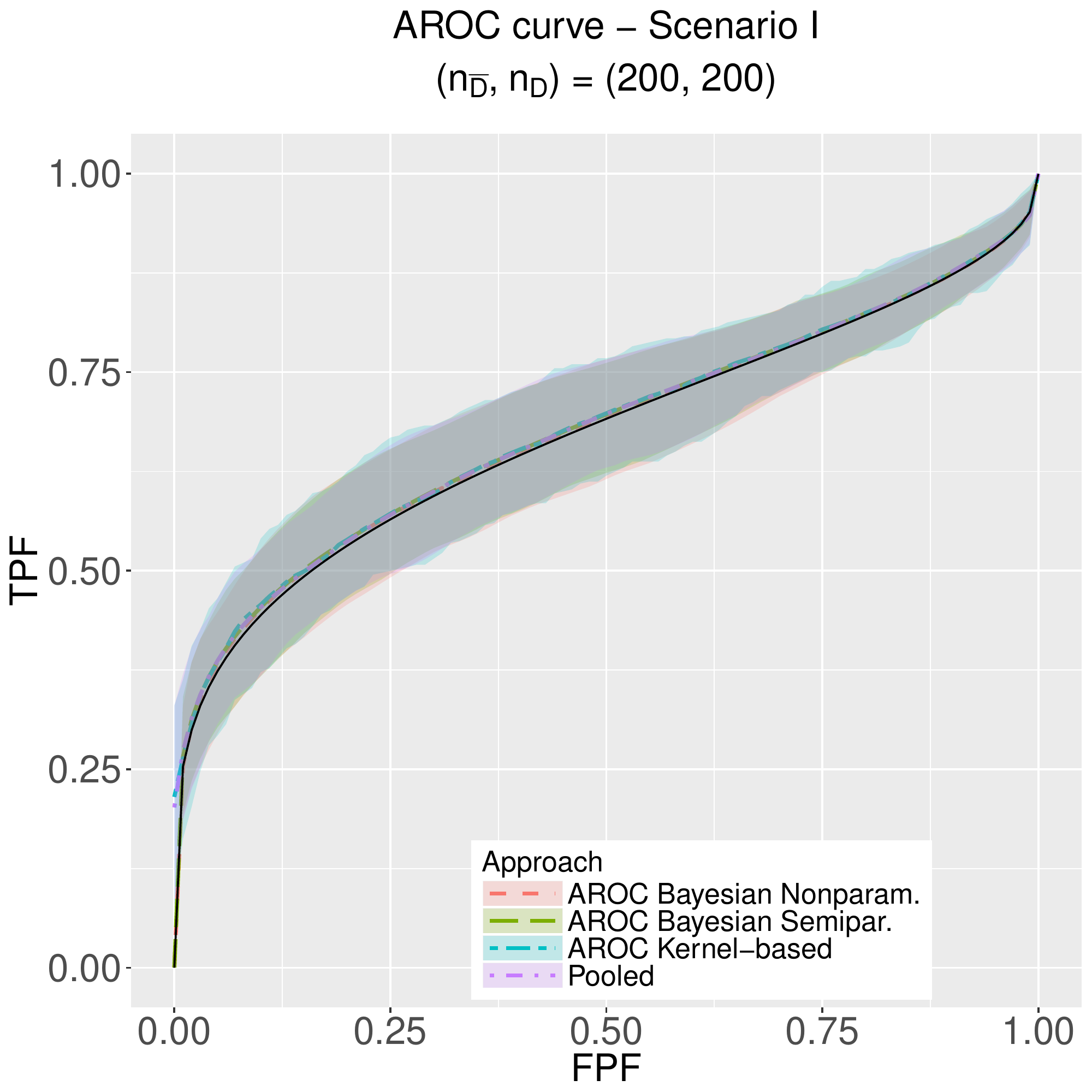}
		\includegraphics[height=5.95cm]{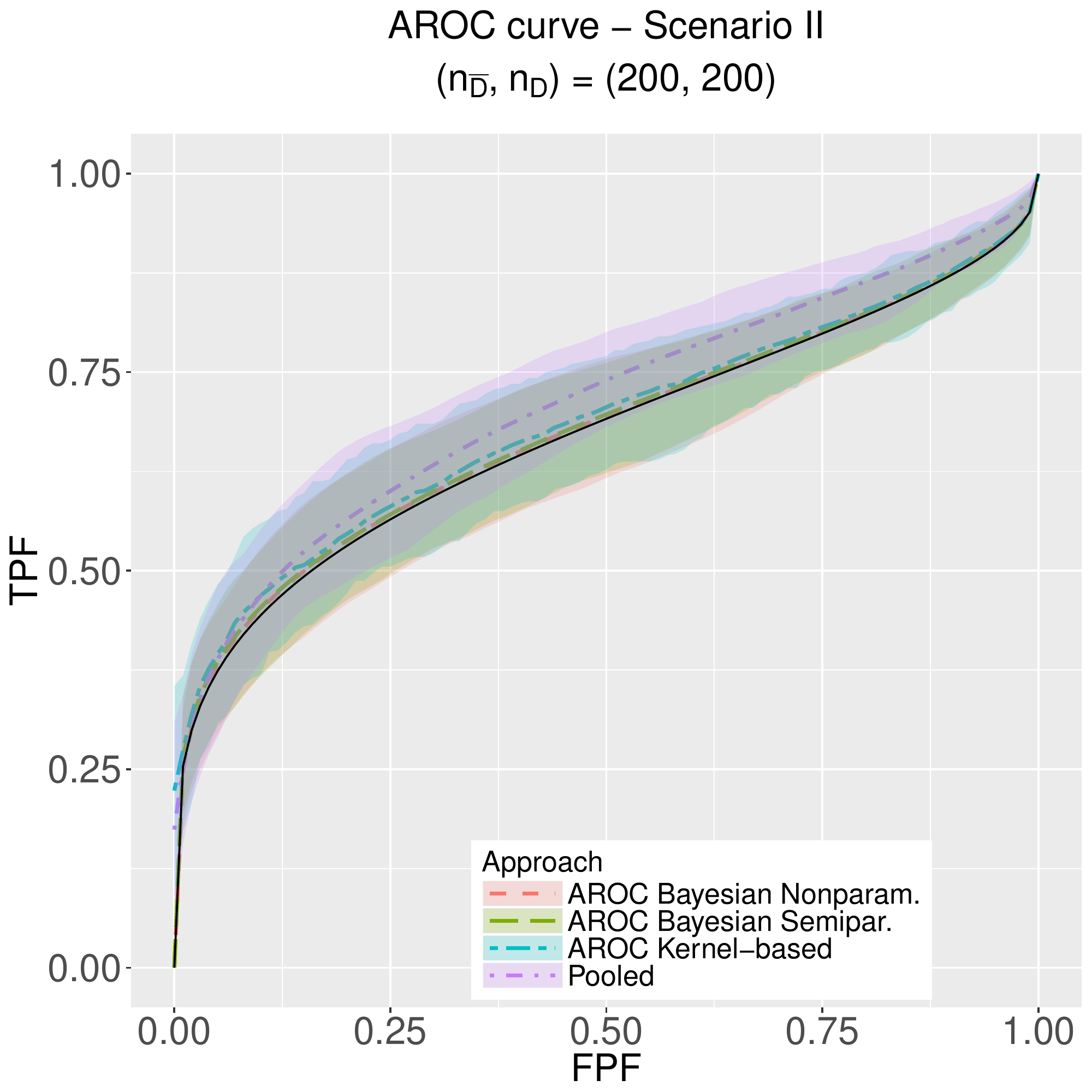}
		\includegraphics[width=5.95cm]{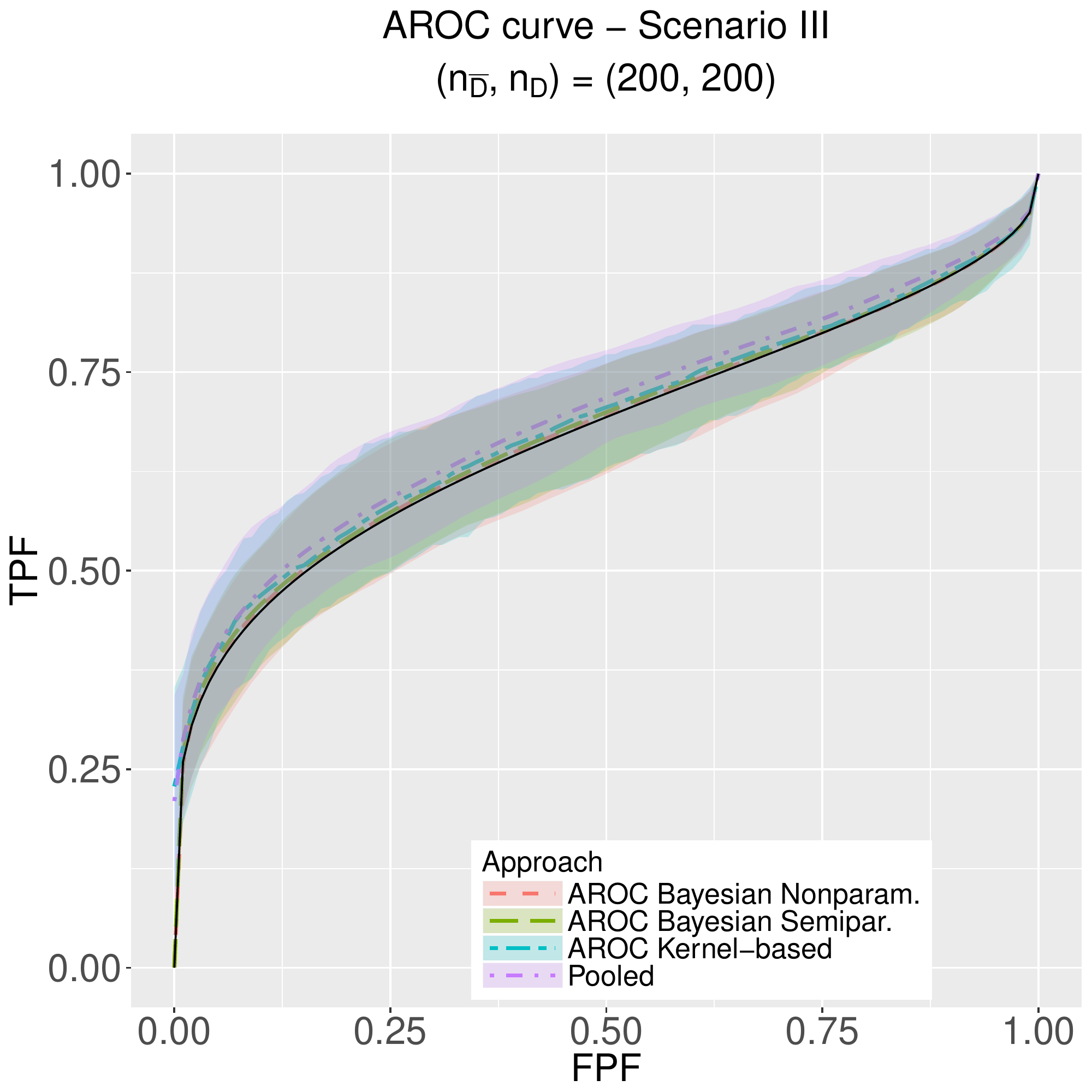}\vspace{0.3cm}
		\includegraphics[height=5.95cm]{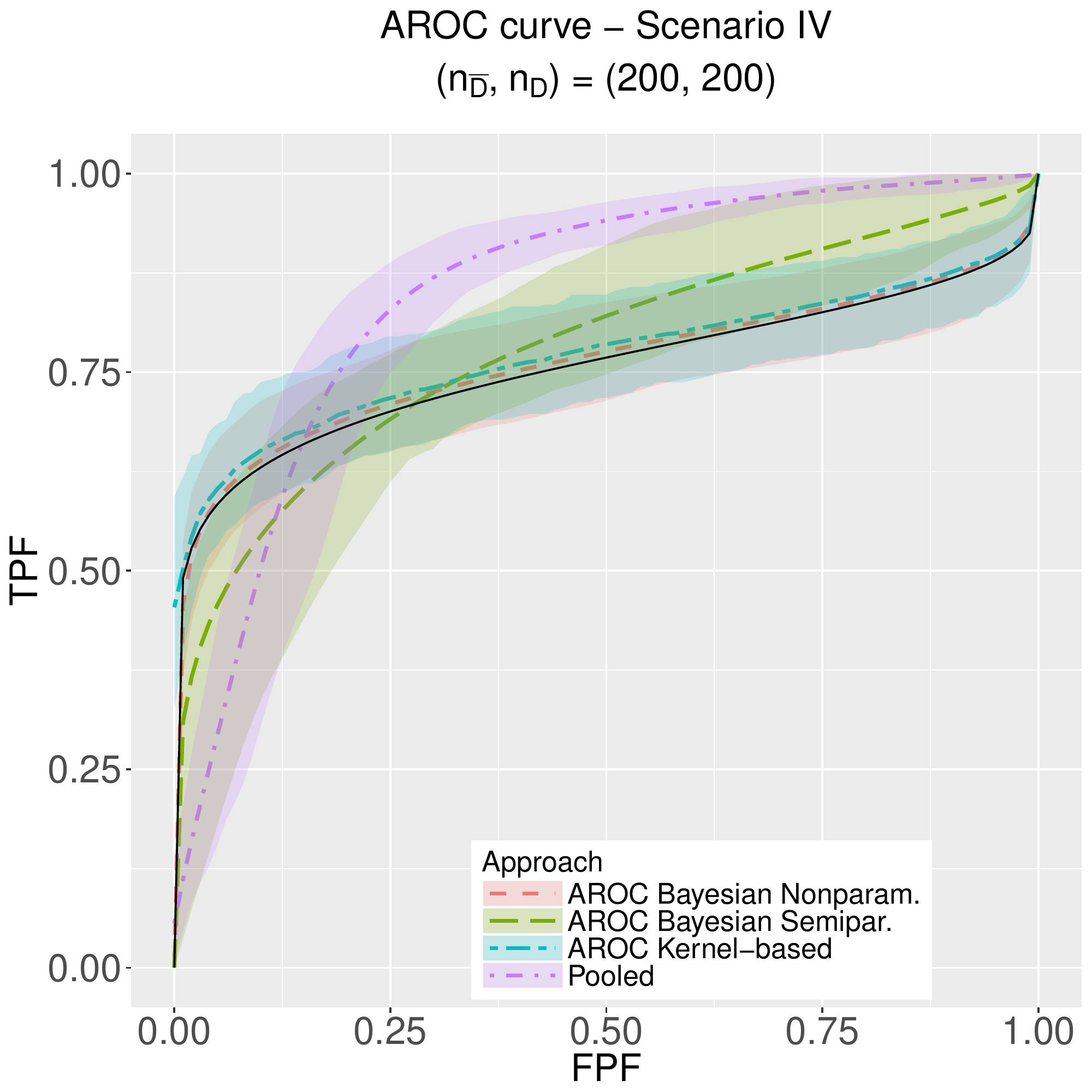}
		\includegraphics[height=5.95cm]{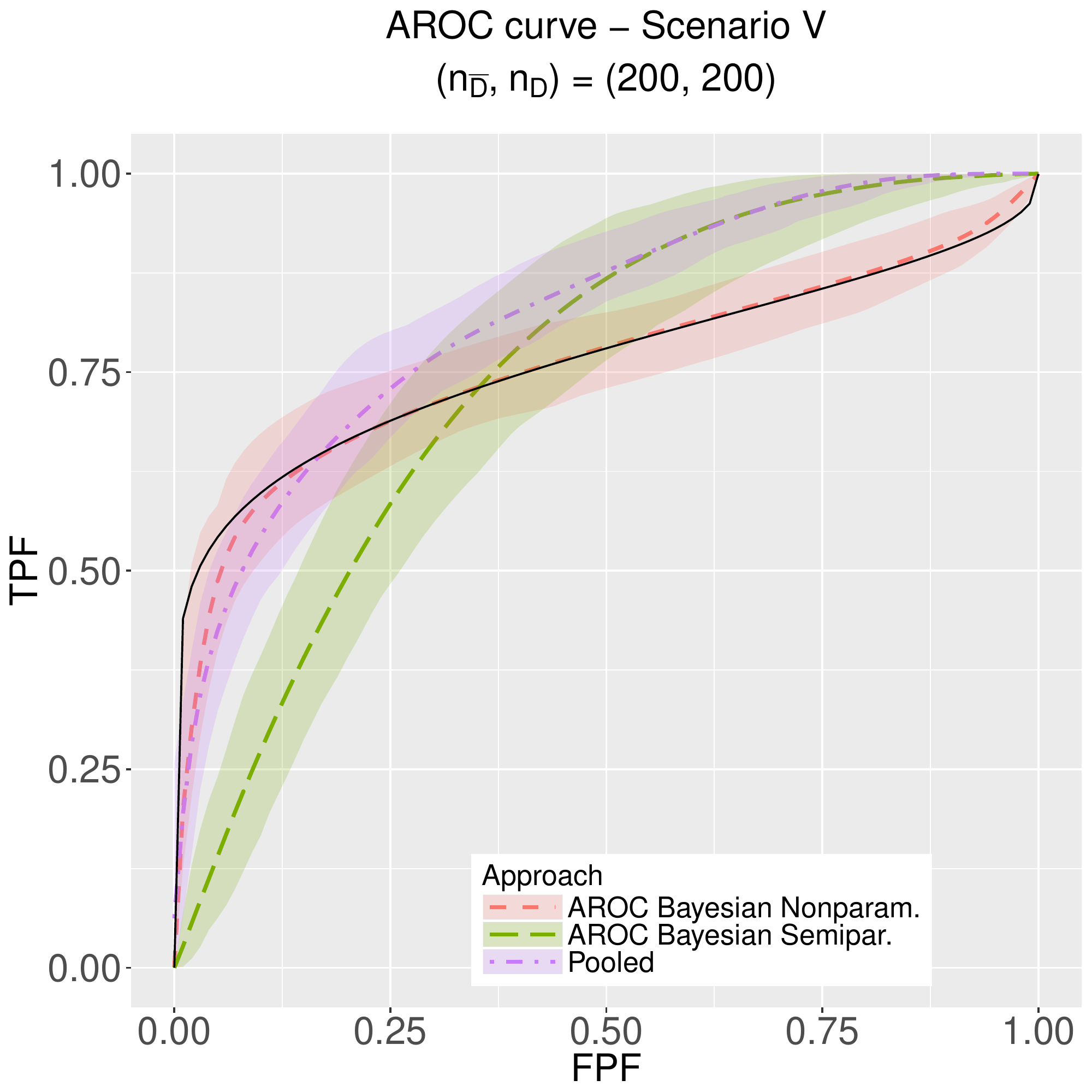}
		\includegraphics[height=5.95cm]{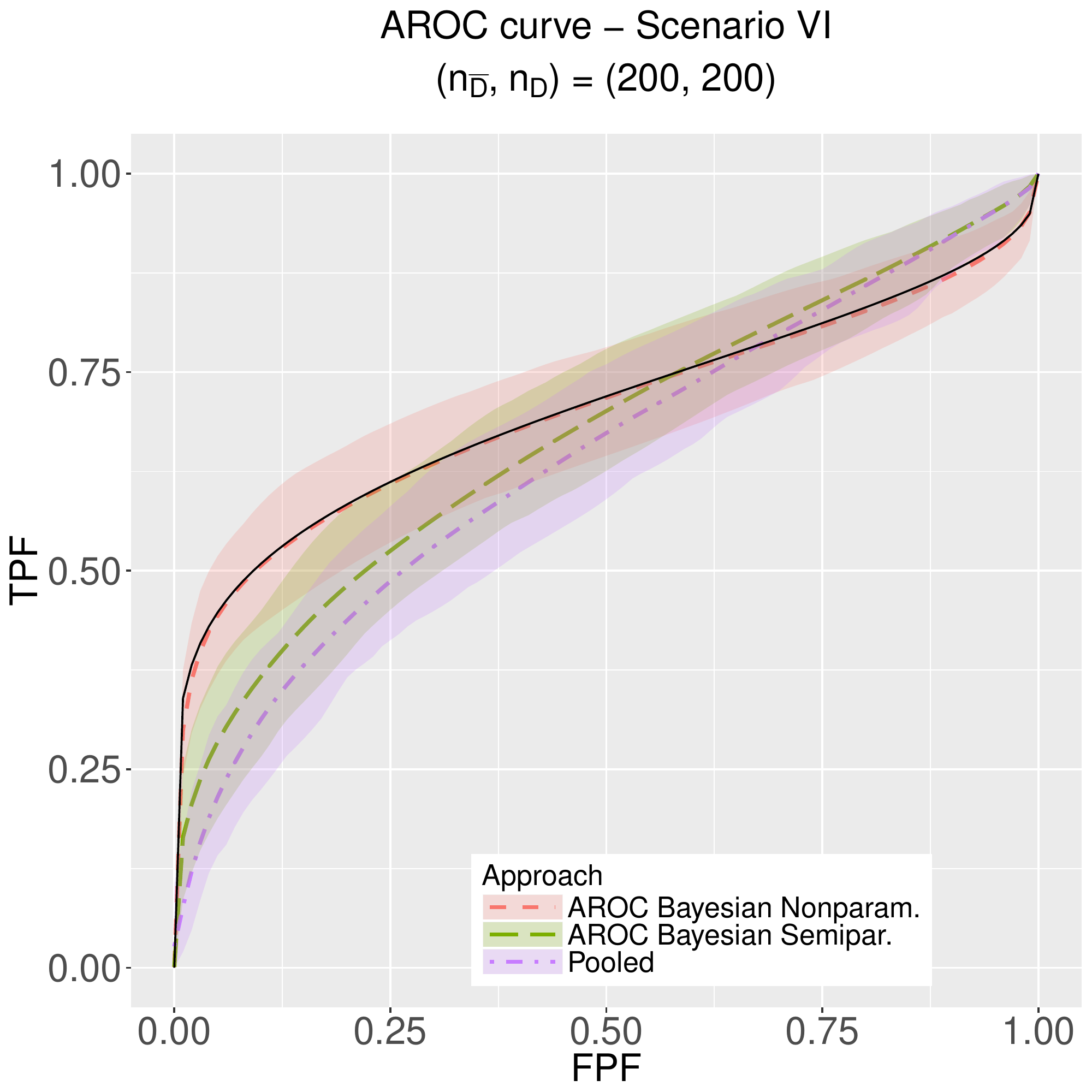}
	\end{center}
		 \caption{True (solid black line) and average value of 100 simulated datasets (dashed lines) of the posterior mean (for the Bayesian estimators) of the covariate adjusted ROC curve/pooled ROC curve for each of the scenarios and sample size $(n_{\bar{D}},n_D)=(200,200)$. The shaded area are bands constructed using the pointwise $2.5\%$ and $97.5\%$ quantiles across simulations.}
		\label{aROC_sim}
\end{figure}
\begin{table}[H]
\caption{$95\%$ coverage probabilities for the AROC (average over all FPFs) for the different approaches under consideration and for $K=4$. The results are presented for each of the simulated scenarios and sample sizes.}\label{covAROC_K_4}
 \begin{center}
 \footnotesize	
	\begin{tabular}{cccccc}
	& & \multicolumn{4}{c}{Sample size}\\
	& & \multicolumn{4}{c}{$(n_{\bar{D}},n_{D})$}\\\hline
	Scenario & Approach & $(50,50)$ & $(200,70)$ & $(200,200)$ &  $(300,100)$ \\\hline
	\multirow{2}{*}{I} &	
Bayesian nonparametric & 96.0 & 95.0 & 95.0 & 95.0\\
 & Bayesian semiparametric& 93.0 & 95.0 & 94.0 & 94.0\\ \hline
	\multirow{2}{*}{II} &	
Bayesian nonparametric & 96.0 & 96.0 & 94.0 & 95.0\\
 & Bayesian semiparametric & 93.0 & 95.0 & 94.0 & 94.0\\ \hline
	\multirow{2}{*}{III} &	
Bayesian nonparametric & 96.0 & 95.0 & 95.0 & 94.0\\
 & Bayesian semiparametric& 94.0 & 95.0 & 95.0 & 94.0\\ \hline
	\multirow{2}{*}{IV} &	
Bayesian nonparametric & 96.0 & 96.0 & 93.0 & 93.0\\
 & Bayesian semiparametric & 81.0 & 70.0 & 52.0 & 55.0\\ \hline
	\multirow{2}{*}{V} &	
Bayesian nonparametric & 97.0 & 95.0 & 95.0 & 96.0\\
 & Bayesian semiparametric & 55.0 & 35.0 & 27.0 & 28.0\\ \hline
	\multirow{2}{*}{VI} &	
Bayesian nonparametric & 98.0 & 97.0 & 96.0 & 95.0\\
 & Bayesian semiparametric & 87.0 & 77.0 & 60.0 & 68.0\\ \hline
	\end{tabular}
 \end{center}
\end{table}

\begin{figure}[H]
\begin{center}
\subfigure[Age-specific AUC]{
\includegraphics[width =5.5cm]{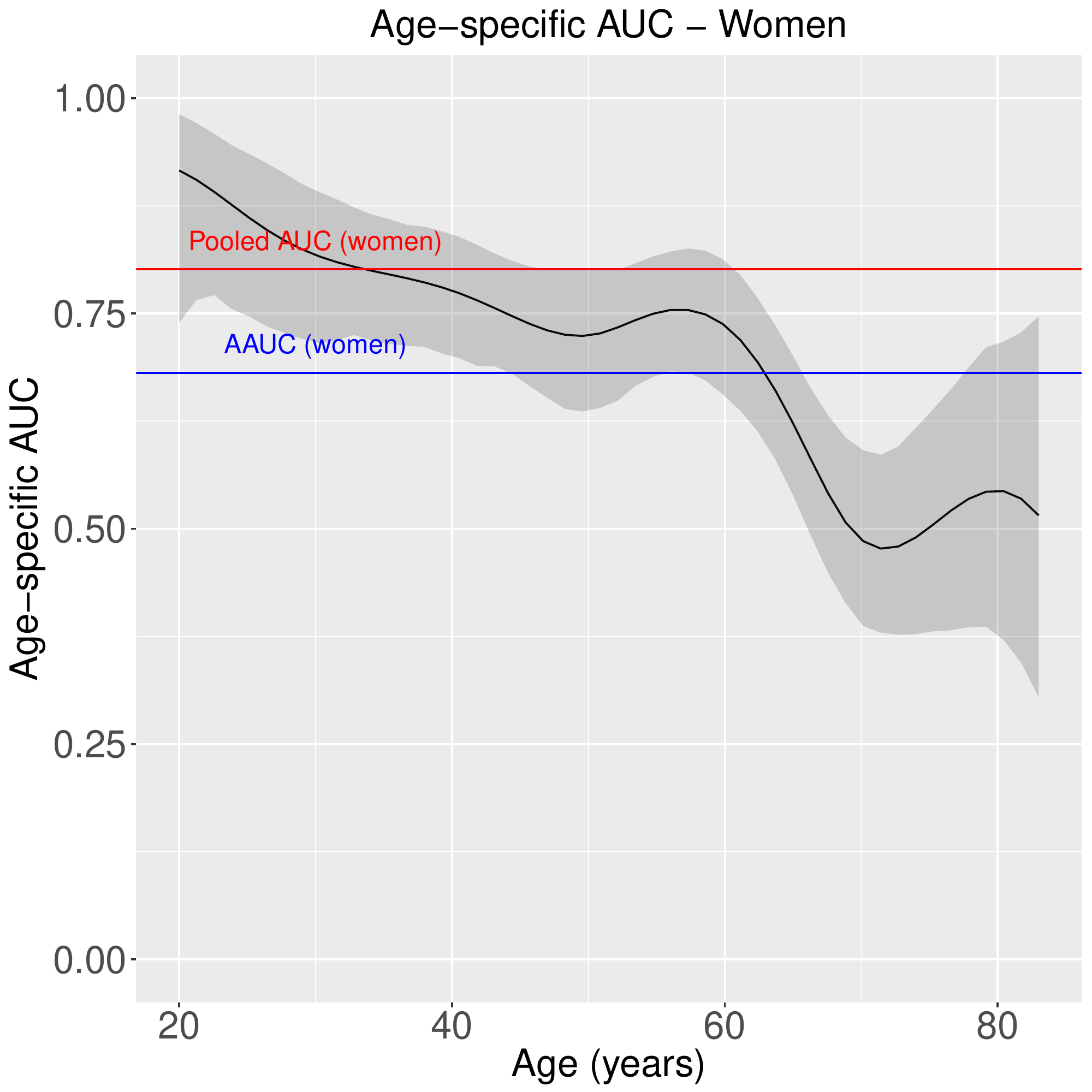}\label{cROC_aROC_gender_a} \hspace{0.4cm}
\includegraphics[width=5.5cm]{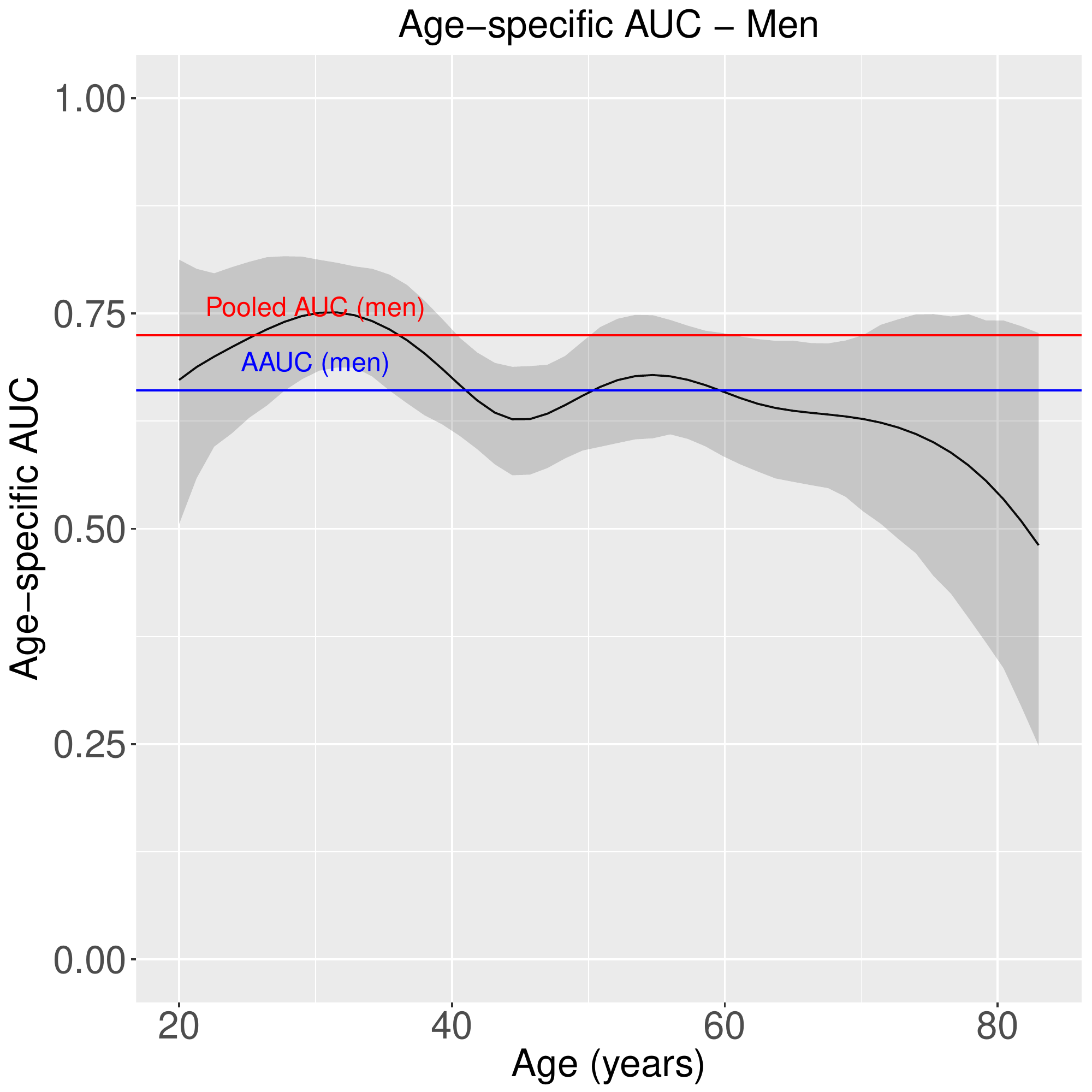}
}
\subfigure[Age-adjusted ROC curve]{
\includegraphics[height=5.5cm]{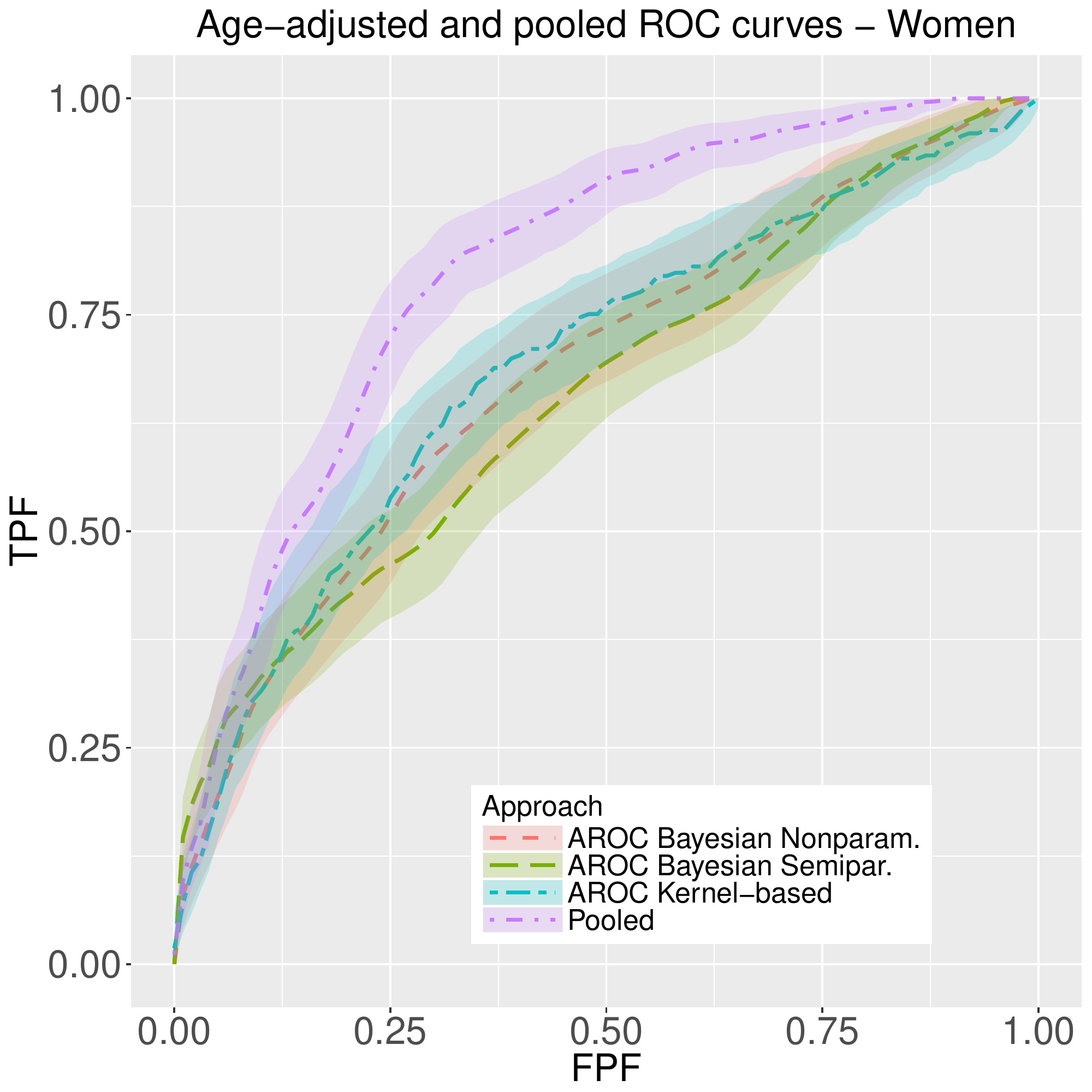} \hspace{0.4cm}
\includegraphics[height=5.5cm]{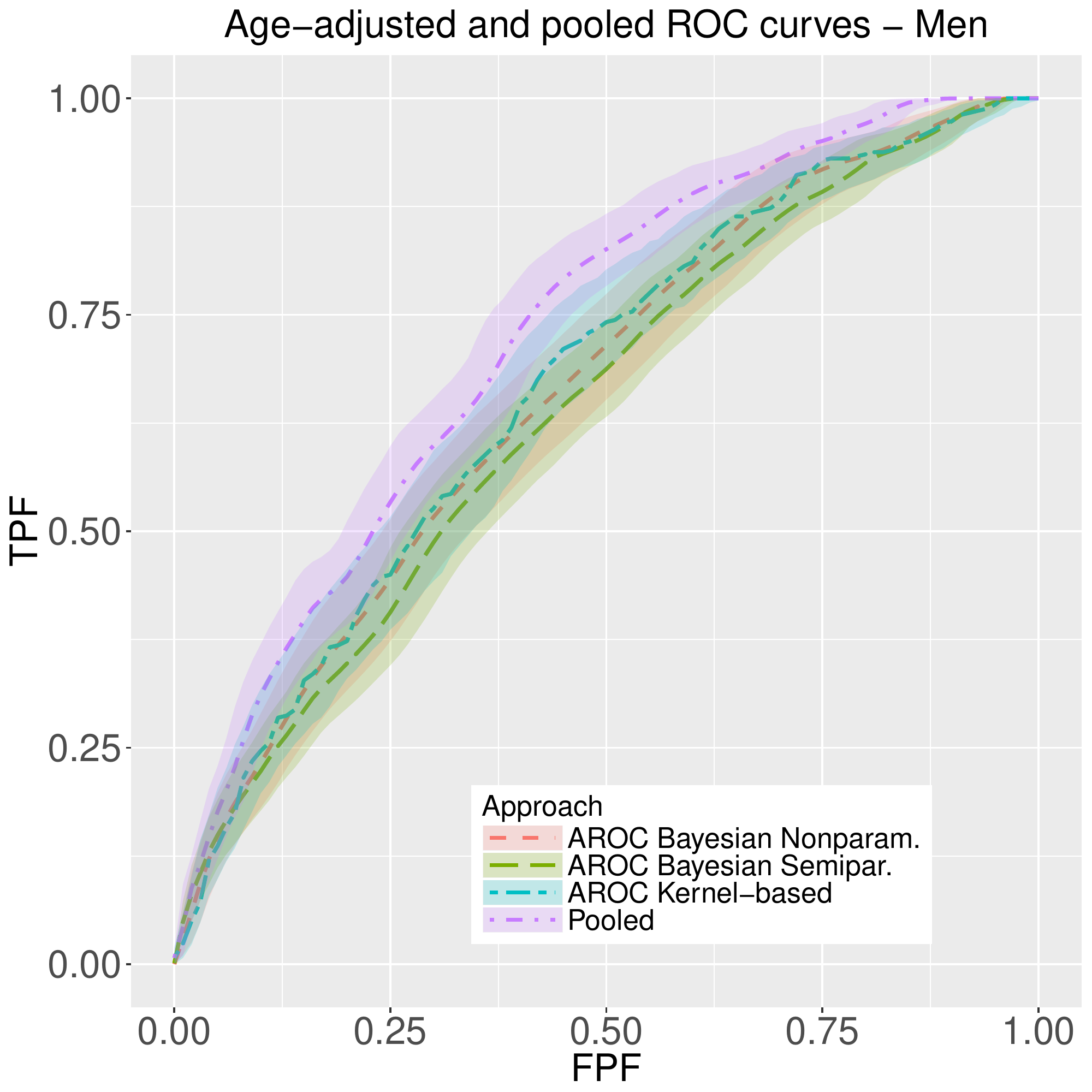}\label{cROC_aROC_gender_b}
}
\end{center}
\caption{\footnotesize{Top row: Age-specific AUC: posterior mean (black line) and 95\% pointwise credible band (grey area) for women (left) and men (right). The red line corresponds to the estimated pooled AUC and the blue line to the estimated area under the age-adjusted ROC curve (AAUC). Bottom row: estimated age-adjusted ROC curve (AROC) and 95\% intervals obtained under our Bayesian nonparametric estimator, the Bayesian semiparametric method and the kernel-based approach. In all cases, the analyses were done separately in women and men with age as the single covariate to adjust for. For comparison purposes, the estimated pooled ROC curves are also depicted.}
\label{cROC_aROC_gender}}
\end{figure}

\begin{figure}[H]
\begin{center}
 \subfigure[AROC and pooled ROC curves]{
\includegraphics[width =5cm]{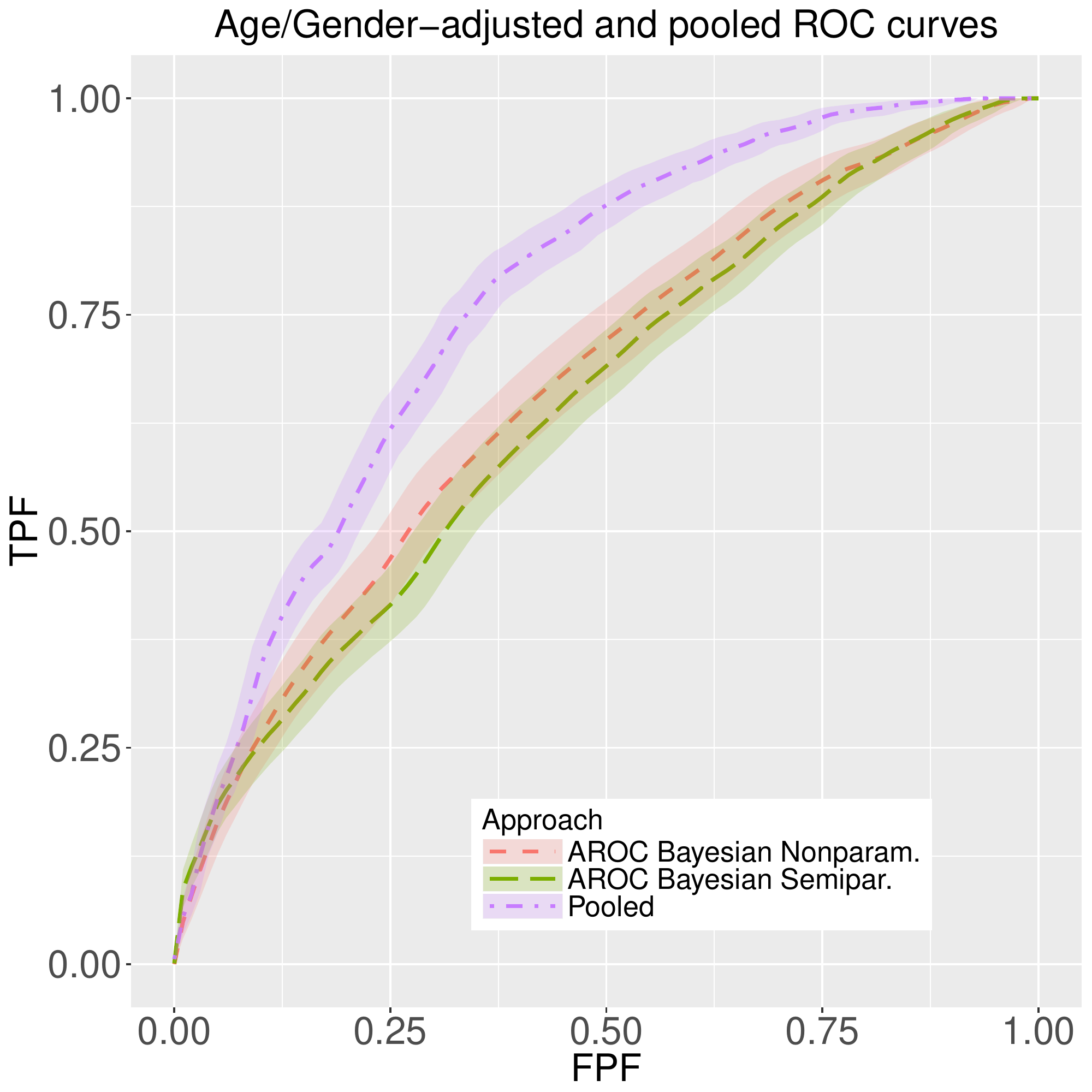}
\label{aROC_a}}
\subfigure[AROC-based threshold values]{
\includegraphics[width=5cm]{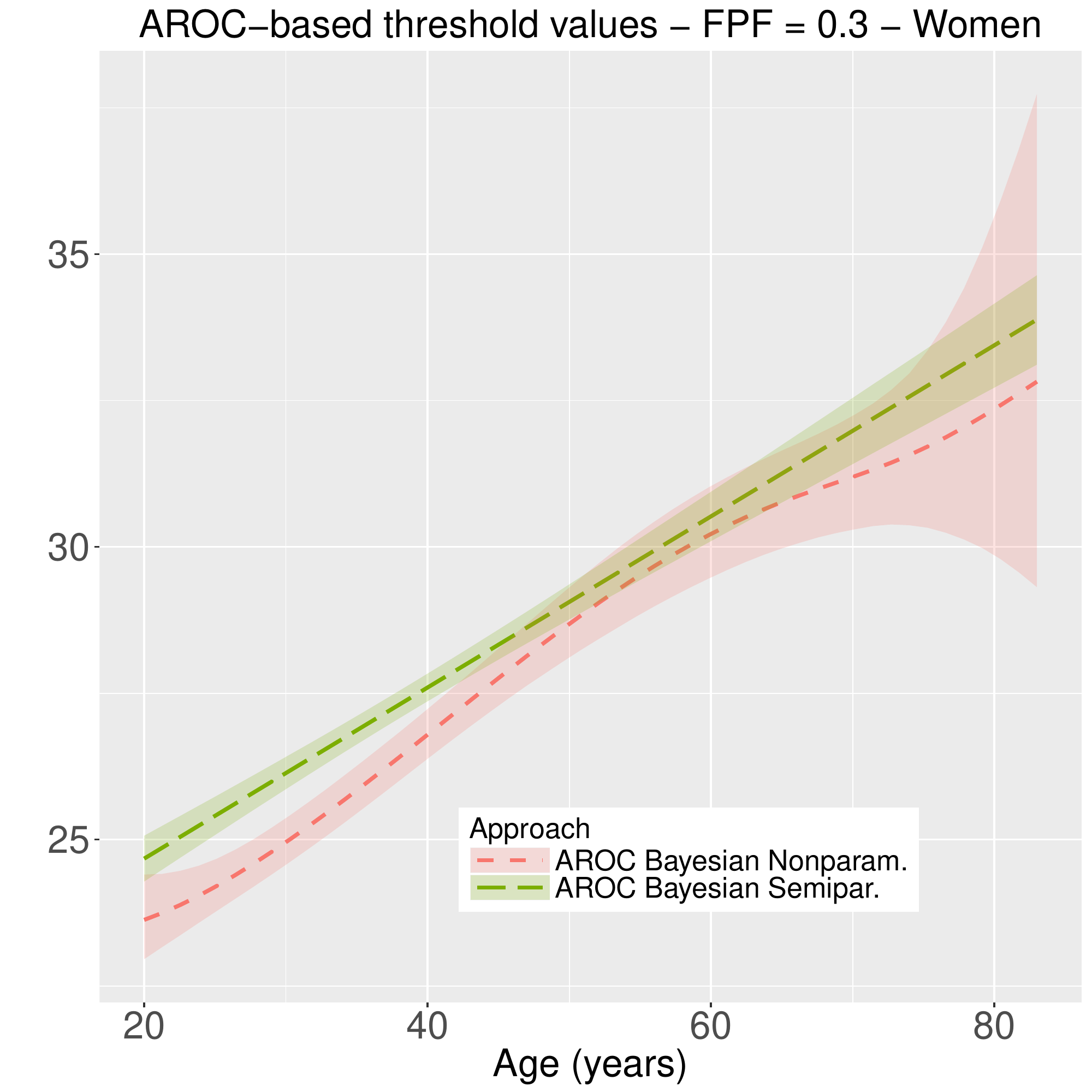}
\includegraphics[height=5cm]{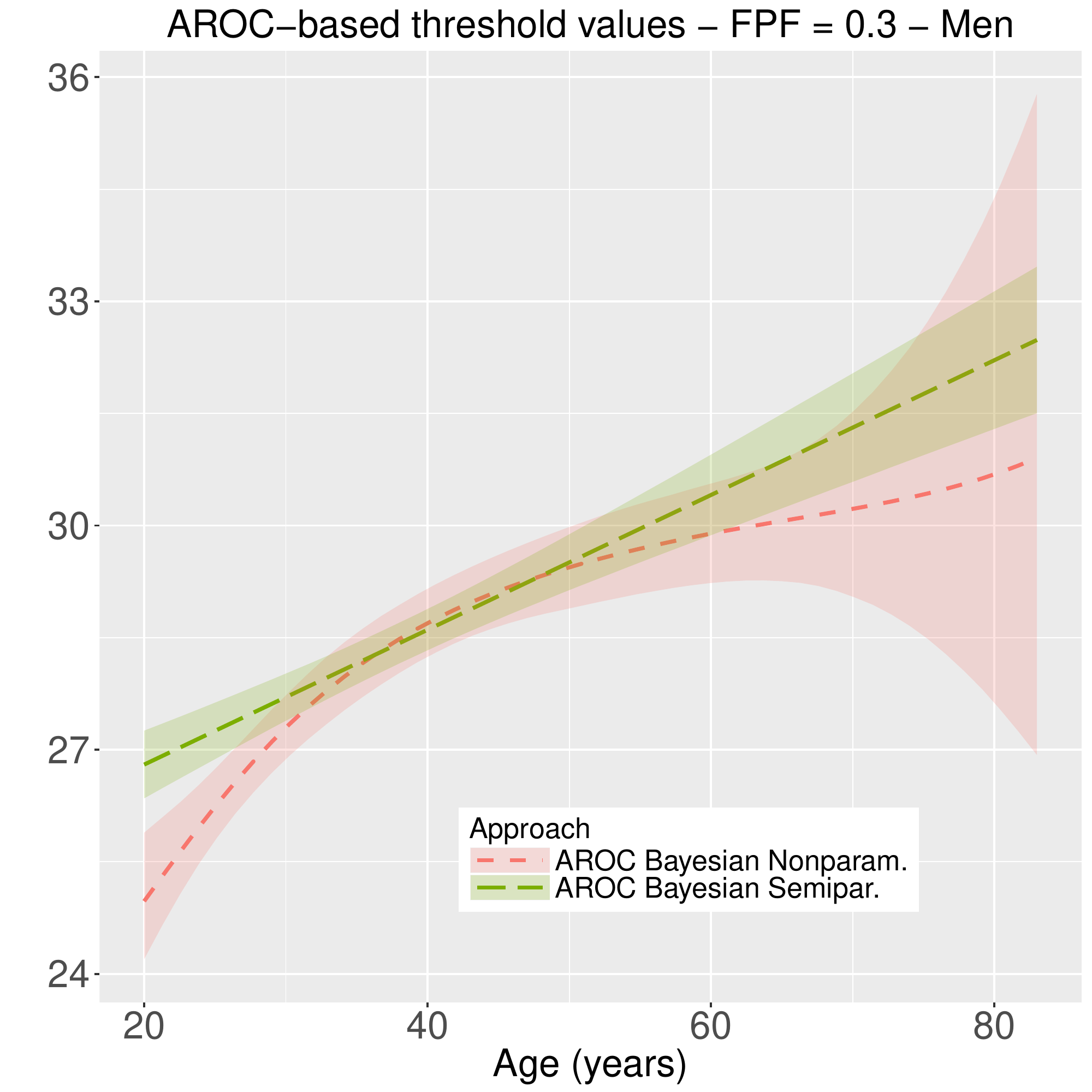}\label{aROC_b}}
\end{center}
\caption{\footnotesize{(a) Age/gender-adjusted ROC curves (AROC): posterior means and 95\% pointwise credible bands under our Bayesian nonparametric estimator and the Bayesian semiparametric estimator. In both cases, the analyses were done including the age-by-gender interaction. For comparison purposes, the estimated pooled ROC based on the global sample is also depicted. (b) Posterior mean and 95\% pointwise credible band of the AROC-based threshold values along age for women and men, for a FPF = 0.3.}}
\label{aROC}
\end{figure}

\begin{figure}[H]
\begin{center}
\subfigure[Predictive checks: Bayesian nonparametric model (B-splines DDP)]{
\includegraphics[width =5cm]{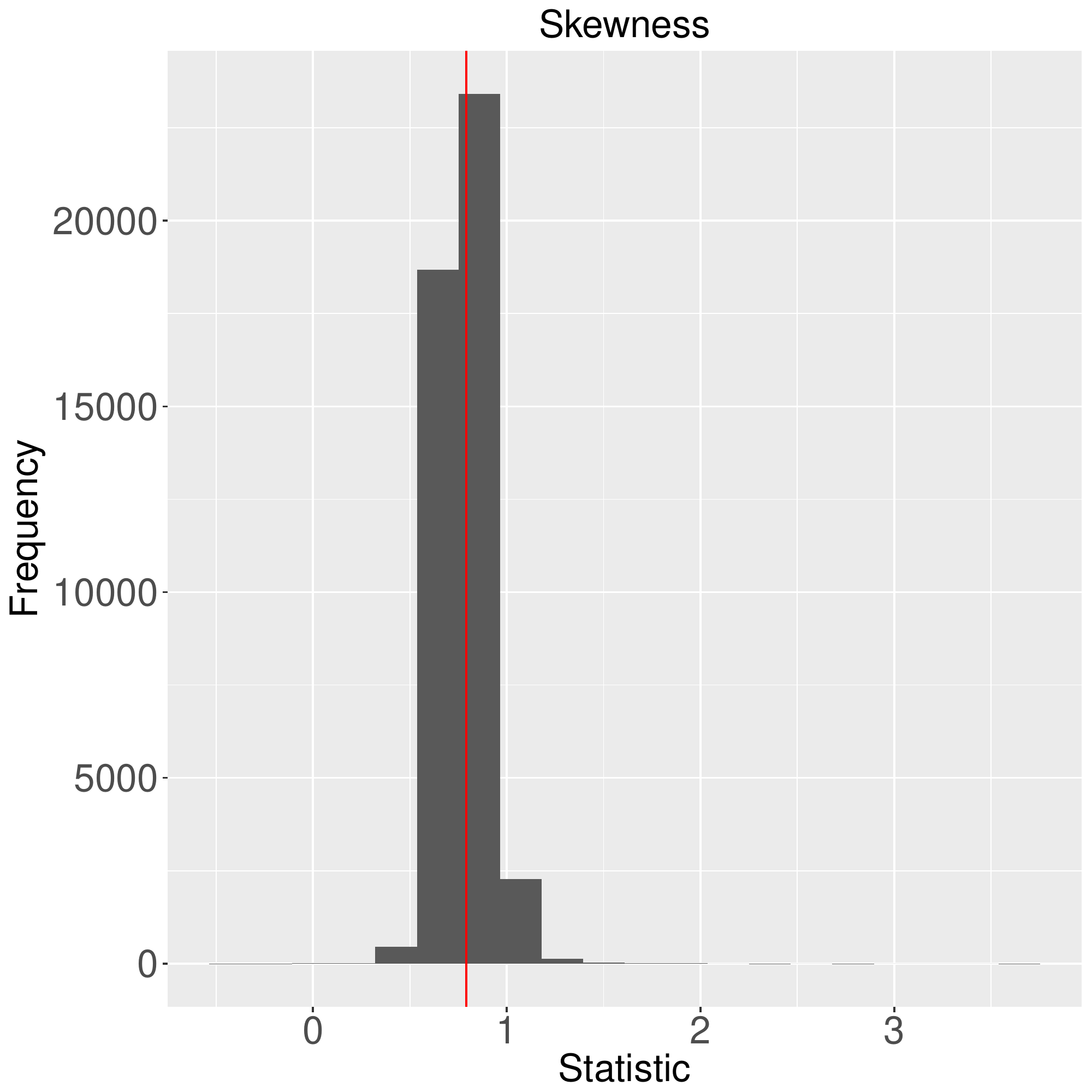}
\includegraphics[height=5cm]{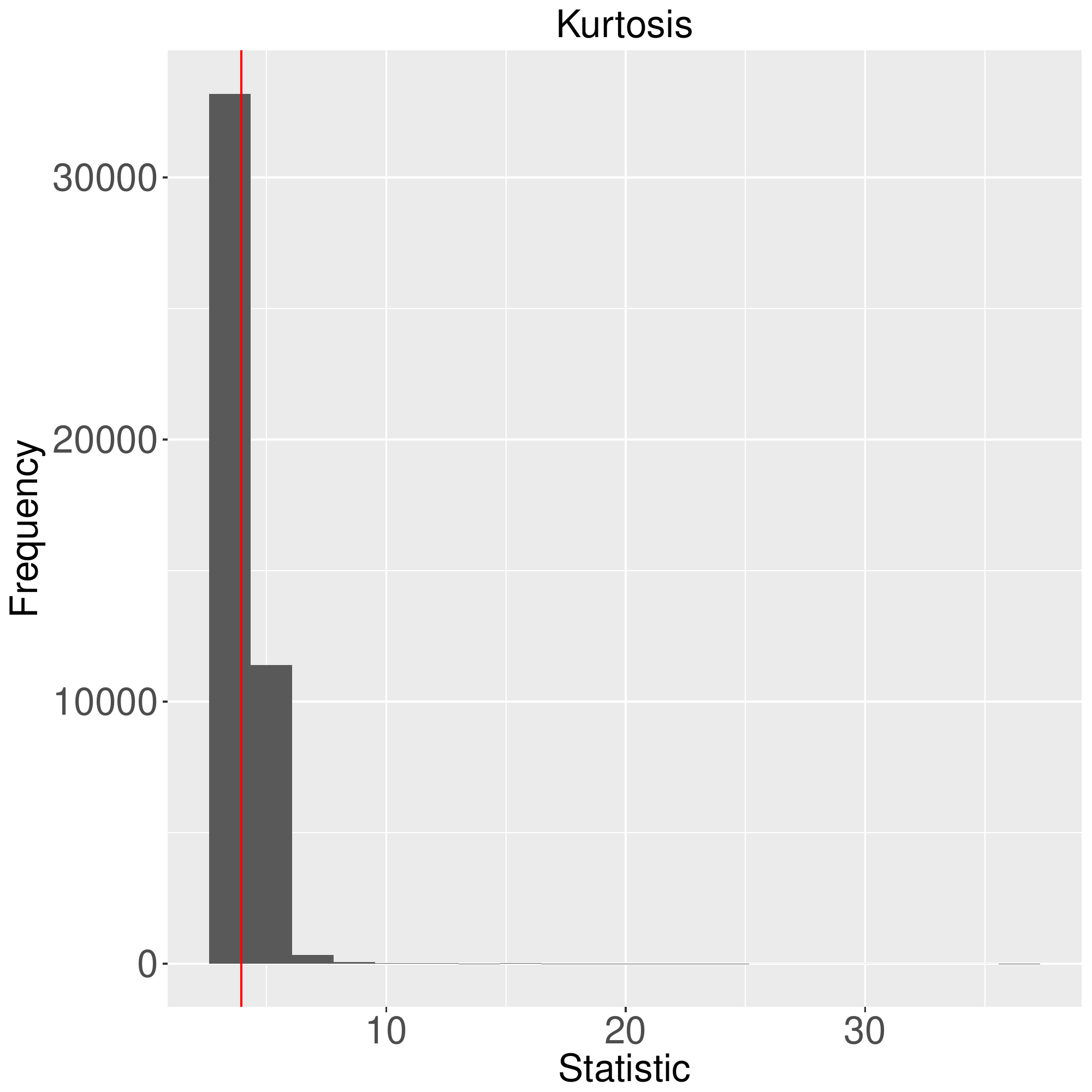}
\includegraphics[height=5cm]{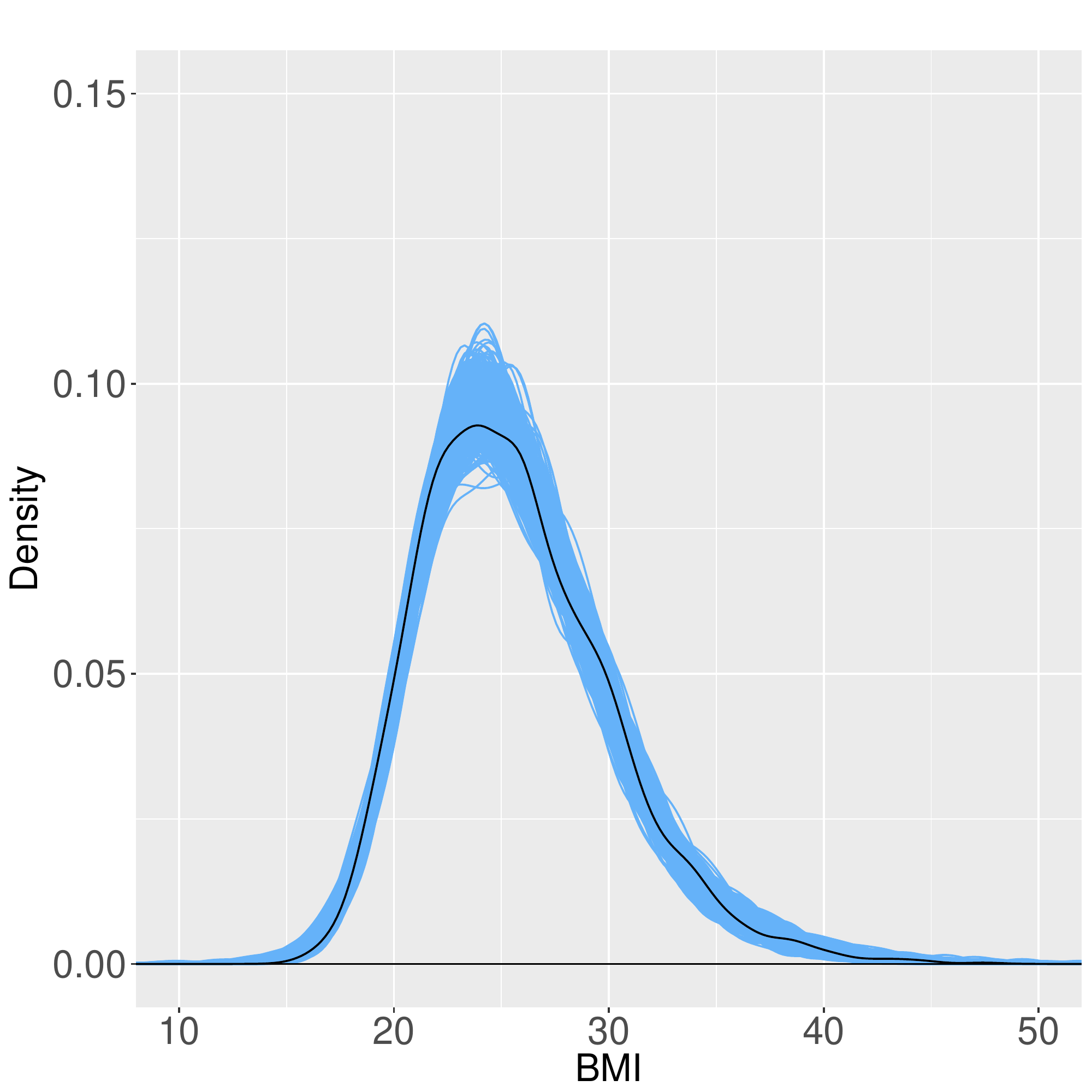}
\label{predchecks_a}}
\subfigure[Predictive checks: Bayesian normal-linear model]{
\includegraphics[width =5cm]{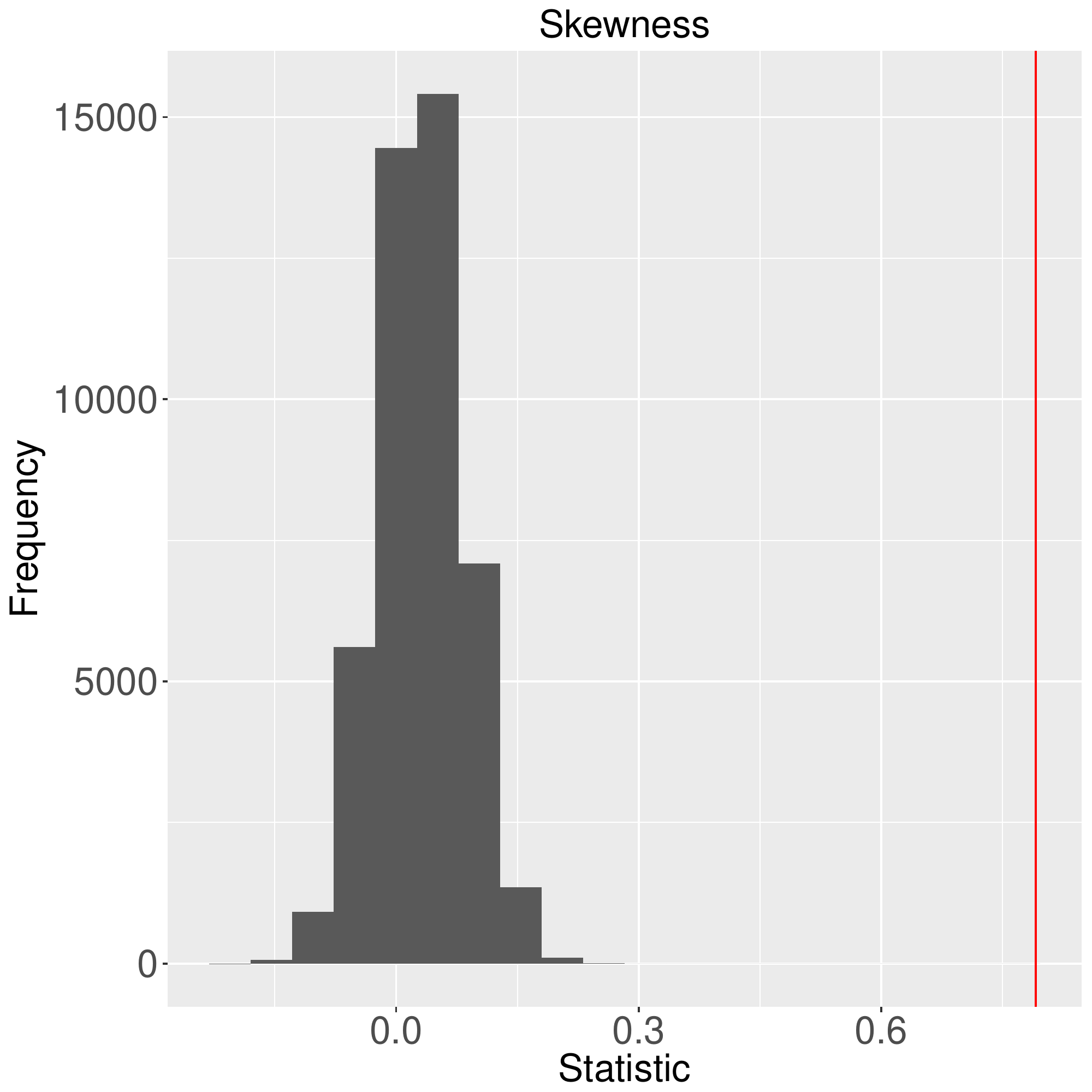}
\includegraphics[height=5cm]{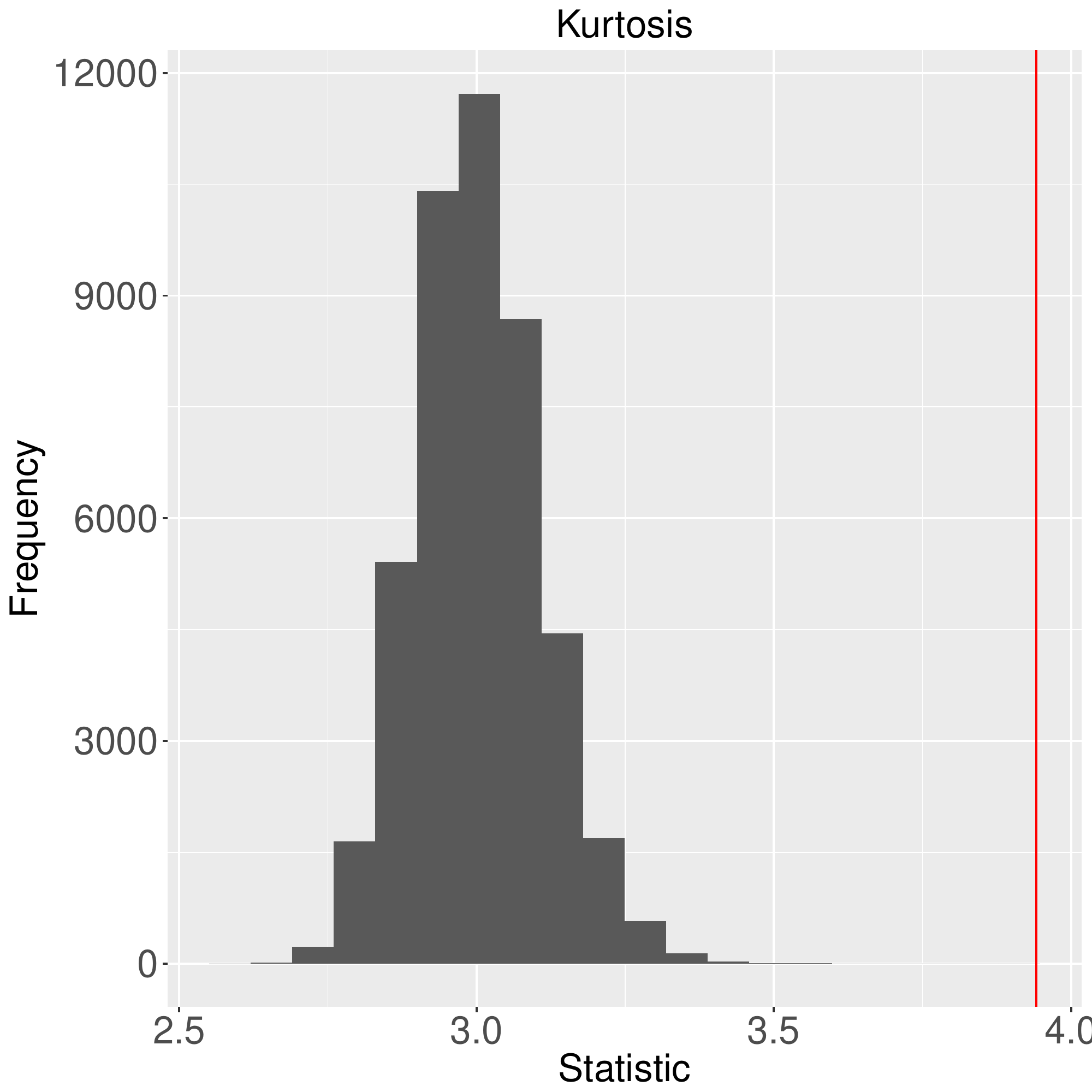}
\includegraphics[height=5cm]{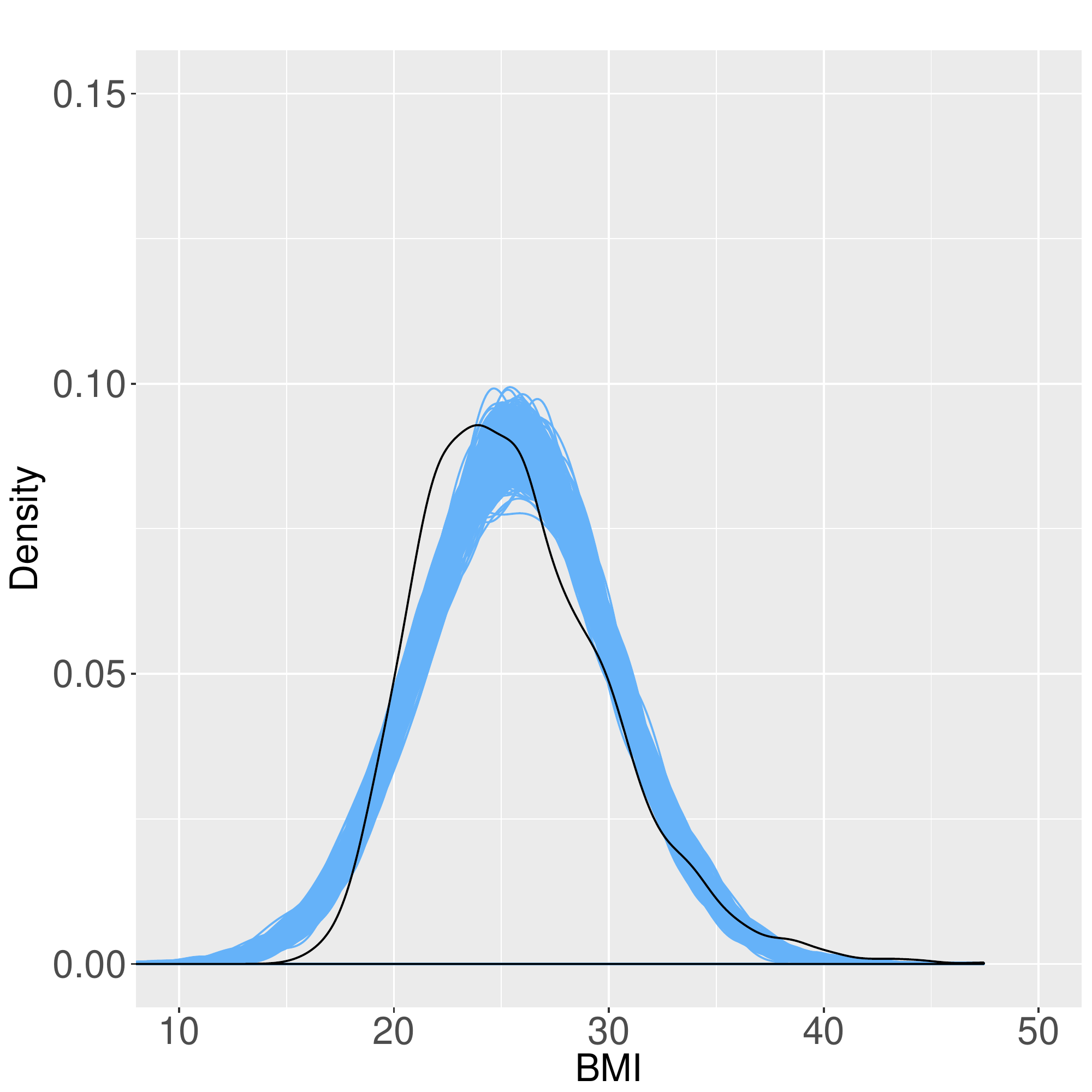}
\label{predchecks_b}}
\subfigure[Conditional predictive ordinate]{
\includegraphics[height=5cm]{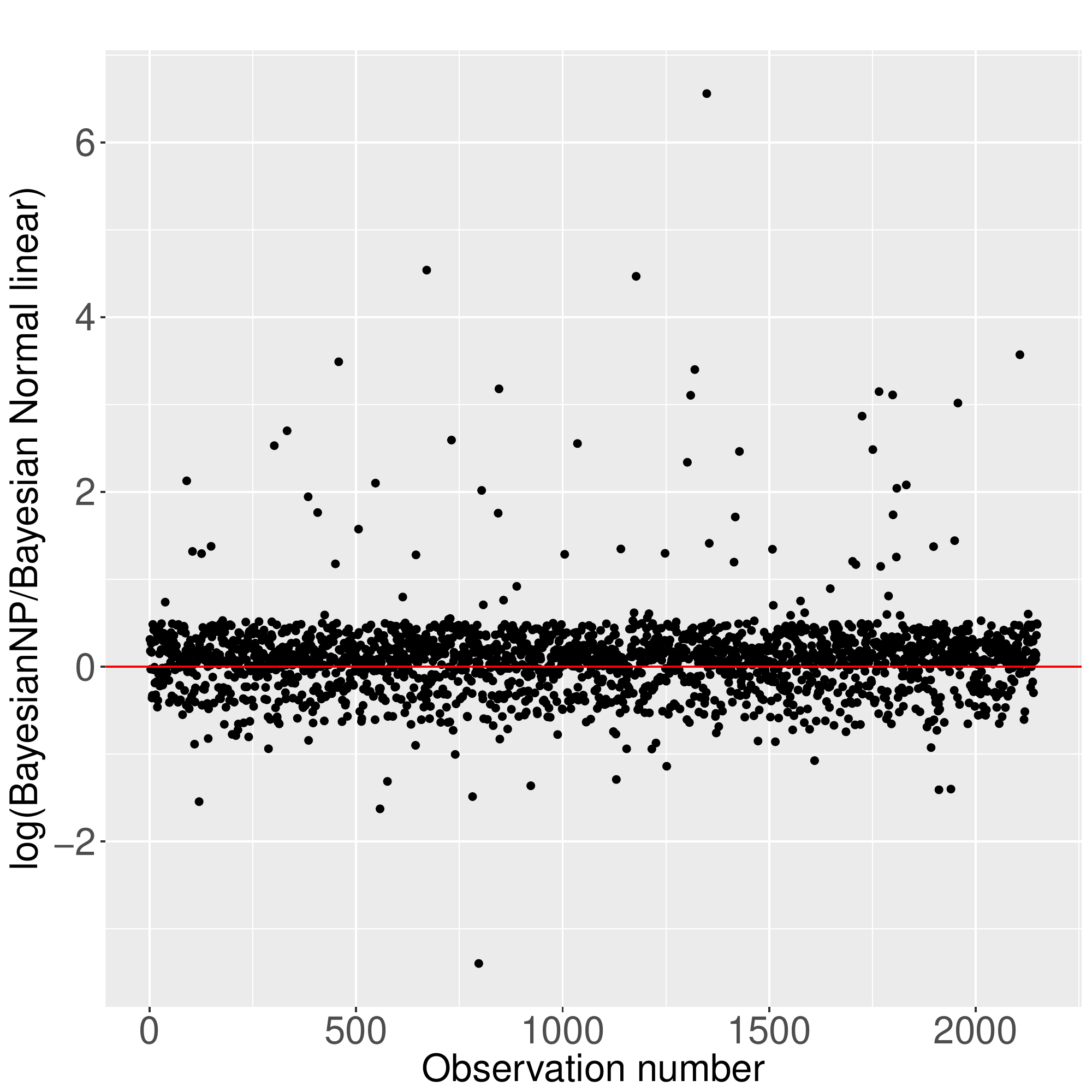}
\label{predchecks_c}}
\end{center}
\caption{\footnotesize{Posterior predictive checks. Histograms of statistics \textit{skewness} and \textit{kurtosis} computed from $45000$ draws from the posterior predictive distribution in the nondiseased group. The red line is the estimated statistic from the observed BMI (nondiseased group). The right-hand side plots show the kernel density estimate of the observed BMI (solid black line), jointly with the kernel density estimates for $500$ simulated datasets drawn from the posterior predictive distribution. Shown in (c) is the logarithm of the ratio of the CPO from the Bayesian nonparametric approach to the CPO from the Bayesian normal linear model. In all cases the results are for the analyses including the age-by-gender interaction.}}
\label{predchecks}
\end{figure}

\begin{table}[H]
\centering
\begin{tabular}{lccc}
& \textbf{Global} & \multicolumn{2}{c}{\textbf{Gender}}\\
& \textbf{sample} & Female & Male \\
\hline
\textbf{Pooled AUC} & 0.766 (0.747, 0.783) &  0.800 (0.772, 0.826) &  0.725 (0.696, 0.751)\\
\textbf{AAUC} & 0.667 (0.639, 0.695) & 0.684 (0.642, 0.724)  & 0.661 (0.624, 0.696)\\
\hline
\end{tabular}
\vspace{0.4cm}
\caption{\footnotesize{Estimated areas under the pooled (AUC) and AROC (AAUC) curves jointly with 95\% credible intervals. The pooled AUCs were estimated using the global sample and in men and women separately. The AAUCs were estimated on the basis of our Bayesian nonparametric estimator. For the global sample, the AAUC was estimated incorporating into the B-splines DDP mixture model the age-by-gender interaction. For the results in men and women, the analyses were done separately in each group with age as the single covariate. \label{AUCDescriptive}}}
\end{table}
\newpage
\begin{bibunit}[hapalike]
\makeatletter
\def\maxwidth{ %
  \ifdim\Gin@nat@width>\linewidth
    \linewidth
  \else
    \Gin@nat@width
  \fi
}
\makeatother

\definecolor{fgcolor}{rgb}{0.345, 0.345, 0.345}
\newcommand{\hlnum}[1]{\textcolor[rgb]{0.686,0.059,0.569}{#1}}%
\newcommand{\hlstr}[1]{\textcolor[rgb]{0.192,0.494,0.8}{#1}}%
\newcommand{\hlcom}[1]{\textcolor[rgb]{0.678,0.584,0.686}{\textit{#1}}}%
\newcommand{\hlopt}[1]{\textcolor[rgb]{0,0,0}{#1}}%
\newcommand{\hlstd}[1]{\textcolor[rgb]{0.345,0.345,0.345}{#1}}%
\newcommand{\hlkwa}[1]{\textcolor[rgb]{0.161,0.373,0.58}{\textbf{#1}}}%
\newcommand{\hlkwb}[1]{\textcolor[rgb]{0.69,0.353,0.396}{#1}}%
\newcommand{\hlkwc}[1]{\textcolor[rgb]{0.333,0.667,0.333}{#1}}%
\newcommand{\hlkwd}[1]{\textcolor[rgb]{0.737,0.353,0.396}{\textbf{#1}}}%
\let\hlipl\hlkwb

\makeatletter
\newenvironment{kframe}{%
 \def\at@end@of@kframe{}%
 \ifinner\ifhmode%
  \def\at@end@of@kframe{\end{minipage}}%
  \begin{minipage}{\columnwidth}%
 \fi\fi%
 \def\FrameCommand##1{\hskip\@totalleftmargin \hskip-\fboxsep
 \colorbox{shadecolor}{##1}\hskip-\fboxsep
     \hskip-\linewidth \hskip-\@totalleftmargin \hskip\columnwidth}%
 \MakeFramed {\advance\hsize-\width
   \@totalleftmargin\z@ \linewidth\hsize
   \@setminipage}}%
 {\par\unskip\endMakeFramed%
 \at@end@of@kframe}
\makeatother

\definecolor{shadecolor}{rgb}{.97, .97, .97}
\definecolor{messagecolor}{rgb}{0, 0, 0}
\definecolor{warningcolor}{rgb}{1, 0, 1}
\definecolor{errorcolor}{rgb}{1, 0, 0}
\newenvironment{knitrout}{}{} 

\setcounter{section}{0}
\setcounter{equation}{0}
\setcounter{figure}{0}
\setcounter{table}{0}
\renewcommand{\theequation}{\Alph{section}\arabic{equation}}
\addto\captionsenglish{\renewcommand{\figurename}{Web Figure}}
\addto\captionsenglish{\renewcommand{\tablename}{Web Table}}
\renewcommand\thesection{Web Appendix \Alph{section}}

\makeatletter
\renewcommand{\fnum@figure}{Web Figure \thefigure}
\renewcommand{\fnum@table}{Web Table \thetable}
\makeatother

\makeatletter
\def\verbatim@font{\linespread{1}\normalfont\ttfamily}
\makeatother

\title{Supplementary materials for ``Bayesian nonparametric inference for the covariate-adjusted ROC curve''}
\author{\textsc{Vanda~In\'acio de Carvalho} and \textsc{Mar\'ia Xos\'e Rodr\'iguez-\'Alvarez}}
\date{}
\maketitle 

This document contains supplementary materials to the paper ``Bayesian nonparametric inference for the covariate-adjusted ROC curve''. In \ref{gibbs_sampler} the complete Gibbs sampler used to fit our B-splines DDP mixture model for the conditional distribution of test outcomes in the nondiseased group is reported. \ref{kernel_est} describes the details of the kernel-based estimator for the AROC proposed by \cite{MX11a}. In \ref{threshold_est} we describe the computation of covariate-specific threshold values based on the B-splines DDP mixture model. In \ref{Rpackage} we present a brief usage description of the \texttt{AROC} package that accompanies this paper. Supporting figures and tables for the simulation study discussed in Section \ref{simulation} of the main document are provided in \ref{supp_sim}. Specifically, in \ref{sim_k_4} we show the results when the B-splines DDP mixture model was fitted considering four interior knots ($K = 4$), and in and \ref{sim_k_0} the comparisons between the results using four interior knots ($K = 4$) and no interior knots ($K = 0$). Finally, \ref{supp_app} contains extra results for the real data application described in Section \ref{application} of the main document.

\let\thefootnote\relax\footnotetext{Vanda In\'acio de Carvalho, School of Mathematics, University of Edinburgh, Scotland, UK (\textit{Vanda.Inacio@ed.ac.uk}). Mar\'ia Xos\'e Rodr\'iguez-\'Alvarez, BCAM-Basque Center for Applied Mathematics \& IKERBASQUE, Basque Foundation for Science, Bilbao, Basque Country, Spain (\textit{mxrodriguez@bcamath.org}).}

\newpage
\section{\large{\textsf{BLOCKED GIBBS SAMPLER ALGORITHM}}\label{gibbs_sampler}}
We report the complete Gibbs sampler used to fit our B-splines DDP mixture model for $F_{\bar{D}(\cdot\mid \mathbf{X}_{\bar{D}}\mid\mathbf{x})}$, which is essentially a covariate-dependent version of the Blocked Gibbs sampler algorithm in \cite{Ishwaran2001}. We use configuration variables to identify the label of the mixture component to which the $i$th nondiseased subject belongs to. Let $S_i=l$ denotes that the $i$th nondiseased subject is allocated to component $l$, $i=1,\ldots,n_{\bar{D}}$, $l=1,\ldots,L$. The Gibbs sampling procedure alternates through the following steps. 

\begin{description}[align=left]
\item [\textbf{Step 1:}] Update $S_i$, for $i=1,\ldots,n_{\bar{D}}$, by sampling from its full conditional distribution, which is a multinomial distribution with probabilities
\begin{equation*}
\Pr(S_i=l\mid\text{else})=\frac{\omega_l\phi(y_{\bar{D}i}\mid\mathbf{z}_{\bar{D}i}^{\prime}\boldsymbol{\beta}_l,\sigma_l^2)}{\sum_{k=1}^{L}\omega_k\phi(y_{\bar{D}i}\mid\mathbf{z}_{\bar{D}i}^{\prime}\boldsymbol{\beta}_k,\sigma_k^2)},\qquad \omega_l=v_l\prod_{r<l}(1-v_r),\qquad l=1,\ldots,L.
\end{equation*}
\item [\textbf{Step 2:}]  Update the stick-breaking weights from their conjugate beta posterior distribution
\begin{equation*}
v_l\mid\text{else}\sim\text{Beta}\left(n_l+1,\alpha+\sum_{r=l+1}^{L}n_r\right),\qquad l=1,\ldots, L-1,
\end{equation*}
with $n_l=\sum_{i=1}^{n_{\bar{D}}}I(S_i=l)$ being the number of observations from component $l$.
\item [\textbf{Step 3:}] Update the component specific parameters, $\boldsymbol{\beta}_l$ and $\sigma_l^2$ ($l=1,\ldots,L$), as in a standard Bayesian normal linear regression model, using observations from component $l$, i.e., 
\begin{align*}
&\boldsymbol{\beta}_l\mid\text{else}\sim\text{N}_Q\left(\mathbf{V}_l\left(\mathbf{S}^{-1}\mathbf{m}+\sigma_l^{-2}\sum_{\{i:S_i=l\}}\mathbf{z}_{\bar{D}i}y_{\bar{D}i}\right),\mathbf{V}_l\right),\quad \mathbf{V}_l=\left(\mathbf{S}^{-1}+\sigma_l^{-2}\sum_{\{ i:S_i=l\}}\mathbf{z}_{\bar{D}i}\mathbf{z}_{\bar{D}i}^{\prime}\right)^{-1},\\
&\sigma_l^{-2}\mid\text{else}\sim\Gamma\left(a+\frac{n_l}{2},b+\frac{1}{2}\sum_{\{i:S_i=l\}}\left(y_{\bar{D}i}-\mathbf{z}_{\bar{D}i}^{\prime}\boldsymbol{\beta}_l\right)^2\right).
\end{align*}
\item [\textbf{Step 4:}] Update the hyperparameters, $\mathbf{m}$ and $\mathbf{S}$, by sampling from their full conditional distributions, namely
\begin{align*}
&\mathbf{m}\mid\text{else}\sim\text{N}_Q\left(\mathbf{V}\left(\mathbf{S}_0^{-1}\mathbf{m}_0+\mathbf{S}^{-1}\sum_{l=1}^{L}\boldsymbol{\beta}_l\right),\mathbf{V}\right),\quad \mathbf{V}=(\mathbf{S}_0^{-1}+L\mathbf{S}^{-1})^{-1},\\
&\mathbf{S}^{-1}\mid\text{else}\sim\text{W}_Q\left(\nu+L,\left(\nu\Psi+\sum_{l=1}^{L}(\boldsymbol{\beta}_l-\mathbf{m})(\boldsymbol{\beta}_l-\mathbf{m})^{\prime}\right)^{-1}\right).
\end{align*}
\end{description}

\newpage
\section{\large{\textsf{DETAILS ON THE KERNEL APPROACH OF \cite{MX11a}}}\label{kernel_est}}
We start by assuming that we have random samples $\{(y_{\bar{D}i},x_{\bar{D}i})\}_{i=1}^{n_{\bar{D}}}$ and $\{(y_{Dj},x_{Dj})\}_{j=1}^{n_{D}}$ and that test outcomes in the nondiseased group are modelled through a location-scale regression model
\begin{equation*} 
Y_{\bar{D}}=\mu_{\bar{D}}(X_{\bar{D}})+\sigma_{\bar{D}}(X_{\bar{D}})\varepsilon_{\bar{D}}.
\end{equation*}
Here $\mu_{\bar{D}}$ is the regression function, $\sigma_{\bar{D}}^2$ is the variance function, and $\varepsilon_{\bar{D}}$ has zero mean, variance one, and distribution function $F_{\varepsilon_{\bar{D}}}$. Both the regression and variance functions are estimated using local estimators, particularly, using Nadaraya--Watson estimators \citep[][Chapter 2]{Fan1996}. Once we have estimates of the mean and variance functions, which we denote by $\widehat{\mu}_{\bar{D}}$ and $\widehat{\sigma}_{\bar{D}}^2$, we can compute the standardised residuals and estimate their distribution through the empirical distribution function, i.e., 
\begin{equation*} 
\widehat{\varepsilon}_{\bar{D}i}=\frac{y_{\bar{D}i}-\widehat{\mu}_{\bar{D}}(x_{\bar{D}i})}{\widehat{\sigma}_{\bar{D}}(x_{\bar{D}i})},\qquad \widehat{F}_{\varepsilon_{\bar{D}}}(y)=\frac{1}{n_{\bar{D}}}\sum_{i=1}^{n_{\bar{D}}}I\{\widehat{\varepsilon}_{\bar{D}i}\leq y\}.
\end{equation*}
Then, we can model $U_{Dj}$ as
\begin{equation*}
U_{Dj}=1-\widehat{F}_{\varepsilon_{\bar{D}}}\left(\frac{y_{Dj}-\widehat{\mu}_{\bar{D}}(x_{Dj})}{\widehat{\sigma}_{\bar{D}}(x_{Dj})}\right),\qquad j=1,\ldots,n_{D},
\end{equation*}
and therefore, the AROC curve is estimated as
\begin{equation*}
\widehat{\text{AROC}}(t)=\frac{1}{n_{D}}\sum_{j=1}^{n_{D}}I\left\{1-\widehat{F}_{\varepsilon_{\bar{D}}}\left(\frac{y_{Dj}-\widehat{\mu}_{\bar{D}}(x_{Dj})}{\widehat{\sigma}_{\bar{D}}(x_{Dj})}\right)\leq t\right\},\qquad 0\leq t \leq 1.
\end{equation*}
Uncertainty estimation is done via the bootstrap \citep{Davison1997}, and the scheme we present relies on a combination of a bootstrap of the residuals in the nondiseased group and a resampling cases approach in the diseased group. Let $B$ denotes the number of resamples; the algorithm is as follows.
\begin{description}[align=left]
\item [\textbf{Step 1:}] In the \underline{nondiseased} group sample with replacement from the standardised residuals $(\widehat{\varepsilon}_{\bar{D}1},\ldots,\widehat{\varepsilon}_{\bar{D}n_{\bar{D}}})$ to form bootstrap samples $(\widehat{\varepsilon}_{\bar{D}1}^{(b)},\ldots,\widehat{\varepsilon}_{\bar{D}n_{\bar{D}}}^{(b)})$, then obtaining
\begin{equation*}
y_{\bar{D}i}^{(b)}=\widehat{\mu}_{\bar{D}}(x_{\bar{D}i})+\widehat{\sigma}_{\bar{D}}(x_{\bar{D}i})\widehat{\varepsilon}_{\bar{D}i}^{(b)},\qquad b=1,\ldots, B,
\end{equation*}
where $\widehat{\mu}$ and $\widehat{\sigma}$ are the estimates based on the original sample. Our bootstrap samples in the nondiseased group are thus of the form $\{(y_{\bar{D}1}^{(b)},x_{\bar{D}1}),\ldots,(y_{\bar{D}n_{\bar{D}}}^{(b)},x_{\bar{D}n_{\bar{D}}})\}$. Based on these samples, estimate the mean and variance functions, respectively denoted by $\widehat{\mu}_{\bar{D}}^{(b)}$ and $\widehat{\sigma}_{\bar{D}}^{2(b)}$, compute the standardised residuals $\widehat{\varepsilon}_{\bar{D}i}^{*(b)}=(y_{\bar{D}i}-\mu_{\bar{D}}^{(b)}(x_{\bar{D}i}))/\sigma_{\bar{D}}^{(b)}(x_{\bar{D}i})$, and denote the empirical distribution function of these by $\widehat{F}_{\varepsilon_{\bar{D}}}^{(b)}$. The use of this bootstrap is due to the well-known result of \citet[][p.~781]{Hardle1991} that for nonparametric regression, bootstrapping the original data is not a valid approach and an approach based on bootstrapping the residuals should be preferred.
\item [\textbf{Step 2:}] In the \underline{diseased} group, bootstrap samples $\{(y_{D1}^{(b)},x_{D1}^{(b)}),\ldots,(y_{Dn_{D}}^{(b)},x_{Dn_{D}}^{(b)})\}$ are obtained by resampling, with replacement, from the original diseased sample--resampling cases bootstrap. Note that Steps 1 and 2 are interchangeable. 
\item [\textbf{Step 3:}] Compute 
\begin{equation*}
U_{Dj}^{(b)}=1-\widehat{F}_{\varepsilon_{\bar{D}}}^{(b)}\left(\frac{y_{Dj}^{(b)}-\widehat{\mu}_{\bar{D}}^{(b)}(x_{Dj}^{(b)})}{\widehat{\sigma}_{\bar{D}}^{(b)}(x_{Dj}^{(b)})}\right),
\end{equation*}
and then compute the corresponding AROC curve
\begin{equation*}
\widehat{\text{AROC}}^{(b)}(t)=\frac{1}{n_{D}}\sum_{j=1}^{n_{D}}I\left\{U_{Dj}^{(b)}\leq t\right\},\qquad 0\leq t \leq 1.
\end{equation*}
Once this process has been completed, and according to the percentile method, a bootstrap confidence interval for $\text{AROC}(t)$ of confidence level $1-\alpha$ is given by the quantiles $\alpha/2$ and $1-\alpha/2$ of the ensemble $\left(\widehat{\text{AROC}}^{(1)}(t),\ldots,\widehat{\text{AROC}}^{(B)}(t)\right)$.
\end{description}
\newpage
\section{\large{\textsf{COMPUTATION OF THE COVARIATE-SPECIFIC THRESHOLD VALUES}}\label{threshold_est}}
Let $\mathbf{x}=(x_1,\ldots,x_p)$ be the vector of covariates for which we are interested in computing the threshold value, $c_{\mathbf{x}}$, that gives rise to a FPF of $t$. From step 1 of our modelling procedure described in Section 3.1 of the main paper, we can easily compute
\begin{equation}\label{optthres}
c_{\mathbf{x}}^{(s)}=F_{\bar{D}}^{-1{(s)}}(1-t\mid \mathbf{X}_{\bar{D}}=\mathbf{x}),\qquad 0\leq t \leq 1,
\end{equation}
where
\begin{equation*}
F_{\bar{D}}^{(s)}(1-t\mid \mathbf{X}_{\bar{D}}=\mathbf{x})=\sum_{l=1}^{L}\omega_{l}^{(s)}\Phi\left(1-t\mid \mathbf{z}^{\prime}\boldsymbol{\beta}^{(s)},\left(\sigma_l^{(s)}\right)^{2}\right),\qquad s=1,\ldots,S,
\end{equation*}
with $\mathbf{z}^{\prime}=\left(1,\mathbf{B}^{\prime}_{\mathbf{\xi}_1}(x_{1}), \ldots, \mathbf{B}^{\prime}_{\mathbf{\xi}_p}(x_{p})\right)$ and $S$ denotes the number of posterior samples after burn-in. Note that the inversion in \eqref{optthres} is done numerically. The $S$ posterior samples originate an ensemble of optimal thresholds $\left(c_{\mathbf{x}}^{(1)},\ldots, c_{\mathbf{x}}^{(S)}\right)$ from which the posterior mean can be computed
\begin{equation*}
\hat{c}_{\mathbf{x}}=\frac{1}{S}\sum_{s=1}^{S}c_{\mathbf{x}}^{(s)},
\end{equation*}
with credible intervals derived from the percentiles of the ensemble.
\newpage
\section{\large{\textsf{\texttt{AROC} PACKAGE}}}\label{Rpackage}
This section contains a brief description of the \texttt{R}-package we developed to accompany this paper. The package can be freely downloaded from \texttt{https://CRAN.R-project.org/package=AROC}, where a more detailed explanation of its use can be found. The package implements four different methods for the estimation of the covariate-adjusted ROC (AROC) curve, and two methods for the estimation of the pooled ROC curve. Specifically, the implemented methods  and associated \texttt{R}-functions are:
\begin{itemize}
\item \texttt{AROC.bnp}: Estimates the AROC curve (and AAUC/pAAUC) using our proposed Bayesian nonparametric approach.
\item \texttt{AROC.bsp}: Estimates the AROC curve (and AAUC/pAAUC) using the Bayesian semiparametric approach discussed in Section \ref{simulation} of the main document.
\item \texttt{AROC.kernel}: Estimates the AROC curve (and AAUC) using the nonparametric kernel-based method proposed by \cite{MX11a}. The method, as it stands now, can only deal with one continuous covariate. 
\item \texttt{AROC.sp}: Estimates the AROC curve (and AAUC) using the semiparametric approach proposed by \cite{Janes09a}.
\item \texttt{pooledROC.BB}: Estimates the pooled ROC curve (and AUC) using the Bayesian bootstrap estimator proposed by \cite{Gu2008}.
\item \texttt{pooledROC.emp}: Estimates the pooled ROC curve (and AUC) using the empirical estimator proposed by \cite{Hsieh1996}.
\end{itemize}
In all cases, numerical and graphical summaries can be obtained by calling the functions \texttt{print}, \texttt{summary} and \texttt{plot}. In addition, and for the four methods that estimate the AROC, the package provides functions for estimating the associated AROC-based threshold values (functions \texttt{compute.threshold.AROC.bnp}, \texttt{compute.threshold.AROC.bsp}, \texttt{compute.threshold.AROC.kernel} and \texttt{compute.threshold.AROC.sp}). Finally, for the Bayesian approaches, it is also possible to conduct posterior predictive checks by means of the functions \texttt{predictive.checks.AROC.bnp} and \texttt{predictive.checks.AROC.bsp}. A detailed description of each function and the input arguments they require can be found in the documentation files of the package, where usage examples are also provided.  

By way of example, and since we are not allowed to make the endocrine dataset publicly available, we present here the syntax for the CARET PSA data discussed in \cite{Etzioni99}. The dataset contains 71 cases of prostate cancer and 71 age- and serum samples- matched controls (nondiseased individuals) who participated in a lung cancer prevention trial (CARET, Beta-carotene and retinol trial). For each subject on the study, serum samples were drawn at baseline and at two-year intervals after that. All the cases (and thus the matched controls) have, at least, three and up to eight serum samples. The aim here is to determine if the free prostate specific antigen (PSA) is able to discriminate cases from controls, adjusting for age. It has been reported \citep[see][pp 163]{Pepe03} that age does not modify (have an effect on) the accuracy of the free PSA when discriminating between cases and controls. Age is, however, associated with PSA values. The older the individual, the larger the free PSA values. As extensively discussed in Section 2 of the main paper, even in these cases the age-specific ROC curve (which is the same for all ages) may not coincide with the pooled ROC curve. It does coincide, however, with the age-adjusted ROC curve. Also, in a matched study (as this one), the pooled (unadjusted) ROC curve also has a difficult interpretation: since the sample of control individuals is artificially constructed, the FPF axis of the pooled ROC curve does not represent the FPF of the diagnostic test in the real population. In contrast, the FPF axis of the AROC describes the real-world performance of the diagnostic test in a population with fixed covariate value \citep[see][for a detailed discussion]{Janes2008b}. For a simple illustration of the usage of the package, in the subsequent analyses we only consider the last PSA measurement of each subject. The PSA dataset can be found in \texttt{https://research.fhcrc.org/diagnostic-biomarkers-center/en/datasets.html}, and it has been included in the \texttt{R}-package \texttt{AROC}. Here there is a brief summary of the data 

\begin{knitrout}\small
\definecolor{shadecolor}{rgb}{0.969, 0.969, 0.969}\color{fgcolor}\begin{kframe}
\begin{alltt}
\hlkwd{sessionInfo}\hlstd{()}
\end{alltt}
\begin{verbatim}
## R version 3.4.4 (2018-03-15)
## Platform: x86_64-pc-linux-gnu (64-bit)
## Running under: Ubuntu 16.04.4 LTS
## 
## Matrix products: default
## BLAS: /usr/lib/libblas/libblas.so.3.6.0
## LAPACK: /usr/lib/lapack/liblapack.so.3.6.0
## 
## locale:
##  [1] LC_CTYPE=en_US.UTF-8       LC_NUMERIC=C              
##  [3] LC_TIME=es_ES.UTF-8        LC_COLLATE=en_US.UTF-8    
##  [5] LC_MONETARY=es_ES.UTF-8    LC_MESSAGES=en_US.UTF-8   
##  [7] LC_PAPER=es_ES.UTF-8       LC_NAME=C                 
##  [9] LC_ADDRESS=C               LC_TELEPHONE=C            
## [11] LC_MEASUREMENT=es_ES.UTF-8 LC_IDENTIFICATION=C       
## 
## attached base packages:
## [1] stats     graphics  grDevices utils     datasets  methods   base     
## 
## other attached packages:
## [1] knitr_1.20
## 
## loaded via a namespace (and not attached):
## [1] compiler_3.4.4  magrittr_1.5    tools_3.4.4     stringi_1.2.2  
## [5] stringr_1.3.0   evaluate_0.10.1 tcltk_3.4.4
\end{verbatim}
\begin{alltt}
\hlkwd{library}\hlstd{(AROC)}
\hlkwd{data}\hlstd{(psa)}
\hlkwd{summary}\hlstd{(psa)}
\end{alltt}
\begin{verbatim}
##        id            marker1          marker2           status      
##  Min.   :  1.00   Min.   : 0.030   Min.   :0.0000   Min.   :0.0000  
##  1st Qu.: 39.00   1st Qu.: 1.050   1st Qu.:0.1588   1st Qu.:0.0000  
##  Median : 73.00   Median : 1.800   Median :0.2124   Median :0.0000  
##  Mean   : 72.69   Mean   : 4.803   Mean   :0.2200   Mean   :0.3353  
##  3rd Qu.:107.00   3rd Qu.: 3.960   3rd Qu.:0.2732   3rd Qu.:1.0000  
##  Max.   :141.00   Max.   :99.980   Max.   :1.0003   Max.   :1.0000  
##       age              t          
##  Min.   :46.75   Min.   :-9.0080  
##  1st Qu.:61.07   1st Qu.:-3.7290  
##  Median :65.16   Median :-1.4210  
##  Mean   :64.86   Mean   :-1.3714  
##  3rd Qu.:69.00   3rd Qu.: 0.8255  
##  Max.   :80.83   Max.   : 7.4200
\end{verbatim}
\end{kframe}
\end{knitrout}
\noindent The dataset contains the patient identifier (\texttt{id}), the free PSA (variable \texttt{marker1}) and total PSA (variable \texttt{marlker2}), the variable that indicates the presence/absence of prostate cancer (\texttt{status}), the age at which the serum samples were drawn (variable \texttt{age}), and the time (in years) relative to prostate cancer diagnosis (\texttt{t}). For our analyses, we focus on the last measurement, and the free PSA is log-transformed

\begin{knitrout}\small
\definecolor{shadecolor}{rgb}{0.969, 0.969, 0.969}\color{fgcolor}\begin{kframe}
\begin{alltt}
\hlcom{# Select the last measurement}
\hlstd{newpsa} \hlkwb{<-} \hlstd{psa[}\hlopt{!}\hlkwd{duplicated}\hlstd{(psa}\hlopt{$}\hlstd{id,} \hlkwc{fromLast} \hlstd{=} \hlnum{TRUE}\hlstd{),]}
\hlcom{# Log-transform the biomarker}
\hlstd{newpsa}\hlopt{$}\hlstd{l_marker1} \hlkwb{<-} \hlkwd{log}\hlstd{(newpsa}\hlopt{$}\hlstd{marker1)}
\end{alltt}
\end{kframe}
\end{knitrout}

We start our analyses by estimating the pooled ROC curve using the Bayesian bootstrap estimator proposed \cite{Gu2008}

\begin{knitrout}\small
\definecolor{shadecolor}{rgb}{0.969, 0.969, 0.969}\color{fgcolor}\begin{kframe}
\begin{alltt}
\hlkwd{set.seed}\hlstd{(}\hlnum{123}\hlstd{)} \hlcom{# for reproducibility}
\hlstd{psa_pooled} \hlkwb{<-} \hlkwd{pooledROC.BB}\hlstd{(}\hlkwc{y0} \hlstd{= newpsa}\hlopt{$}\hlstd{l_marker1[newpsa}\hlopt{$}\hlstd{status} \hlopt{==} \hlnum{0}\hlstd{],}
                            \hlkwc{y1} \hlstd{= newpsa}\hlopt{$}\hlstd{l_marker1[newpsa}\hlopt{$}\hlstd{status} \hlopt{==} \hlnum{1}\hlstd{],}
                             \hlkwc{p} \hlstd{=} \hlkwd{seq}\hlstd{(}\hlnum{0}\hlstd{,}\hlnum{1}\hlstd{,}\hlkwc{l}\hlstd{=}\hlnum{101}\hlstd{),}  \hlkwc{B} \hlstd{=} \hlnum{5000}\hlstd{)}
\end{alltt}
\end{kframe}
\end{knitrout}

\noindent Through \texttt{y0} and \texttt{y1} arguments, the user specifies the vector of diagnostic test outcomes in, respectively, the nondiseased and diseased groups. The set of FPFs at which to estimate the pooled ROC curve is specified in \texttt{p}, and argument \texttt{B} allows to indicate the the number of Bayesian bootstrap resamples. A numerical summary of the fitted model can be obtained by calling the function \texttt{summary}, that provides, among other things, the estimated area under pooled ROC curve (AUC) and 95\% credible intervals 

\begin{knitrout}\small
\definecolor{shadecolor}{rgb}{0.969, 0.969, 0.969}\color{fgcolor}\begin{kframe}
\begin{alltt}
\hlkwd{summary}\hlstd{(psa_pooled)}
\end{alltt}
\begin{verbatim}
## 
## Call:
## pooledROC.BB(y0 = newpsa$l_marker1[newpsa$status == 0], y1 = newpsa$l_marker1[newpsa$status == 
##     1], p = seq(0, 1, l = 101), B = 5000)
## 
## Approach: Pooled ROC curve - Bayesian bootstrap
## ----------------------------------------------
## Area under the pooled ROC curve: 0.878 (0.815, 0.928)
\end{verbatim}
\end{kframe}
\end{knitrout}

\noindent To complement those numerical results, the \texttt{AROC} package furnishes graphical results that can be used to further explore the fitted model. Specifically, the function \texttt{plot} depicts the estimated pooled ROC curve and AUC, jointly with 95\% credible intervals

\begin{knitrout}\small
\definecolor{shadecolor}{rgb}{0.969, 0.969, 0.969}\color{fgcolor}\begin{kframe}
\begin{alltt}
\hlkwd{plot}\hlstd{(psa_pooled)}
\end{alltt}
\end{kframe}
\end{knitrout}

\noindent The result of the above code is shown in Web Figures \ref{psa_pooled}. We now turn our attention to the estimation of the age-ajusted ROC curve (AROC) using the Bayesian nonparametric approach proposed in this paper. The code for the PSA data is as follows

\begin{knitrout}\small
\definecolor{shadecolor}{rgb}{0.969, 0.969, 0.969}\color{fgcolor}\begin{kframe}
\begin{alltt}
\hlkwd{set.seed}\hlstd{(}\hlnum{123}\hlstd{)} \hlcom{# for reproducibility}
\hlstd{psa_aroc_bnp} \hlkwb{<-} \hlkwd{AROC.bnp}\hlstd{(}\hlkwc{formula.healthy} \hlstd{= l_marker1} \hlopt{~} \hlkwd{f}\hlstd{(age,} \hlkwc{K} \hlstd{=} \hlnum{0}\hlstd{),}
                         \hlkwc{group} \hlstd{=} \hlstr{"status"}\hlstd{,} \hlkwc{tag.healthy} \hlstd{=} \hlnum{0}\hlstd{,} \hlkwc{data} \hlstd{= newpsa,} \hlkwc{scale} \hlstd{=} \hlnum{TRUE}\hlstd{,}
                         \hlkwc{p} \hlstd{=} \hlkwd{seq}\hlstd{(}\hlnum{0}\hlstd{,}\hlnum{1}\hlstd{,}\hlkwc{l}\hlstd{=}\hlnum{101}\hlstd{),} \hlkwc{compute.lpml} \hlstd{=} \hlnum{TRUE}\hlstd{,} \hlkwc{compute.WAIC} \hlstd{=} \hlnum{TRUE}\hlstd{,}
                         \hlkwc{m0} \hlstd{=} \hlkwd{rep}\hlstd{(}\hlnum{0}\hlstd{,}\hlnum{4}\hlstd{),} \hlkwc{S0} \hlstd{=} \hlnum{100}\hlopt{*}\hlkwd{diag}\hlstd{(}\hlnum{4}\hlstd{),} \hlkwc{nu} \hlstd{=} \hlnum{6}\hlstd{,} \hlkwc{Psi} \hlstd{=} \hlkwd{diag}\hlstd{(}\hlnum{4}\hlstd{),}
                         \hlkwc{a} \hlstd{=} \hlnum{2}\hlstd{,} \hlkwc{b} \hlstd{=} \hlnum{0.5}\hlstd{,} \hlkwc{L} \hlstd{=} \hlnum{10}\hlstd{,} \hlkwc{nsim} \hlstd{=} \hlnum{50000}\hlstd{,}\hlkwc{nburn} \hlstd{=} \hlnum{5000}\hlstd{)}
\end{alltt}
\end{kframe}
\end{knitrout}

\noindent The argument \verb|formula.healthy| is a \verb|formula| object specifying the B-splines DDP mixture of normals model for estimation of the conditional distribution function of the diagnostic test outcomes in the nondiseased/healthy population. This formula is similar to that used for the \verb|glm()| function, except that nonparametric functions (modelled by means of B-spline basis functions) can be added by means of function \verb|f()|. For instance, specification \texttt{y $\sim$ x1 + f(x2, K = 0)} assumes a linear effect of \verb|x1| and a nonparametric effect of \verb|x2|. The argument \verb|K = 0| indicates that \verb|0| internal knots will be used. Categorical variables (factors) can be also incorporated, as well as factor-by-curve interaction terms. For example, to include the interaction between \verb|age| and \verb|gender| we need to specify \texttt{y $\sim$ gender + s(age, by = gender, K = 0)}. The name of the variable that distinguishes nondiseased from diseased individuals is represented by argument \verb|group|, and the value codifying the nondiseased individuals in \verb|group| is specified by \verb|tag.healthy|. The \verb|data| argument is a data frame containing the data and all needed variables. The argument \verb|scale = TRUE| is used to indicate that test outcomes will be scaled, i.e., divided by the standard deviation (and then transformed to the original scale when presenting the results). The set of FPFs at which to estimate the pooled ROC curve is indicated in \verb|p|.  The log pseudo marginal likelihood (LPML) and the widely applicable information criterion (WAIC) can be computed by setting the arguments \verb|compute.lpml| and \verb|compute.WAIC| to \verb|TRUE|. The arguments \verb|m0, S0, nu, Psi, a, b| are the hyperparameters associated with the centering distribution. Note that in this case the hyperparameter specification is made on the scaled data (\verb|scale = TRUE|). Finally, in argument \verb|L| the user specifies the maximum number of mixture components for the B-splines DDP mixture model, and \verb|nsim| and \verb|nburn| are, respectively, the total number of Gibbs sampler iterates (including the burn-in period) and the number of burn-in iterations. As before, a numerical summary can be obtained by calling the function \verb|summary|

\begin{knitrout}\small
\definecolor{shadecolor}{rgb}{0.969, 0.969, 0.969}\color{fgcolor}\begin{kframe}
\begin{alltt}
\hlkwd{summary}\hlstd{(psa_aroc_bnp)}
\end{alltt}
\begin{verbatim}
## 
## Call:
## AROC.bnp(formula.healthy = l_marker1 ~ f(age, K = 0), group = "status", 
##     tag.healthy = 0, data = newpsa, scale = TRUE, p = seq(0, 
##         1, l = 101), compute.lpml = TRUE, compute.WAIC = TRUE, 
##     m0 = rep(0, 4), S0 = 100 * diag(4), nu = 6, Psi = diag(4), 
##     a = 2, b = 0.5, L = 10, nsim = 50000, nburn = 5000)
## 
## Approach: AROC Bayesian nonparametric
## ----------------------------------------------
## Area under the covariate-adjusted ROC curve: 0.88 (0.812, 0.933)
## 
## 
## Model selection criteria - Healthy population
## ----------------------------------------------
## Widely applicable information criterion (WAIC):  193.165 
## Pseudo marginal likelihood (LPML):  -97
\end{verbatim}
\end{kframe}
\end{knitrout}

\noindent Note that the \verb|summary| function provides the WAIC and LPML. Graphical results in this case can be obtained using functions \verb|plot| and \verb|predictive.checks.AROC.bnp|. Specifically, the estimated AROC curve and AAUC, jointly with 95\% credible intervals, is obtained using the function \verb|plot|. The function \linebreak \verb|predictive.checks.AROC.bnp| depicts the histograms of several test statistics computed from datasets drawn from the posterior predictive distribution in the nondiseased group, jointly with the estimated test statistics from the observed diagnostic test outcomes in the nondiseased group. The function also provides the kernel density estimate of the observed diagnostic test outcomes in the nondiseased group and the kernel density estimates for $500$ simulated datasets drawn from the posterior predictive distribution in the nondiseased group.

\begin{knitrout}\small
\definecolor{shadecolor}{rgb}{0.969, 0.969, 0.969}\color{fgcolor}\begin{kframe}
\begin{alltt}
\hlkwd{plot}\hlstd{(psa_aroc_bnp)}

\hlkwd{predictive.checks.AROC.bnp}\hlstd{(psa_aroc_bnp,} \hlkwc{statistics} \hlstd{=} \hlkwd{c}\hlstd{(}\hlstr{"skewness"}\hlstd{,}\hlstr{"kurtosis"}\hlstd{))}
\end{alltt}
\end{kframe}
\end{knitrout}

The results are shown in Web Figures \ref{psa_AROC_bnp} and \ref{psa_bnp_predictive_checks}. We finish this Web Appendix by showing the code needed for estimating the age-adjusted ROC (AROC) curve using the Bayesian semiparametric model discussed in Section \ref{simulation} of the main document (function \verb|AROC.bsp|), and the kernel-based method proposed by \cite{MX11a} (function \verb|AROC.kernel|). For the sake of simplicity, we do not describe here either the usage of the functions or the input arguments they require. This information can be found in the documentation files of the package.

\begin{knitrout}\small
\definecolor{shadecolor}{rgb}{0.969, 0.969, 0.969}\color{fgcolor}\begin{kframe}
\begin{alltt}
\hlcom{# Bayesian semiparametric approach}
\hlkwd{set.seed}\hlstd{(}\hlnum{123}\hlstd{)} \hlcom{# for reproducibility}
\hlstd{psa_aroc_bsp} \hlkwb{<-} \hlkwd{AROC.bsp}\hlstd{(}\hlkwc{formula.healthy} \hlstd{= l_marker1} \hlopt{~} \hlstd{age,}
                         \hlkwc{group} \hlstd{=} \hlstr{"status"}\hlstd{,} \hlkwc{tag.healthy} \hlstd{=} \hlnum{0}\hlstd{,} \hlkwc{data} \hlstd{= newpsa,} \hlkwc{scale} \hlstd{=} \hlnum{TRUE}\hlstd{,}
                         \hlkwc{p} \hlstd{=} \hlkwd{seq}\hlstd{(}\hlnum{0}\hlstd{,}\hlnum{1}\hlstd{,}\hlkwc{l}\hlstd{=}\hlnum{101}\hlstd{),} \hlkwc{compute.lpml} \hlstd{=} \hlnum{TRUE}\hlstd{,} \hlkwc{compute.WAIC} \hlstd{=} \hlnum{TRUE}\hlstd{,}
                         \hlkwc{m0} \hlstd{=} \hlkwd{rep}\hlstd{(}\hlnum{0}\hlstd{,}\hlnum{2}\hlstd{),} \hlkwc{S0} \hlstd{=} \hlnum{100}\hlopt{*}\hlkwd{diag}\hlstd{(}\hlnum{2}\hlstd{),} \hlkwc{nu}\hlstd{=}\hlnum{4}\hlstd{,}
                         \hlkwc{Psi}\hlstd{=}\hlkwd{diag}\hlstd{(}\hlnum{2}\hlstd{),} \hlkwc{a} \hlstd{=} \hlnum{2}\hlstd{,} \hlkwc{b} \hlstd{=} \hlnum{0.5}\hlstd{,} \hlkwc{nsim} \hlstd{=} \hlnum{50000}\hlstd{,} \hlkwc{nburn} \hlstd{=} \hlnum{5000}\hlstd{)}

\hlcom{# Kernel based approach}
\hlkwd{set.seed}\hlstd{(}\hlnum{123}\hlstd{)} \hlcom{# for reproducibility}
\hlstd{psa_aroc_kernel} \hlkwb{<-} \hlkwd{AROC.kernel}\hlstd{(}\hlkwc{marker} \hlstd{=} \hlstr{"l_marker1"}\hlstd{,} \hlkwc{covariate} \hlstd{=} \hlstr{"age"}\hlstd{,}
                               \hlkwc{group} \hlstd{=} \hlstr{"status"}\hlstd{,} \hlkwc{tag.healthy} \hlstd{=} \hlnum{0}\hlstd{,} \hlkwc{data} \hlstd{= newpsa,}
                               \hlkwc{p} \hlstd{=} \hlkwd{seq}\hlstd{(}\hlnum{0}\hlstd{,}\hlnum{1}\hlstd{,}\hlkwc{l}\hlstd{=}\hlnum{101}\hlstd{),} \hlkwc{B} \hlstd{=} \hlnum{500}\hlstd{)}
\end{alltt}
\end{kframe}
\end{knitrout}

\noindent Numerical and graphical summaries are obtained as follows
\begin{knitrout}\small
\definecolor{shadecolor}{rgb}{0.969, 0.969, 0.969}\color{fgcolor}\begin{kframe}
\begin{alltt}
\hlkwd{summary}\hlstd{(psa_aroc_bsp)}
\end{alltt}
\begin{verbatim}
## 
## Call:
## AROC.bsp(formula.healthy = l_marker1 ~ age, group = "status", 
##     tag.healthy = 0, data = newpsa, scale = TRUE, p = seq(0, 
##         1, l = 101), compute.lpml = TRUE, compute.WAIC = TRUE, 
##     m0 = rep(0, 2), S0 = 100 * diag(2), nu = 4, Psi = diag(2), 
##     a = 2, b = 0.5, nsim = 50000, nburn = 5000)
## 
## Approach: AROC Bayesian semiparametric
## ----------------------------------------------
## Area under the covariate-adjusted ROC curve: 0.887 (0.826, 0.934)
## 
## 
## Model selection criteria - Healthy population
## ----------------------------------------------
## Widely applicable information criterion (WAIC):  198.447 
## Pseudo marginal likelihood (LPML):  -99
\end{verbatim}
\begin{alltt}
\hlkwd{summary}\hlstd{(psa_aroc_kernel)}
\end{alltt}
\begin{verbatim}
## 
## Call:
## AROC.kernel(marker = "l_marker1", covariate = "age", group = "status", 
##     tag.healthy = 0, data = newpsa, p = seq(0, 1, l = 101), B = 500)
## 
## Approach: AROC Kernel-based
## ----------------------------------------------
## Area under the covariate-adjusted ROC curve: 0.891 (0.835, 0.942)
\end{verbatim}
\end{kframe}
\end{knitrout}

\begin{knitrout}\small
\definecolor{shadecolor}{rgb}{0.969, 0.969, 0.969}\color{fgcolor}\begin{kframe}
\begin{alltt}
\hlkwd{plot}\hlstd{(psa_aroc_bsp)}

\hlkwd{predictive.checks.AROC.bsp}\hlstd{(psa_aroc_bsp,} \hlkwc{statistics} \hlstd{=} \hlkwd{c}\hlstd{(}\hlstr{"skewness"}\hlstd{,}\hlstr{"kurtosis"}\hlstd{))}

\hlkwd{plot}\hlstd{(psa_aroc_kernel)}
\end{alltt}
\end{kframe}
\end{knitrout}

\noindent The graphical results are shown in Web Figures \ref{psa_AROC_bsp}, \ref{psa_bsp_predictive_checks} and \ref{psa_AROC_kernel}. The code for plotting all results using the \verb|R|-package \verb|ggplot2| \citep{Wickham09} (similarly to how results are presented in the main paper) is

\begin{knitrout}\small
\definecolor{shadecolor}{rgb}{0.969, 0.969, 0.969}\color{fgcolor}\begin{kframe}
\begin{alltt}
\hlkwd{library}\hlstd{(ggplot2)}
\hlstd{p} \hlkwb{<-} \hlkwd{seq}\hlstd{(}\hlnum{0}\hlstd{,} \hlnum{1}\hlstd{,} \hlkwc{l} \hlstd{=} \hlnum{101}\hlstd{)}

\hlstd{df} \hlkwb{<-} \hlkwd{data.frame}\hlstd{(}\hlkwc{Approach} \hlstd{=} \hlkwd{rep}\hlstd{(}\hlkwd{c}\hlstd{(}\hlstr{"AROC Bayesian Nonparam."}\hlstd{,} \hlstr{"AROC Bayesian Semipar"}\hlstd{,}
             \hlstr{"AROC Kernel-based"}\hlstd{,} \hlstr{"Pooled"}\hlstd{),} \hlkwc{each} \hlstd{=} \hlkwd{length}\hlstd{(p)),} \hlkwc{x} \hlstd{=} \hlkwd{rep}\hlstd{(p,} \hlnum{4}\hlstd{),}
              \hlkwc{y} \hlstd{=} \hlkwd{c}\hlstd{(psa_aroc_bnp}\hlopt{$}\hlstd{ROC[,}\hlnum{1}\hlstd{], psa_aroc_bsp}\hlopt{$}\hlstd{ROC[,}\hlnum{1}\hlstd{],}
                       \hlstd{psa_aroc_kernel}\hlopt{$}\hlstd{ROC[,}\hlnum{1}\hlstd{], psa_pooled}\hlopt{$}\hlstd{ROC[,}\hlnum{1}\hlstd{]),}
             \hlkwc{ql} \hlstd{=} \hlkwd{c}\hlstd{(psa_aroc_bnp}\hlopt{$}\hlstd{ROC[,}\hlnum{2}\hlstd{], psa_aroc_bsp}\hlopt{$}\hlstd{ROC[,}\hlnum{2}\hlstd{],}
                       \hlstd{psa_aroc_kernel}\hlopt{$}\hlstd{ROC[,}\hlnum{2}\hlstd{], psa_pooled}\hlopt{$}\hlstd{ROC[,}\hlnum{2}\hlstd{]),}
             \hlkwc{qh} \hlstd{=} \hlkwd{c}\hlstd{(psa_aroc_bnp}\hlopt{$}\hlstd{ROC[,}\hlnum{3}\hlstd{], psa_aroc_bsp}\hlopt{$}\hlstd{ROC[,}\hlnum{3}\hlstd{],}
                       \hlstd{psa_aroc_kernel}\hlopt{$}\hlstd{ROC[,}\hlnum{3}\hlstd{], psa_pooled}\hlopt{$}\hlstd{ROC[,}\hlnum{3}\hlstd{]))}

\hlstd{g0} \hlkwb{<-} \hlkwd{ggplot}\hlstd{(df,} \hlkwd{aes}\hlstd{(}\hlkwc{x} \hlstd{= x,} \hlkwc{y} \hlstd{= y,} \hlkwc{ymin} \hlstd{= ql,} \hlkwc{ymax} \hlstd{= qh))} \hlopt{+}
         \hlkwd{geom_line}\hlstd{(}\hlkwd{aes}\hlstd{(}\hlkwc{colour} \hlstd{= Approach,} \hlkwc{linetype} \hlstd{= Approach),} \hlkwc{size} \hlstd{=} \hlnum{1}\hlstd{)} \hlopt{+}
         \hlkwd{scale_color_manual}\hlstd{(}\hlkwc{values} \hlstd{=} \hlkwd{c}\hlstd{(}\hlstr{"#F8766D"}\hlstd{,} \hlstr{"#7CAE00"}\hlstd{,} \hlstr{"#00BFC4"}\hlstd{,} \hlstr{"#C77CFF"}\hlstd{))} \hlopt{+}
         \hlkwd{geom_ribbon}\hlstd{(}\hlkwd{aes}\hlstd{(}\hlkwc{fill} \hlstd{= Approach),} \hlkwc{alpha} \hlstd{=} \hlnum{0.2}\hlstd{)} \hlopt{+}
         \hlkwd{scale_fill_manual}\hlstd{(}\hlkwc{values} \hlstd{=} \hlkwd{c}\hlstd{(}\hlstr{"#F8766D"}\hlstd{,} \hlstr{"#7CAE00"}\hlstd{,} \hlstr{"#00BFC4"}\hlstd{,} \hlstr{"#C77CFF"}\hlstd{))} \hlopt{+}
         \hlkwd{scale_linetype_manual}\hlstd{(}\hlkwc{values}\hlstd{=}\hlkwd{c}\hlstd{(}\hlstr{"dashed"}\hlstd{,} \hlstr{"longdash"}\hlstd{,}\hlstr{"twodash"}\hlstd{,} \hlstr{"dotdash"}\hlstd{))} \hlopt{+}
         \hlkwd{guides}\hlstd{(}\hlkwc{colour}\hlstd{=}\hlkwd{guide_legend}\hlstd{(}\hlkwc{keywidth} \hlstd{=} \hlnum{3}\hlstd{,} \hlkwc{keyheight} \hlstd{=} \hlnum{1}\hlstd{))} \hlopt{+}
         \hlkwd{labs}\hlstd{(}\hlkwc{title} \hlstd{=} \hlstr{"Age-adjusted and pooled ROC curves"}\hlstd{,} \hlkwc{x} \hlstd{=} \hlstr{"FPF"}\hlstd{,} \hlkwc{y} \hlstd{=} \hlstr{"TPF"}\hlstd{)} \hlopt{+}
         \hlkwd{theme}\hlstd{(}\hlkwc{legend.position} \hlstd{=} \hlkwd{c}\hlstd{(}\hlnum{0.7}\hlstd{,} \hlnum{0.15}\hlstd{),}
               \hlkwc{plot.title} \hlstd{=} \hlkwd{element_text}\hlstd{(}\hlkwc{hjust} \hlstd{=} \hlnum{0.5}\hlstd{,} \hlkwc{size} \hlstd{=} \hlnum{20}\hlstd{),}
               \hlkwc{axis.text} \hlstd{=} \hlkwd{element_text}\hlstd{(}\hlkwc{size}\hlstd{=}\hlnum{20}\hlstd{),}
               \hlkwc{axis.title} \hlstd{=} \hlkwd{element_text}\hlstd{(}\hlkwc{size}\hlstd{=}\hlnum{20}\hlstd{),}
               \hlkwc{legend.title} \hlstd{=} \hlkwd{element_text}\hlstd{(}\hlkwc{size}\hlstd{=}\hlnum{15}\hlstd{),}
               \hlkwc{legend.text} \hlstd{=} \hlkwd{element_text}\hlstd{(}\hlkwc{size}\hlstd{=}\hlnum{15}\hlstd{))}

 \hlkwd{print}\hlstd{(g0)}
\end{alltt}
\end{kframe}
\end{knitrout}
The result is shown in Web Figure \ref{cROC_aROC_psa}.

\begin{figure}[H]
    \begin{center}
		\subfigure[\texttt{pooledROC.BB}]{\includegraphics[width=5cm]{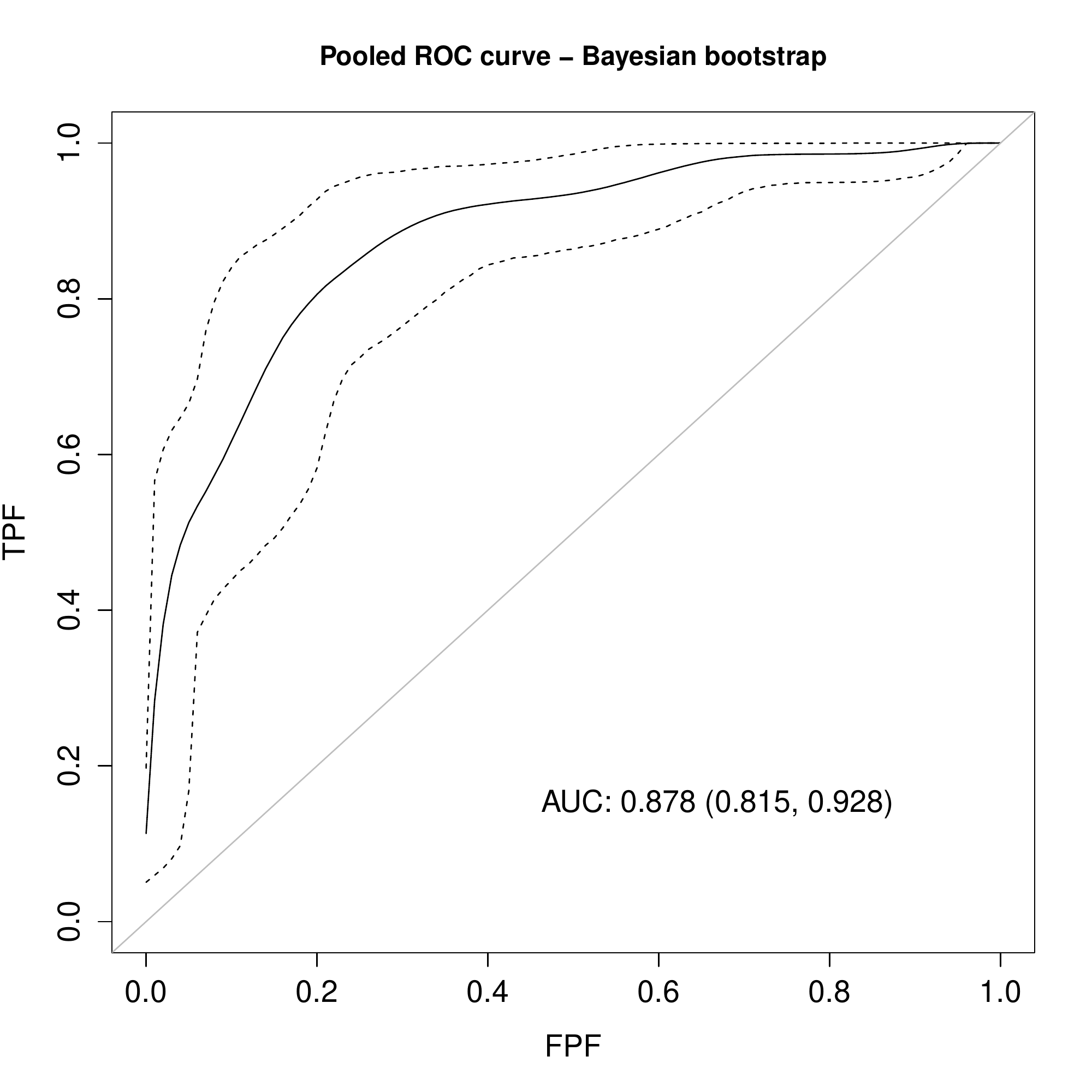}\label{psa_pooled}}
		\subfigure[\texttt{AROC.bnp}]{\includegraphics[width=5cm]{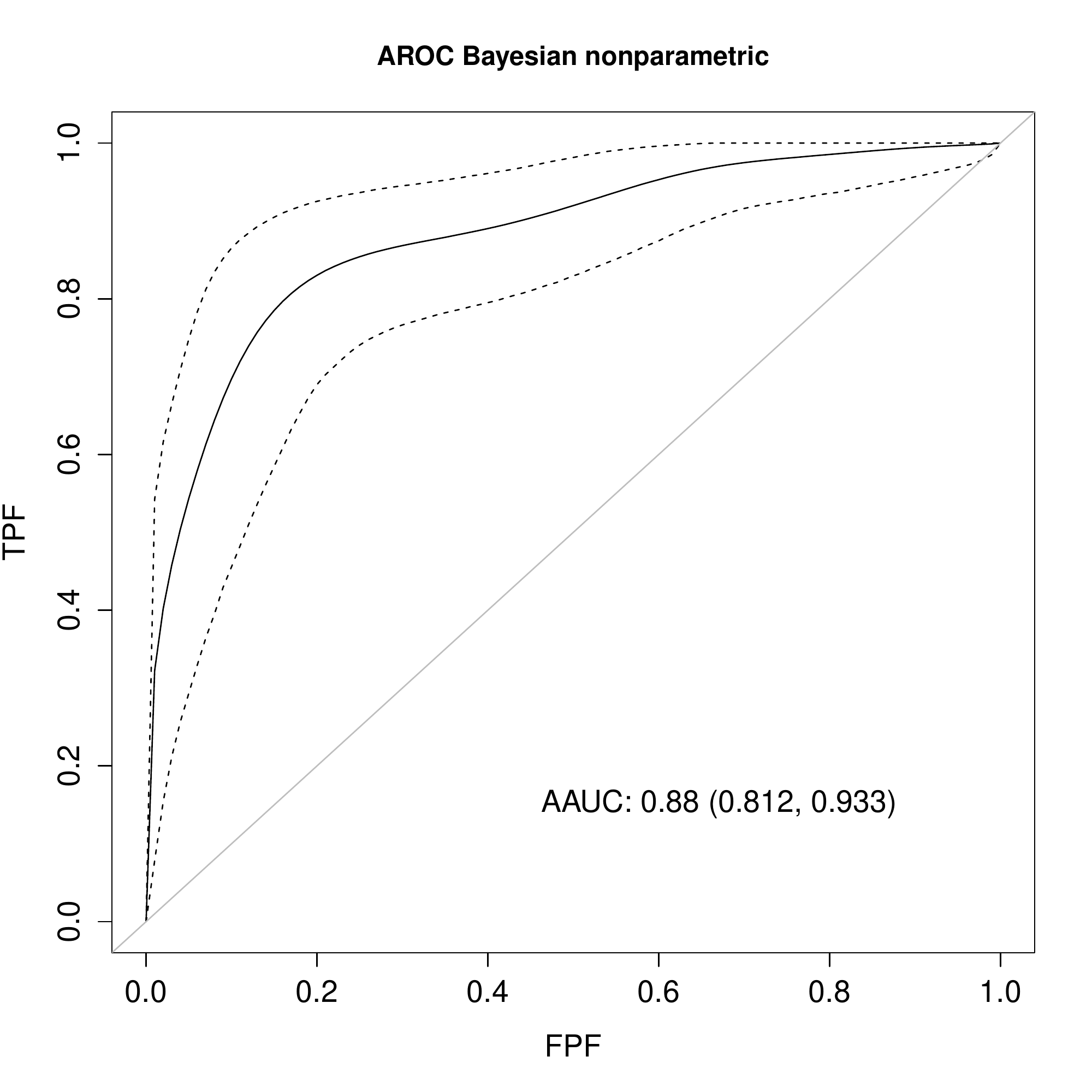}\label{psa_AROC_bnp}}\\
		\subfigure[\texttt{AROC.bsp}]{\includegraphics[width=5cm]{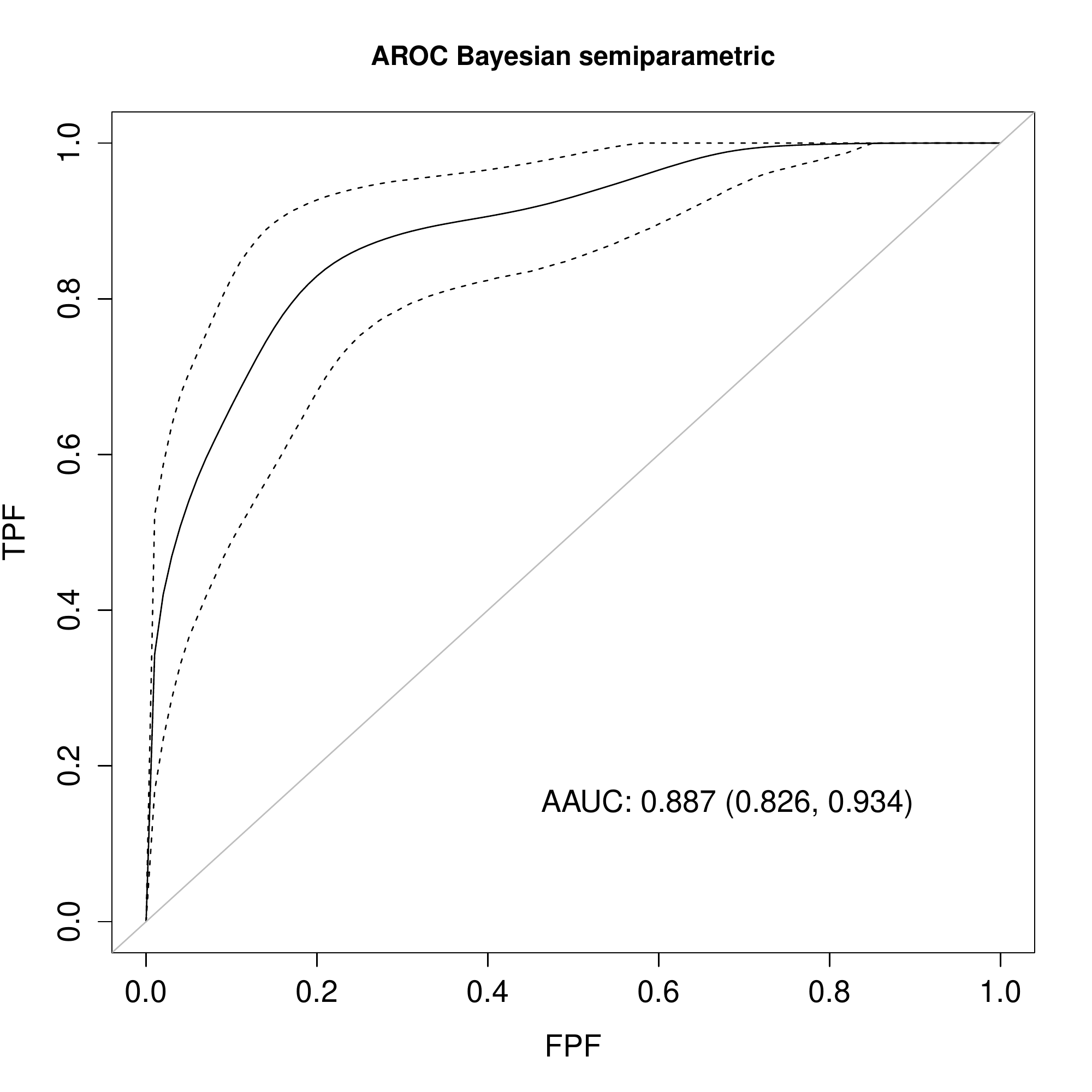}\label{psa_AROC_bsp}}
		\subfigure[\texttt{AROC.kernel}]{\includegraphics[width=5cm]{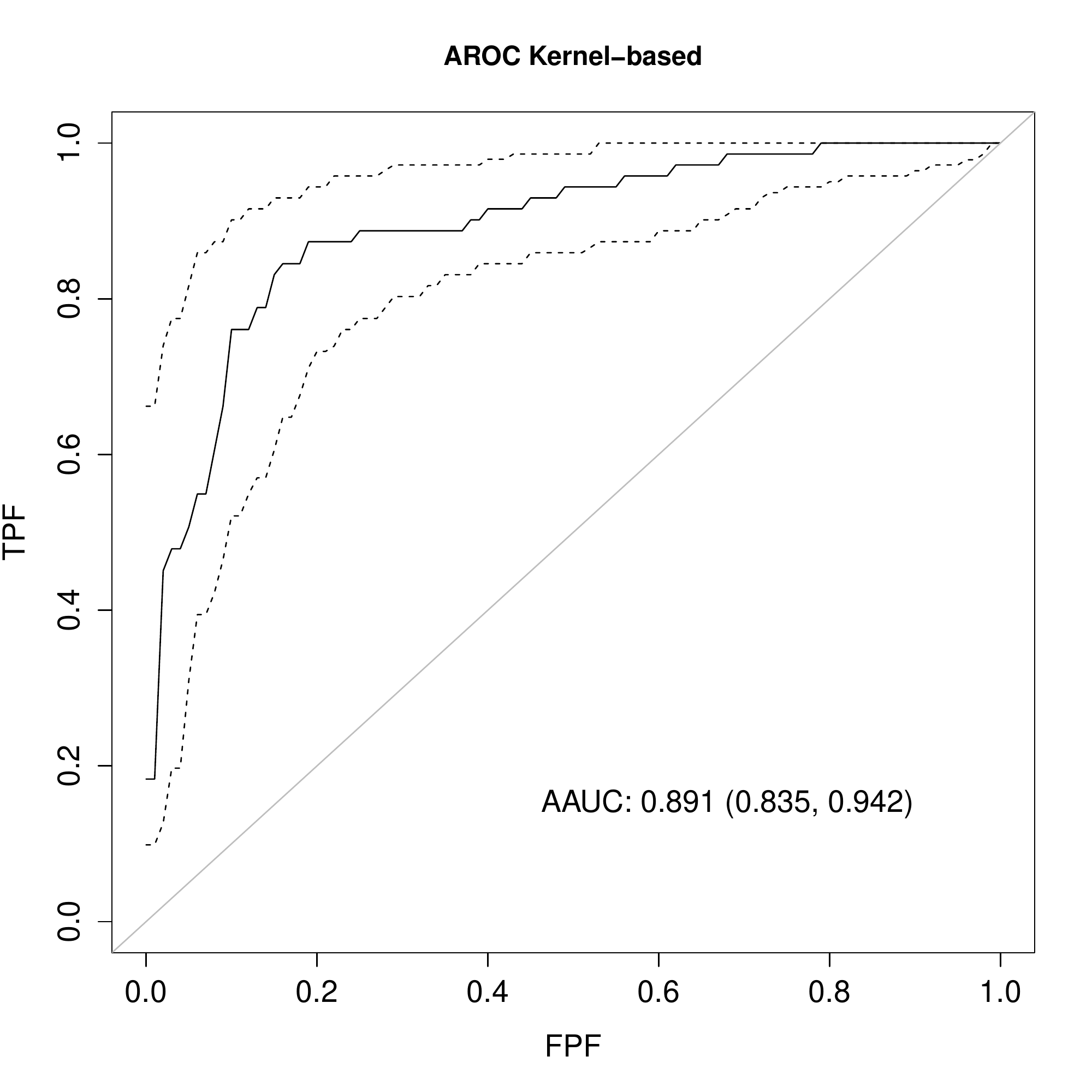}\label{psa_AROC_kernel}}
\end{center}
		 \caption{PSA CARET data: graphical results as provided by the \texttt{AROC} package.}
	\label{psa_AROC_pooled}
\end{figure}

\begin{figure}[H]
    \begin{center}
    \subfigure[Predictive checks: Bayesian nonparametric]{
			\includegraphics[width =5cm, page = 1]{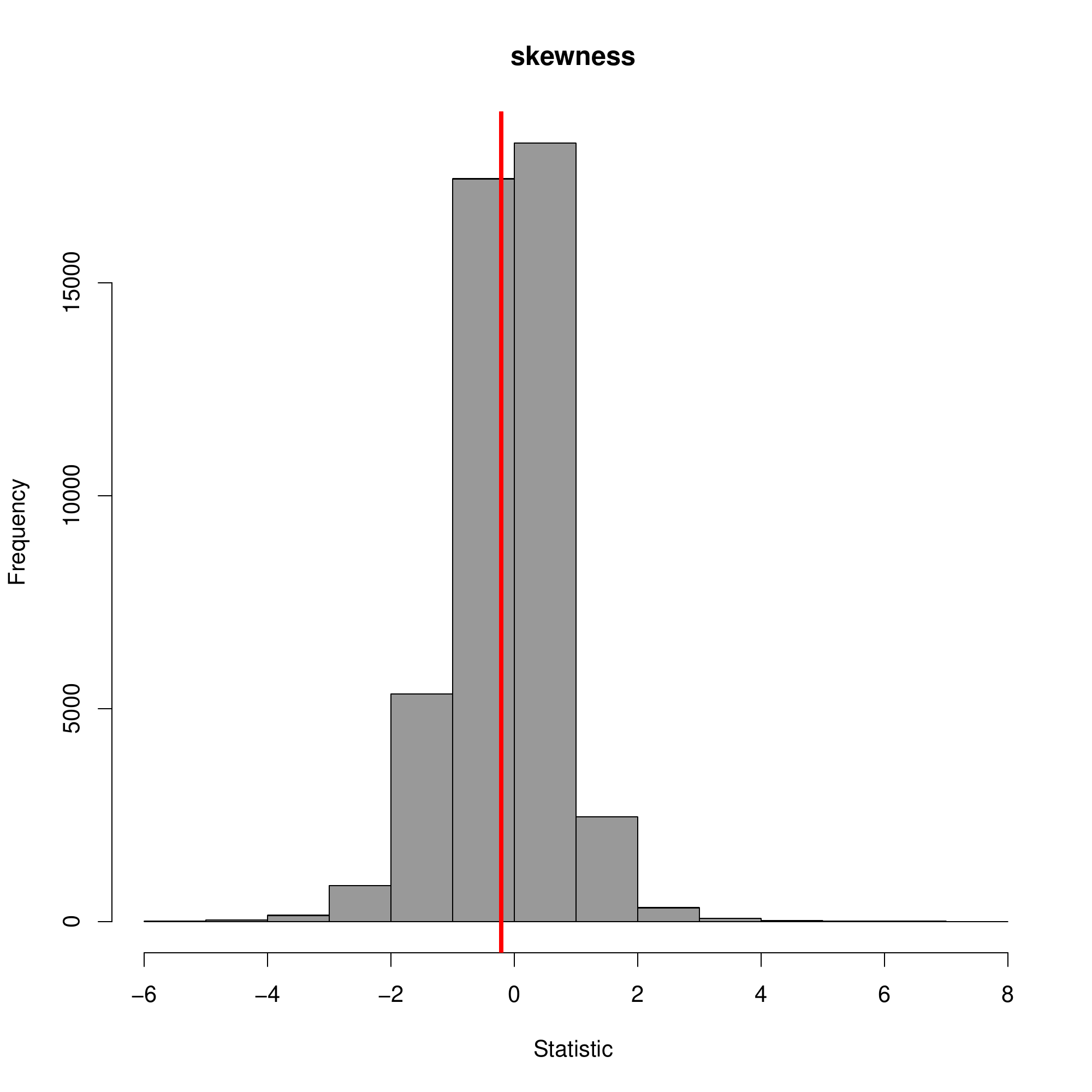}
			\includegraphics[height=5cm, page = 2]{psa_pred_checks_bnp.pdf}
			\includegraphics[height=5cm, page = 3]{psa_pred_checks_bnp.pdf}
			\label{psa_bnp_predictive_checks}
			}
	\subfigure[Predictive checks: Bayesian semiparametric]{
		 	\includegraphics[width =5cm, page = 1]{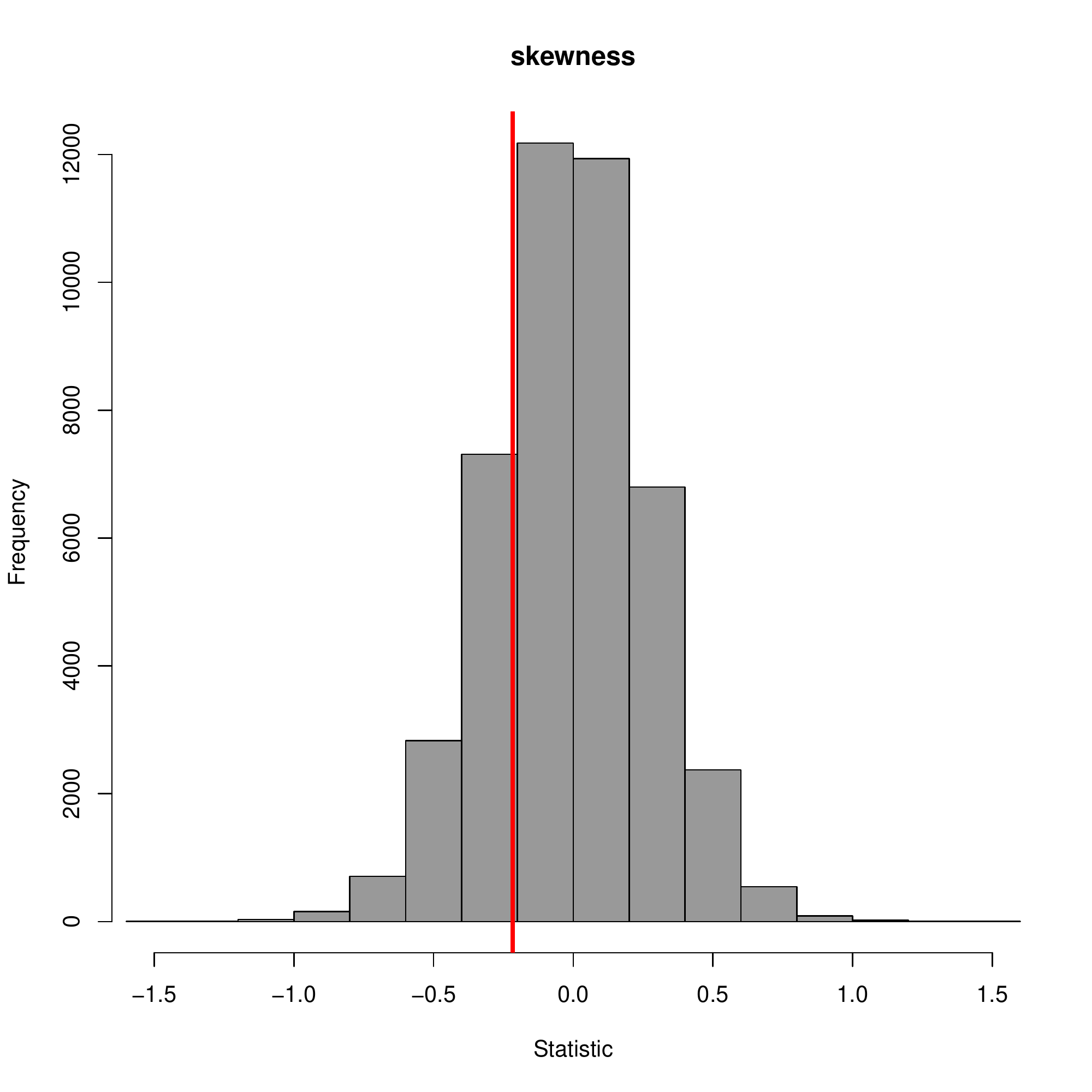}
			\includegraphics[height=5cm, page = 2]{psa_pred_checks_bsp.pdf}
			\includegraphics[height=5cm, page = 3]{psa_pred_checks_bsp.pdf}
			\label{psa_bsp_predictive_checks}
		}
		\end{center}
		 \caption{PSA CARET data: posterior predictive checks as provided by the \texttt{AROC} package}
		\label{predchecks_psa}
\end{figure}

\begin{figure}[H]
    \begin{center}
		\includegraphics[height=5cm]{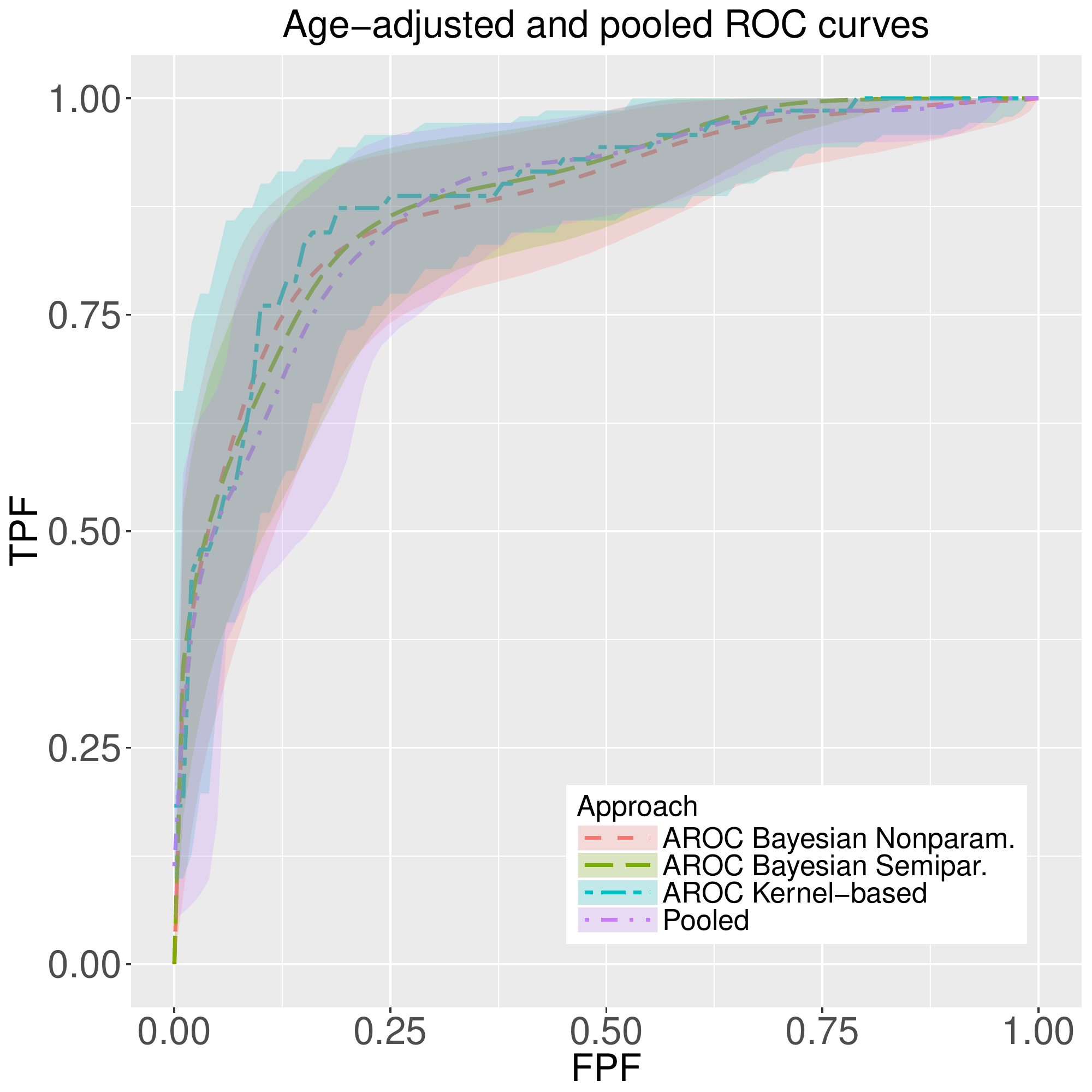}
\end{center}
		 \caption{PSA CARET data: estimated age-adjusted ROC (AROC) curve and 95\% intervals by means of our Bayesian nonparametric proposal (Bayesian Nonparam.), the Bayesian semiparametric model (Bayesian Semipar.), and the kernel-based approach \citep{MX11a}. For comparison purposes, the estimated pooled ROC curve is also depicted.} \label{cROC_aROC_psa}
\end{figure}

\newpage
\section{\large{\textsf{SUPPORTING FIGURES AND TABLES FOR THE SIMULATION STUDY}}\label{supp_sim}}
\begin{figure}[H]
\begin{center}
 \subfigure[]{
\includegraphics[width =5cm]{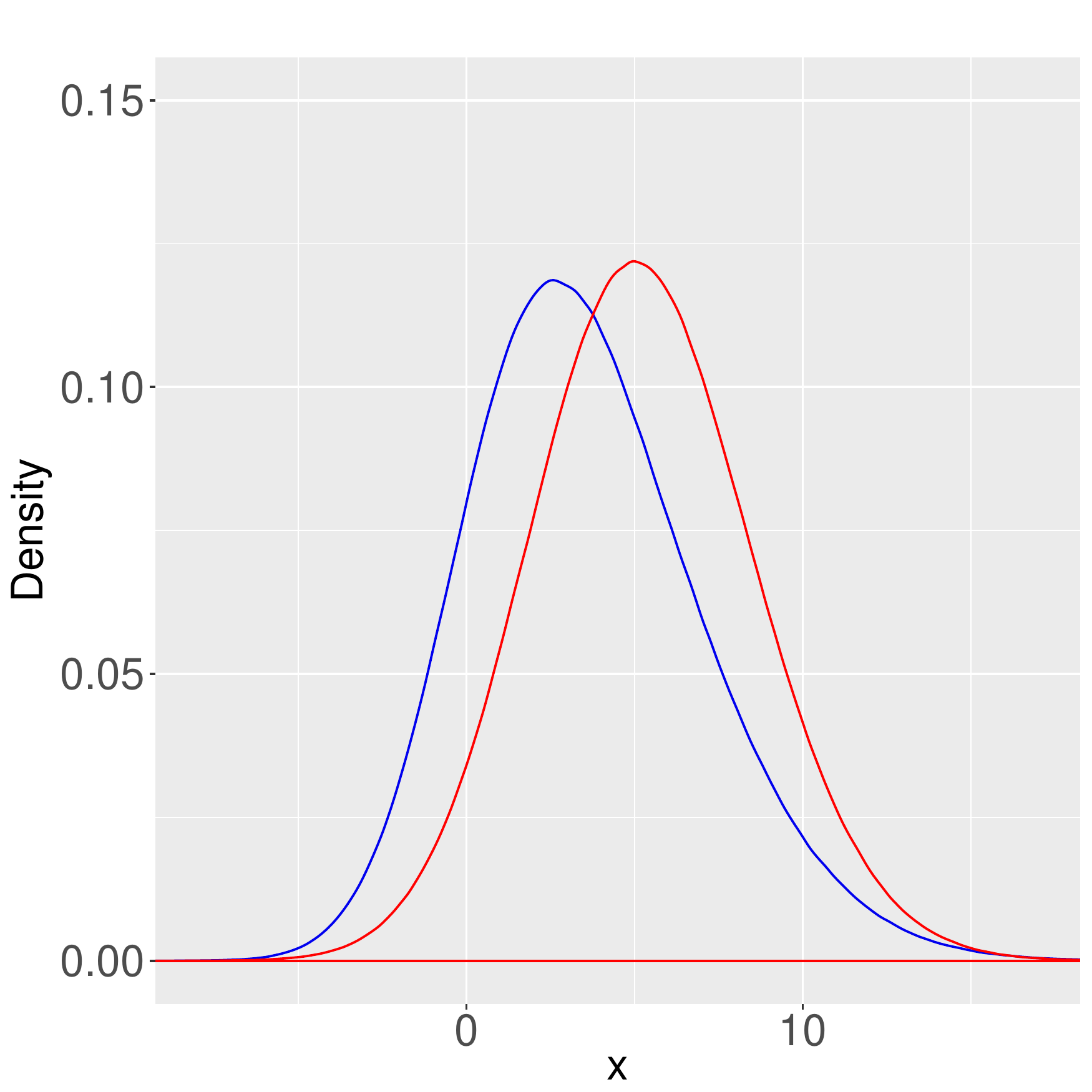}}
\subfigure[]{
\includegraphics[width=5cm]{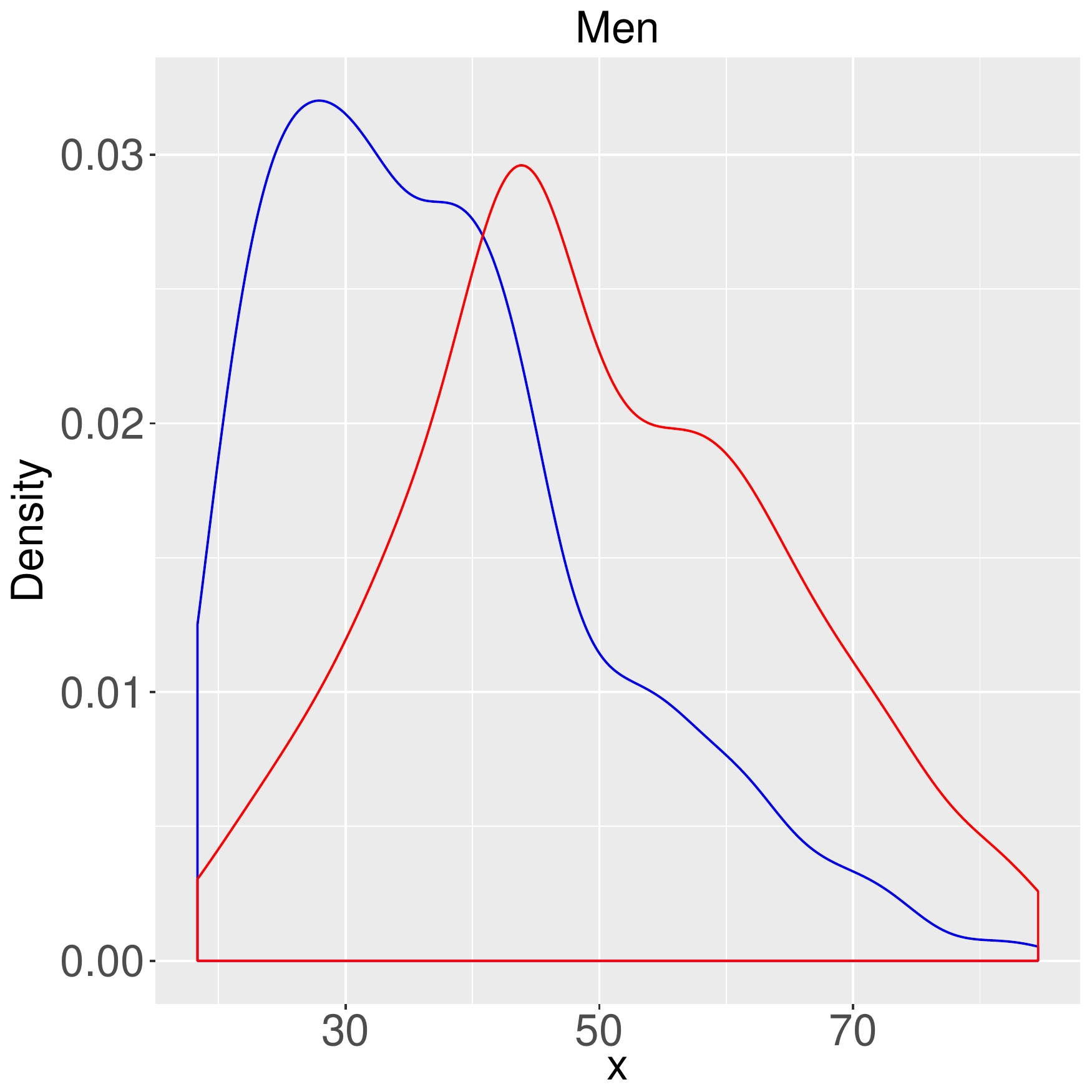}}
\subfigure[]{
\includegraphics[height=5cm]{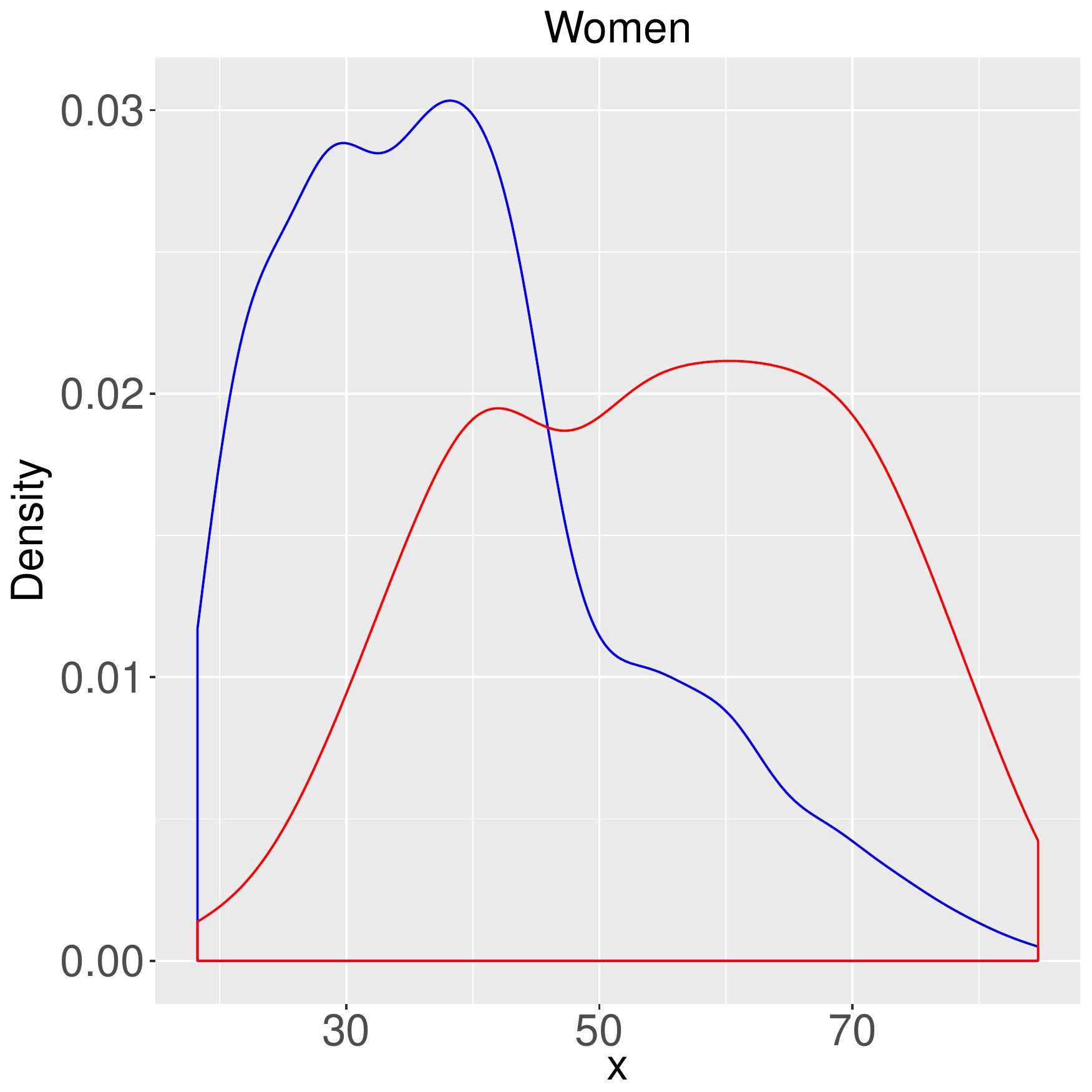}}
\end{center}
\caption{(a) Simulated covariate distribution in the nondiseased group (blue line) and in the diseased group (red line). Endocrine data: kernel density estimate of age in men (panel (b)) and in women (panel (c)).}
\label{covariates}
\end{figure}

\subsection{Results for four interior knots ($K=4$)\label{sim_k_4}}

\begin{table}[H]
\caption{Bias (standard deviation) ($\times$ 100), across simulations, of the AUC for the different approaches under consideration and for $K=4$. The results are presented for each of the simulated scenarios and sample sizes.}\label{AUC_K_4}
 \begin{center}
 \footnotesize	
	\begin{tabular}{cccccc}
	& & \multicolumn{4}{c}{Sample size}\\
	& & \multicolumn{4}{c}{$(n_{\bar{D}},n_{D})$}\\\hline
	Scenario & Approach & $(50,50)$ & $(200,70)$ & $(200,200)$ &  $(300,100)$ \\\hline
	\multirow{4}{*}{I} &	
Bayesian nonparametric & -0.620 (6.140) & -0.228 (4.293) & 0.383 (3.104) & 0.474 (3.944)\\
 & Bayesian semiparametric & 0.422 (6.264) & -0.133 (4.224) & 0.556 (2.955) & 0.525 (3.974)\\
 & Kernel & 0.680 (6.346) & 0.010 (4.261) & 0.676 (2.947) & 0.650 (3.963)\\
 & Pooled & 0.632 (6.040) & -0.053 (4.249) & 0.609 (2.895) & 0.599 (3.897)\\ \hline
	\multirow{4}{*}{II} &	
Bayesian nonparametric & -0.625 (6.114) & -0.235 (4.294) & 0.380 (3.090) & 0.472 (3.946)\\
 & Bayesian semiparametric & 0.404 (6.245) & -0.127 (4.206) & 0.550 (2.958) & 0.524 (3.970)\\
 & Kernel & 1.697 (6.378) & 0.655 (4.314) & 1.343 (3.043) & 1.173 (3.981)\\
 & Pooled & 4.134 (5.354) & 3.270 (4.131) & 3.828 (2.767) & 3.924 (3.577)\\ \hline
	\multirow{4}{*}{III} &	
Bayesian nonparametric & -0.749 (6.110) & -0.212 (4.347) & 0.311 (3.092) & 0.456 (3.926)\\
 & Bayesian semiparametric & 0.448 (6.139) & -0.089 (4.247) & 0.503 (2.955) & 0.501 (3.947)\\
 & Kernel & 1.578 (6.230) & 0.569 (4.342) & 1.139 (2.992) & 1.056 (3.961)\\
 & Pooled & 2.557 (5.595) & 1.715 (4.251) & 2.206 (2.848) & 2.293 (3.764)\\ \hline
	\multirow{4}{*}{IV} &	
Bayesian nonparametric & -0.648 (5.769) & -0.071 (3.902) & 0.670 (2.797) & 0.126 (4.004)\\
 & Bayesian semiparametric & 1.869 (5.618) & 1.922 (3.487) & 2.383 (2.782) & 2.359 (3.804)\\
 & Kernel & 1.483 (5.771) & 0.957 (4.099) & 1.521 (3.023) & 0.948 (4.142)\\
 & Pooled & 9.262 (3.785) & 8.728 (2.522) & 9.130 (2.014) & 9.281 (2.270)\\ \hline
	\multirow{4}{*}{V} &	
Bayesian nonparametric & -1.616 (4.845) & -0.740 (3.956) & -0.558 (2.370) & -0.723 (3.385)\\
 & Bayesian semiparametric & -1.942 (5.853) & -1.731 (3.479) & -1.609 (3.133) & -1.573 (2.767)\\
 & Kernel & - & - & - & -\\
 & Pooled & 5.921 (4.077) & 5.451 (3.150) & 5.786 (1.840) & 5.396 (2.318)\\ \hline
	\multirow{4}{*}{VI} &	
Bayesian nonparametric & -2.881 (5.090) & -0.759 (4.086) & -0.314 (3.175) & -0.105 (3.733)\\
 & Bayesian semiparametric & -3.867 (5.757) & -4.067 (3.805) & -3.357 (3.183) & -3.183 (3.744)\\
 & Kernel & - & - & - & -\\
 & Pooled & -7.565 (4.812) & -6.870 (3.850) & -6.173 (3.092) & -5.813 (3.662)\\ \hline
 \end{tabular}
 \end{center}
\end{table}

\begin{table}[H]
\caption{$95\%$ coverage probabilities for the AAUC for the different approaches under consideration and for $K=4$. The results are presented for each of the simulated scenarios and sample sizes.}\label{covAAROC_K_4}
 \begin{center}
 \footnotesize	
	\begin{tabular}{cccccc}
	& & \multicolumn{4}{c}{Sample size}\\
	& & \multicolumn{4}{c}{$(n_{\bar{D}},n_{D})$}\\\hline
	Scenario & Approach & $(50,50)$ & $(200,70)$ & $(200,200)$ &  $(300,100)$ \\\hline
	\multirow{2}{*}{I} &	
Bayesian nonparametric & 94.0 & 96.0 & 94.0 & 94.0\\
 & Bayesian semiparametric & 91.0 & 96.0 & 94.0 & 94.0\\ \hline
	\multirow{2}{*}{II} &	
Bayesian nonparametric & 94.0 & 97.0 & 93.0 & 94.0\\
 & Bayesian semiparametric & 91.0 & 96.0 & 94.0 & 94.0\\ \hline
	\multirow{2}{*}{III} &	
Bayesian nonparametric & 94.0 & 95.0 & 93.0 & 94.0\\
 & Bayesian semiparametric & 92.0 & 95.0 & 94.0 & 95.0\\ \hline
	\multirow{2}{*}{IV} &	
Bayesian nonparametric & 92.0 & 96.0 & 93.0 & 93.0\\
 & Bayesian semiparametric & 89.0 & 91.0 & 81.0 & 80.0\\ \hline
	\multirow{2}{*}{V} &	
Bayesian nonparametric & 98.0 & 96.0 & 97.0 & 97.0\\
 & Bayesian semiparametric & 95.0 & 93.0 & 90.0 & 92.0\\ \hline
	\multirow{2}{*}{VI} &	
Bayesian nonparametric & 98.0 & 96.0 & 94.0 & 96.0\\
 & Bayesian semiparametric & 87.0 & 82.0 & 71.0 & 84.0\\ \hline
	\end{tabular}
 \end{center}
\end{table}

\begin{table}[H]
\caption{Percentage of time the WAIC/LPML is smaller/higher for the B-splines dependent Dirichlet process (DDP) mixture of normals model with four interior knots ($K=4$) against no interior knots ($K=0$). The results are presented for each of the simulated scenarios and sample sizes (in the nondiseased group).}\label{WAIC_NPML_ndx_5_ndx_1}
 \begin{center}
 \footnotesize	
	\begin{tabular}{cccccc}
	& & \multicolumn{3}{c}{Sample size}\\
	& & \multicolumn{3}{c}{$n_{\bar{D}}$}\\\hline
	Scenario & Approach & $50$ & $200$ &  $300$ \\\hline
	\multirow{2}{*}{I} &	
WAIC ($K= 4$) $<$ WAIC ($K= 0$) & 15\%  & 5\% & 6\%\\
 & LPML ($K= 4$) $>$ LPML ($K= 0$) & 13\%  & 5\% & 4\%\\ \hline
	\multirow{2}{*}{II} &	
WAIC ($K= 4$) $<$ WAIC ($K= 0$) & 11\%  & 5\% & 5\%\\
 & LPML ($K= 4$) $>$ LPML ($K= 0$) & 9\%  & 6\% & 4\%\\ \hline
	\multirow{2}{*}{III} &	
WAIC ($K= 4$) $<$ WAIC ($K= 0$) & 13\%  & 4\% & 6\%\\
 & LPML ($K= 4$) $>$ LPML ($K= 0$) & 10\%  & 4\% & 3\%\\ \hline
	\multirow{2}{*}{IV} &	
WAIC ($K= 4$) $<$ WAIC ($K= 0$) & 45\%  & 98\% & 100\%\\
 & LPML ($K= 4$) $>$ LPML ($K= 0$) & 39\%  & 97\% & 100\%\\ \hline
 	\multirow{2}{*}{V} &	
WAIC ($K= 4$) $<$ WAIC ($K= 0$) & 3\% & 1\% & 1\%\\
 & LPML ($K= 4$) $>$ LPML ($K= 0$) & 1\%  & 1\% & 1\%\\ \hline
 	\multirow{2}{*}{VI} &	
WAIC ($K= 4$) $<$ WAIC ($K= 0$) & 15\%  & 58\% & 80\%\\
 & LPML ($K= 4$) $>$ LPML ($K= 0$) & 9\%  & 50\% & 76\%\\ \hline
	\end{tabular}
 \end{center}
\end{table}

\begin{table} [H]
\caption{For $K = 4$: percentage of time the WAIC/LPML is smaller/higher for the Bayesian normal linear (BNL) model against our B-splines dependent Dirichlet process (DDP) mixture of normals model with $K=4$. The results are presented for each of the simulated scenarios and sample sizes (in the nondiseased group).}\label{WAIC_LPML_ndx5}
 \begin{center}
 \footnotesize	
	\begin{tabular}{cccccc}
	& & \multicolumn{3}{c}{Sample size}\\
	& & \multicolumn{3}{c}{$n_{\bar{D}}$}\\\hline
	Scenario & Approach & $50$  & $200$ &  $300$ \\\hline
	\multirow{2}{*}{I} &	
WAIC (BNL) $<$ WAIC (DDP) & 86\%  & 97\% & 99\%\\
 & LPML (BNL) $>$ LPML (DDP) & 91\%  & 97\% & 100\%\\ \hline
	\multirow{2}{*}{II} &	
WAIC (BNL) $<$ WAIC (DDP) & 88\%  & 97\% & 100\%\\
 & LPML (BNL) $>$ LPML (DDP) & 92\%  & 97\% & 100\%\\ \hline
	\multirow{2}{*}{III} &	
WAIC (BNL) $<$ WAIC (DDP) & 88\%  & 97\% & 100\%\\
 & LPML (BNL) $>$ LPML (DDP) & 91\%  & 97\% & 100\%\\ \hline
	\multirow{2}{*}{IV} &	
WAIC (BNL) $<$ WAIC (DDP) & 3\%  & 0\% & 0\%\\
 & LPML (BNL) $>$ LPML (DDP) & 4\%  & 0\% & 0\%\\ \hline
 \multirow{2}{*}{V} &	
WAIC (BNL) $<$ WAIC (DDP) & 0\%  & 0\% & 0\%\\
 & LPML (BNL) $>$ LPML (DDP) & 0\%  & 0\% & 0\%\\ \hline
 \multirow{2}{*}{VI} &	
WAIC (BNL) $<$ WAIC (DDP) & 3\%  & 0\% & 0\%\\
 & LPML (BNL) $>$ LPML (DDP) & 11\%  & 0\% & 0\%\\ \hline
	\end{tabular}
 \end{center}
\end{table}

\begin{figure}[H]
  \begin{center}
    	\subfigure{
		\includegraphics[height=4.5cm, page = 1]{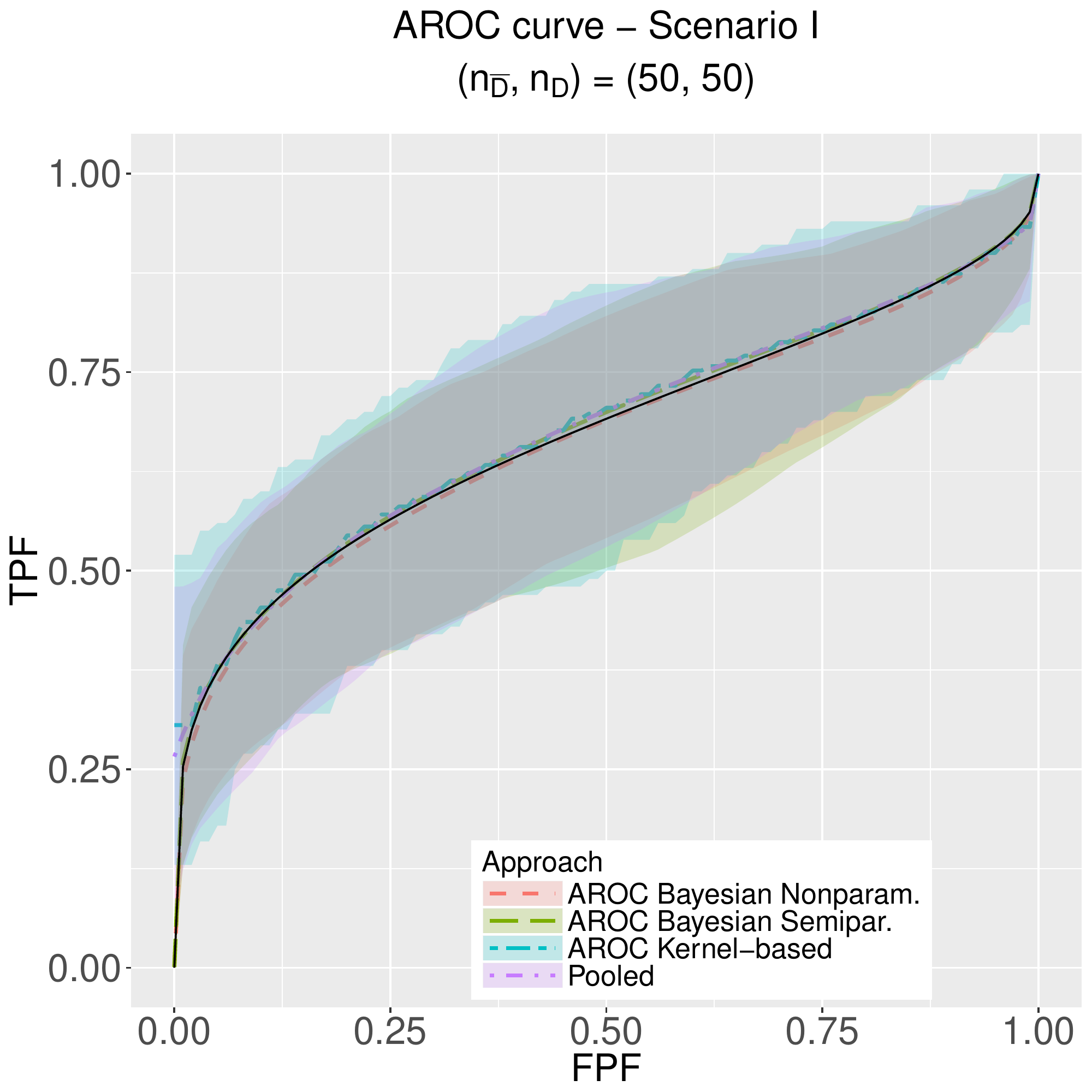}
		\includegraphics[height=4.5cm, page = 1]{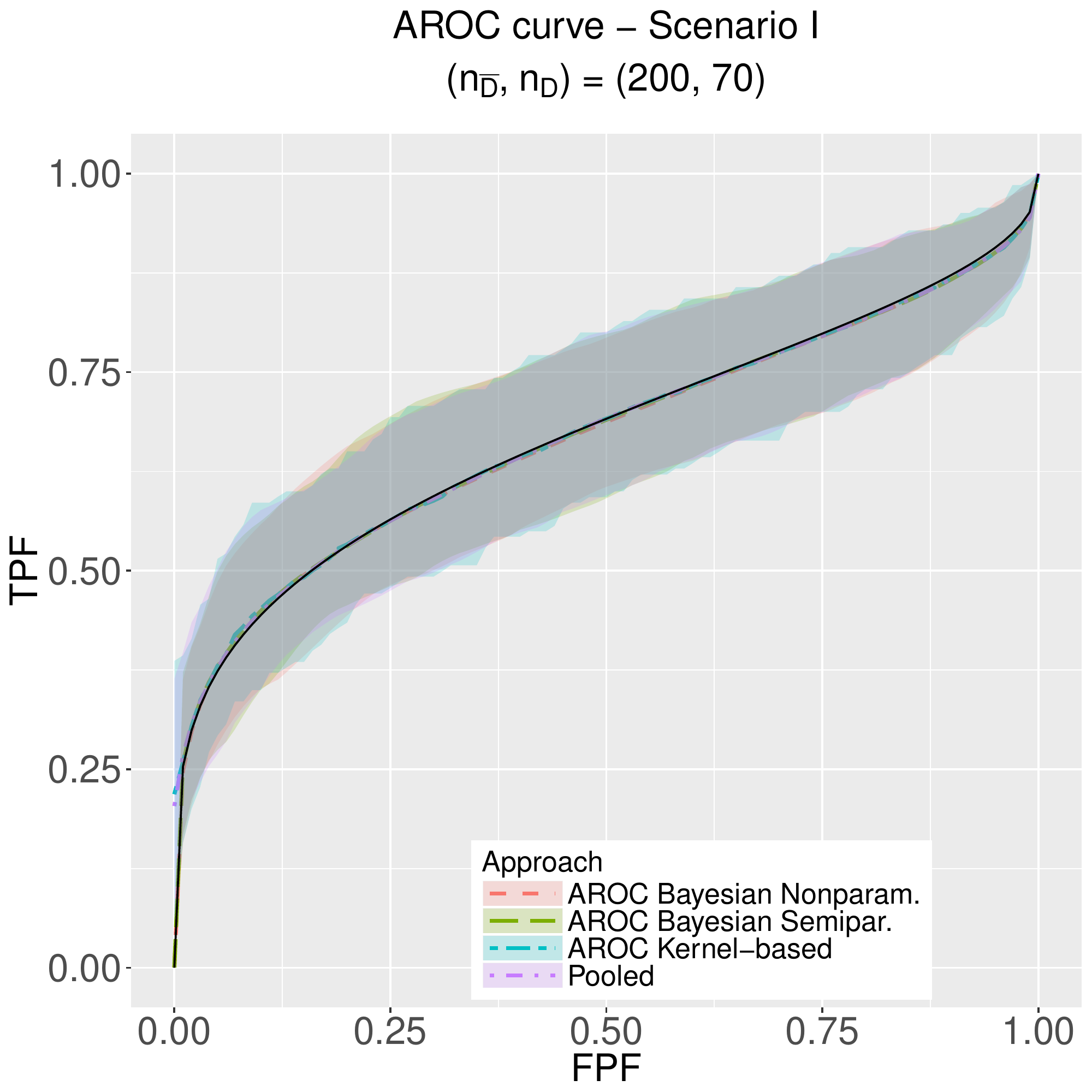}
		\includegraphics[height=4.5cm, page = 1]{sim_I_200_200_ndx_5_bs.pdf}
		\includegraphics[height=4.5cm, page = 1]{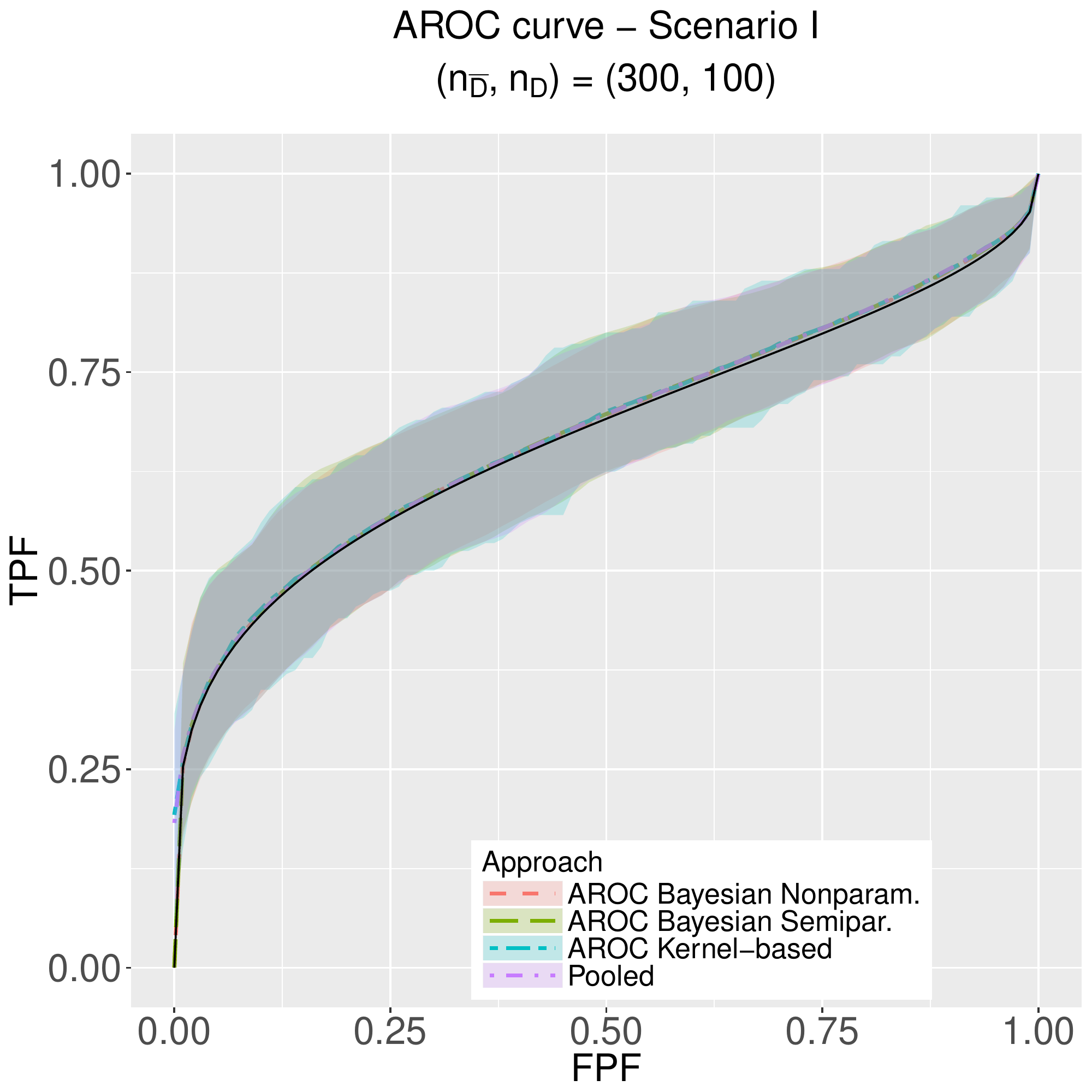}}
\end{center}
		 \caption{Scenario I and $K = 4$: true (solid black line) and average value of 100 simulated datasets (dashed lines) of the posterior mean (for the Bayesian estimators) of the covariate adjusted ROC curve/pooled ROC curve for the different approaches under consideration and sample sizes.}
		\label{sim_ndx_5_I}
\end{figure}

\begin{figure}[H]
  \begin{center}
    	\subfigure{
		\includegraphics[height=4.5cm, page = 1]{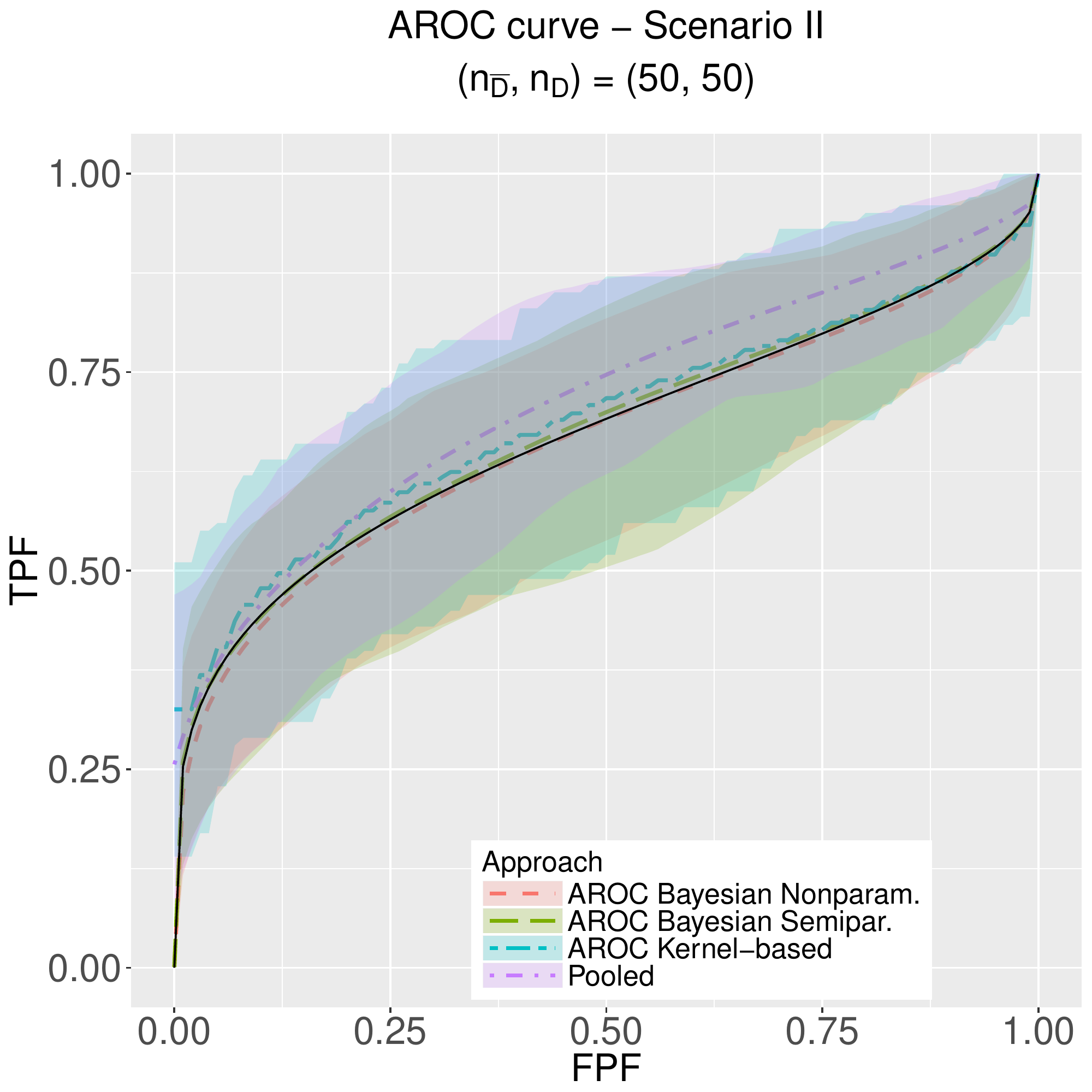}
		\includegraphics[height=4.5cm, page = 1]{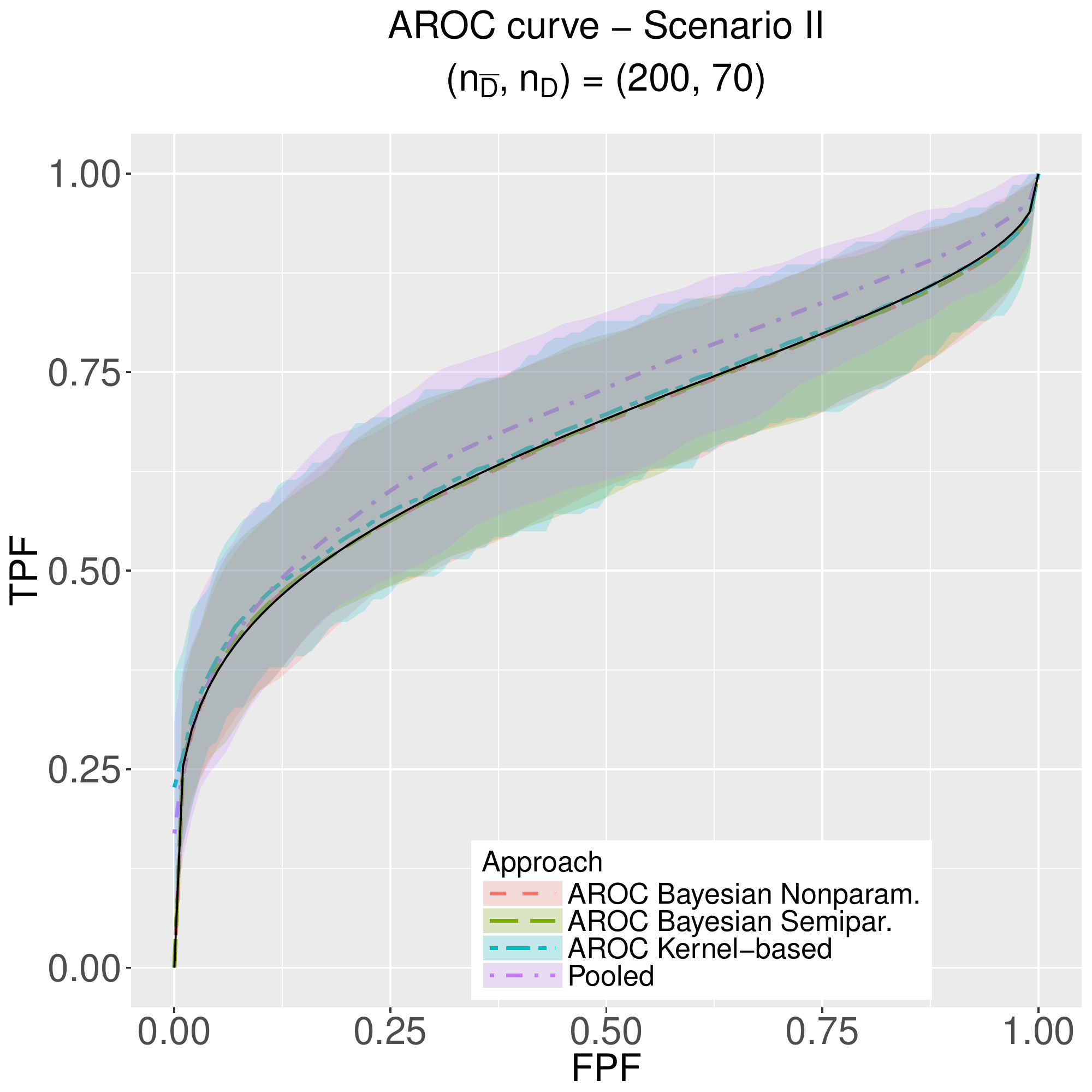}
		\includegraphics[height=4.5cm, page = 1]{sim_II_200_200_ndx_5_bs.pdf}
		\includegraphics[height=4.5cm, page = 1]{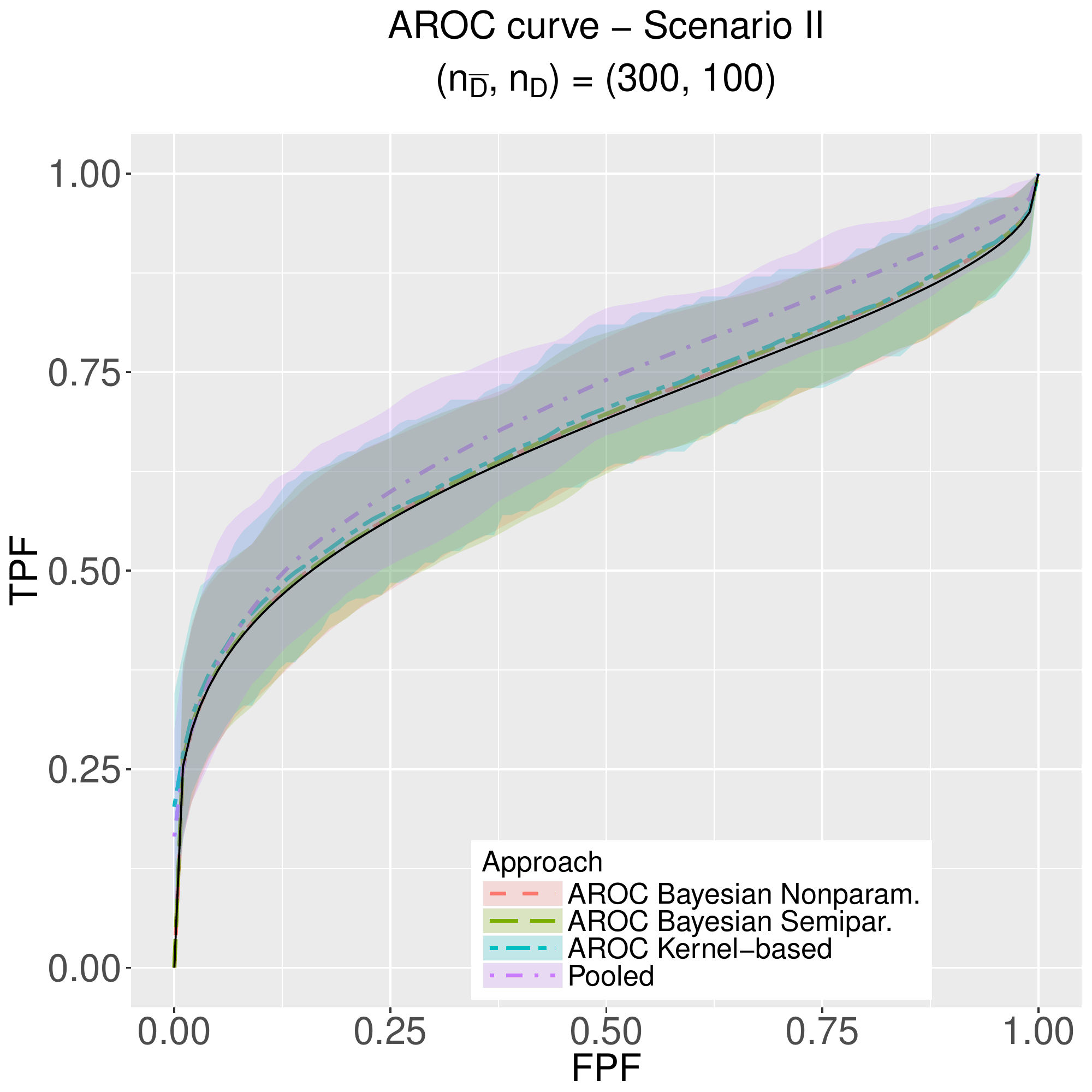}}
\end{center}
		 \caption{Scenario II and $K=4$ : true (solid black line) and average value of 100 simulated datasets (dashed lines) of the posterior mean (for the Bayesian estimators) of the covariate adjusted ROC curve/pooled ROC curve for the different approaches under consideration and sample sizes.}
		\label{sim_ndx_5_II}
\end{figure}

\begin{figure}[H]
  \begin{center}
    	\subfigure{
		\includegraphics[height=4.5cm, page = 1]{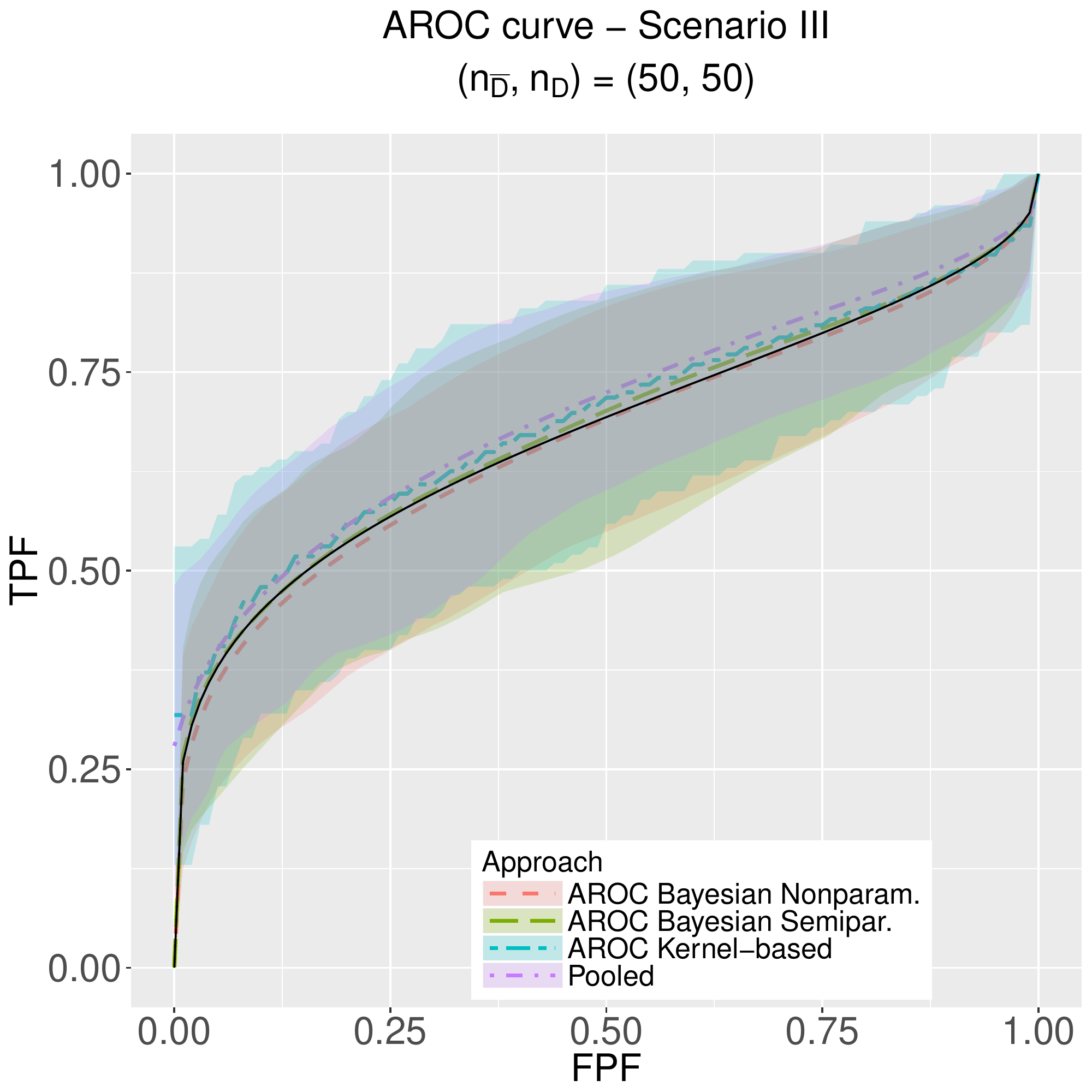}
		\includegraphics[height=4.5cm, page = 1]{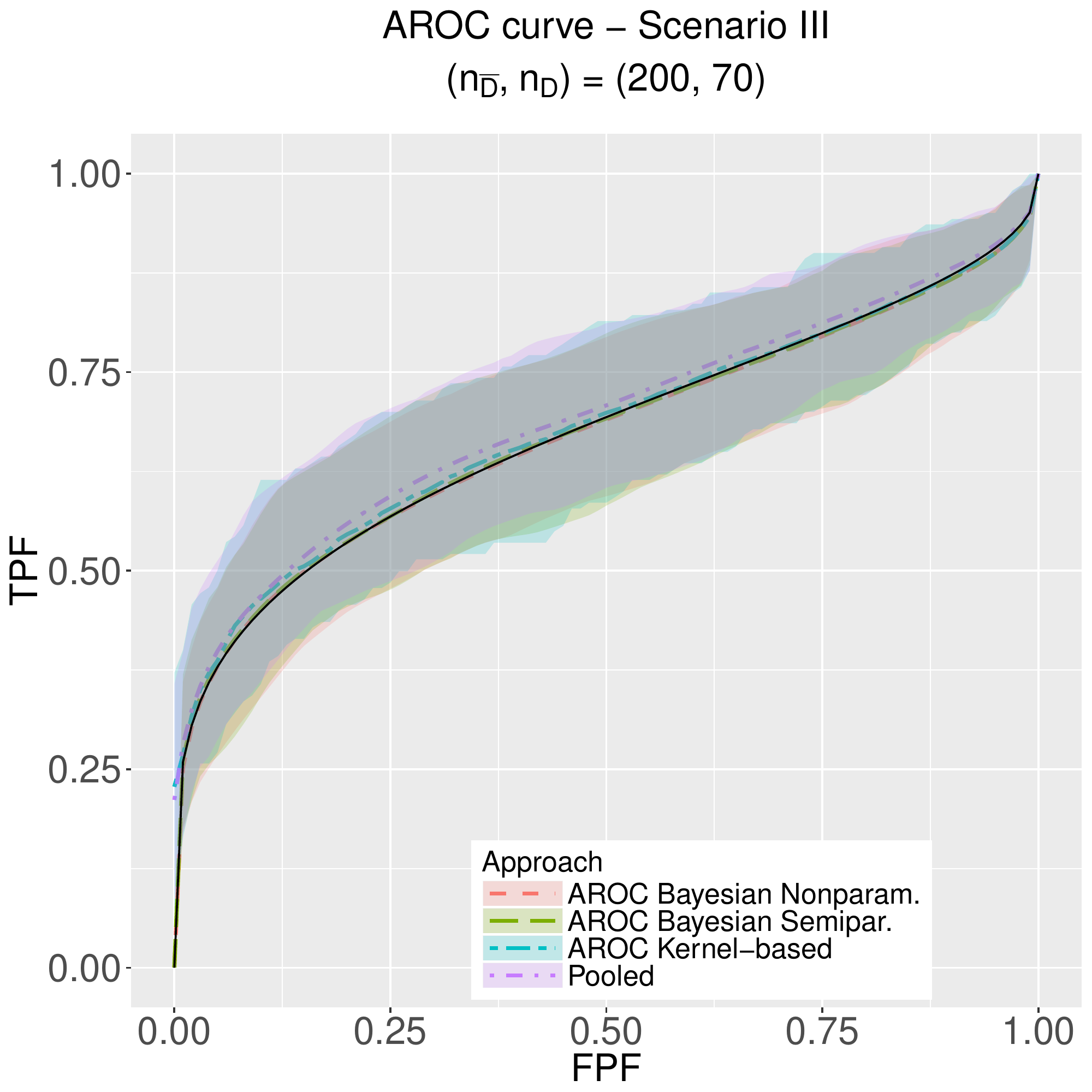}
		\includegraphics[height=4.5cm, page = 1]{sim_III_200_200_ndx_5_bs.pdf}
		\includegraphics[height=4.5cm, page = 1]{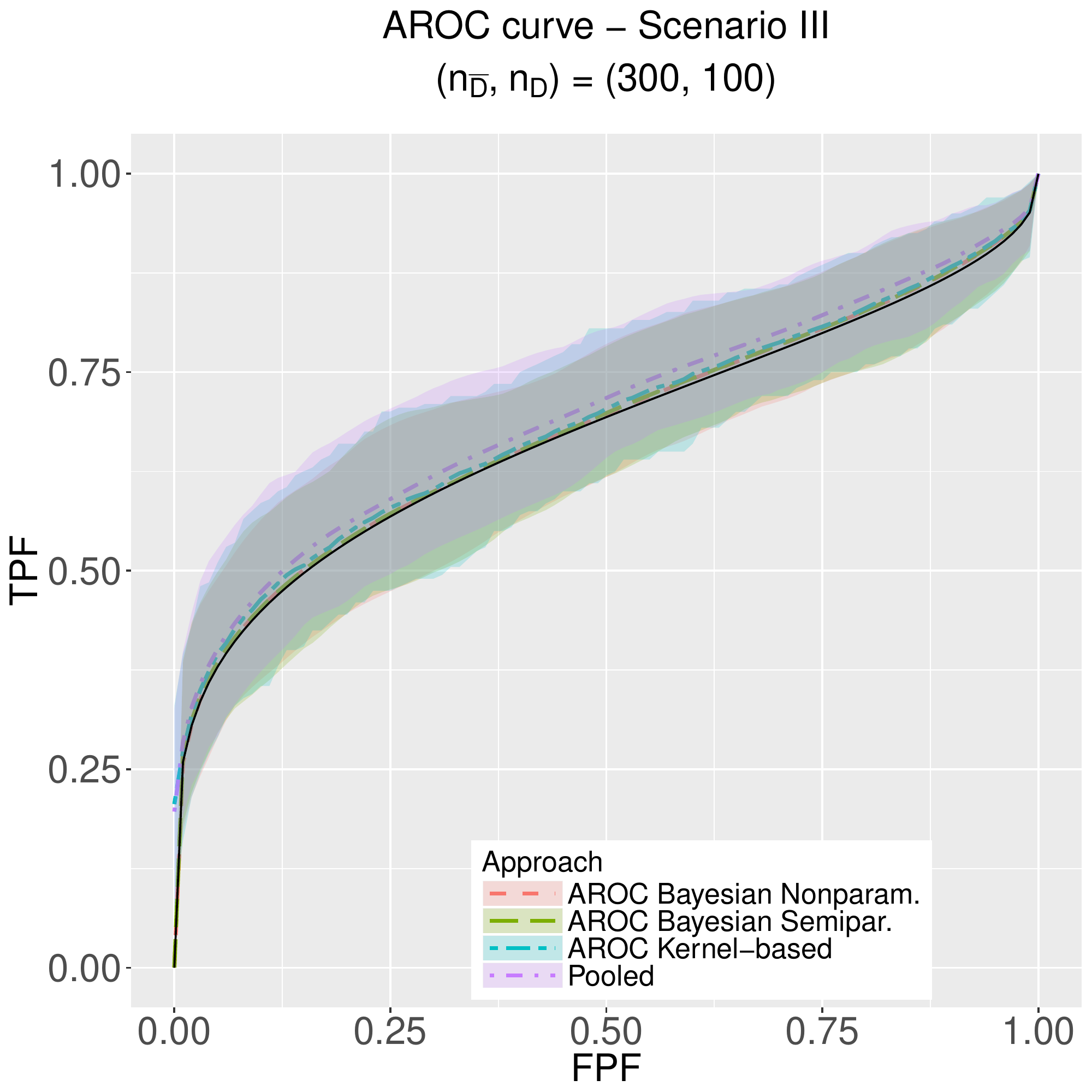}}
\end{center}
		 \caption{Scenario III and $K=4$: true (solid black line) and average value of 100 simulated datasets (dashed lines) of the posterior mean (for the Bayesian estimators) of the covariate adjusted ROC curve/pooled ROC curve for the different approaches under consideration and sample sizes.}
		\label{sim_ndx_5_III}
\end{figure}

\begin{figure}[H]
  \begin{center}
    	\subfigure{
		\includegraphics[height=4.5cm, page = 1]{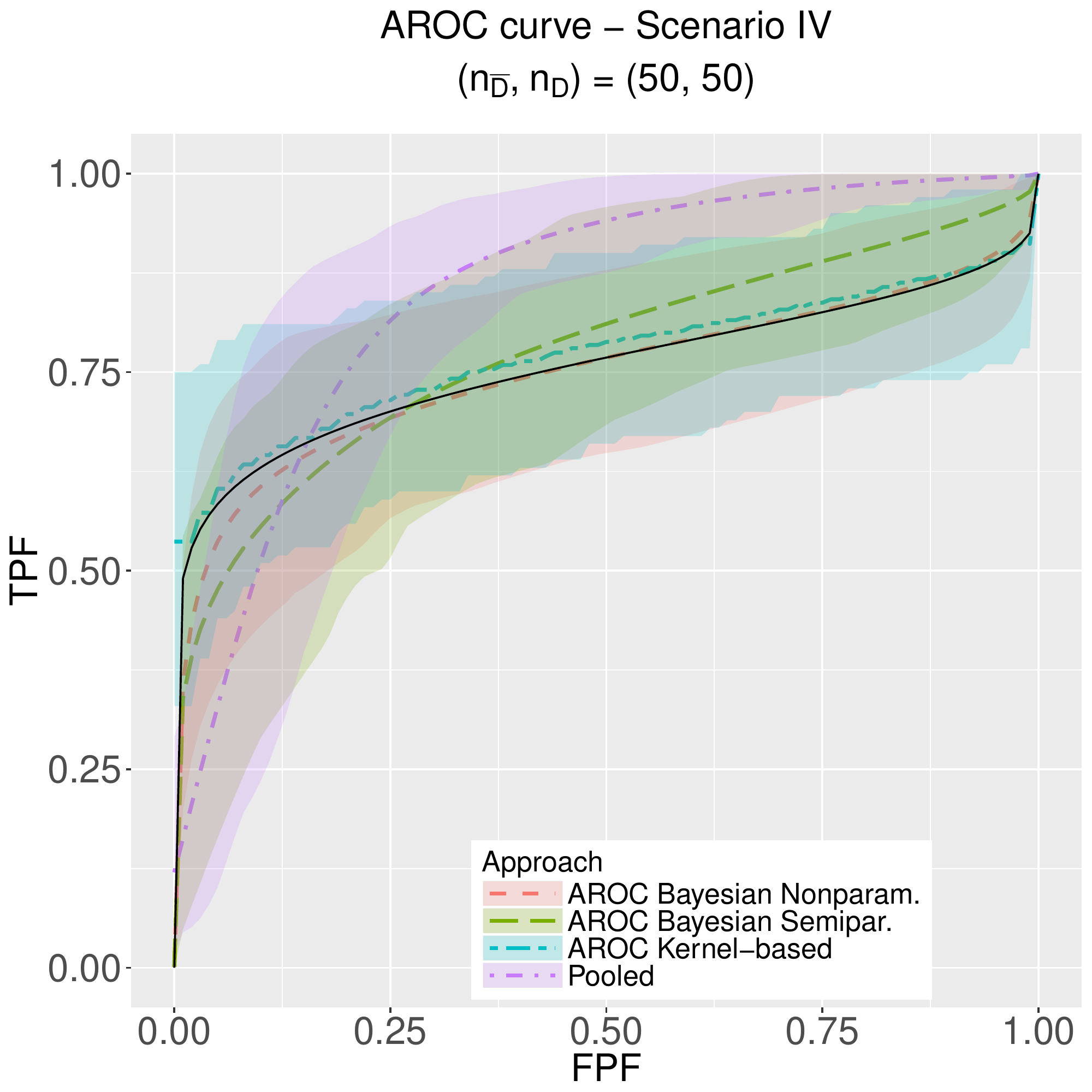}
		\includegraphics[height=4.5cm, page = 1]{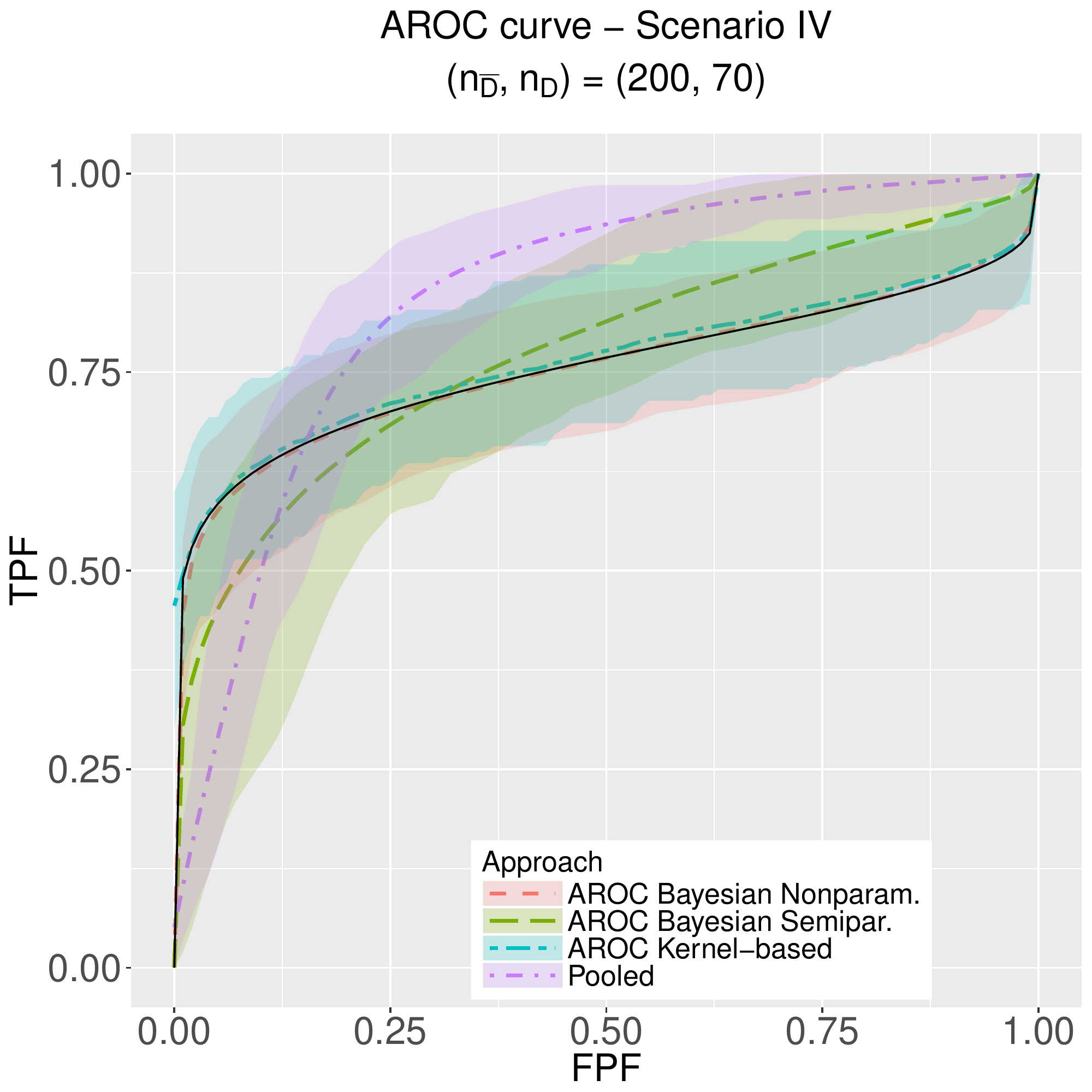}
		\includegraphics[height=4.5cm, page = 1]{sim_IV_200_200_ndx_5_bs.pdf}
		\includegraphics[height=4.5cm, page = 1]{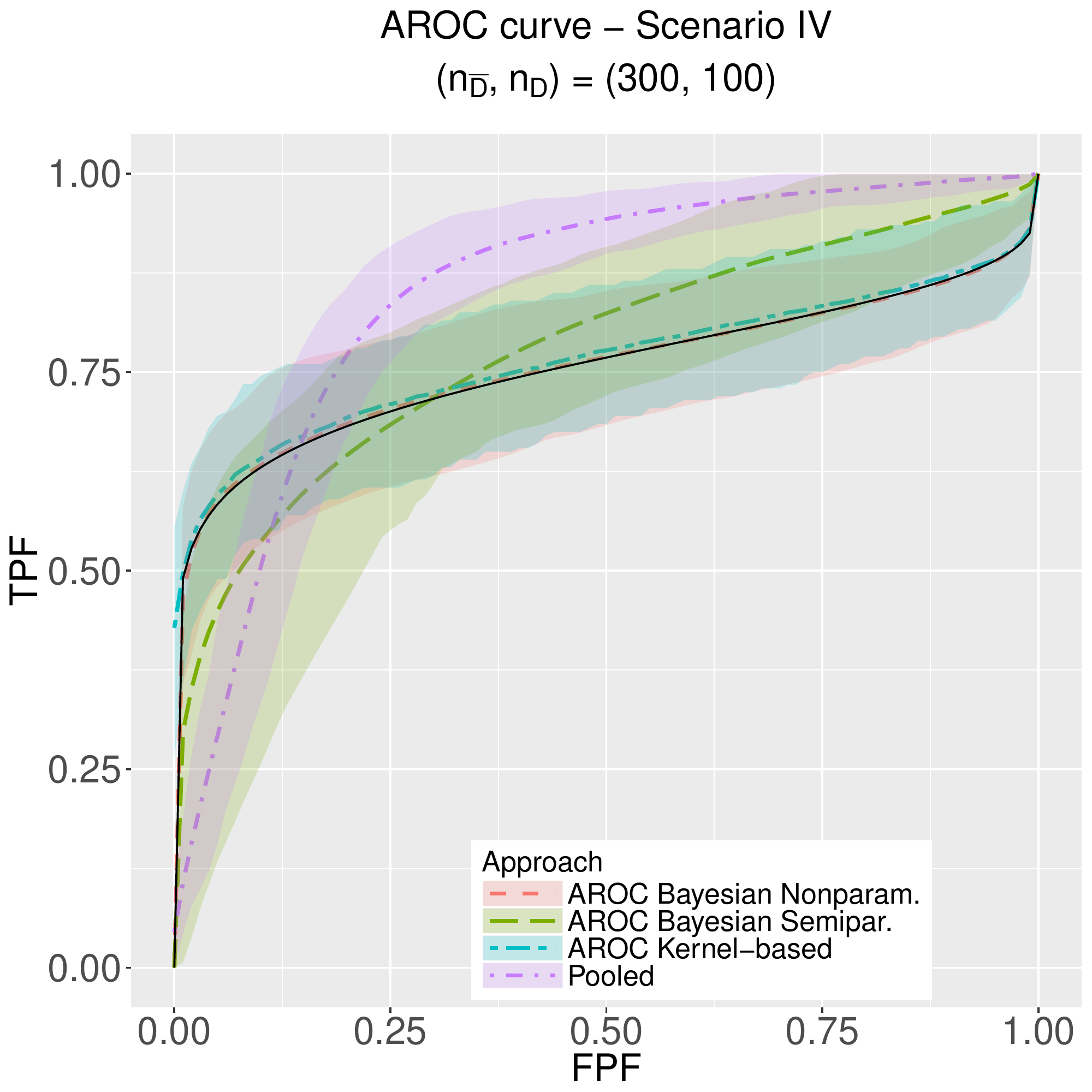}}
\end{center}
		 \caption{Scenario IV and $K=4$: true (solid black line) and average value of 100 simulated datasets (dashed lines) of the posterior mean (for the Bayesian estimators) of the covariate adjusted ROC curve/pooled ROC curve for the different approaches under consideration and sample sizes.}
		\label{sim_ndx_5_IV}
\end{figure}

\begin{figure}[H]
  \begin{center}
    	\subfigure{
		\includegraphics[height=4.5cm, page = 1]{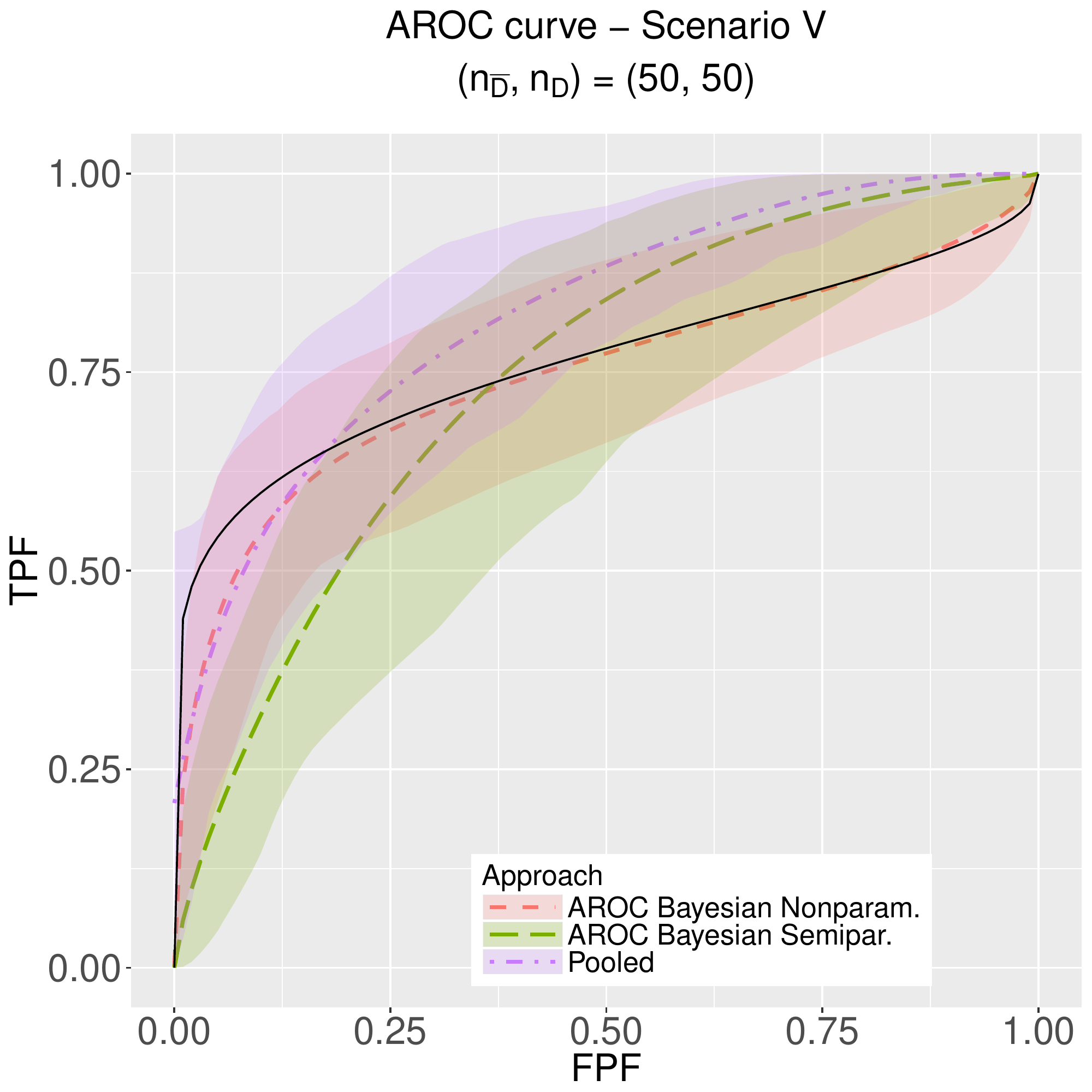}
		\includegraphics[height=4.5cm, page = 1]{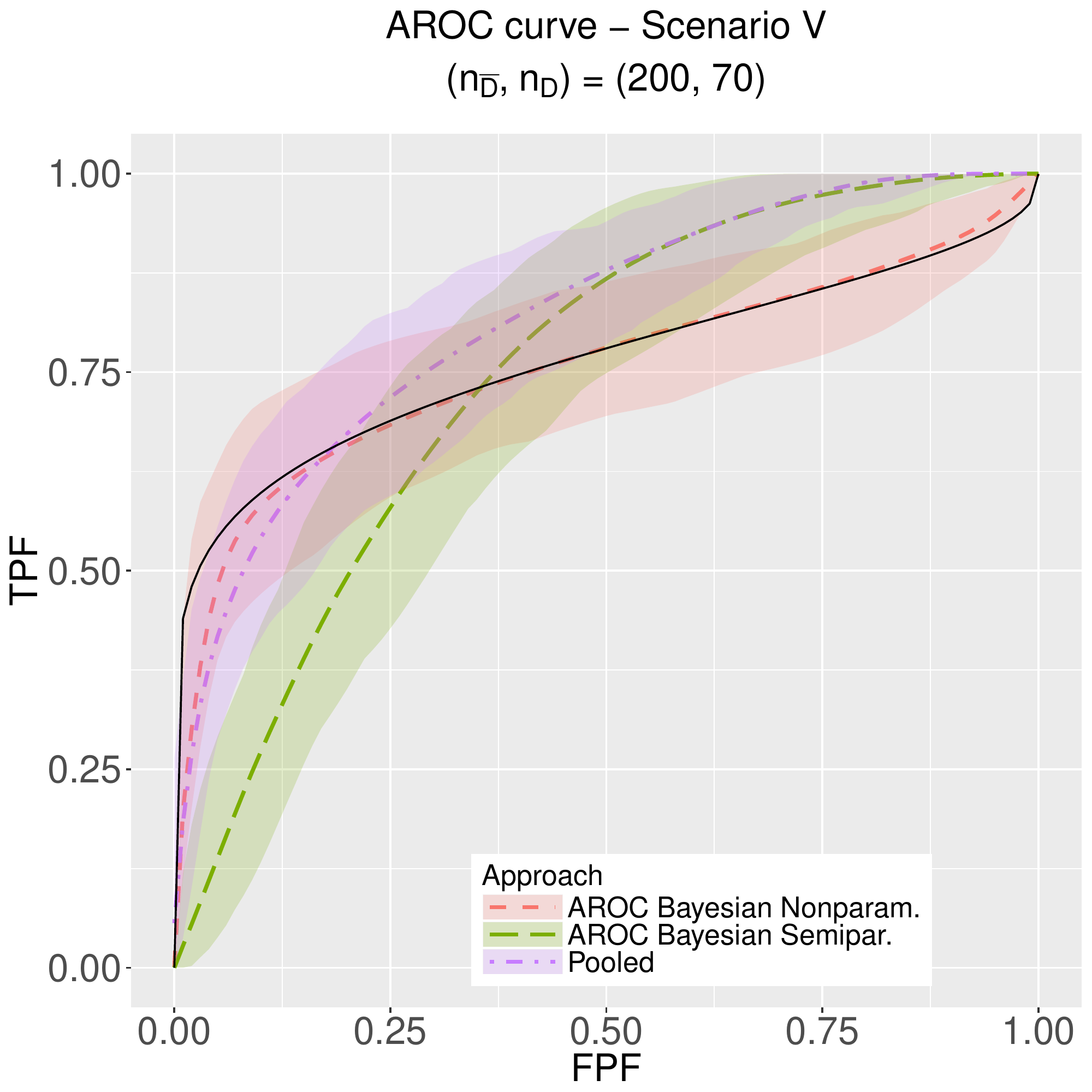}
		\includegraphics[height=4.5cm, page = 1]{sim_V_200_200_ndx_5_bs.pdf}
		\includegraphics[height=4.5cm, page = 1]{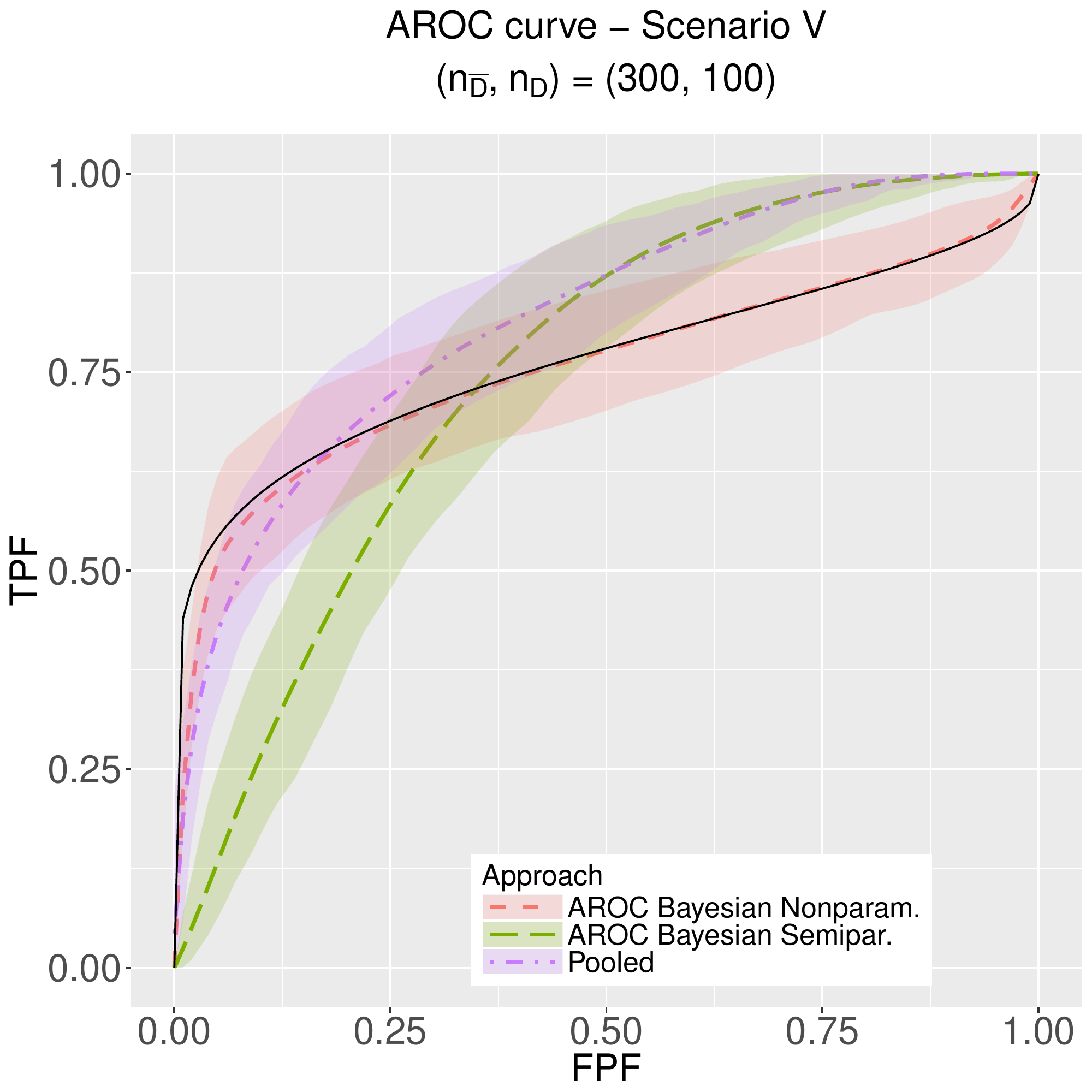}}
\end{center}
		 \caption{Scenario V and $K=4$: true (solid black line) and average value of 100 simulated datasets (dashed lines) of the posterior mean (for the Bayesian estimators) of the covariate adjusted ROC curve/pooled ROC curve for the different approaches under consideration and sample sizes.}
		\label{sim_ndx_5_V}
\end{figure}

\begin{figure}[H]
  \begin{center}
    	\subfigure{
		\includegraphics[height=4.5cm, page = 1]{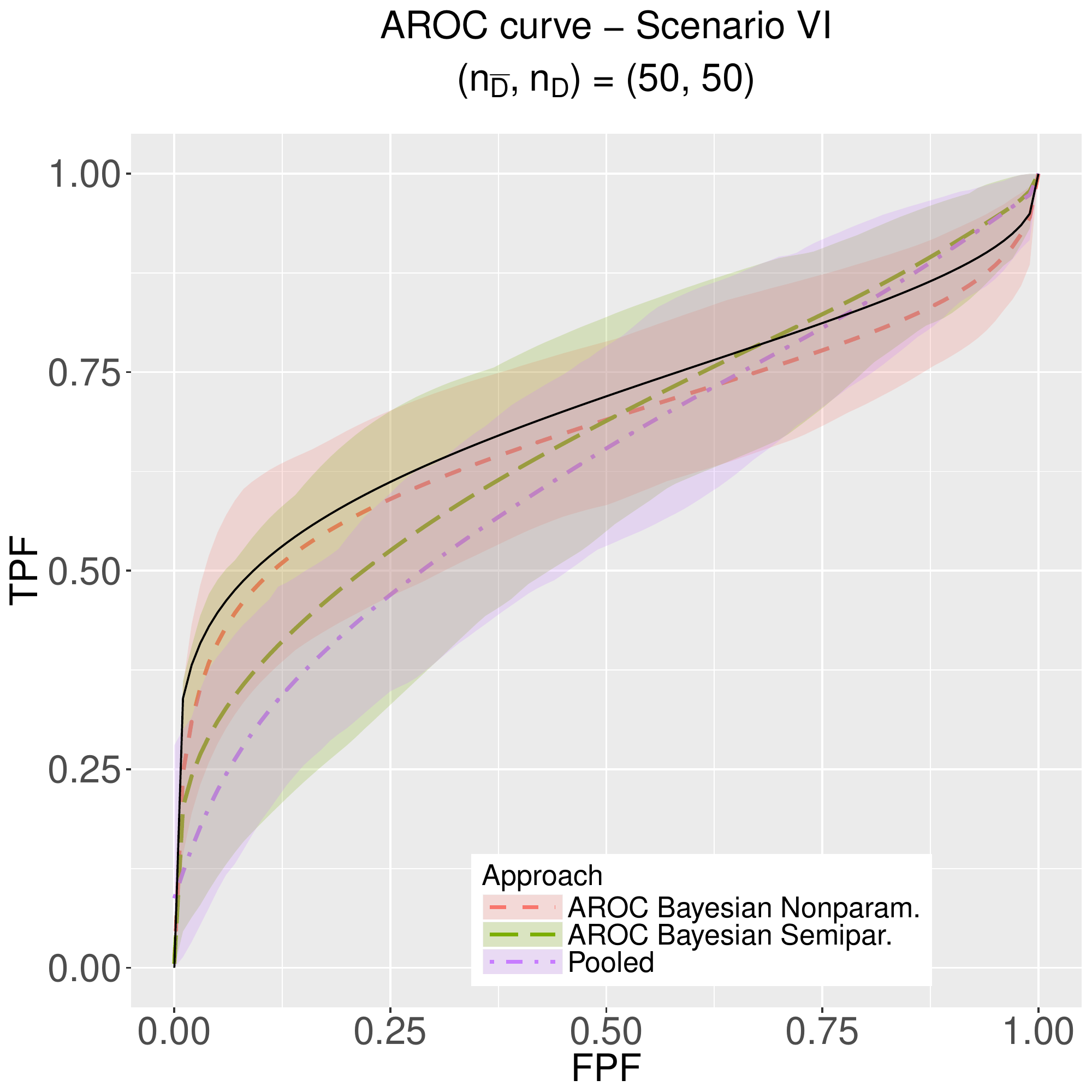}
		\includegraphics[height=4.5cm, page = 1]{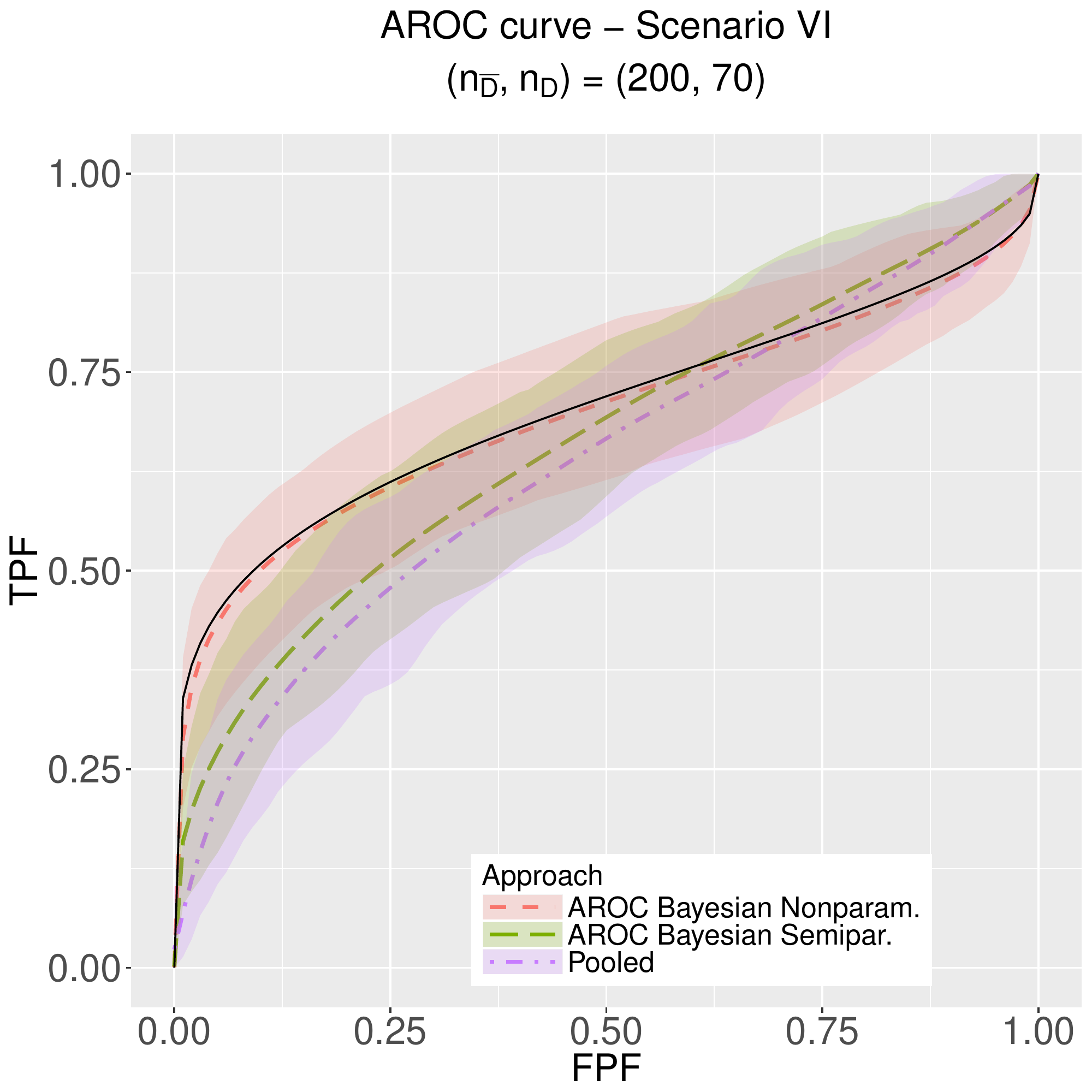}
		\includegraphics[height=4.5cm, page = 1]{sim_VI_200_200_ndx_5_bs.pdf}
		\includegraphics[height=4.5cm, page = 1]{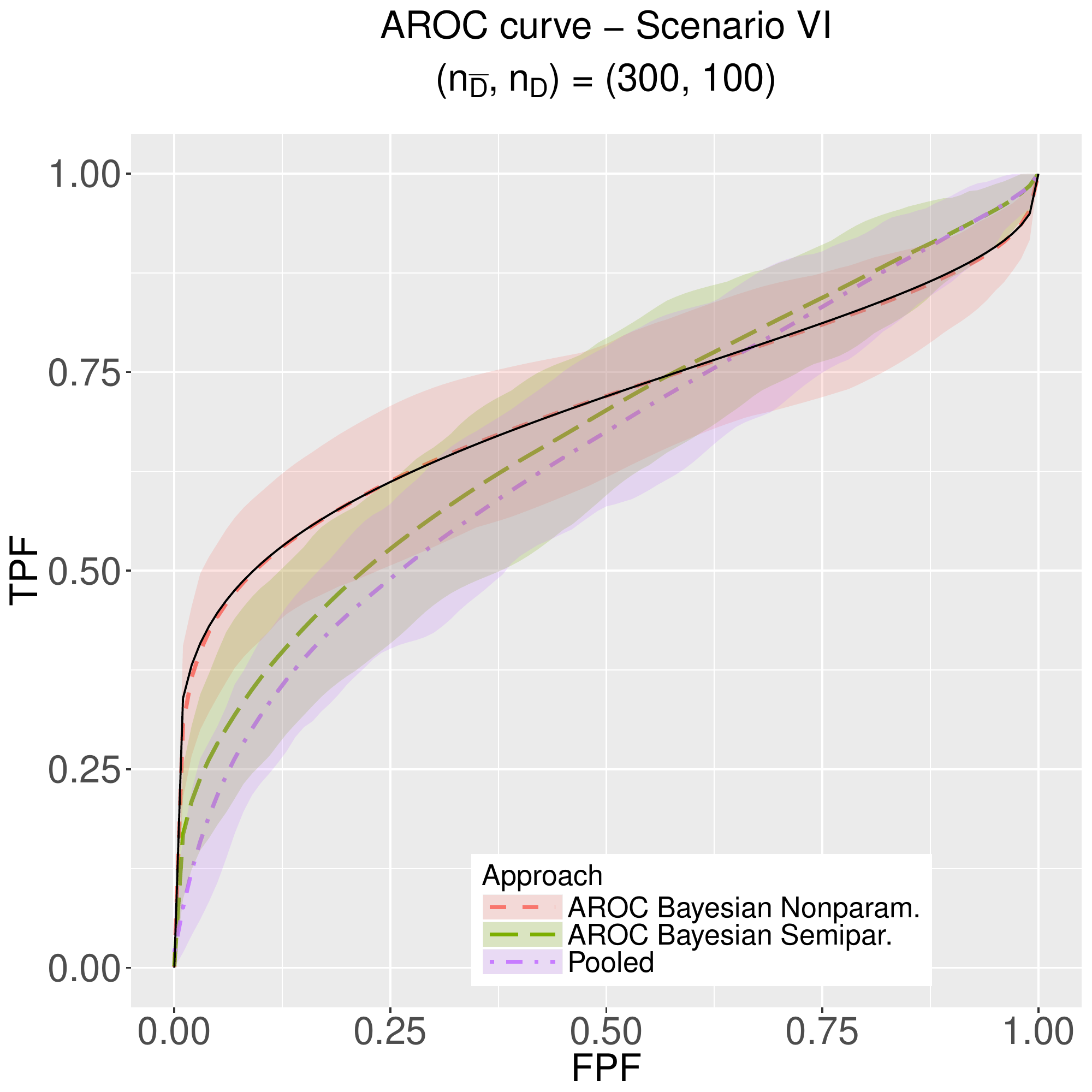}}
\end{center}
		 \caption{Scenario VI and $K=4$: true (solid black line) and average value of 100 simulated datasets (dashed lines) of the posterior mean (for the Bayesian estimators) of the covariate adjusted ROC curve/pooled ROC curve for the different approaches under consideration and sample sizes.}
		\label{sim_ndx_5_VI}
\end{figure}

\begin{figure}[H]
    \begin{center}
    	\subfigure[FPF = 0.1]{
		\includegraphics[height=4.5cm, page = 1]{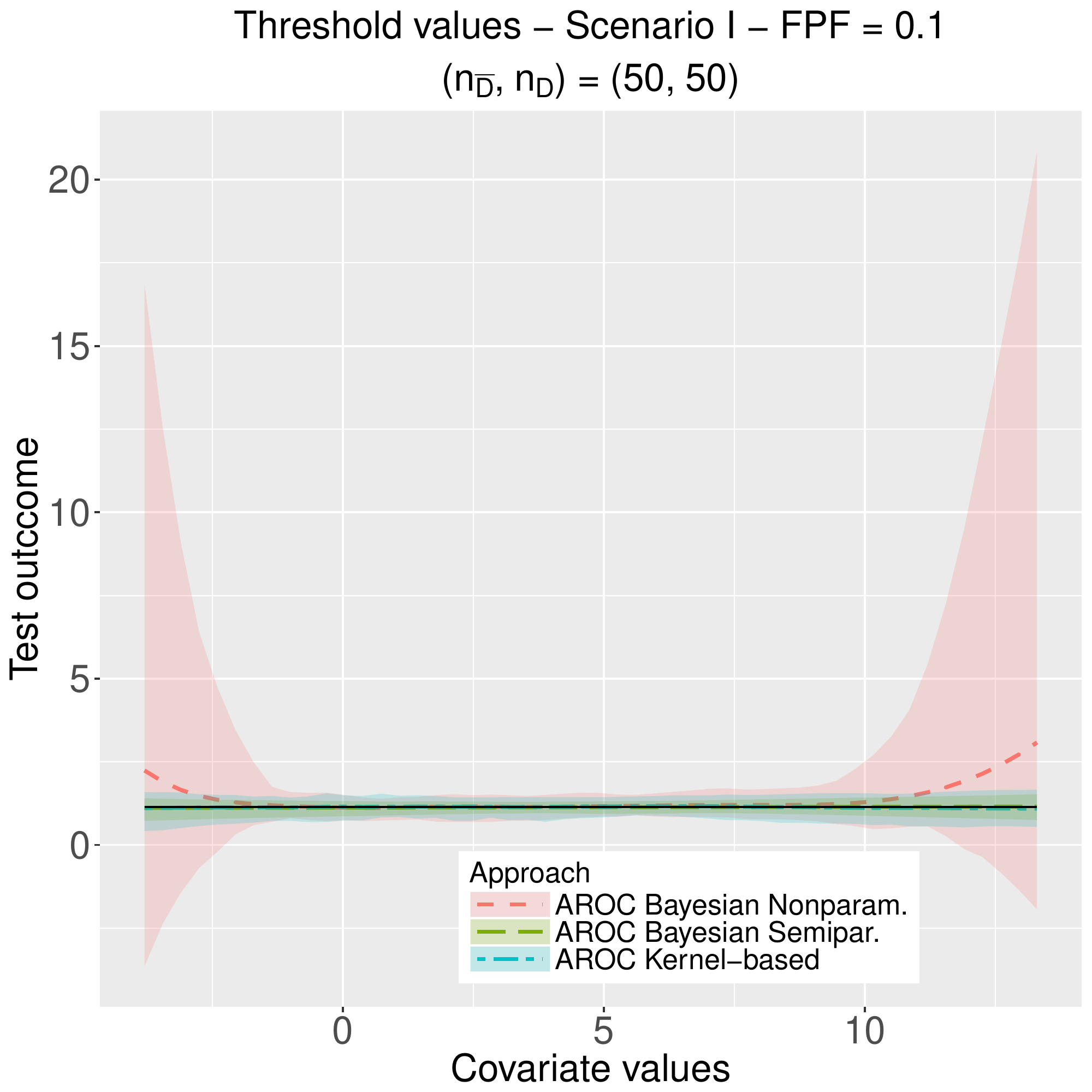}
		\includegraphics[height=4.5cm, page = 1]{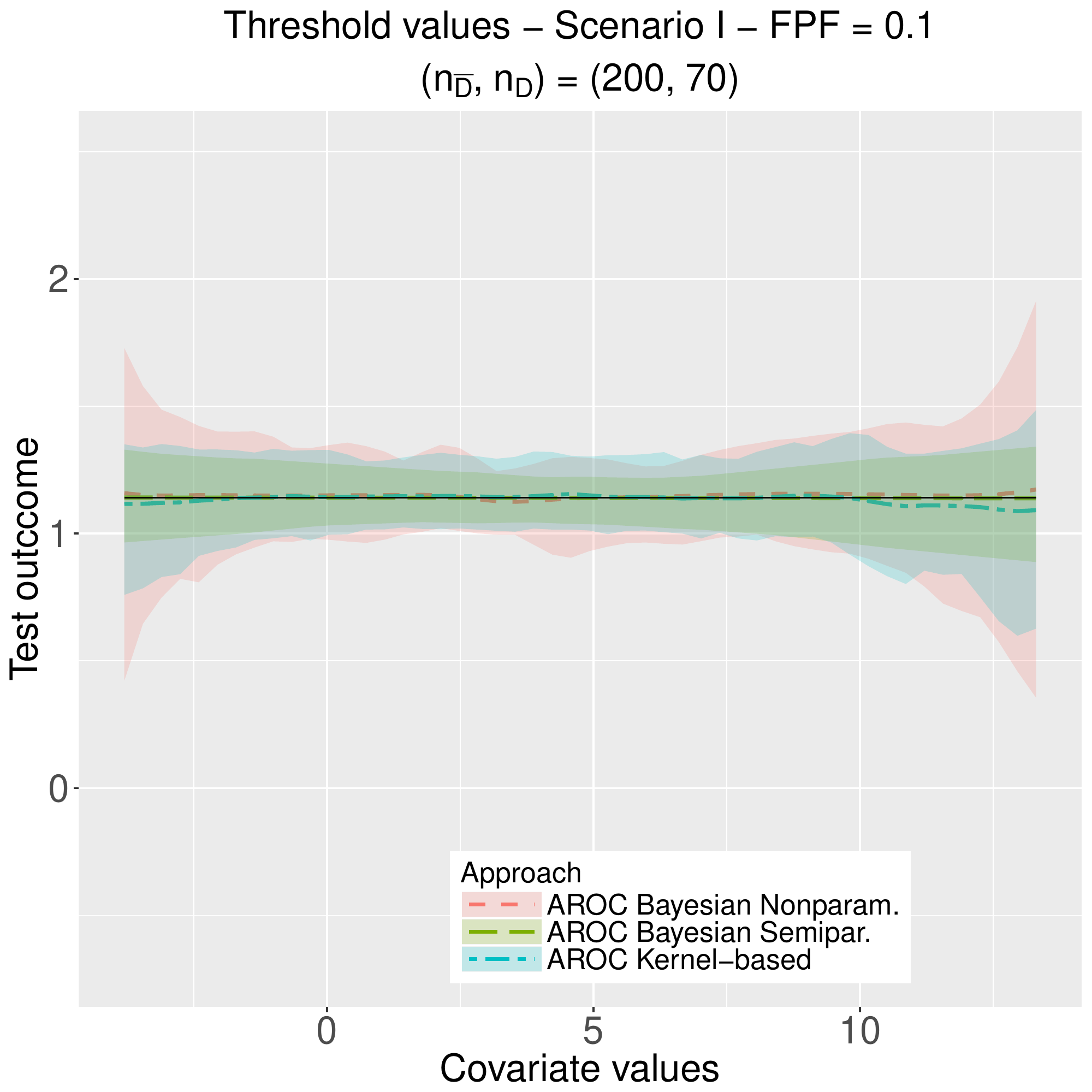}
		\includegraphics[height=4.5cm, page = 1]{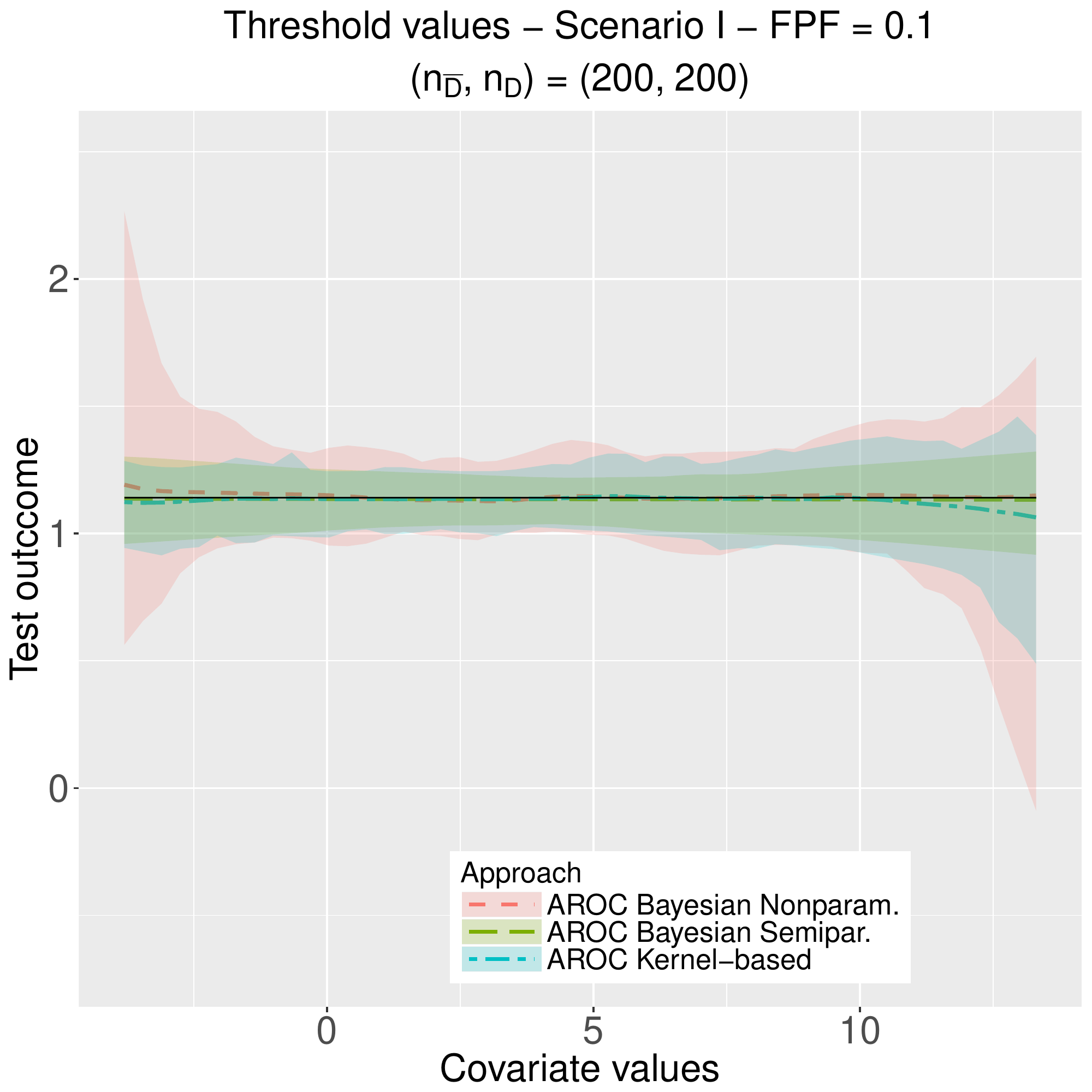}
		\includegraphics[height=4.5cm, page = 1]{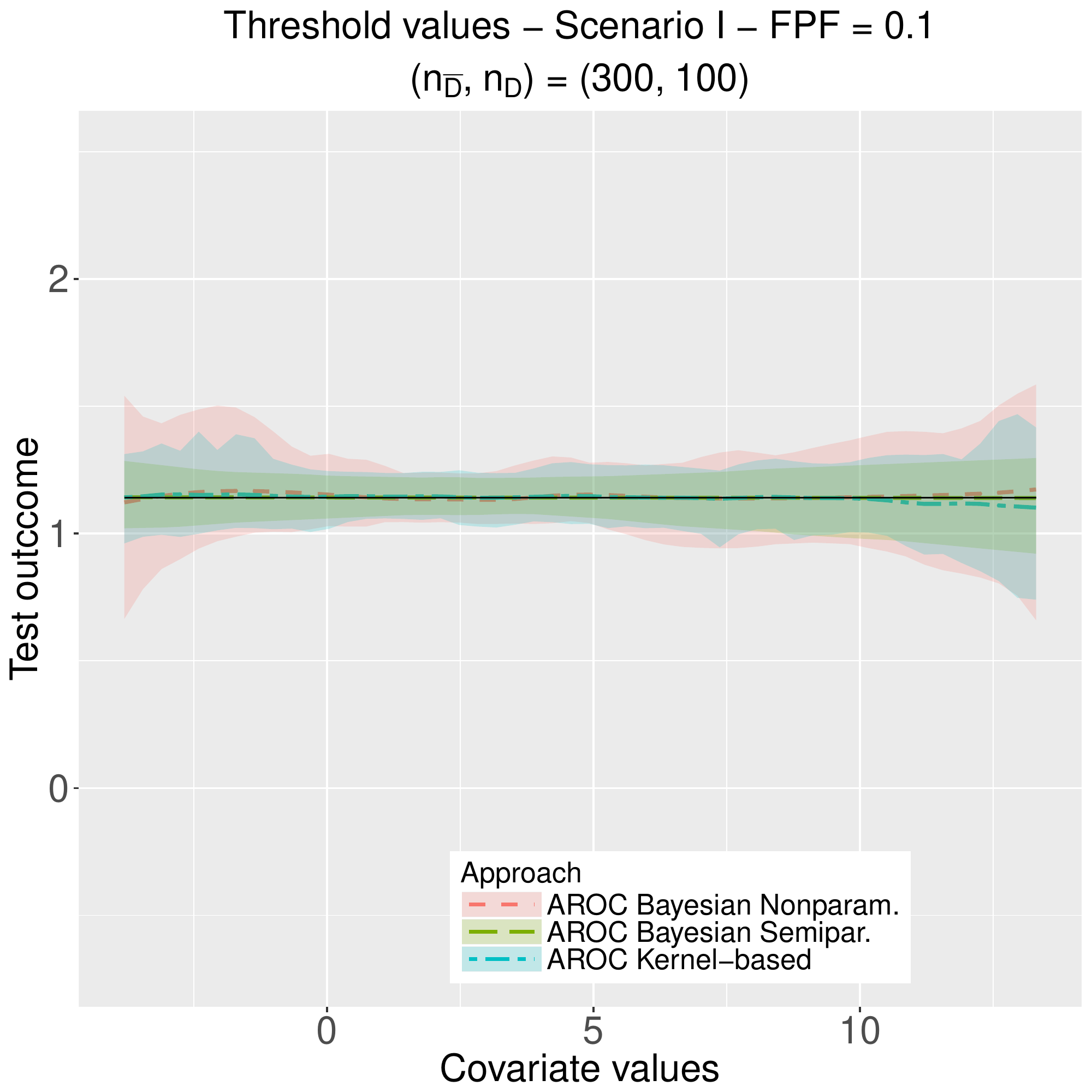}} \vspace{0.3cm}
		\subfigure[FPF = 0.3]{
		\includegraphics[height=4.4cm, page = 2]{sim_thresholds_I_50_50_ndx_5_bs.pdf}
		\includegraphics[height=4.4cm, page = 2]{sim_thresholds_I_200_70_ndx_5_bs.pdf}
		\includegraphics[height=4.4cm, page = 2]{sim_thresholds_I_200_200_ndx_5_bs.pdf}
		\includegraphics[height=4.4cm, page = 2]{sim_thresholds_I_300_100_ndx_5_bs.pdf}}
	\end{center}
		 \caption{Scenario I and $K=4$: true (solid black line) and average value of 100 simulated datasets (dashed lines) of the posterior mean (for the Bayesian estimators) of the thresholds used for defining a positive test result. The shaded area are bands constructed using the pointwise $2.5\%$ and $97.5\%$ quantiles across simulations.}
		\label{thresholds_sim_ndx_5_I}
\end{figure}

\begin{figure}[H]
    \begin{center}
    	\subfigure[FPF = 0.1]{
		\includegraphics[height=4.5cm, page = 1]{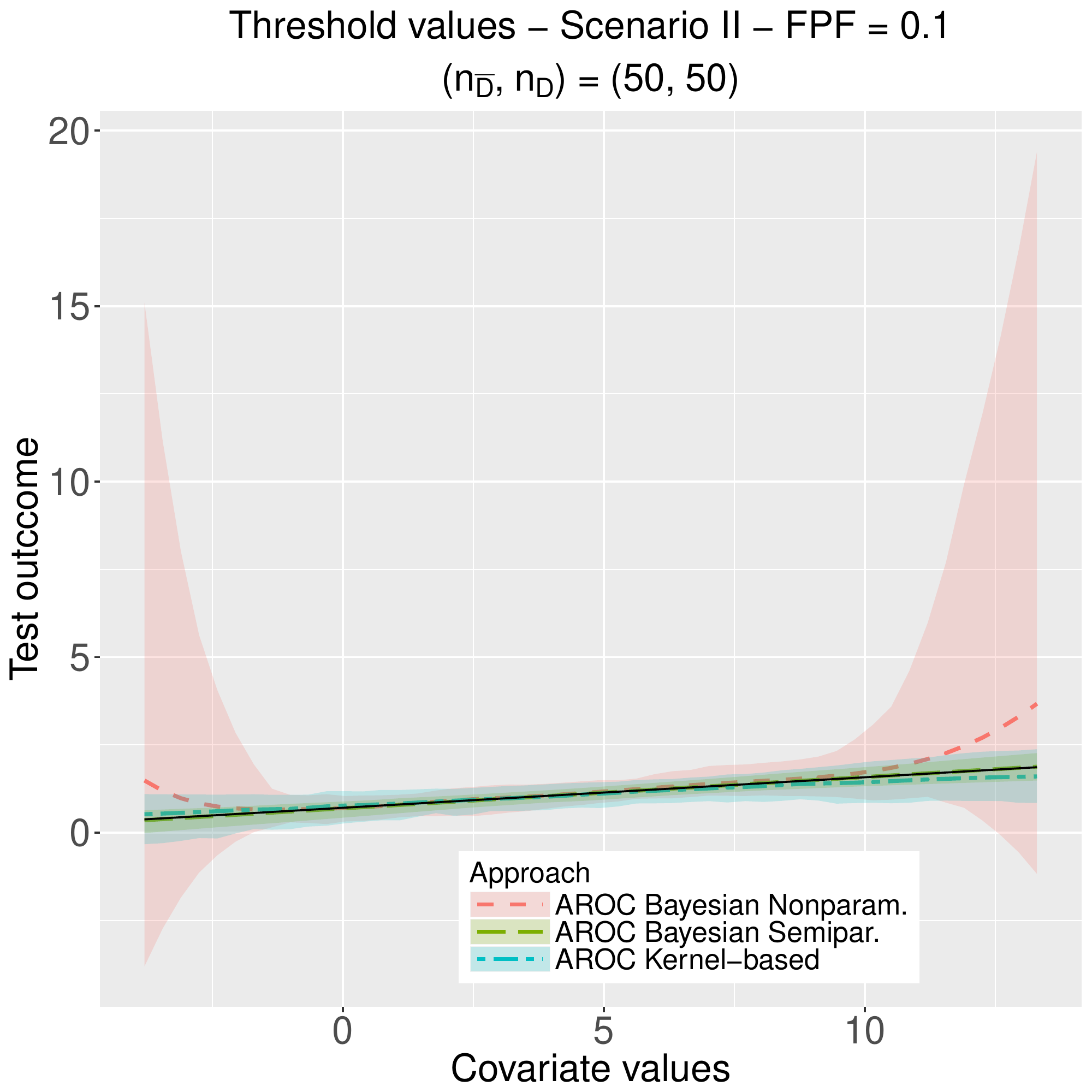}
		\includegraphics[height=4.5cm, page = 1]{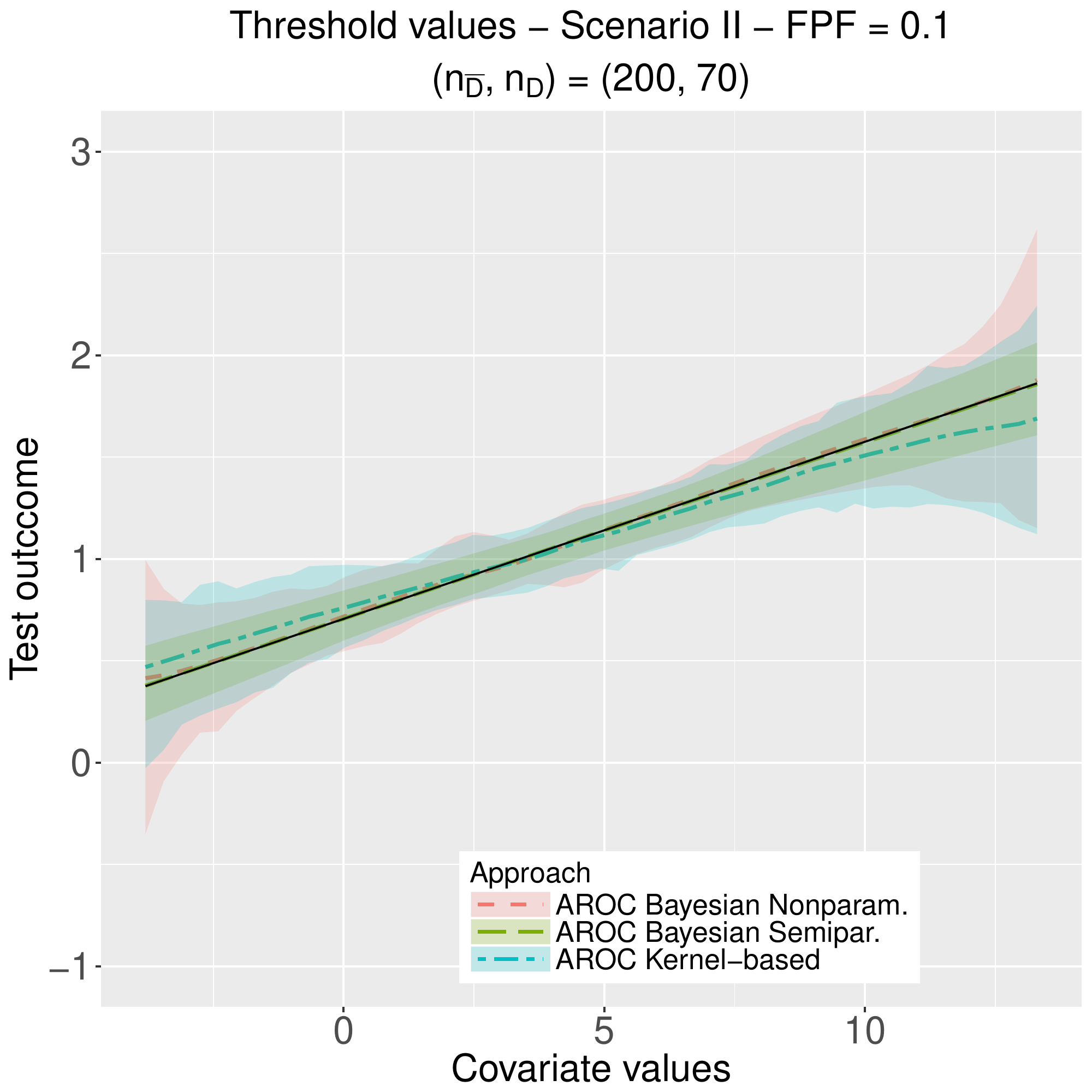}
		\includegraphics[height=4.5cm, page = 1]{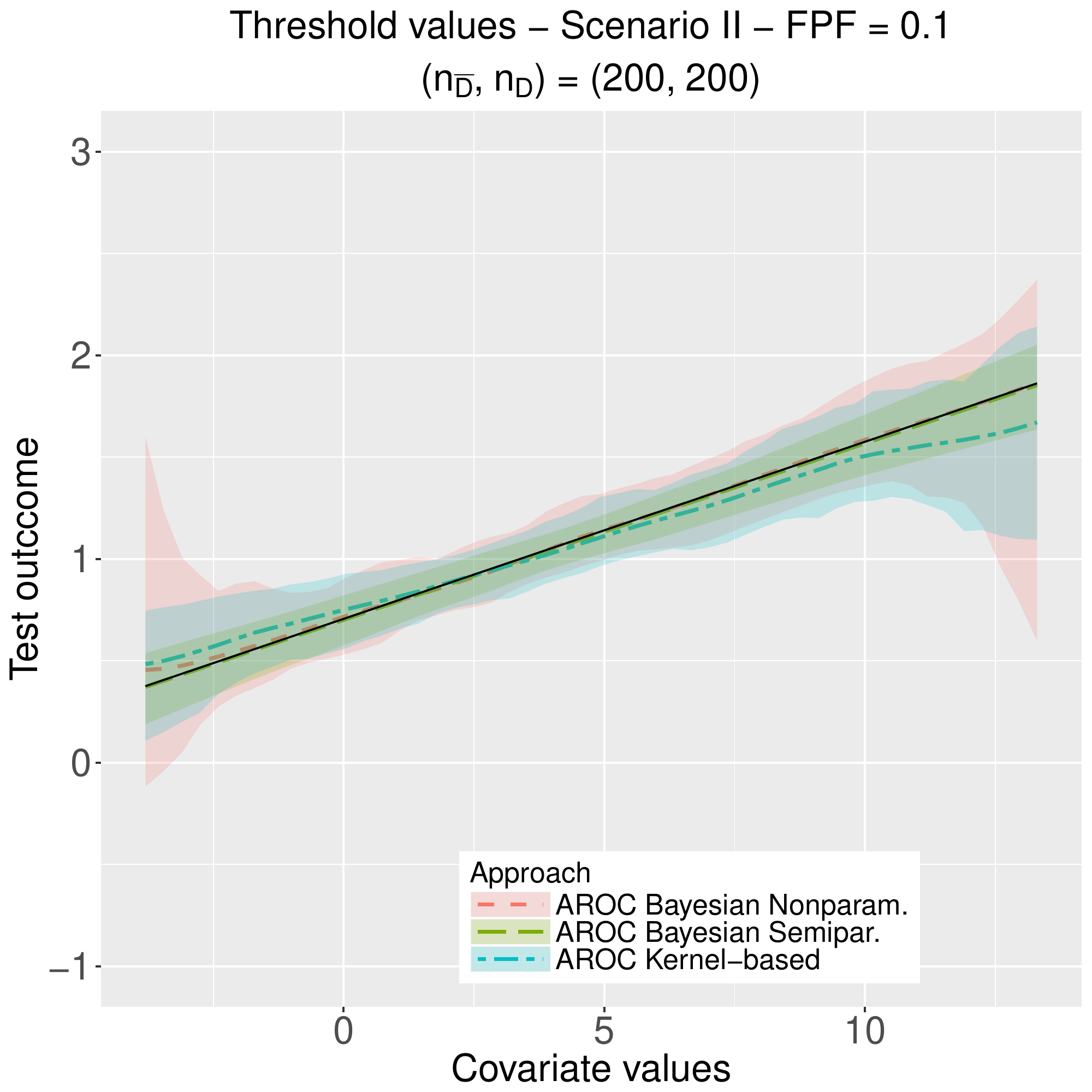}
		\includegraphics[height=4.5cm, page = 1]{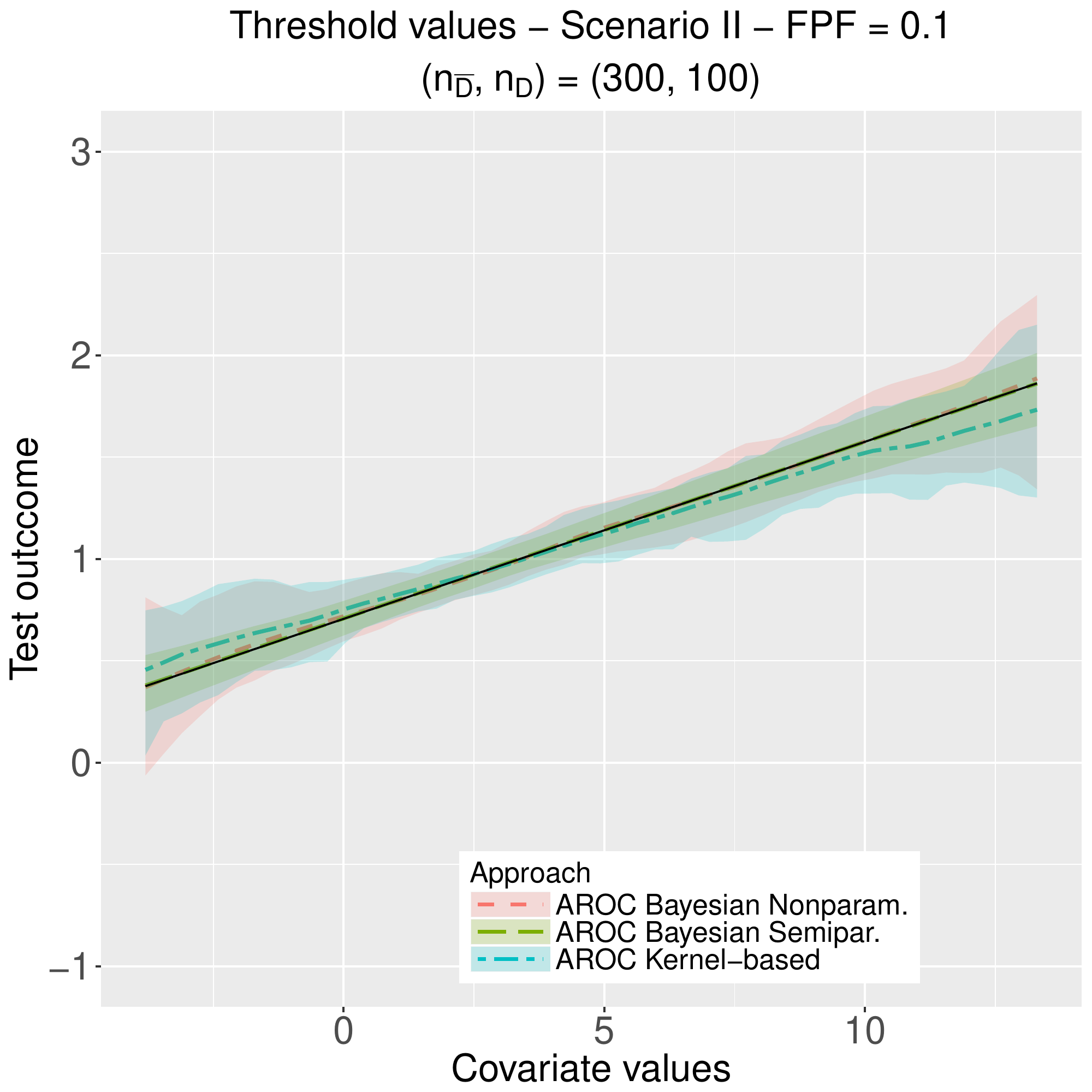}} \vspace{0.3cm}
		\subfigure[FPF = 0.3]{
		\includegraphics[height=4.4cm, page = 2]{sim_thresholds_II_50_50_ndx_5_bs.pdf}
		\includegraphics[height=4.4cm, page = 2]{sim_thresholds_II_200_70_ndx_5_bs.pdf}
		\includegraphics[height=4.4cm, page = 2]{sim_thresholds_II_200_200_ndx_5_bs.pdf}
		\includegraphics[height=4.4cm, page = 2]{sim_thresholds_II_300_100_ndx_5_bs.pdf}}
	\end{center}
		 \caption{Scenario II and $K=4$: true (solid black line) and average value of 100 simulated datasets (dashed lines) of the posterior mean (for the Bayesian estimators) of the thresholds used for defining a positive test result. The shaded area are bands constructed using the pointwise $2.5\%$ and $97.5\%$ quantiles across simulations.}
		\label{thresholds_sim_ndx_5_II}
\end{figure}

\begin{figure}[H]
    \begin{center}
    	\subfigure[FPF = 0.1]{
		\includegraphics[height=4.5cm, page = 1]{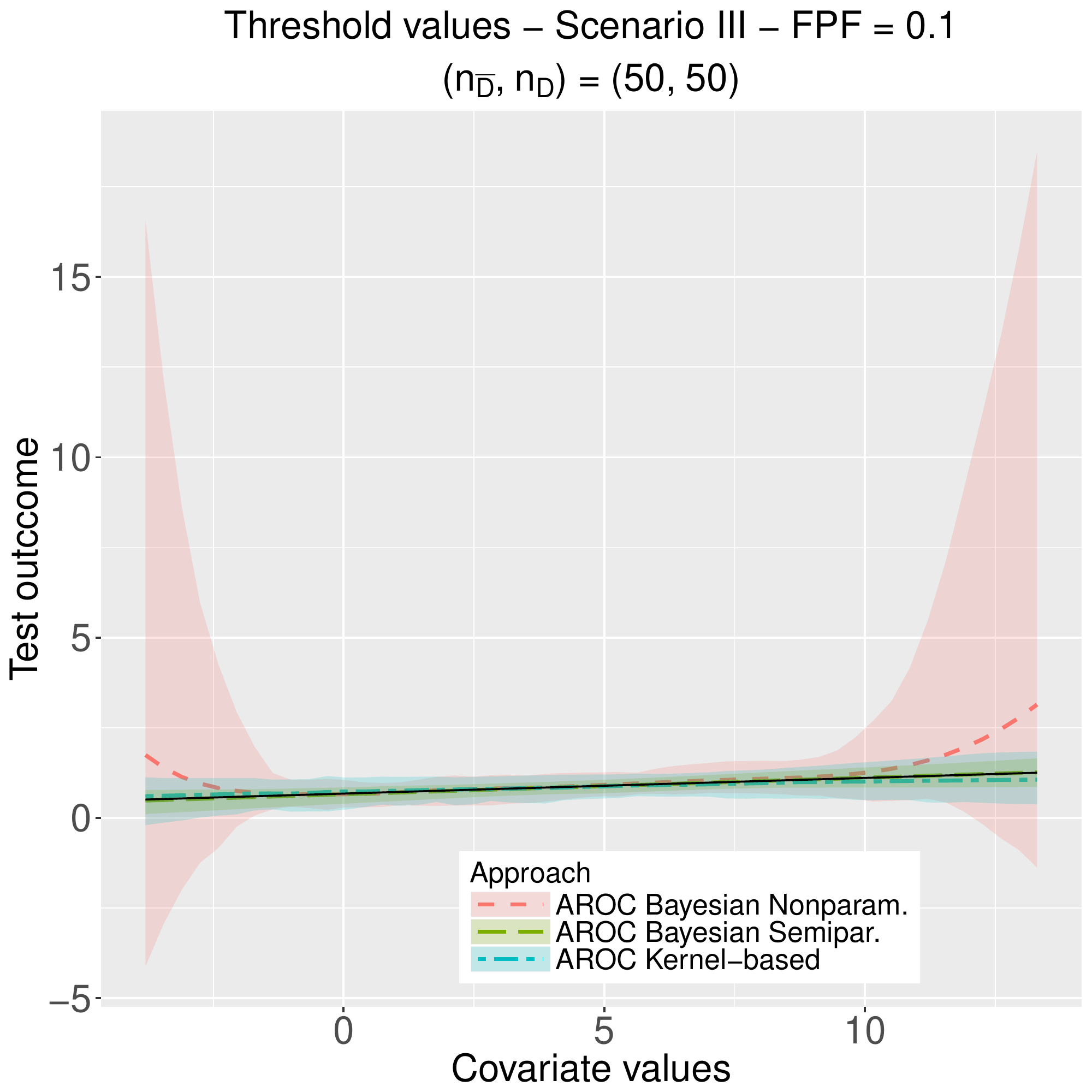}
		\includegraphics[height=4.5cm, page = 1]{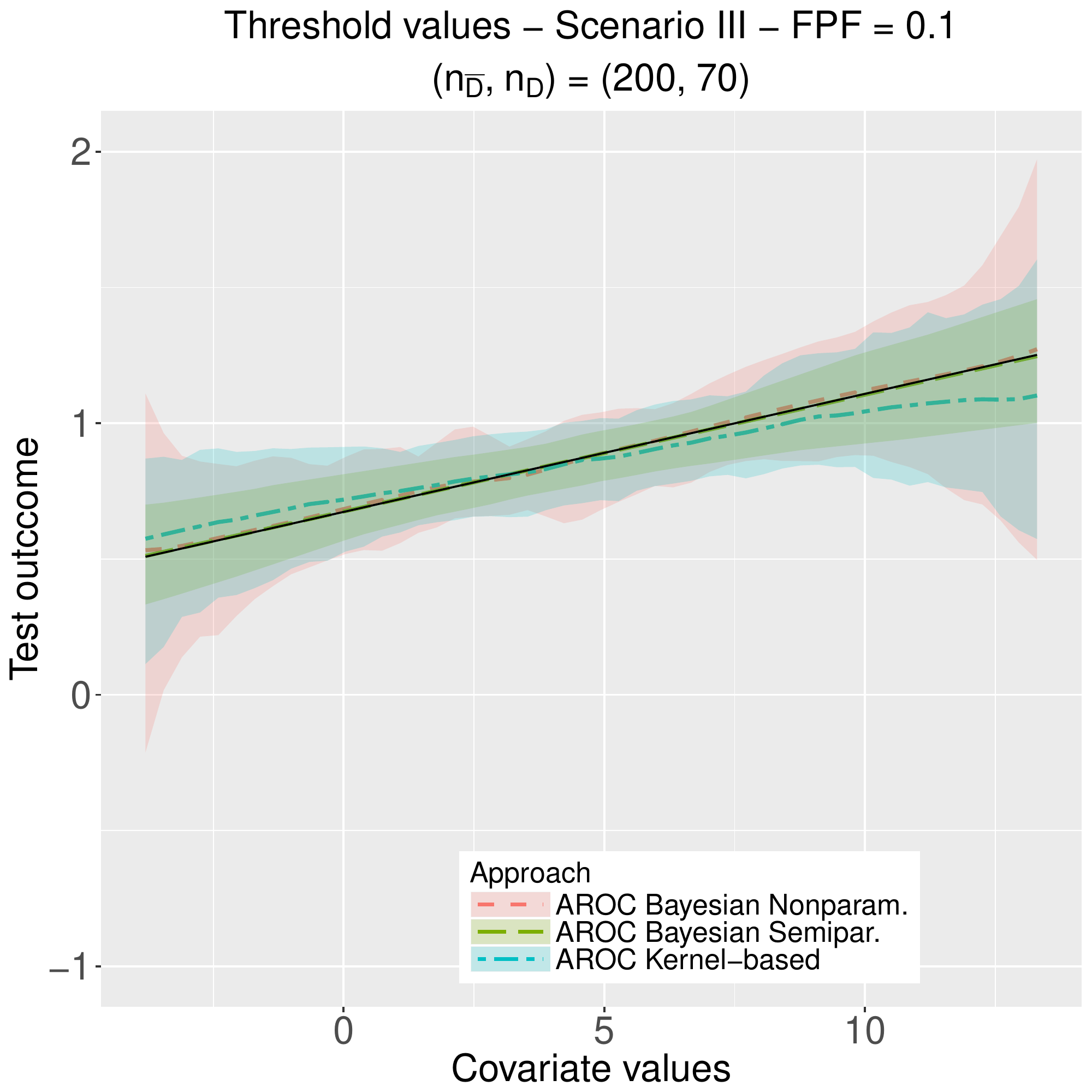}
		\includegraphics[height=4.5cm, page = 1]{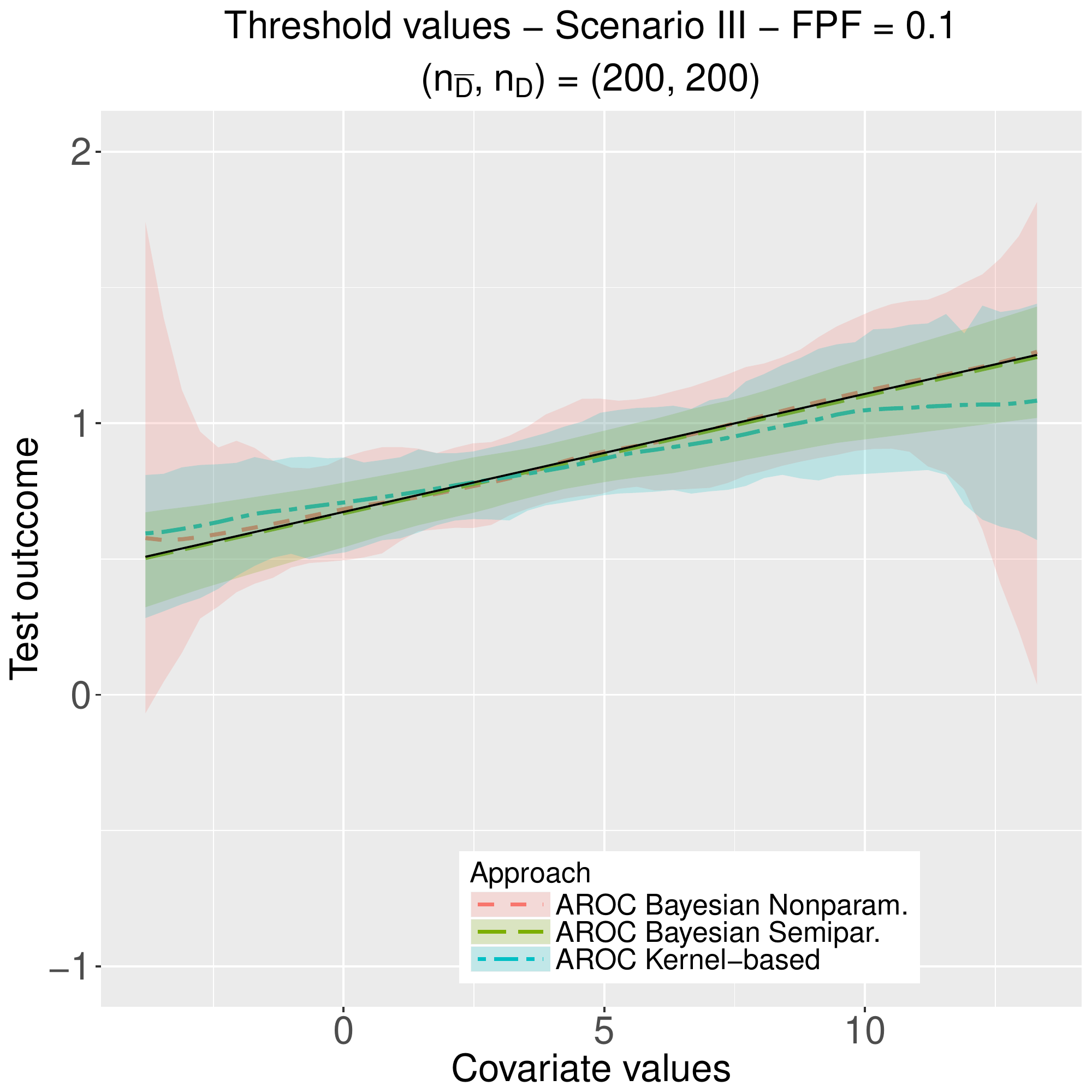}
		\includegraphics[height=4.5cm, page = 1]{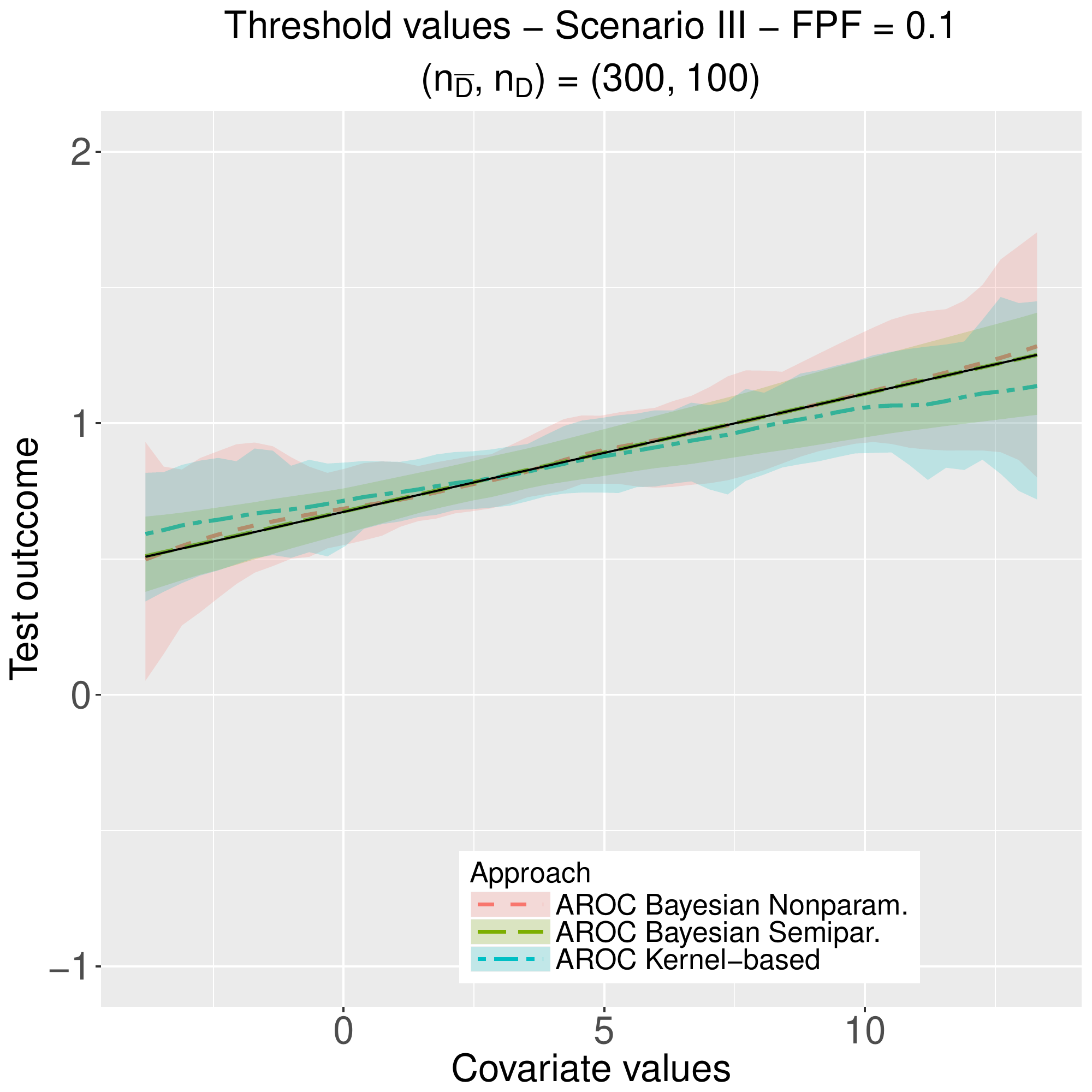}} \vspace{0.3cm}
		\subfigure[FPF = 0.3]{
		\includegraphics[height=4.4cm, page = 2]{sim_thresholds_III_50_50_ndx_5_bs.pdf}
		\includegraphics[height=4.4cm, page = 2]{sim_thresholds_III_200_70_ndx_5_bs.pdf}
		\includegraphics[height=4.4cm, page = 2]{sim_thresholds_III_200_200_ndx_5_bs.pdf}
		\includegraphics[height=4.4cm, page = 2]{sim_thresholds_III_300_100_ndx_5_bs.pdf}}
	\end{center}
		 \caption{Scenario III and $K=4$: true (solid black line) and average value of 100 simulated datasets (dashed lines) of the posterior mean (for the Bayesian estimators) of the thresholds used for defining a positive test result. The shaded area are bands constructed using the pointwise $2.5\%$ and $97.5\%$ quantiles across simulations.}
		\label{thresholds_sim_ndx_5_III}
\end{figure}

\begin{figure}[H]
    \begin{center}
    	\subfigure[FPF = 0.1]{
		\includegraphics[height=4.5cm, page = 1]{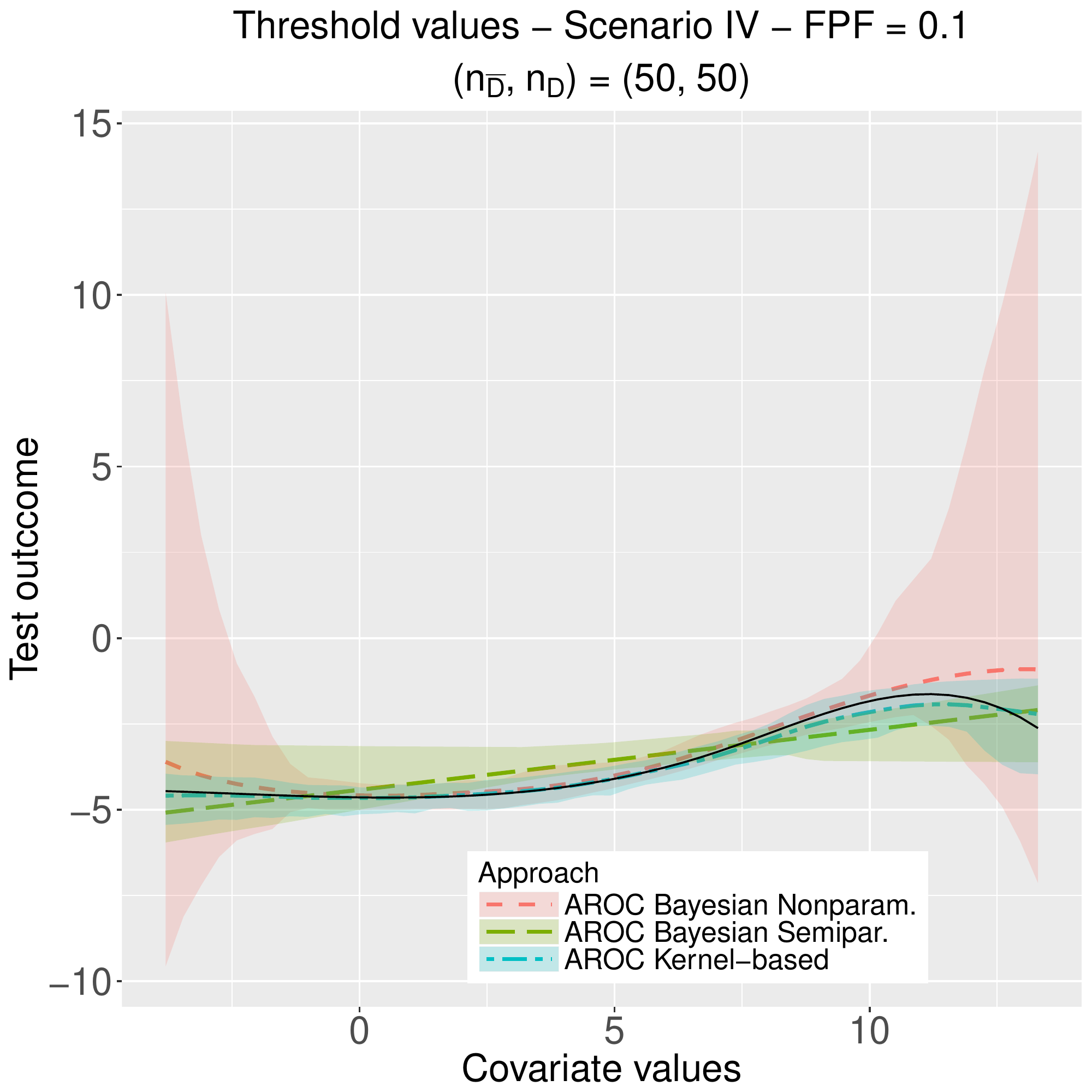}
		\includegraphics[height=4.5cm, page = 1]{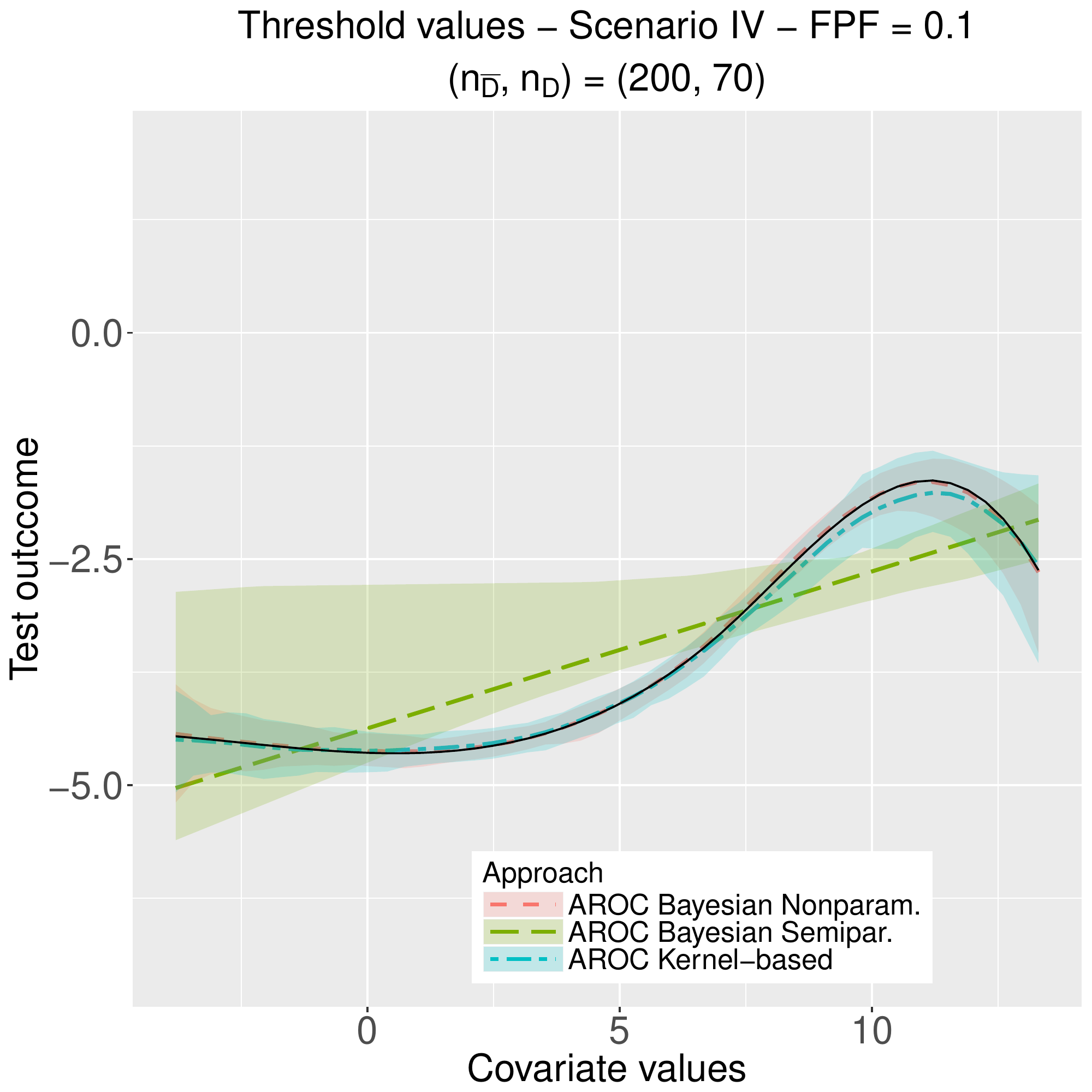}
		\includegraphics[height=4.5cm, page = 1]{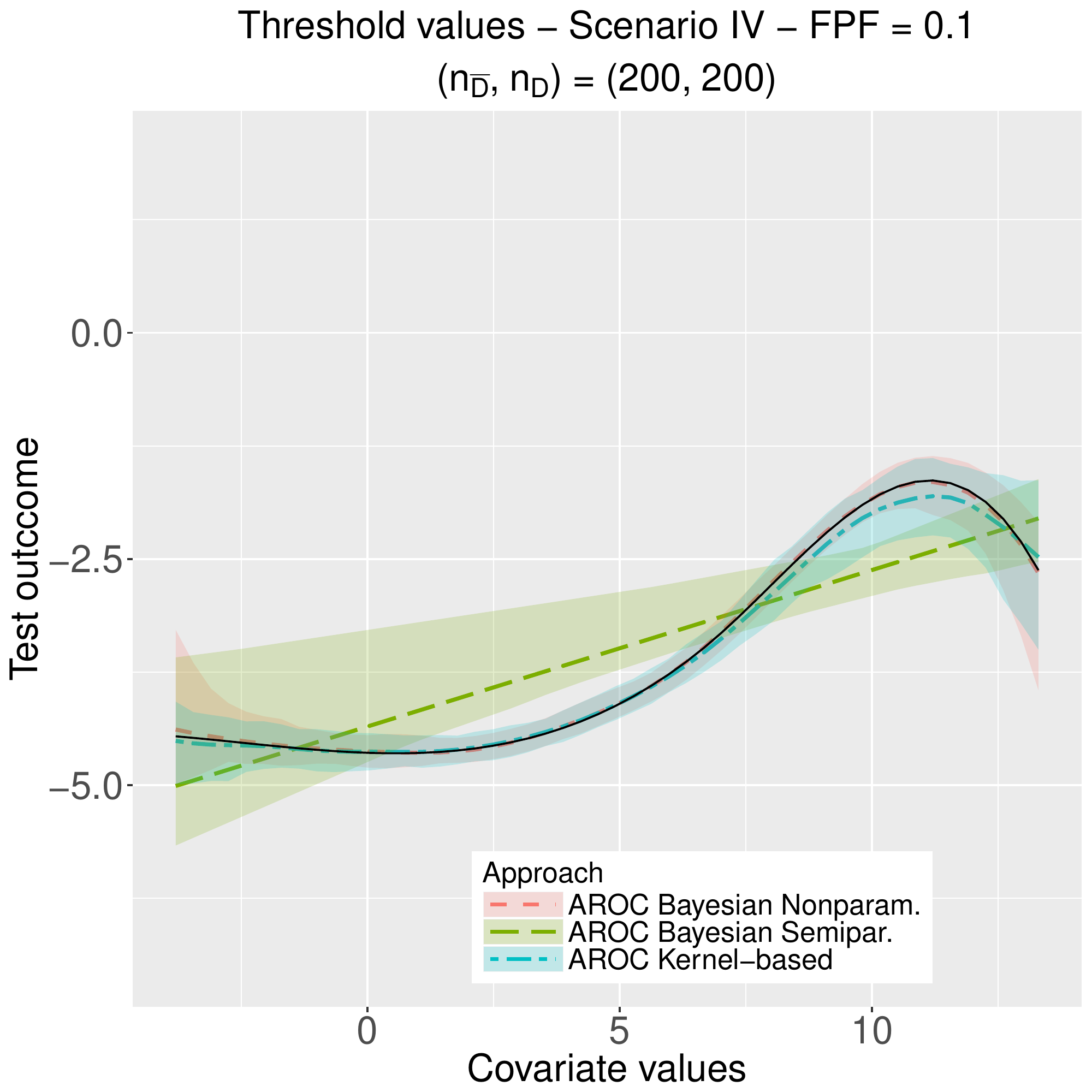}
		\includegraphics[height=4.5cm, page = 1]{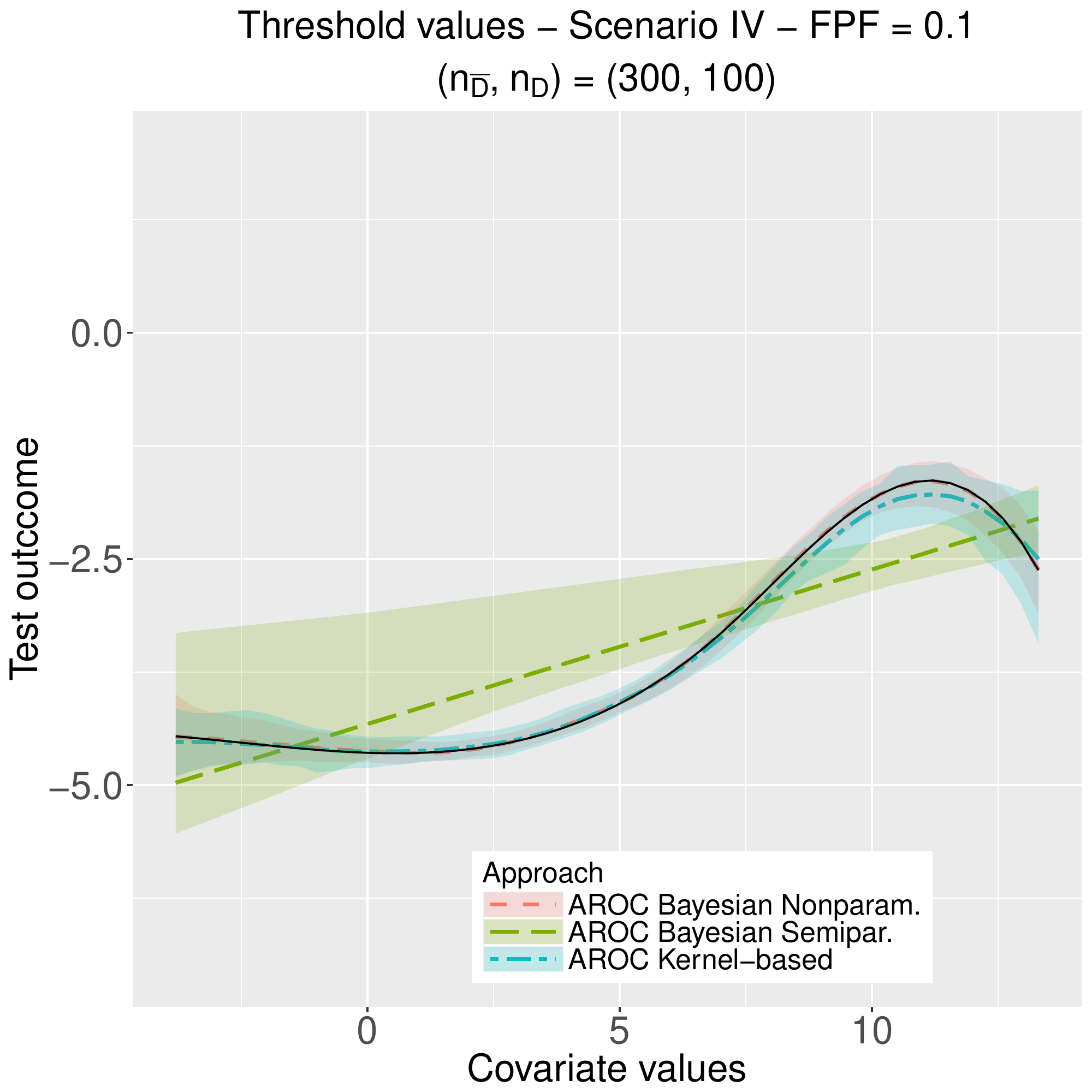}} \vspace{0.3cm}
		\subfigure[FPF = 0.3]{
		\includegraphics[height=4.4cm, page = 2]{sim_thresholds_IV_50_50_ndx_5_bs.pdf}
		\includegraphics[height=4.4cm, page = 2]{sim_thresholds_IV_200_70_ndx_5_bs.pdf}
		\includegraphics[height=4.4cm, page = 2]{sim_thresholds_IV_200_200_ndx_5_bs.pdf}
		\includegraphics[height=4.4cm, page = 2]{sim_thresholds_IV_300_100_ndx_5_bs.pdf}}
	\end{center}
		 \caption{Scenario IV and $K=4$: true (solid black line) and average value of 100 simulated datasets (dashed lines) of the posterior mean (for the Bayesian estimators) of the thresholds used for defining a positive test result. The shaded area are bands constructed using the pointwise $2.5\%$ and $97.5\%$ quantiles across simulations.}
		\label{thresholds_sim_ndx_5_IV}
\end{figure}

\subsection{Comparison results between no interior knots ($K=0$) and four interior knots ($K=4$)\label{sim_k_0}}
\begin{table}[H]
\caption{Average (standard deviation) ($\times$ 100), across simulations, of the empirical root mean squared error for the Bayesian nonparametric approach proposed in this paper for four interior knots ($K = 4$) and no interior knots ($K = 0$). The results are presented for each of the simulated scenarios and sample sizes.}\label{MSE_K_0}
 \begin{center}
 \footnotesize	
	\begin{tabular}{cccccc}
	& & \multicolumn{4}{c}{Sample size}\\
	& & \multicolumn{4}{c}{$(n_{\bar{D}},n_{D})$}\\\hline
	Scenario & Approach & $(50,50)$ & $(200,70)$ & $(200,200)$ &  $(300,100)$ \\\hline
	\multirow{2}{*}{I} &	
Bayesian nonparametric ($K = 0$) & 6.266 (3.443) & 4.443 (2.275) & 3.250 (1.585) & 4.197 (2.150)\\
& Bayesian nonparametric ($K = 4$) & 6.013 (3.494) & 4.438 (2.296) & 3.174 (1.644) & 4.070 (2.121)\\\hline
	\multirow{2}{*}{II} &	
Bayesian nonparametric ($K = 0$) & 6.212 (3.422) & 4.434 (2.278) & 3.237 (1.581) & 4.194 (2.154)\\
& Bayesian nonparametric ($K = 4$) & 5.990 (3.498) & 4.434 (2.293) & 3.164 (1.630) & 4.071 (2.124)\\\hline
	\multirow{2}{*}{III} &	
Bayesian nonparametric ($K = 0$) & 6.124 (3.383) & 4.416 (2.297) & 3.292 (1.524) & 4.205 (2.178)\\
& Bayesian nonparametric ($K = 4$) & 6.021 (3.454) & 4.405 (2.324) & 3.196 (1.592) & 4.118 (2.133)\\ \hline
	\multirow{2}{*}{IV} &	
Bayesian nonparametric ($K = 0$) & 6.218 (2.986) & 4.355 (2.260) & 3.212 (1.489) & 4.195 (1.943)\\
 & Bayesian nonparametric ($K = 4$) & 6.060 (3.249) & 3.973 (2.288) & 2.914 (1.539) & 3.888 (2.085)\\\hline
	\multirow{2}{*}{V} &	
Bayesian nonparametric ($K = 0$) & 5.521 (2.282) & 4.109 (2.206) & 2.588 (1.429) & 3.785 (1.679)\\
 & Bayesian nonparametric ($K = 4$) & 6.361 (2.832) & 5.553 (1.944) & 4.394 (1.316) & 4.646 (1.396)\\ \hline
	\multirow{2}{*}{VI} &	
Bayesian nonparametric ($K = 0$) & 5.463 (3.024) & 4.497 (2.182) & 3.158 (1.669) & 4.021 (2.135)\\
 & Bayesian nonparametric ($K = 4$) & 5.673 (3.056) & 4.304 (2.180) & 3.063 (1.748) & 3.781 (2.120)\\ \hline
	\end{tabular}
 \end{center}
\end{table}

\begin{table}[H]
\caption{Bias (standard deviation) ($\times$ 100), across simulations, of the AAUC for the Bayesian nonparametric approach proposed in this paper for four interior knots ($K = 4$) and no interior knots ($K = 0$). The results are presented for each of the simulated scenarios and sample sizes.}\label{AUC_K_0}
 \begin{center}
 \footnotesize	
	\begin{tabular}{cccccc}
	& & \multicolumn{4}{c}{Sample size}\\
	& & \multicolumn{4}{c}{$(n_{\bar{D}},n_{D})$}\\\hline
	Scenario & Approach & $(50,50)$ & $(200,70)$ & $(200,200)$ &  $(300,100)$ \\\hline
	\multirow{2}{*}{I} &	
Bayesian nonparametric ($K = 0$) & -0.045 (6.169) & -0.149 (4.235) & 0.507 (3.049) & 0.540 (4.015)\\
 & Bayesian nonparametric ($K = 4$) & -0.620 (6.140) & -0.228 (4.293) & 0.383 (3.104) & 0.474 (3.944)\\\hline
	\multirow{2}{*}{II} &	
Bayesian nonparametric ($K = 0$) & -0.081 (6.101) & -0.161 (4.236) & 0.501 (3.050) & 0.530 (4.018)\\
 & Bayesian nonparametric ($K = 4$) & -0.625 (6.114) & -0.235 (4.294) & 0.380 (3.090) & 0.472 (3.946)\\\hline
	\multirow{2}{*}{III} &	
Bayesian nonparametric ($K = 0$) & -0.128 (6.035) & -0.146 (4.288) & 0.435 (3.045) & 0.514 (3.994)\\
 & Bayesian nonparametric ($K = 4$) & -0.749 (6.110) & -0.212 (4.347) & 0.311 (3.092) & 0.456 (3.926)\\\hline
	\multirow{2}{*}{IV} &	
Bayesian nonparametric ($K = 0$) & -0.422 (5.767) & 0.080 (3.886) & 0.756 (2.848) & 0.312 (3.985)\\
 & Bayesian nonparametric ($K = 4$) & -0.648 (5.769) & -0.071 (3.902) & 0.670 (2.797) & 0.126 (4.004)\\\hline
	\multirow{2}{*}{V} &	
Bayesian nonparametric ($K = 0$) & 0.097 (4.949) & 0.062 (3.889) & 0.200 (2.409) & -0.177 (3.535)\\
 & Bayesian nonparametric ($K = 4$) & -1.616 (4.845) & -0.740 (3.956) & -0.558 (2.370) & -0.723 (3.385)\\ \hline
	\multirow{2}{*}{VI} &	
Bayesian nonparametric ($K = 0$) & -1.878 (5.178) & -0.686 (4.151) & -0.209 (3.115) & -0.106 (3.851)\\
 & Bayesian nonparametric ($K = 4$) & -2.881 (5.090) & -0.759 (4.086) & -0.314 (3.175) & -0.105 (3.733)\\\hline
	\end{tabular}
 \end{center}
\end{table}

\begin{table}[H]
\caption{$95\%$ coverage probabilities for the AROC (average over all FPFs) for the Bayesian nonparametric approach proposed in this paper for four interior knots ($K = 4$) and no interior knots ($K = 0$). The results are presented for each of the simulated scenarios and sample sizes.}\label{covAROC_K_0}
 \begin{center}
 \footnotesize	
	\begin{tabular}{cccccc}
	& & \multicolumn{4}{c}{Sample size}\\
	& & \multicolumn{4}{c}{$(n_{\bar{D}},n_{D})$}\\\hline
	Scenario & Approach & $(50,50)$ & $(200,70)$ & $(200,200)$ &  $(300,100)$ \\\hline
	\multirow{2}{*}{I} &	
Bayesian nonparametric ($K = 0$) & 95.0 & 95.0 & 94.0 & 94.0\\
 & Bayesian nonparametric ($K = 4$) & 96.0 & 95.0 & 95.0 & 95.0\\\hline
	\multirow{2}{*}{II} &	
Bayesian nonparametric ($K = 0$) & 95.0 & 95.0 & 95.0 & 95.0\\
 & Bayesian nonparametric ($K = 4$) & 96.0 & 96.0 & 94.0 & 95.0\\\hline
	\multirow{2}{*}{III} &	
Bayesian nonparametric ($K = 0$) & 95.0 & 95.0 & 95.0 & 94.0\\
 & Bayesian nonparametric ($K = 4$) & 96.0 & 95.0 & 95.0 & 94.0\\\hline
	\multirow{2}{*}{IV} &	
Bayesian nonparametric ($K = 0$) & 95.0 & 94.0 & 92.0 & 93.0\\
 & Bayesian nonparametric ($K = 4$) & 96.0 & 96.0 & 93.0 & 93.0\\\hline
	\multirow{2}{*}{V} &	
Bayesian nonparametric ($K = 0$) & 96.0 & 95.0 & 95.0 & 96.0\\
 & Bayesian nonparametric ($K = 4$) & 97.0 & 95.0 & 95.0 & 96.0\\\hline
	\multirow{2}{*}{VI} &	
Bayesian nonparametric ($K = 0$) & 97.0 & 96.0 & 94.0 & 94.0\\
 & Bayesian nonparametric ($K = 4$) & 98.0 & 97.0 & 96.0 & 95.0\\\hline
	\end{tabular}
 \end{center}
\end{table}

\begin{table}[H]
\caption{$95\%$ coverage probabilities for the AAUC for the Bayesian nonparametric approach proposed in this paper for four interior knots ($K = 4$) and no interior knots ($K = 0$). The results are presented for each of the simulated scenarios and sample sizes.}\label{covAAROC_K_0}
 \begin{center}
 \footnotesize	
	\begin{tabular}{cccccc}
	& & \multicolumn{4}{c}{Sample size}\\
	& & \multicolumn{4}{c}{$(n_{\bar{D}},n_{D})$}\\\hline
	Scenario & Approach & $(50,50)$ & $(200,70)$ & $(200,200)$ &  $(300,100)$ \\\hline
	\multirow{2}{*}{I} &	
	Bayesian nonparametric ($K = 0$) & 93.0 & 96.0 & 94.0 & 94.0\\
 & Bayesian nonparametric ($K = 4$) & 94.0 & 96.0 & 94.0 & 94.0\\ \hline
	\multirow{2}{*}{II} &	
Bayesian nonparametric ($K = 0$) & 93.0 & 96.0 & 93.0 & 94.0\\
 & Bayesian nonparametric ($K = 4$) & 94.0 & 97.0 & 93.0 & 94.0\\ \hline
	\multirow{2}{*}{III} &	
Bayesian nonparametric ($K = 0$) & 94.0 & 92.0 & 94.0 & 95.0\\
 & Bayesian nonparametric ($K = 4$) & 94.0 & 95.0 & 93.0 & 94.0\\ \hline
	\multirow{2}{*}{IV} &	
Bayesian nonparametric ($K = 0$) & 95.0 & 97.0 & 91.0 & 92.0\\
 & Bayesian nonparametric ($K = 4$) & 92.0 & 96.0 & 93.0 & 93.0\\ \hline
	\multirow{2}{*}{V} &	
Bayesian nonparametric ($K = 0$) & 97.0 & 97.0 & 95.0 & 96.0\\
 & Bayesian nonparametric ($K = 4$) & 98.0 & 96.0 & 97.0 & 97.0\\ \hline
	\multirow{2}{*}{VI} &	
Bayesian nonparametric ($K = 0$) & 95.0 & 95.0 & 91.0 & 95.0\\
 & Bayesian nonparametric ($K = 4$) & 98.0 & 96.0 & 94.0 & 96.0\\ \hline
	\end{tabular}
 \end{center}
\end{table}

\begin{table}[H]
\caption{For $K=0$: percentage of time the WAIC/LPML is smaller/higher for the Bayesian normal linear (BNL) model against our B-splines dependent Dirichlet process (DDP) mixture of normals model with $K = 0$. The results are presented for each of the simulated scenarios and sample sizes (in the nondiseased group).}\label{WAIC_LPML_ndx1}
 \begin{center}
 \footnotesize	
	\begin{tabular}{cccccc}
	& & \multicolumn{3}{c}{Sample size}\\
	& & \multicolumn{3}{c}{$n_{\bar{D}}$}\\\hline
	Scenario & Approach & $50$ & $200$ & $300$ \\\hline
	\multirow{2}{*}{1} &	
WAIC (BNL) $<$ WAIC (DDP) & 78\% & 89\% & 92\%\\
 & LPML (BNL) $>$ LPML (DDP) & 82\%  & 89\% & 92\%\\ \hline
	\multirow{2}{*}{2} &	
WAIC (BNL) $<$ WAIC (DDP) & 80\%  & 90\% & 94\%\\
 & LPML (BNL) $>$ LPML (DDP) & 83\%  & 93\% & 94\%\\ \hline
	\multirow{2}{*}{3} &	
WAIC (BNL) $<$ WAIC (DDP) & 79\%  & 90\% & 92\%\\
 & LPML (BNL) $>$ LPML (DDP) & 83\% & 92\% & 92\%\\ \hline
	\multirow{2}{*}{4} &	
WAIC (BNL) $<$ WAIC (DDP) & 1\%  & 0\% & 0\%\\
 & LPML (BNL) $>$ LPML (DDP) & 1\%  & 0\% & 0\%\\ \hline
 \multirow{2}{*}{5} &	
WAIC (BNL) $<$ WAIC (DDP) & 0\%  & 0\% & 0\%\\
 & LPML (BNL) $>$ LPML (DDP) & 0\% & 0\% & 0\%\\ \hline
\multirow{2}{*}{6} &	
WAIC (BNL) $<$ WAIC (DDP) & 0\%  & 0\% & 0\%\\
 & LPML (BNL) $>$ LPML (DDP) & 0\%& 0\% & 0\%\\ \hline
	\end{tabular}
 \end{center}
\end{table}

\begin{figure}[H]
  \begin{center}
    	\subfigure{
		\includegraphics[height=4.5cm, page = 1]{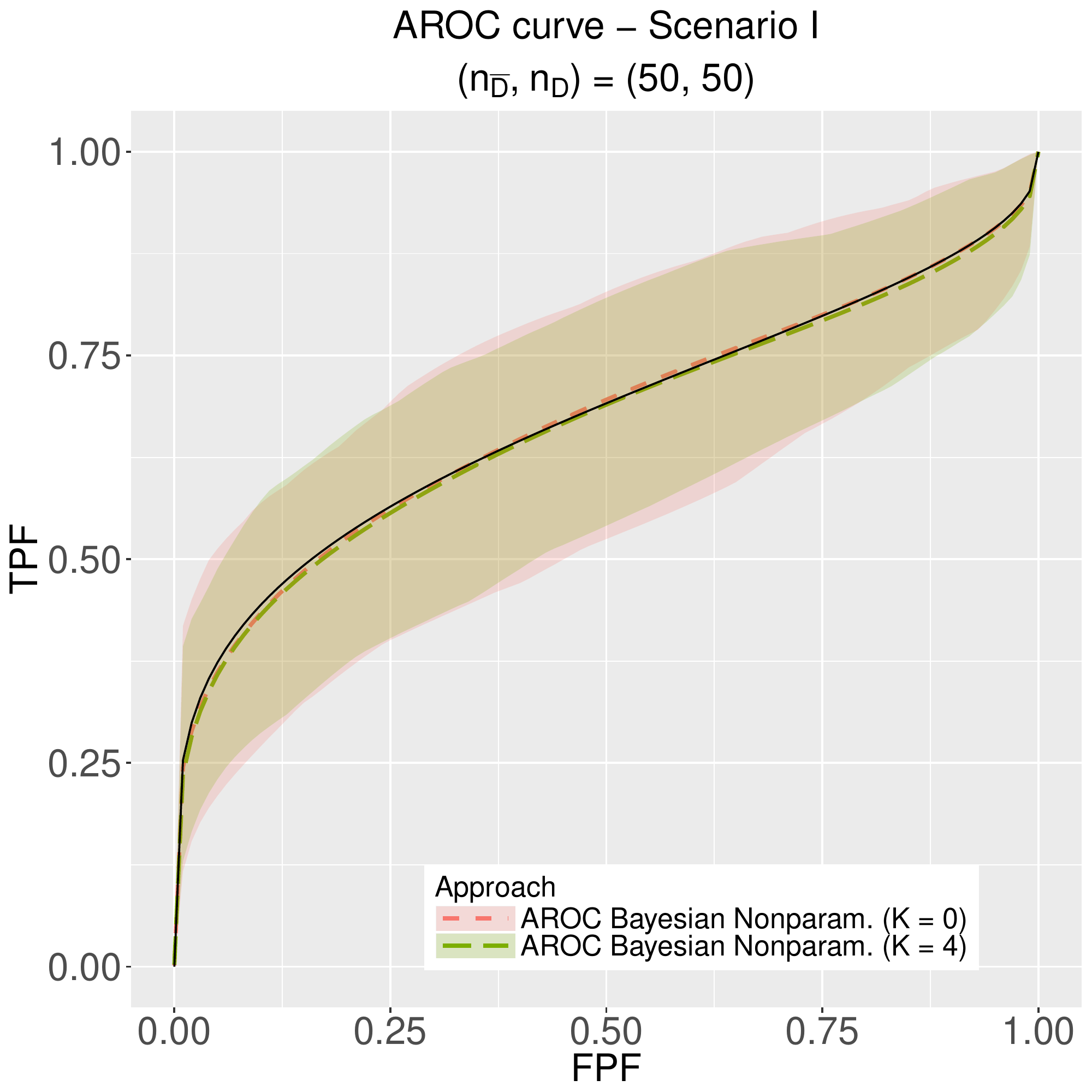}
		\includegraphics[height=4.5cm, page = 1]{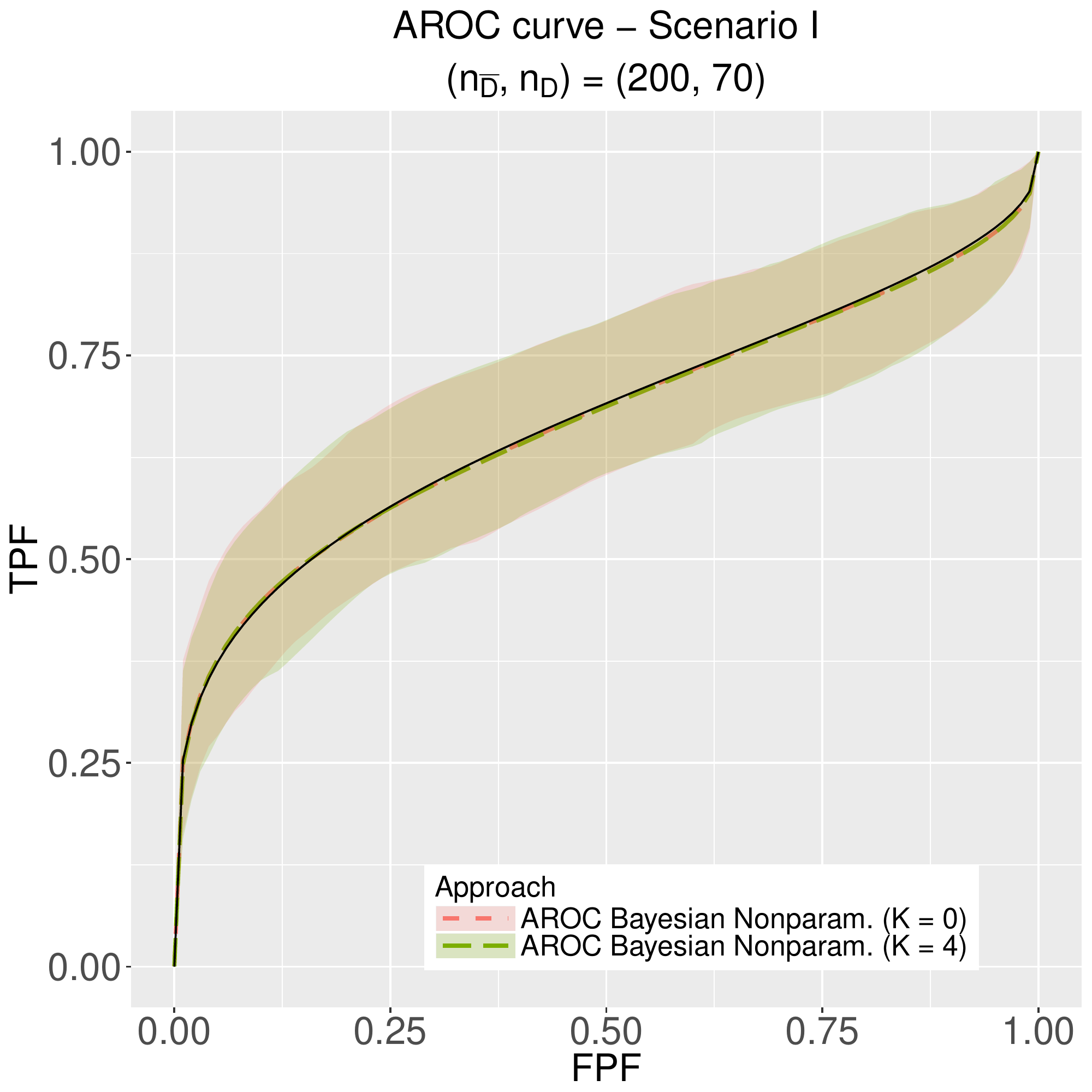}
		\includegraphics[height=4.5cm, page = 1]{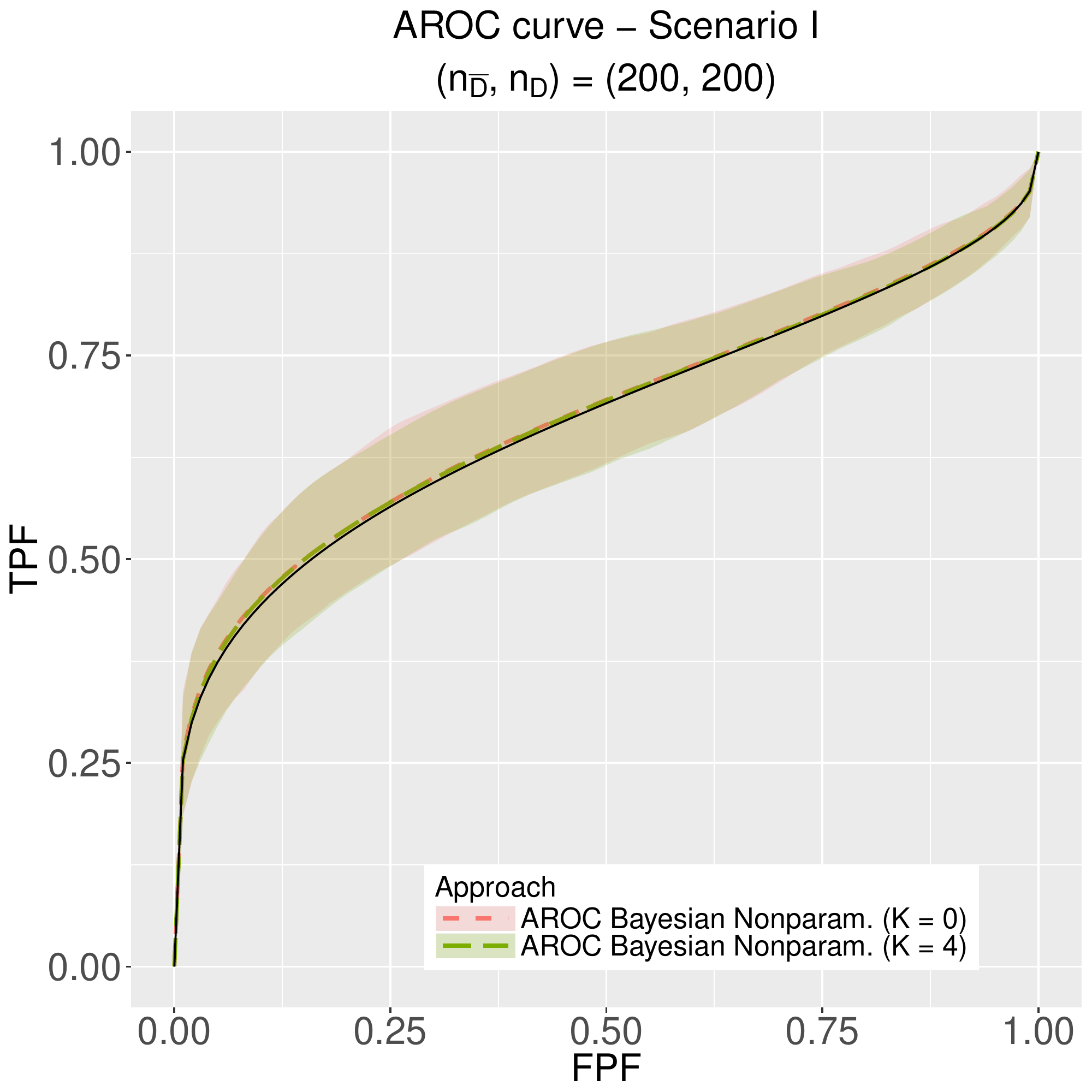}
		\includegraphics[height=4.5cm, page = 1]{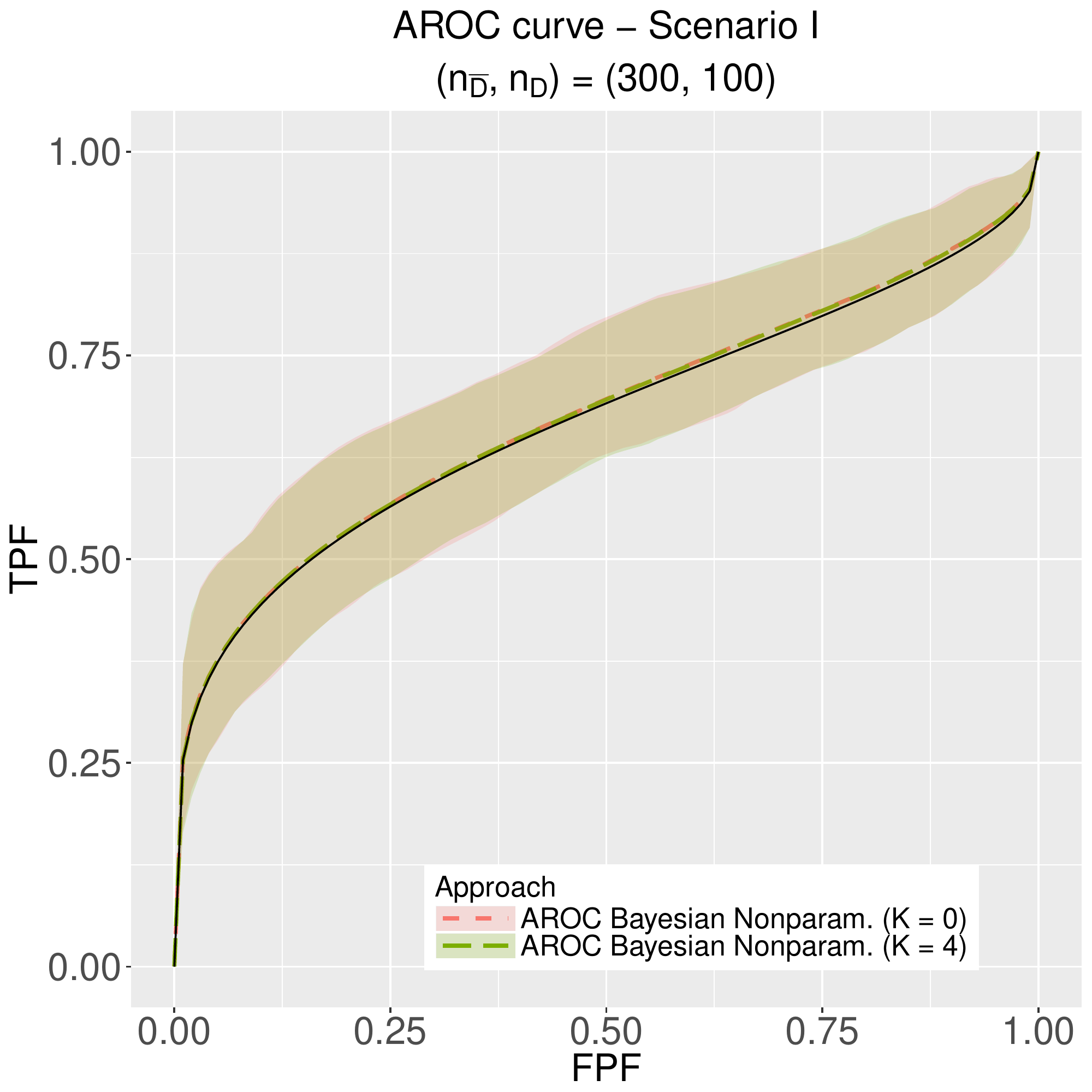}}
\end{center}
		 \caption{Scenario I: true (solid black line) and average value of 100 simulated datasets (dashed lines) of the posterior mean of the covariate adjusted ROC curve/pooled ROC curve for the Bayesian nonparametric approach proposed in this paper. The shaded area are bands constructed using the pointwise $2.5\%$ and $97.5\%$ quantiles across simulations. The results are presented for no interior knots $(K = 0)$ and four interior knots $(K = 4)$ and for each sample size.}
		\label{sim_ndx_1_I}
\end{figure}

\begin{figure}[H]
  \begin{center}
    	\subfigure{
		\includegraphics[height=4.5cm, page = 1]{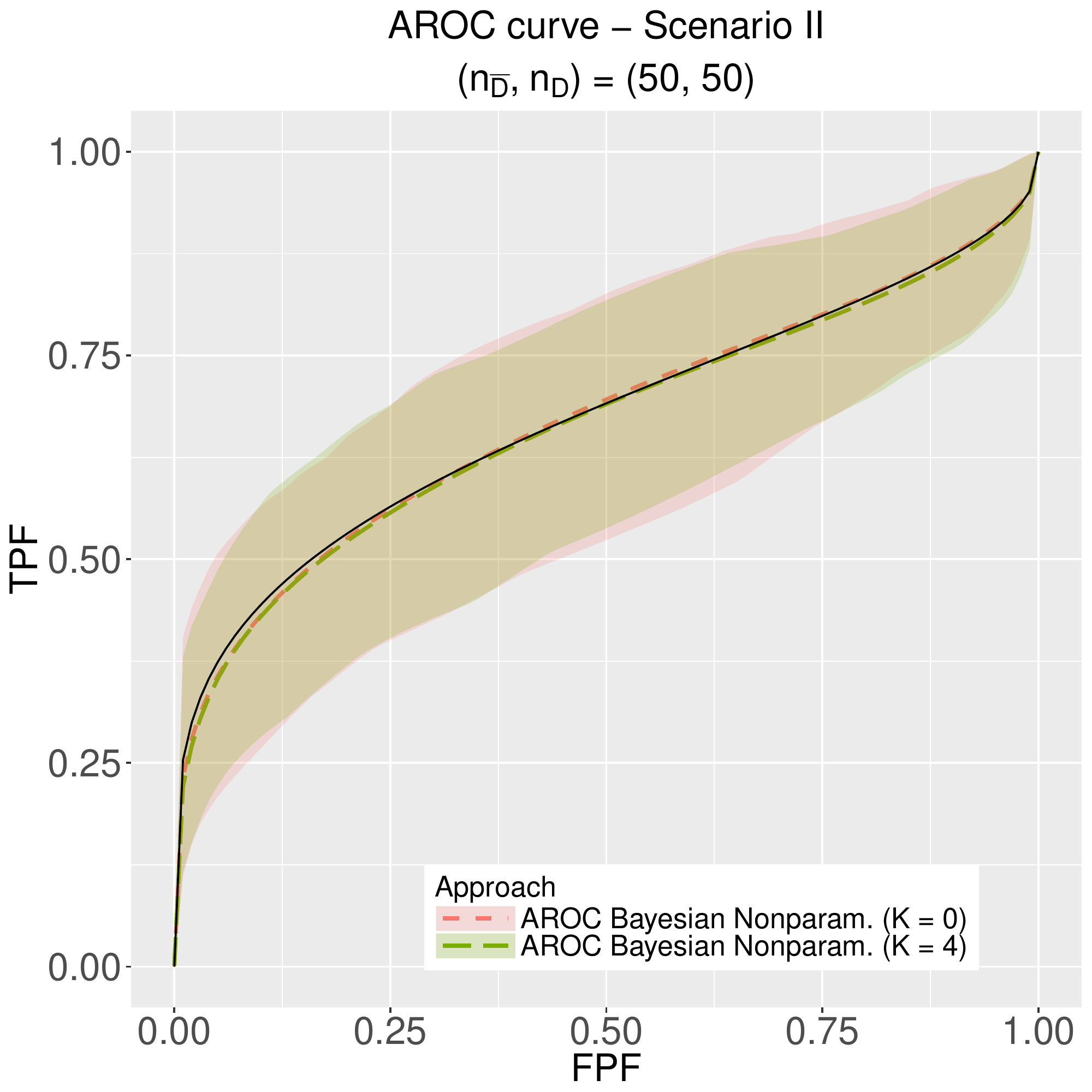}
		\includegraphics[height=4.5cm, page = 1]{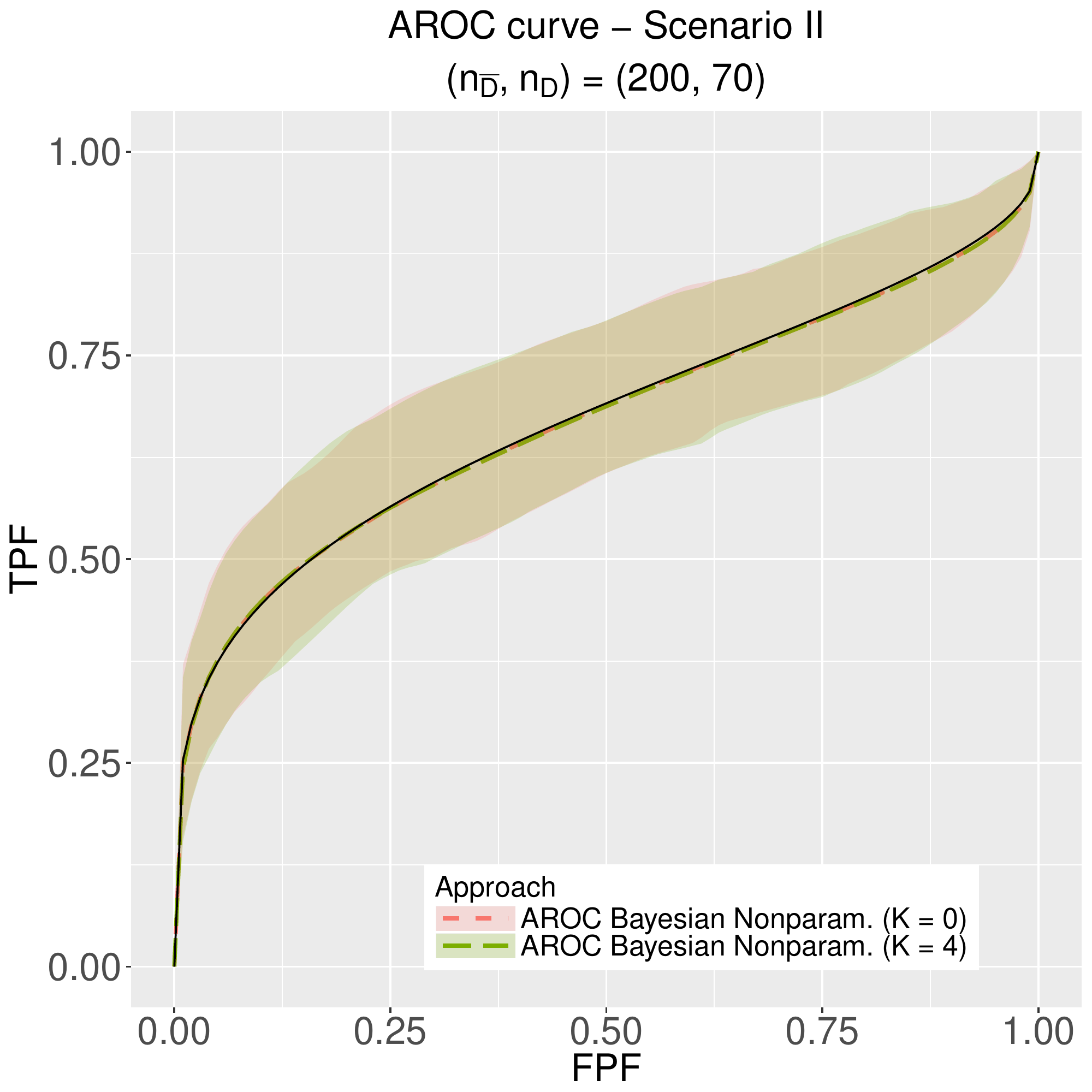}
		\includegraphics[height=4.5cm, page = 1]{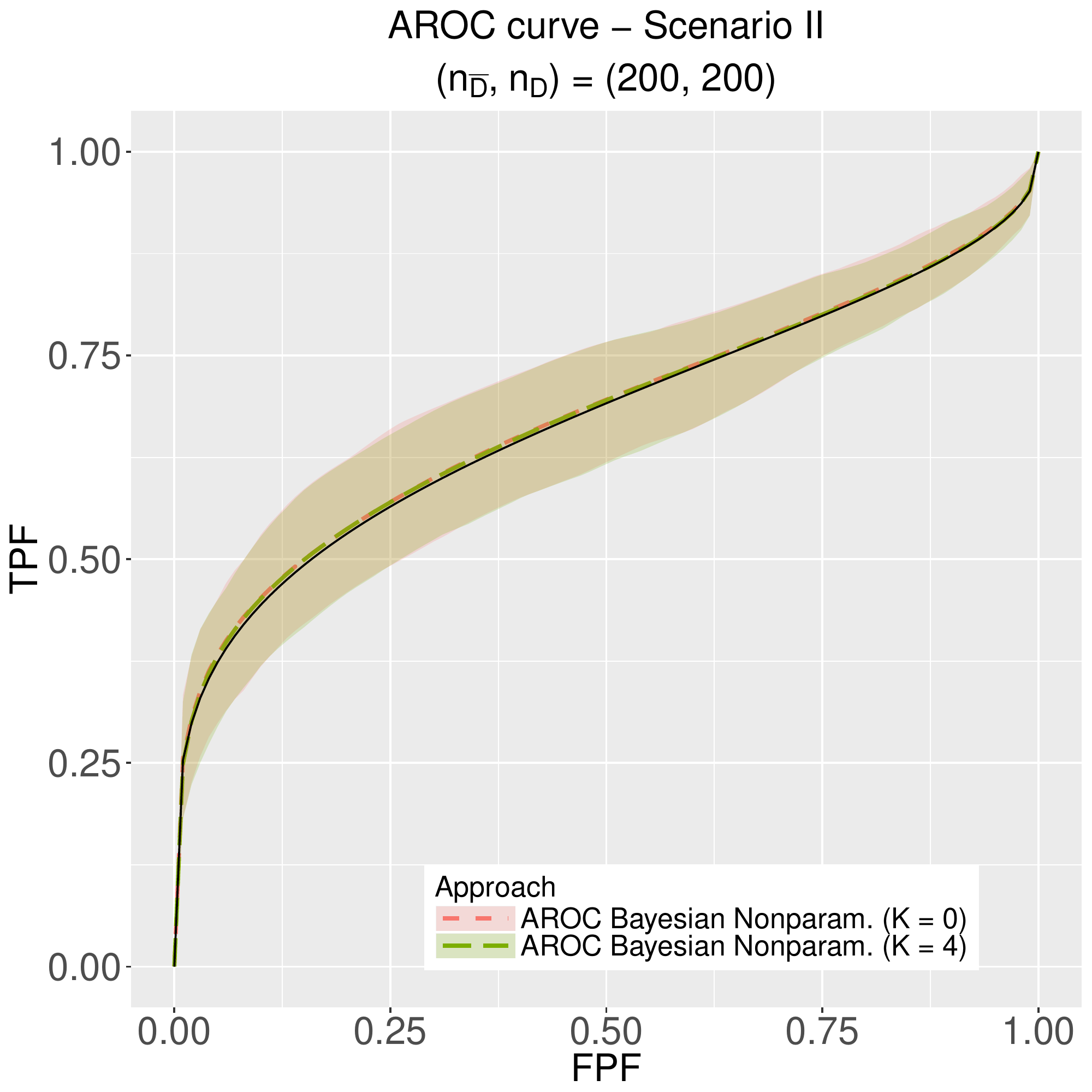}
		\includegraphics[height=4.5cm, page = 1]{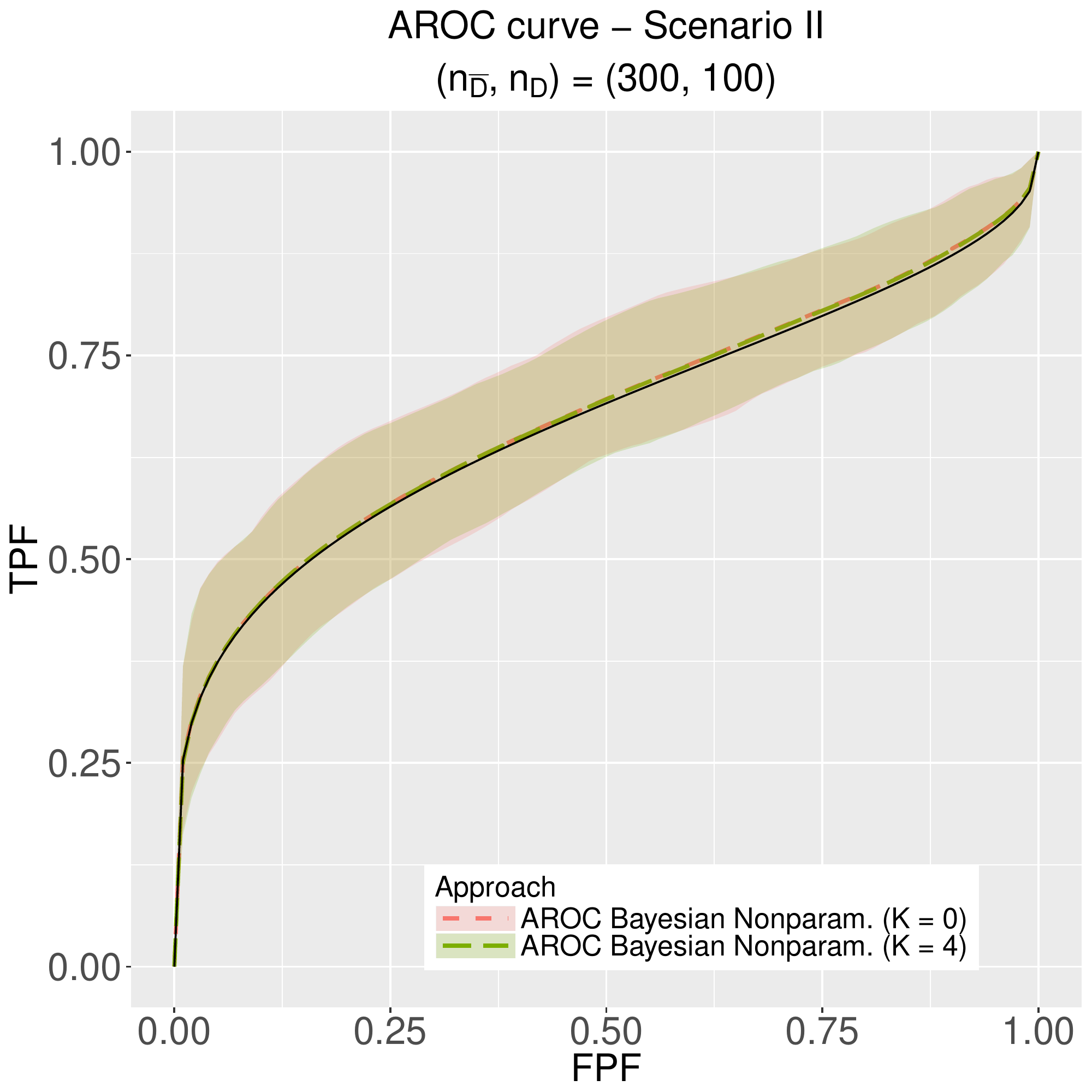}}
\end{center}
		 \caption{Scenario II: true (solid black line) and average value of 100 simulated datasets (dashed lines) of the posterior mean of the covariate adjusted ROC curve/pooled ROC curve for the Bayesian nonparametric approach proposed in this paper. The shaded area are bands constructed using the pointwise $2.5\%$ and $97.5\%$ quantiles across simulations. The results are presented for no interior knots $(K = 0)$ and four interior knots $(K = 4)$ and for each sample size.}
		\label{sim_ndx_1_II}
\end{figure}

\begin{figure}[H]
  \begin{center}
    	\subfigure{
		\includegraphics[height=4.5cm, page = 1]{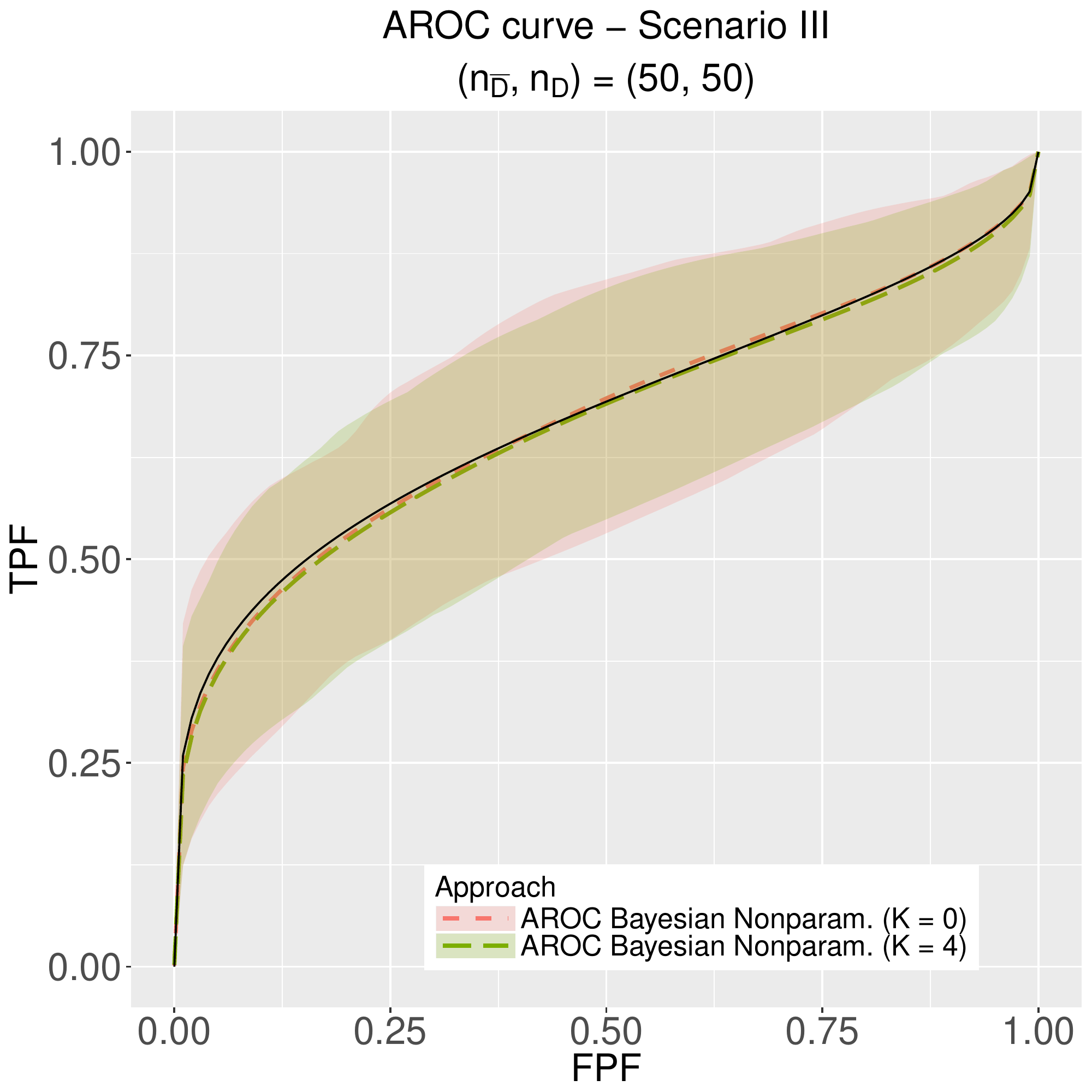}
		\includegraphics[height=4.5cm, page = 1]{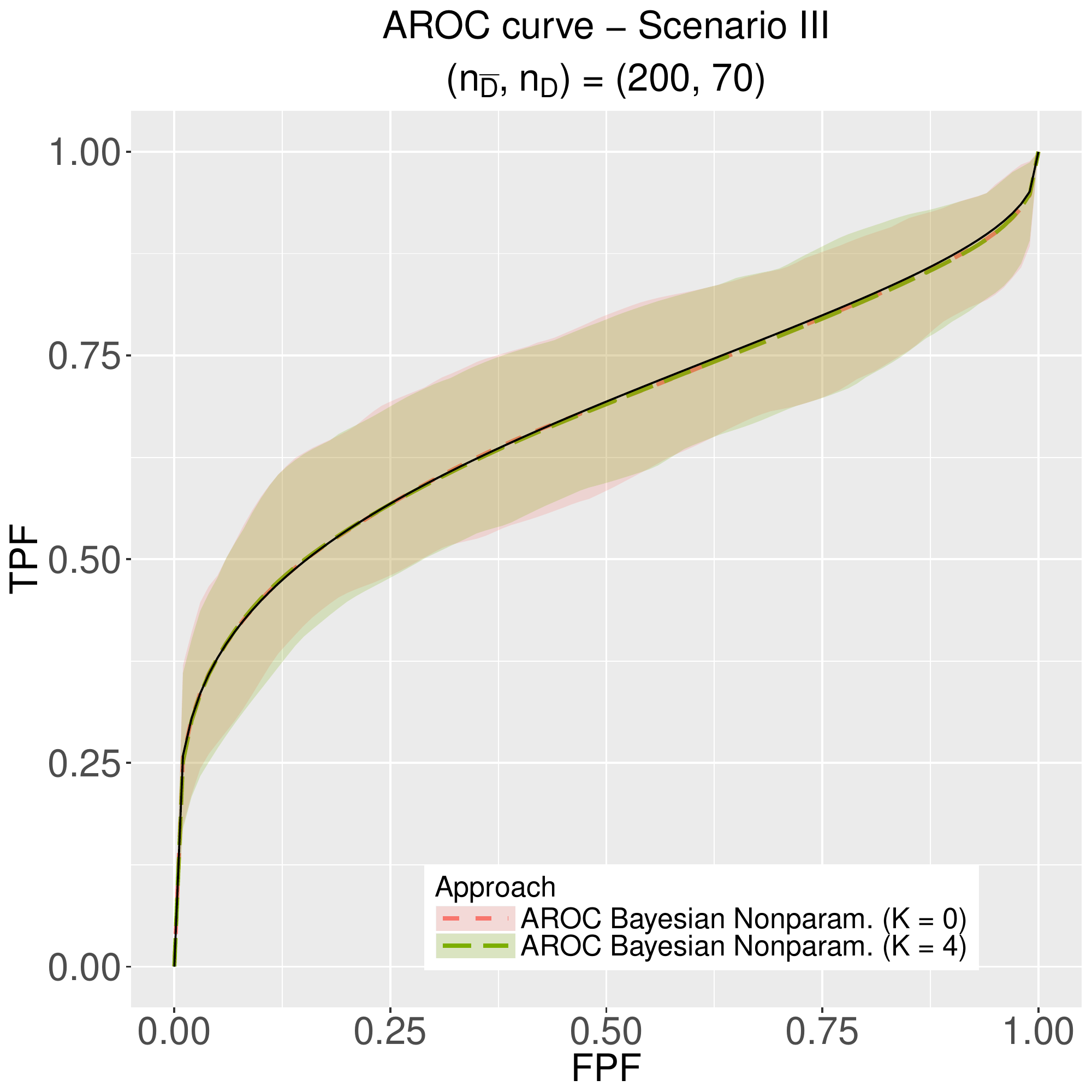}
		\includegraphics[height=4.5cm, page = 1]{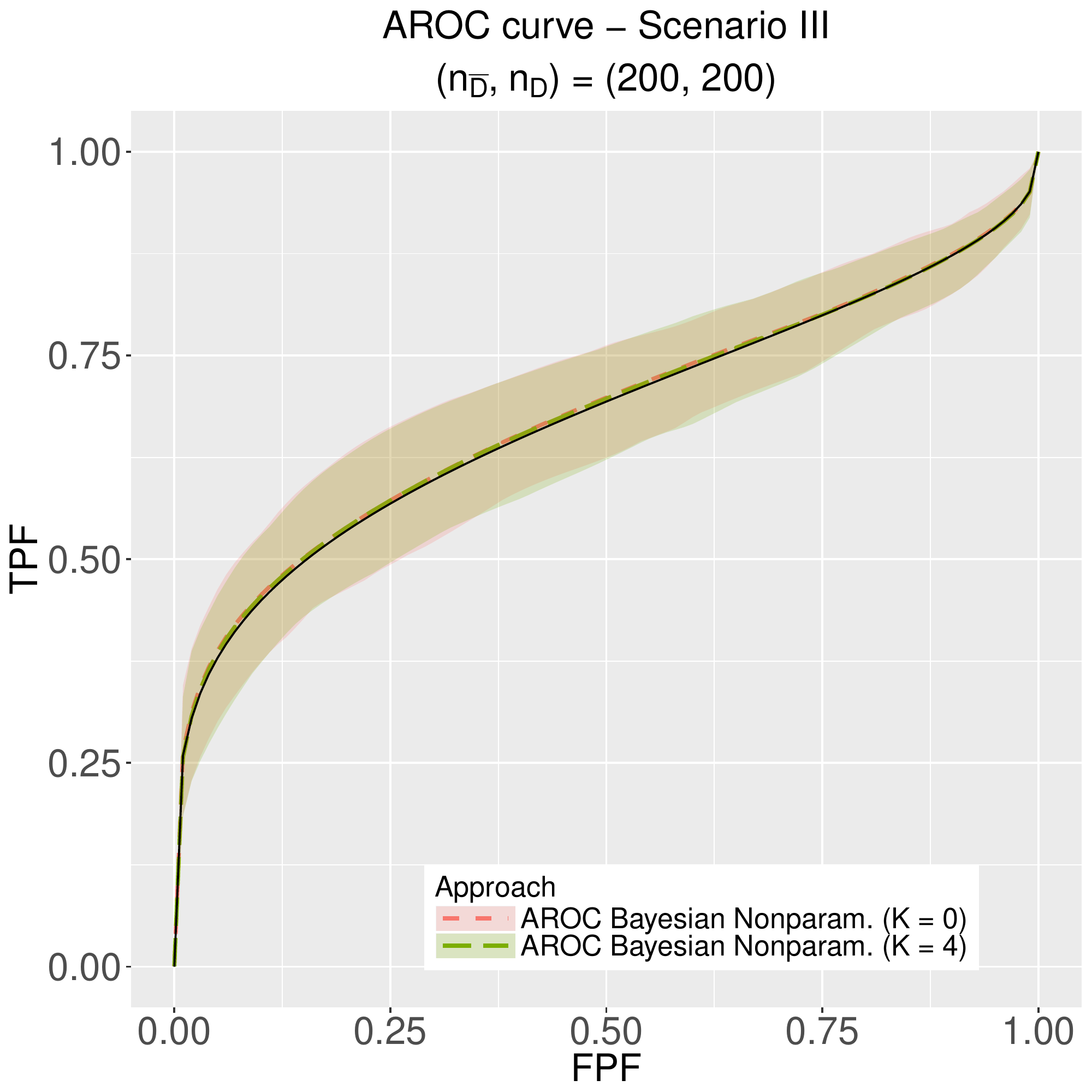}
		\includegraphics[height=4.5cm, page = 1]{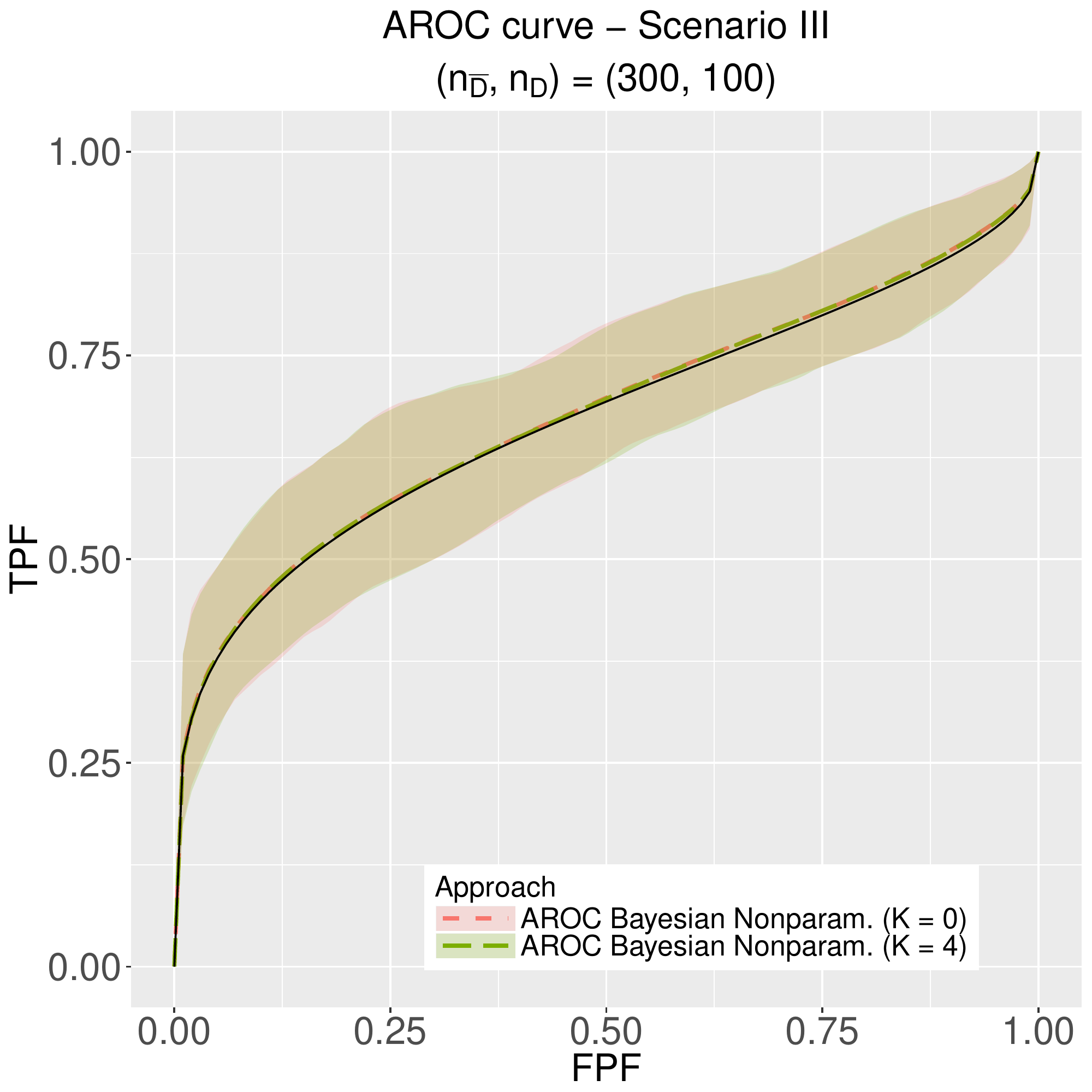}}
\end{center}
		 \caption{Scenario III: true (solid black line) and average value of 100 simulated datasets (dashed lines) of the posterior mean of the covariate adjusted ROC curve/pooled ROC curve for the Bayesian nonparametric approach proposed in this paper. The shaded area are bands constructed using the pointwise $2.5\%$ and $97.5\%$ quantiles across simulations. The results are presented for no interior knots $(K = 0)$ and four interior knots $(K = 4)$ and for each sample size.}
		\label{sim_ndx_1_III}
\end{figure}

\begin{figure}[H]
  \begin{center}
    	\subfigure{
		\includegraphics[height=4.5cm, page = 1]{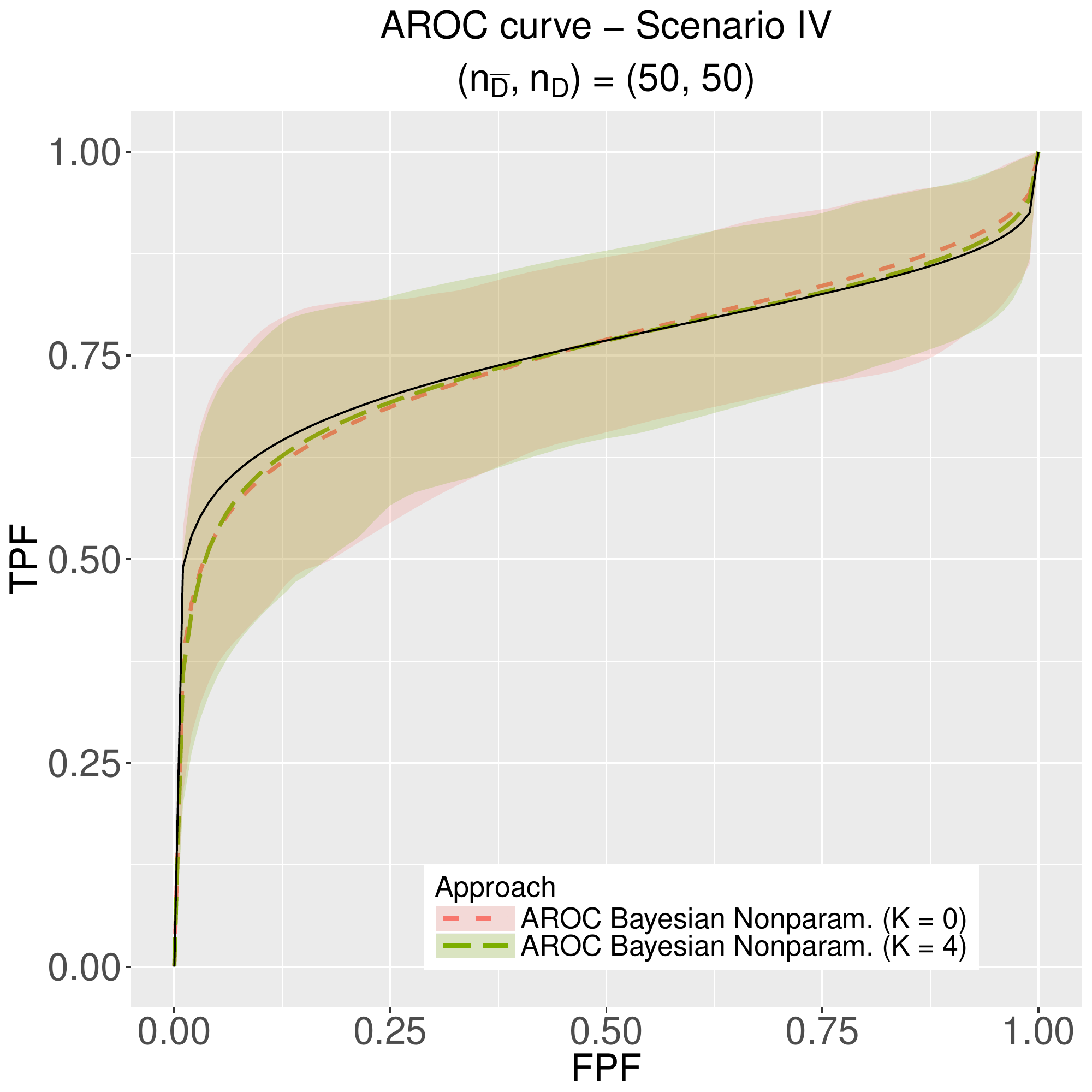}
		\includegraphics[height=4.5cm, page = 1]{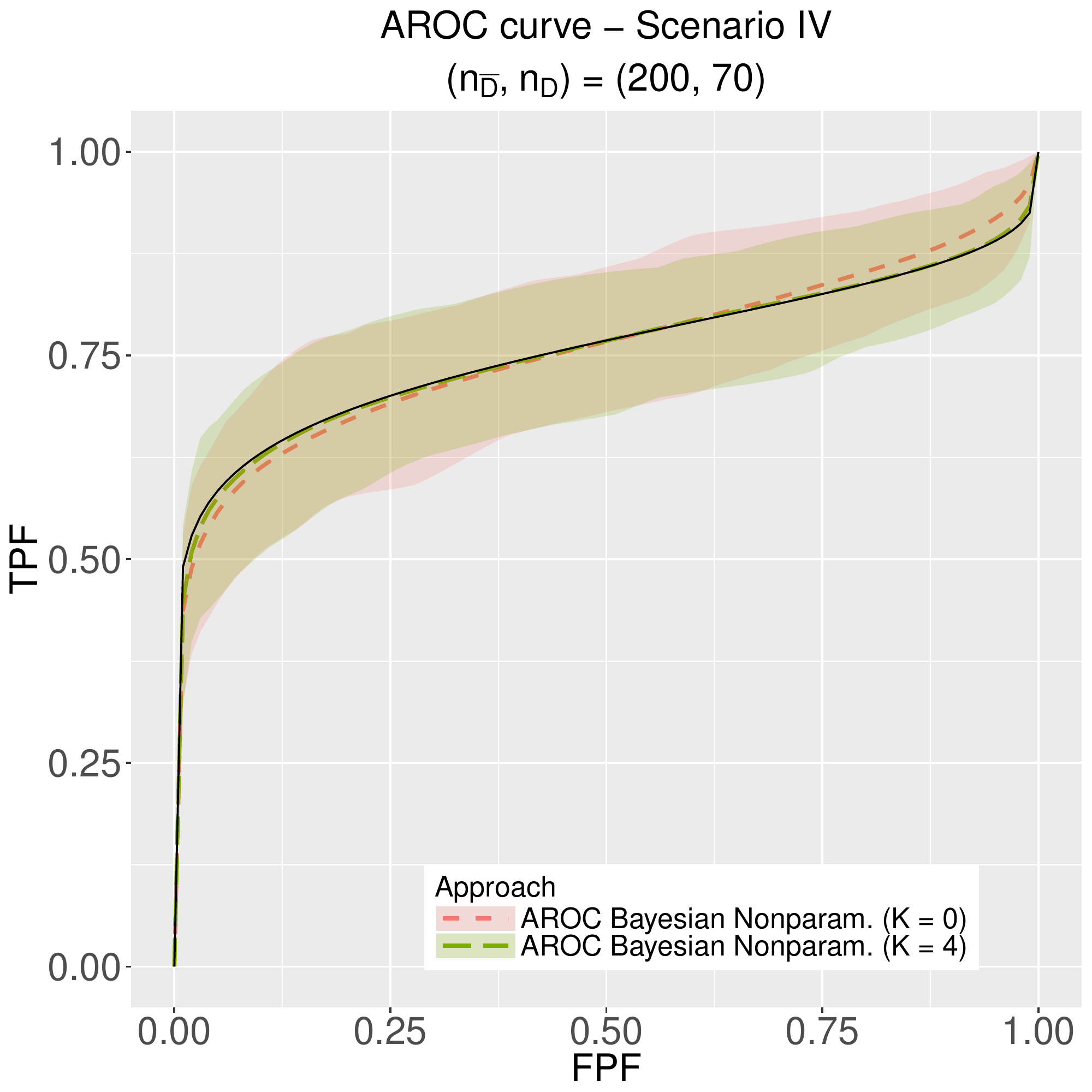}
		\includegraphics[height=4.5cm, page = 1]{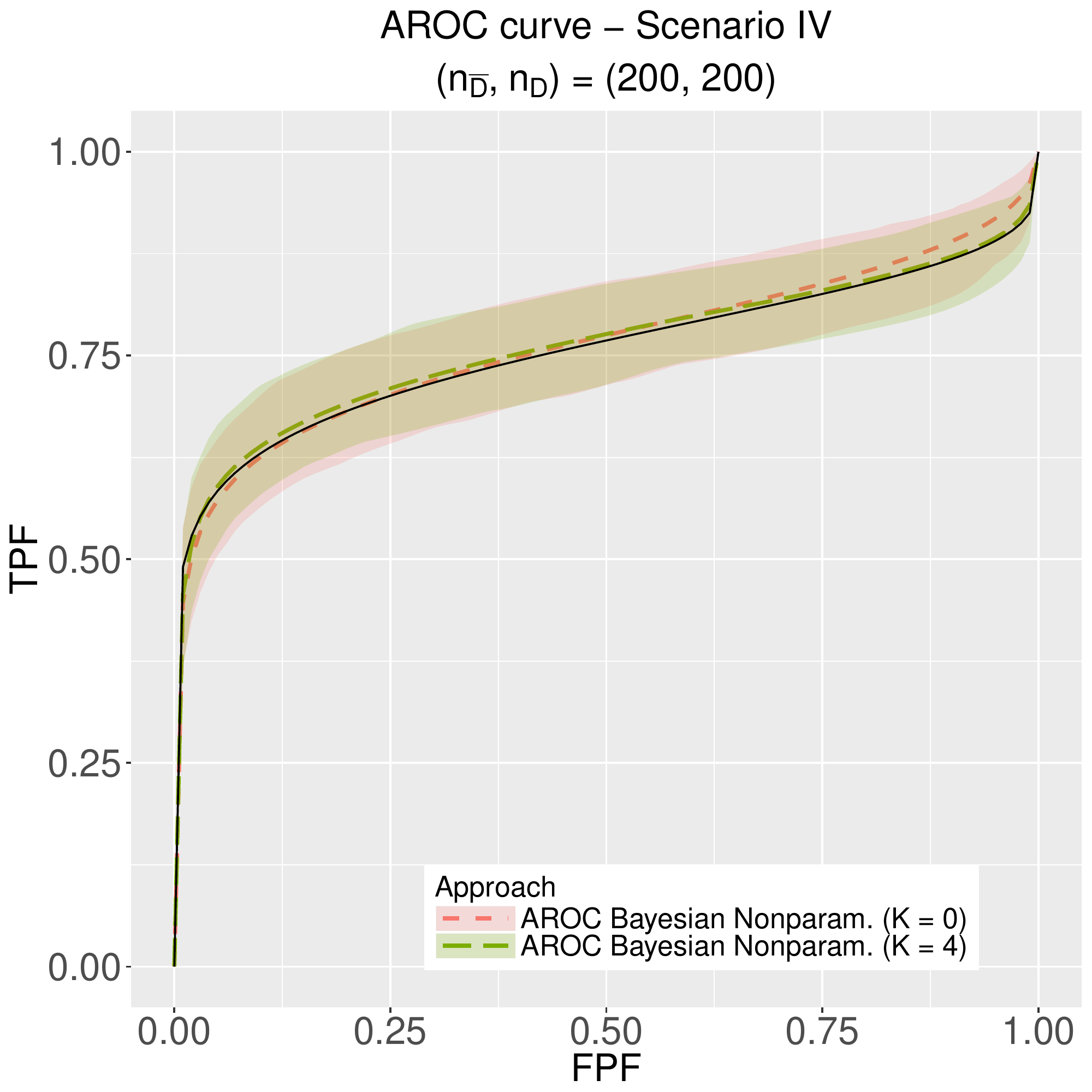}
		\includegraphics[height=4.5cm, page = 1]{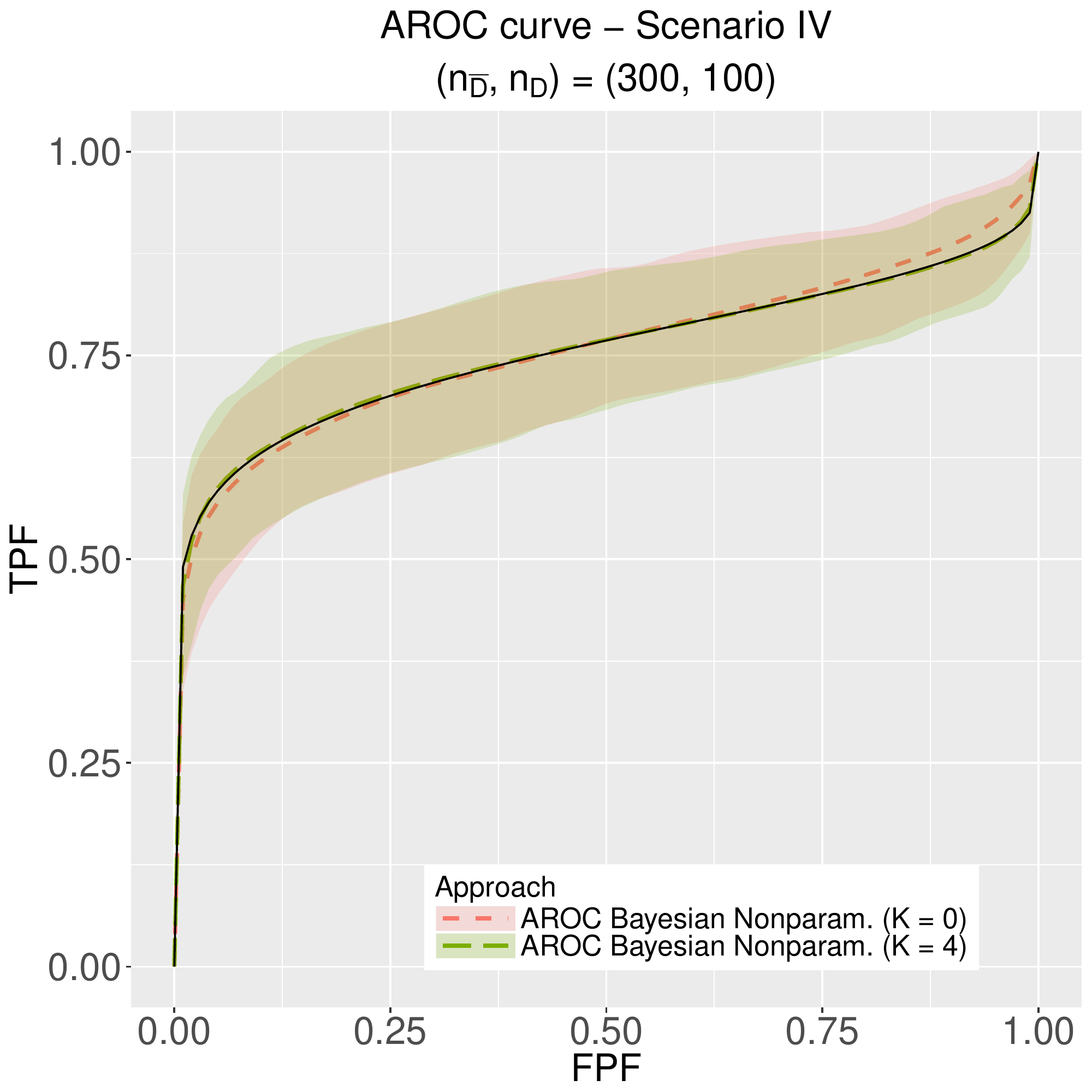}}
\end{center}
		 \caption{Scenario IV: true (solid black line) and average value of 100 simulated datasets (dashed lines) of the posterior mean of the covariate adjusted ROC curve/pooled ROC curve for the Bayesian nonparametric approach proposed in this paper. The shaded area are bands constructed using the pointwise $2.5\%$ and $97.5\%$ quantiles across simulations. The results are presented for no interior knots $(K = 0)$ and four interior knots $(K = 4)$ and for each sample size.}
		\label{sim_ndx_1_IV}
\end{figure}

\begin{figure}[H]
  \begin{center}
    	\subfigure{
		\includegraphics[height=4.5cm, page = 1]{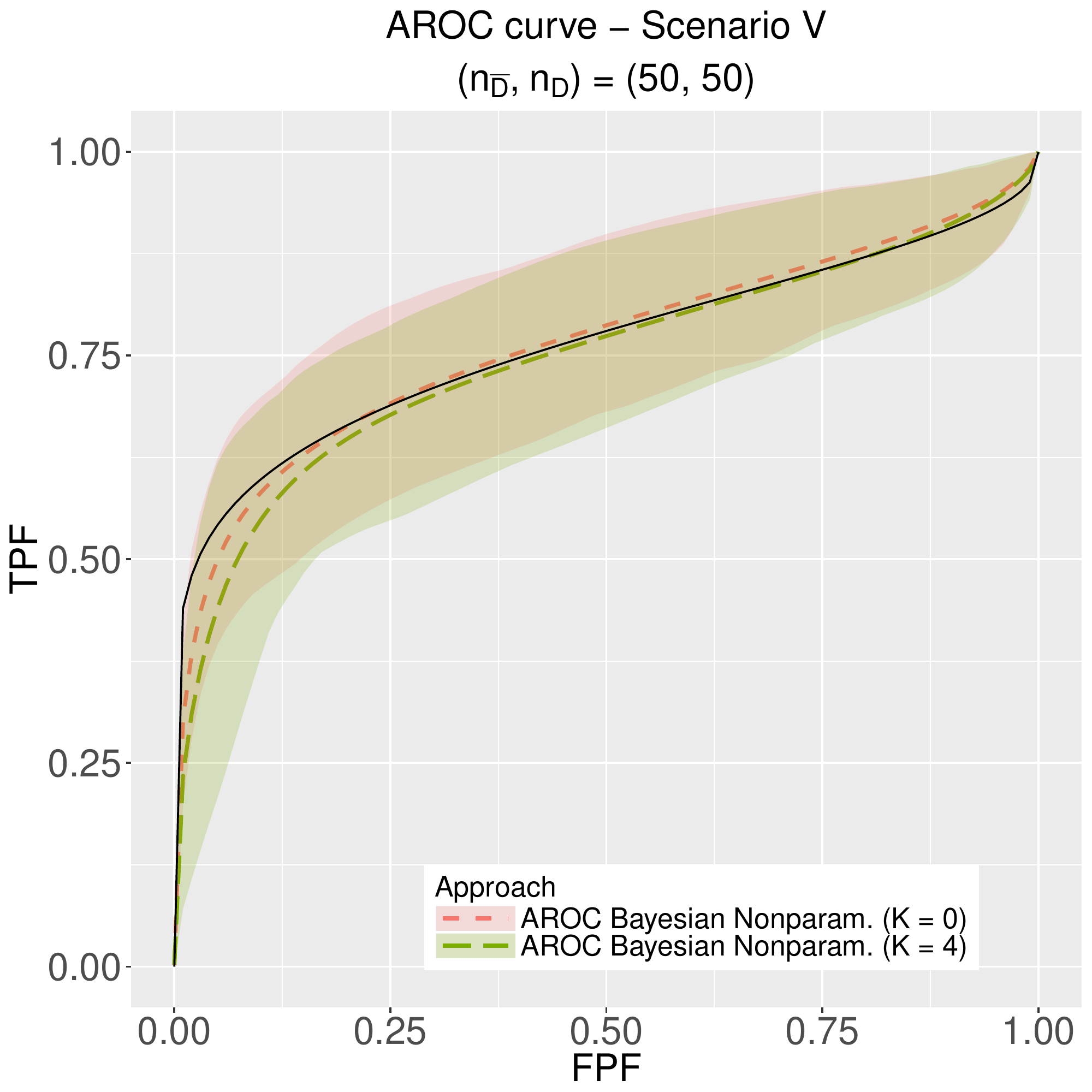}
		\includegraphics[height=4.5cm, page = 1]{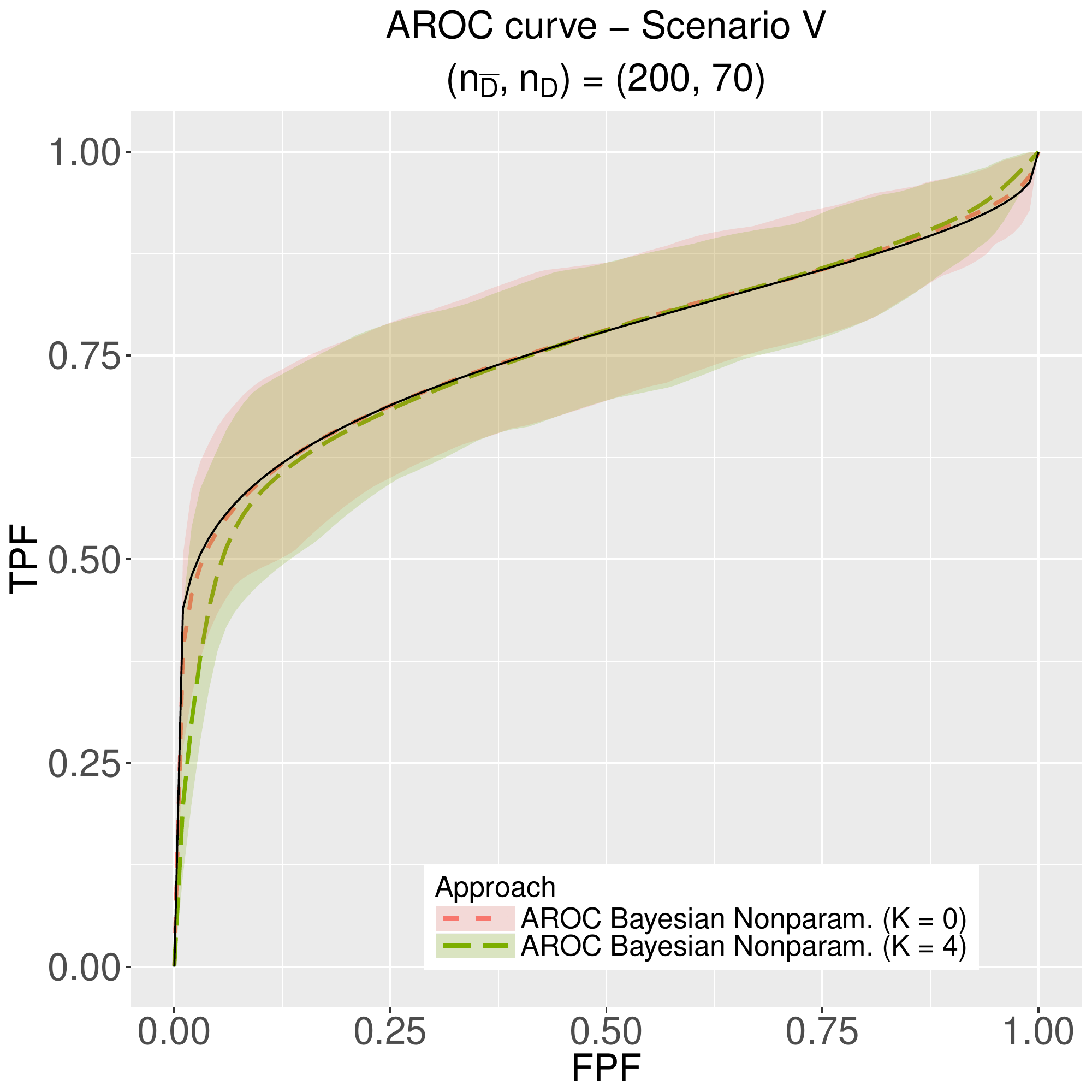}
		\includegraphics[height=4.5cm, page = 1]{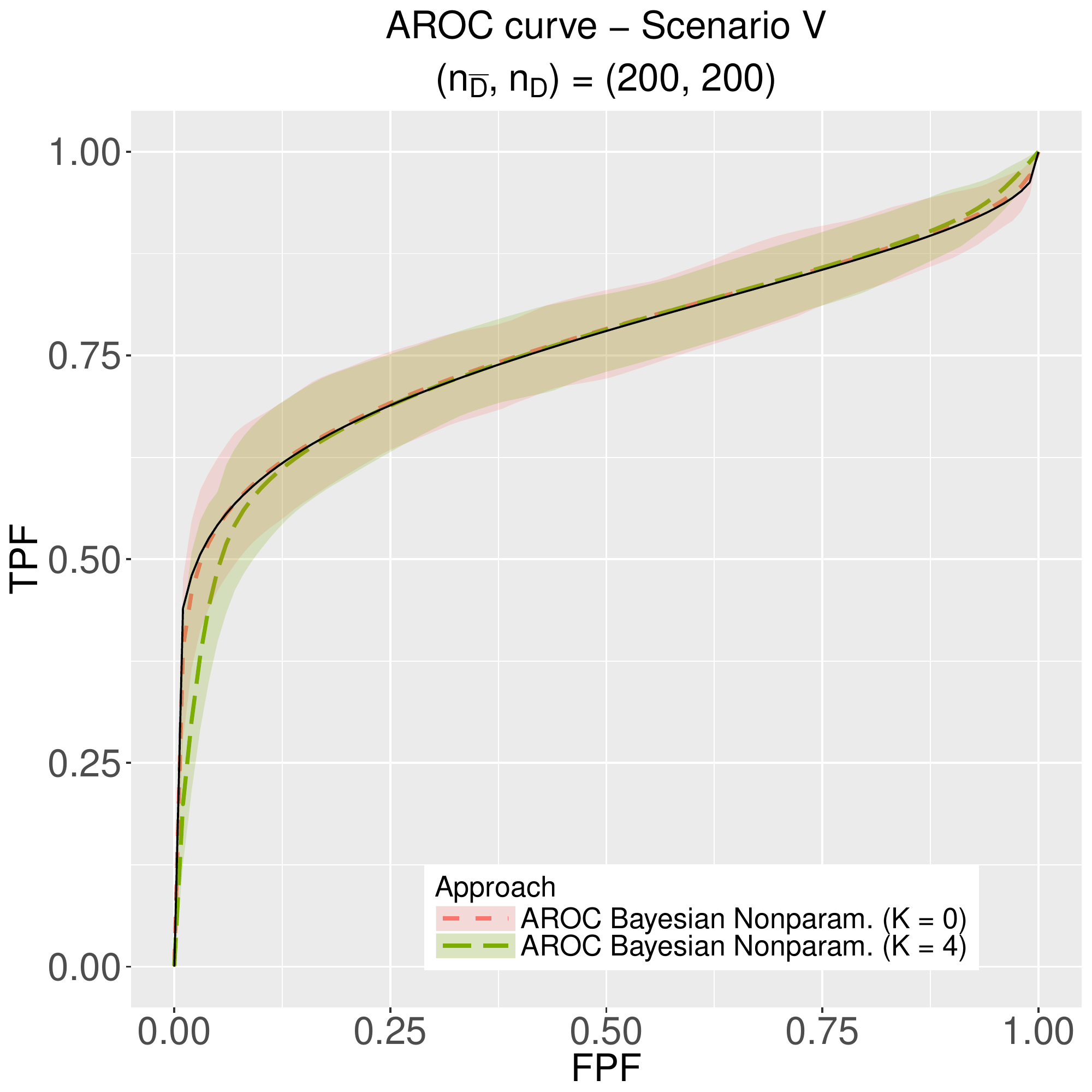}
		\includegraphics[height=4.5cm, page = 1]{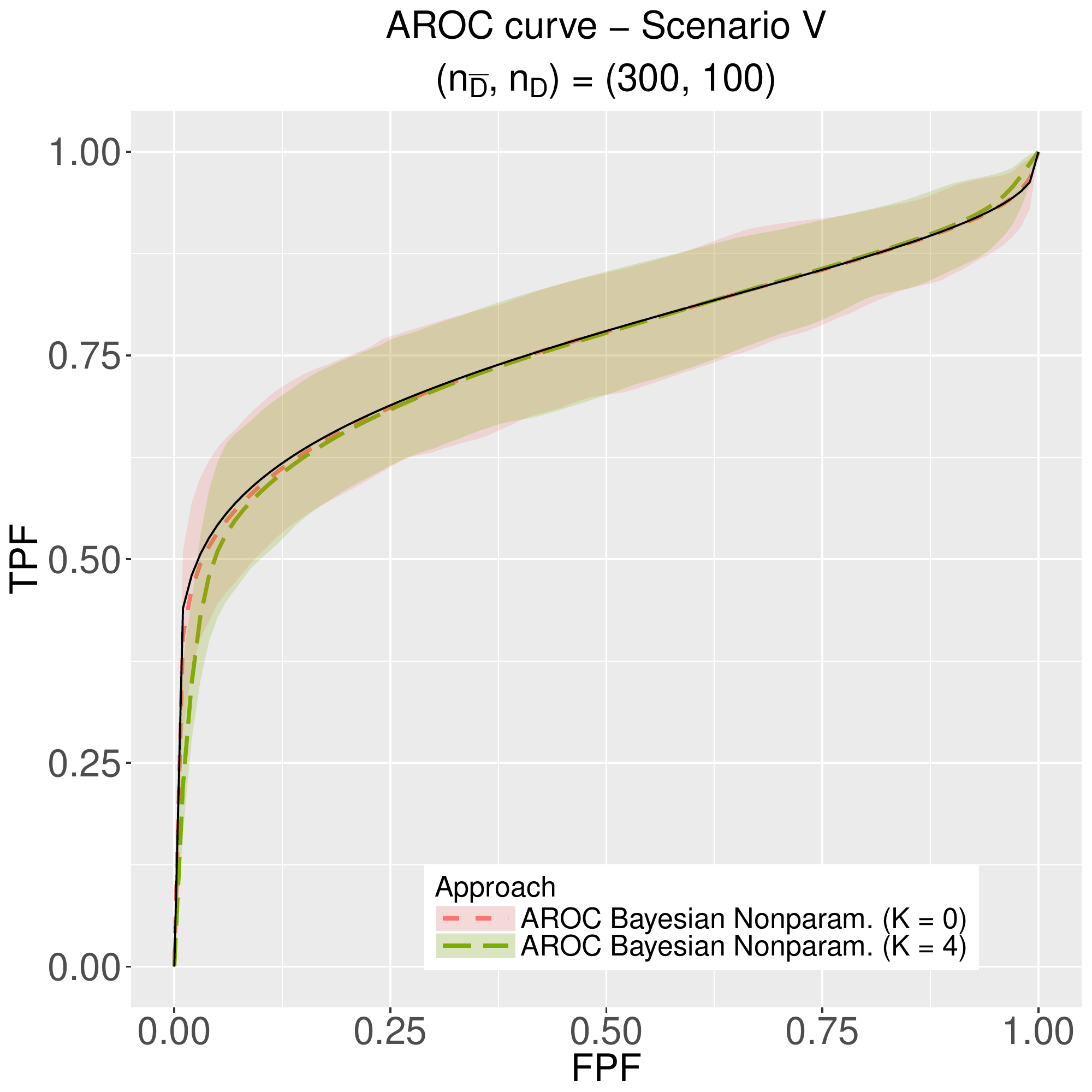}}
\end{center}
		 \caption{Scenario V: true (solid black line) and average value of 100 simulated datasets (dashed lines) of the posterior mean of the covariate adjusted ROC curve/pooled ROC curve for the Bayesian nonparametric approach proposed in this paper. The shaded area are bands constructed using the pointwise $2.5\%$ and $97.5\%$ quantiles across simulations. The results are presented for no interior knots $(K = 0)$ and four interior knots $(K = 4)$ and for each sample size.}
		\label{sim_ndx_1_V}
\end{figure}

\begin{figure}[H]
  \begin{center}
    	\subfigure{
		\includegraphics[height=4.5cm, page = 1]{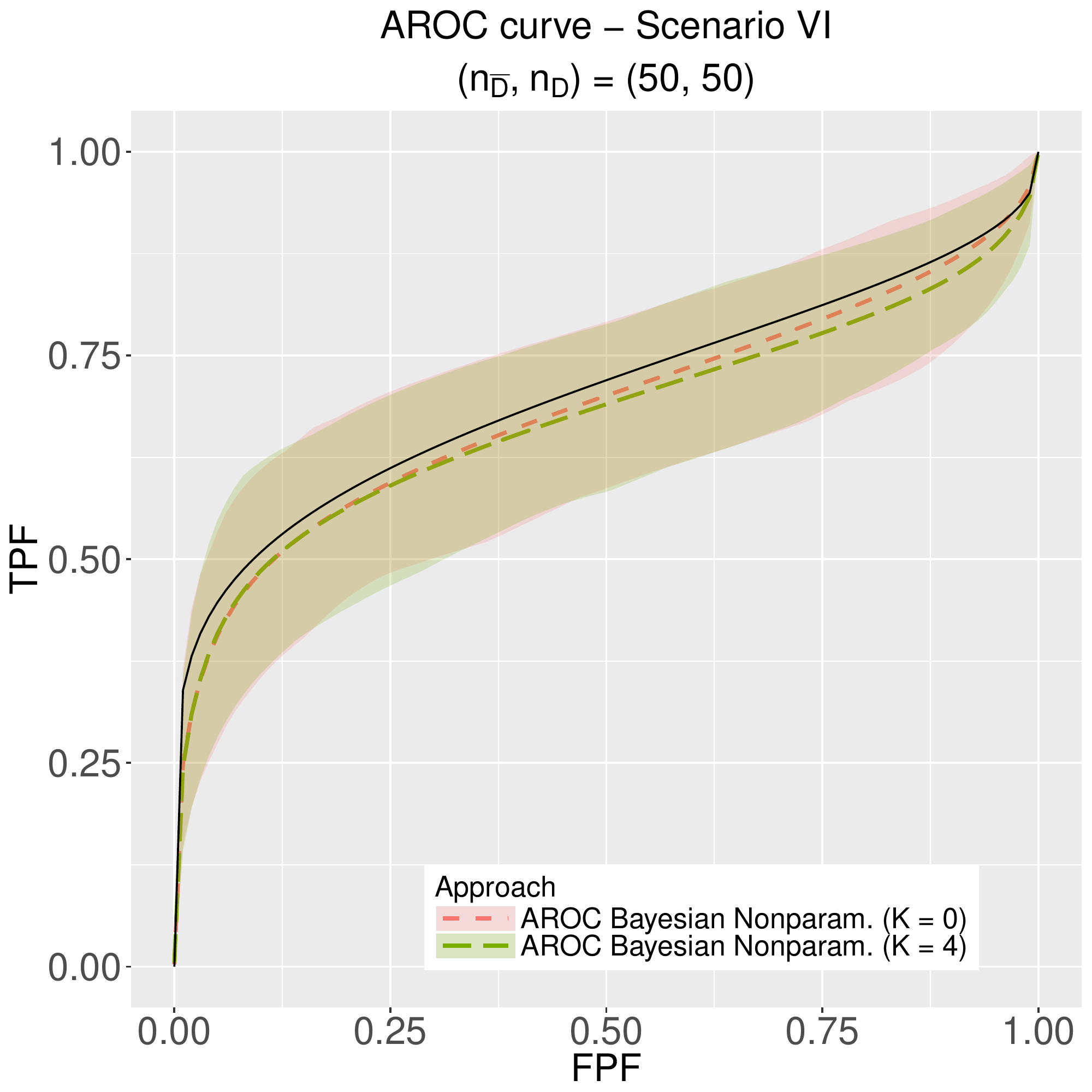}
		\includegraphics[height=4.5cm, page = 1]{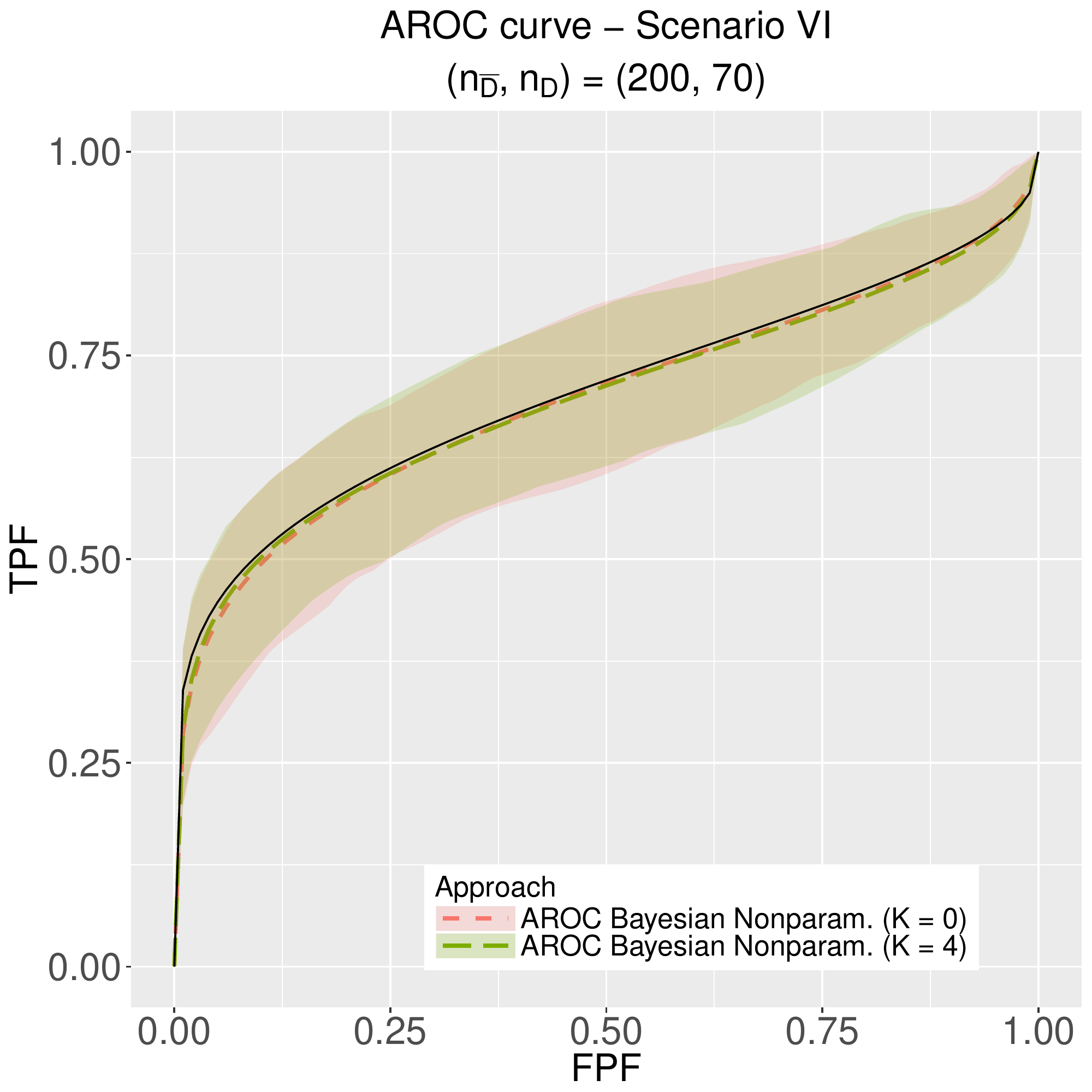}
		\includegraphics[height=4.5cm, page = 1]{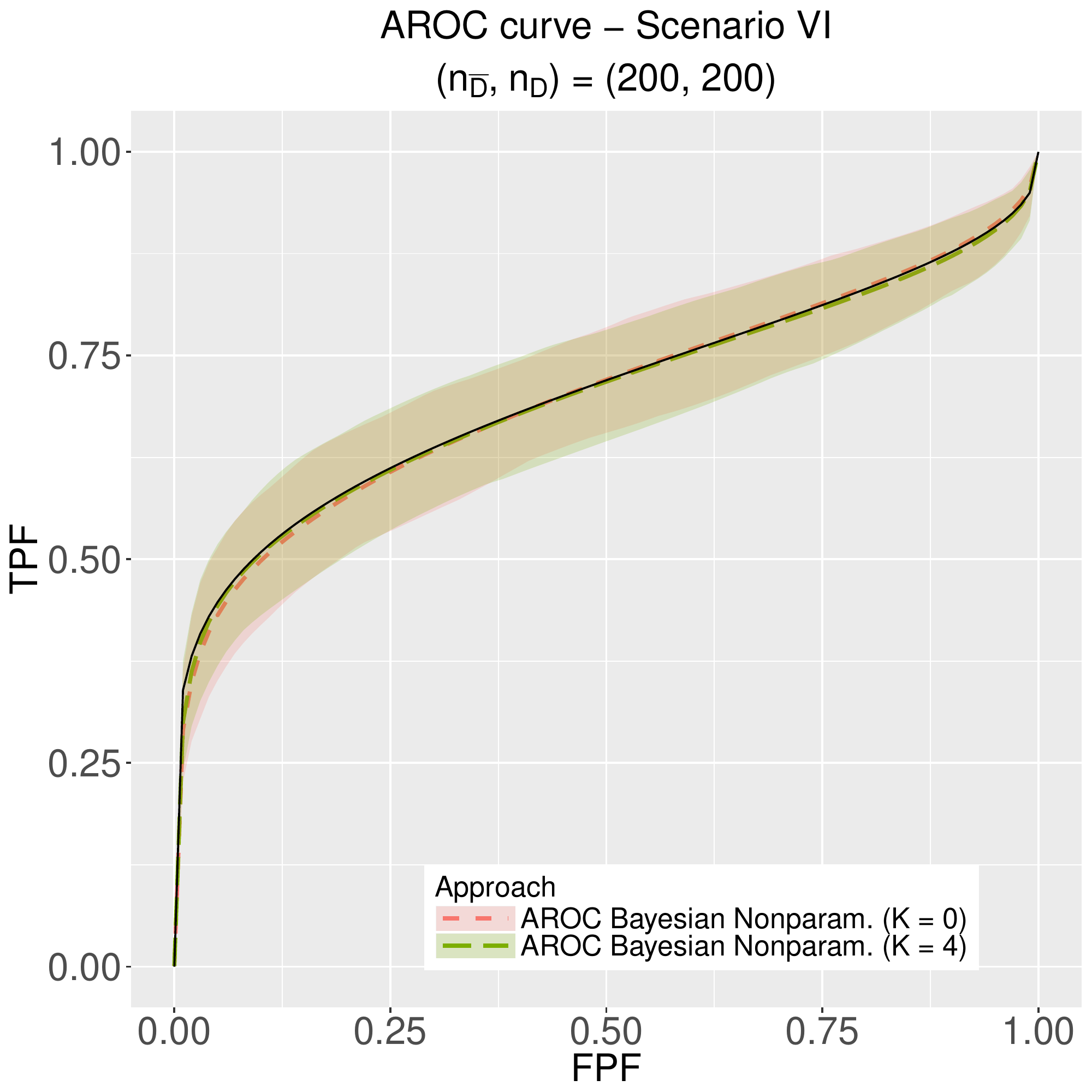}
		\includegraphics[height=4.5cm, page = 1]{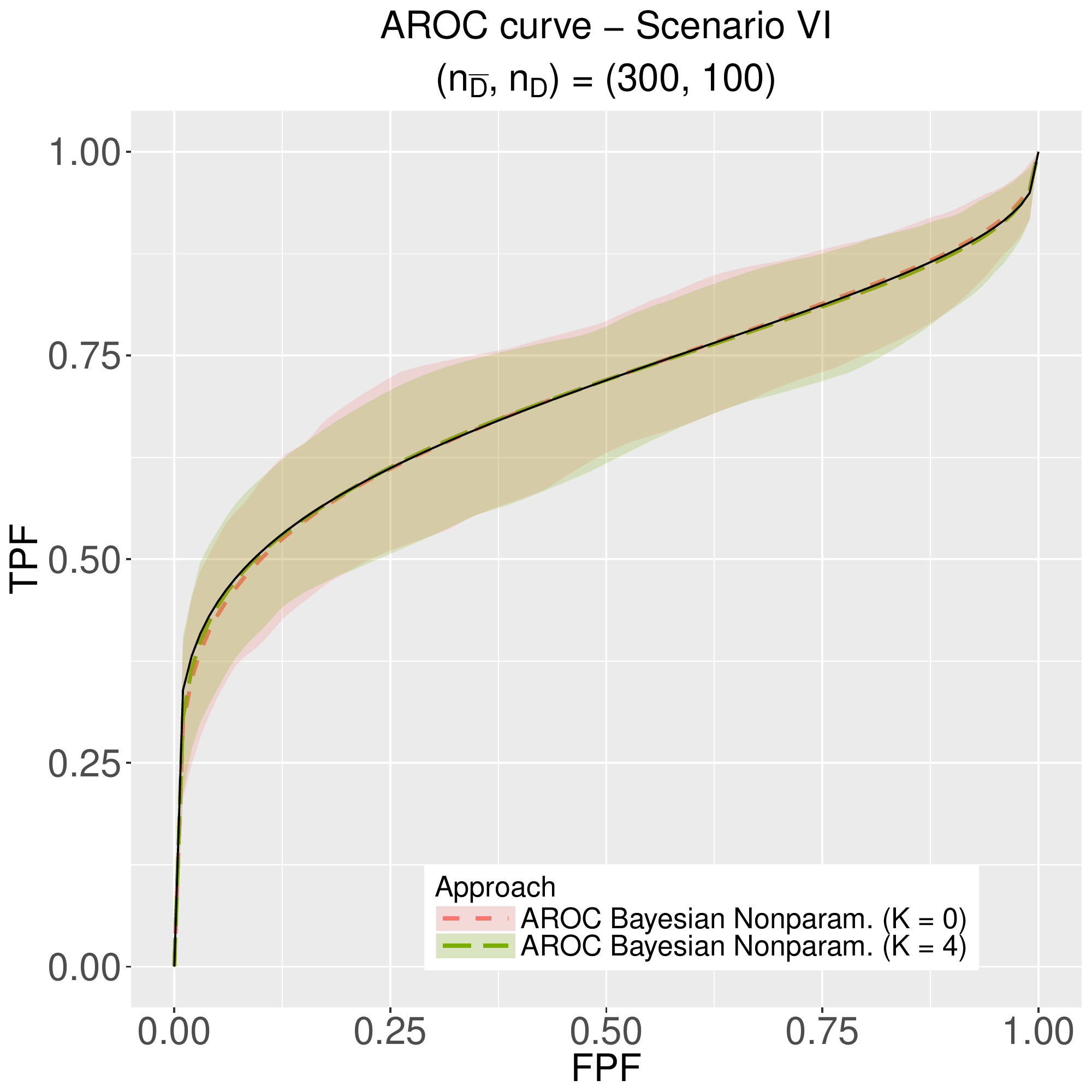}}
\end{center}
		 \caption{Scenario VI: true (solid black line) and average value of 100 simulated datasets (dashed lines) of the posterior mean of the covariate adjusted ROC curve/pooled ROC curve for the Bayesian nonparametric approach proposed in this paper. The shaded area are bands constructed using the pointwise $2.5\%$ and $97.5\%$ quantiles across simulations. The results are presented for no interior knots $(K = 0)$ and four interior knots $(K = 4)$ and for each sample size.}
		\label{sim_ndx_1_VI}
\end{figure}

\begin{figure}[H]
    \begin{center}
    	\subfigure[FPF = 0.1]{
		\includegraphics[height=4.5cm, page = 1]{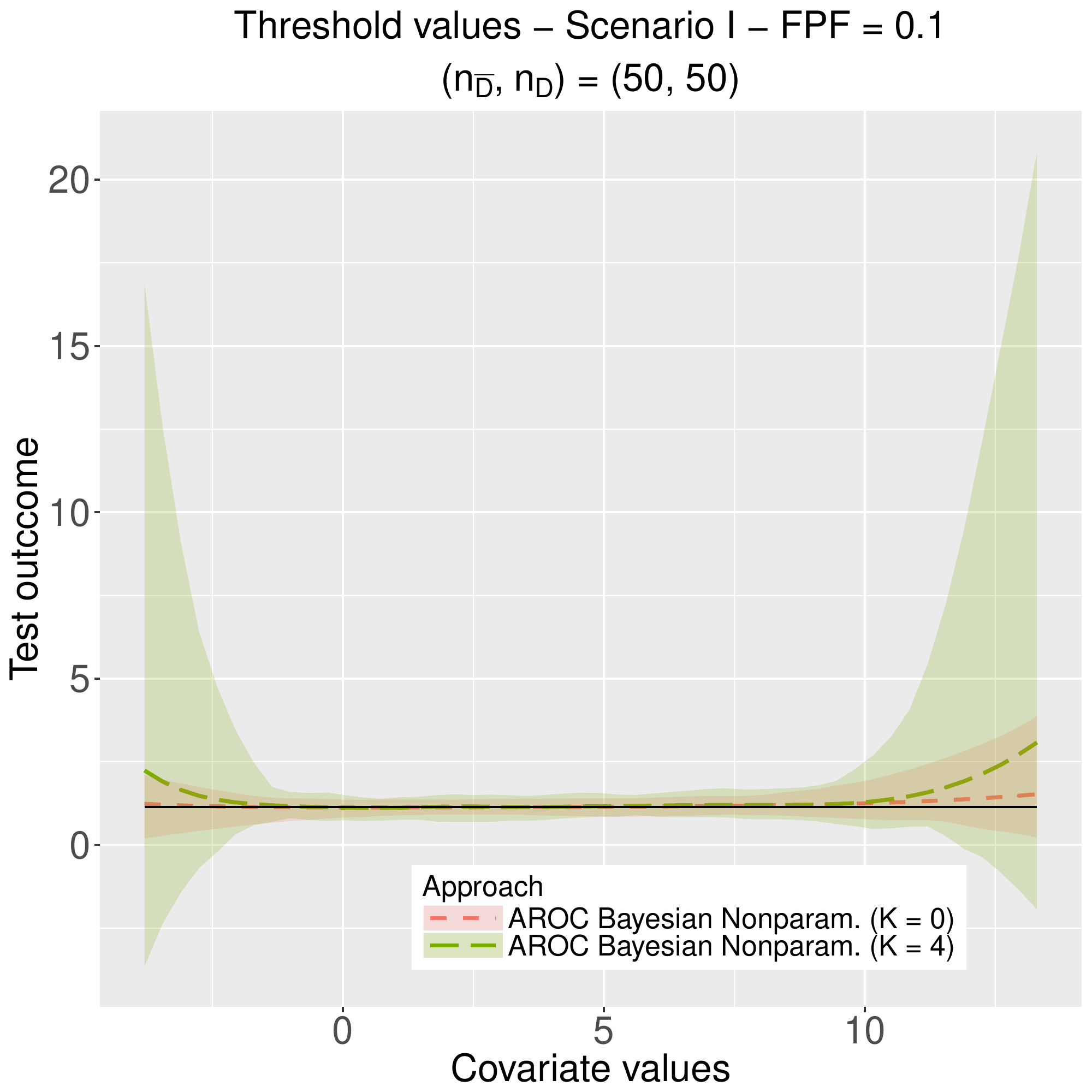}
		\includegraphics[height=4.5cm, page = 1]{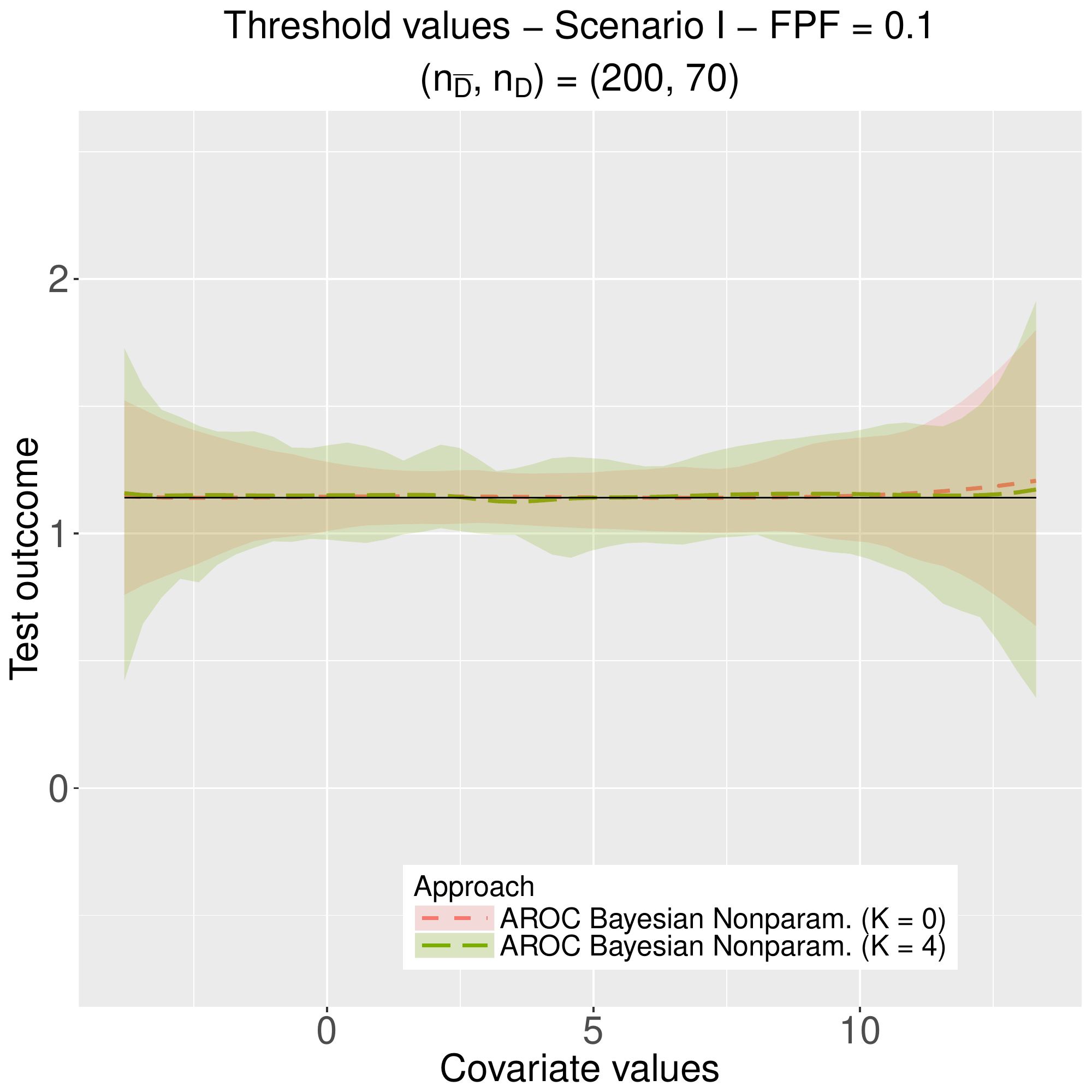}
		\includegraphics[height=4.5cm, page = 1]{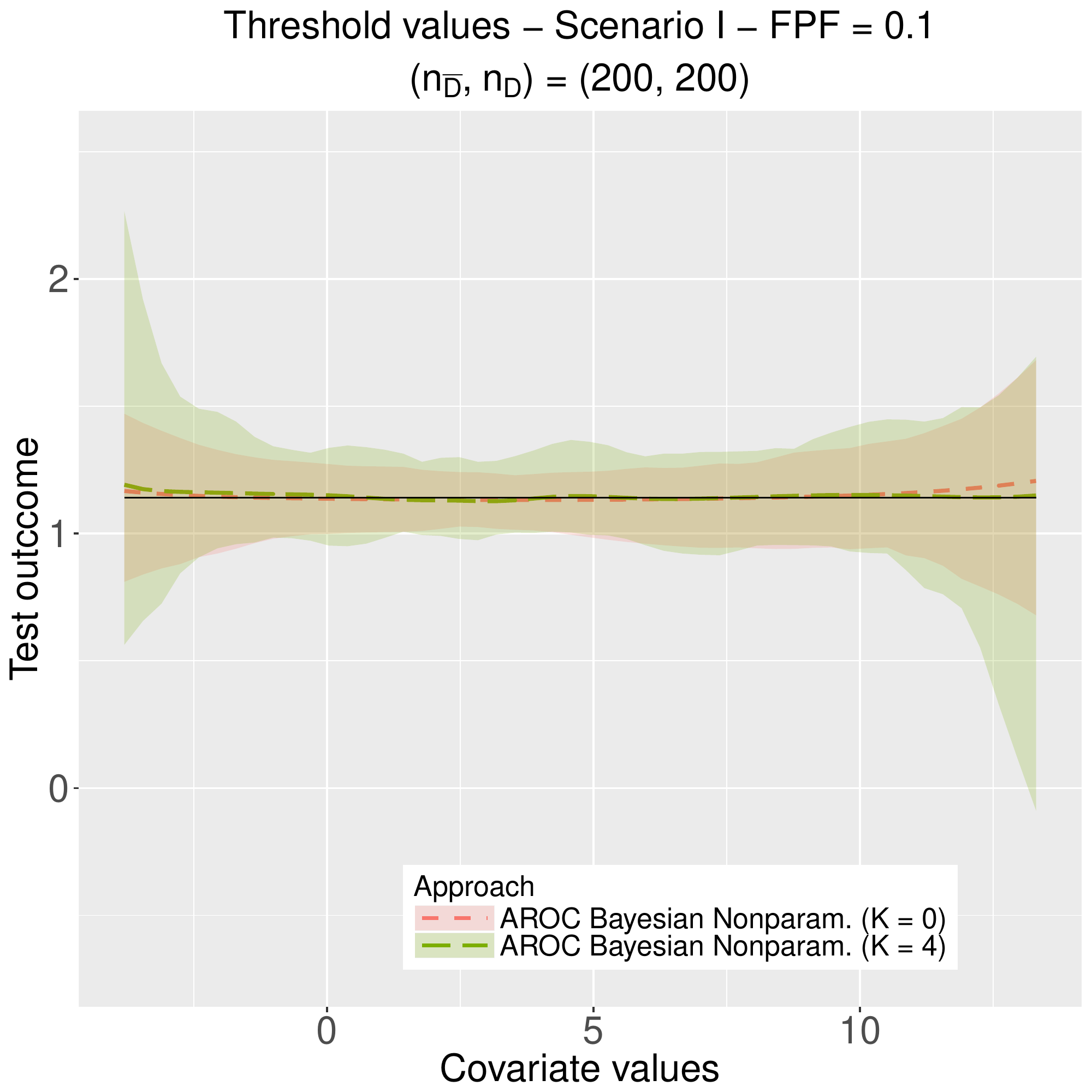}
		\includegraphics[height=4.5cm, page = 1]{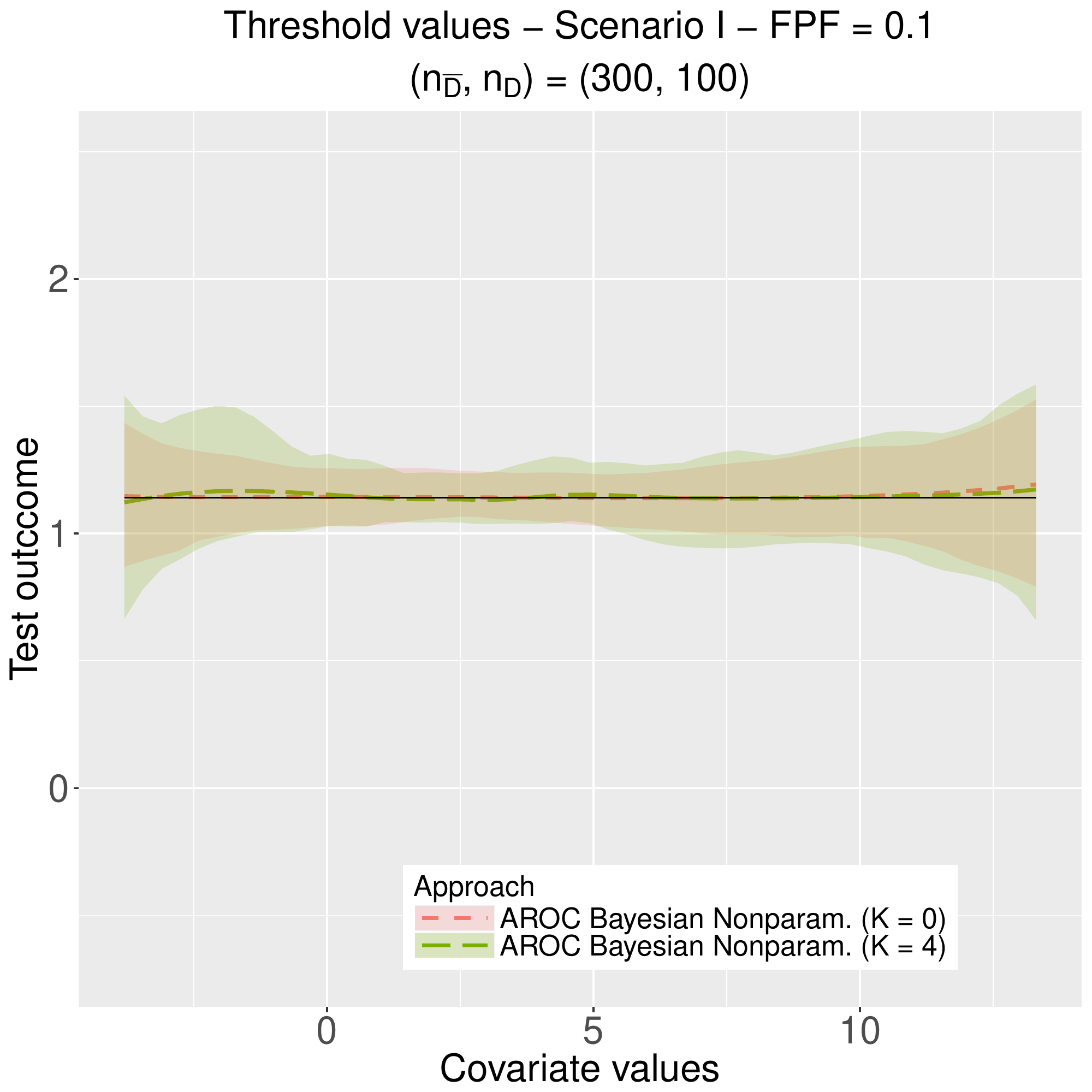}} \vspace{0.3cm}
		\subfigure[FPF = 0.3]{
		\includegraphics[height=4.4cm, page = 2]{sim_thresholds_I_50_50_ndx_1_5_bs.pdf}
		\includegraphics[height=4.4cm, page = 2]{sim_thresholds_I_200_70_ndx_1_5_bs.pdf}
		\includegraphics[height=4.4cm, page = 2]{sim_thresholds_I_200_200_ndx_1_5_bs.pdf}
		\includegraphics[height=4.4cm, page = 2]{sim_thresholds_I_300_100_ndx_1_5_bs.pdf}}
	\end{center}
		 \caption{Scenario I: true (solid black line) and average value of 100 simulated datasets (dashed lines) of the posterior mean of the thresholds used for defining a positive test result for the Bayesian nonparametric approach proposed in this paper. The shaded area are bands constructed using the pointwise $2.5\%$ and $97.5\%$ quantiles across simulations. The results are presented for no interior knots $(K = 0)$ and four interior knots $(K = 4)$ and for each sample size.}
		\label{thresholds_sim_ndx_1_I}
\end{figure}

\begin{figure}[H]
    \begin{center}
    	\subfigure[FPF = 0.1]{
		\includegraphics[height=4.5cm, page = 1]{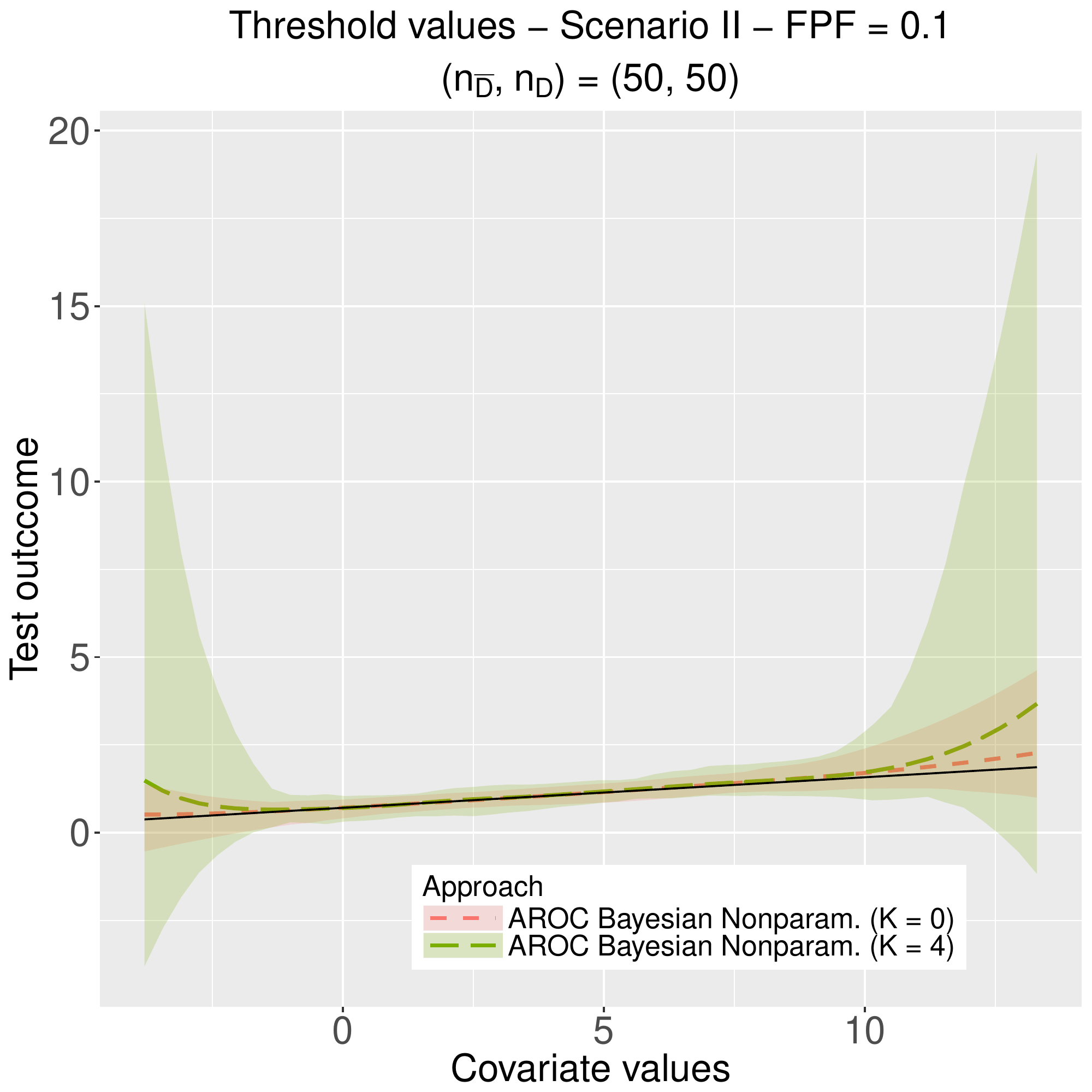}
		\includegraphics[height=4.5cm, page = 1]{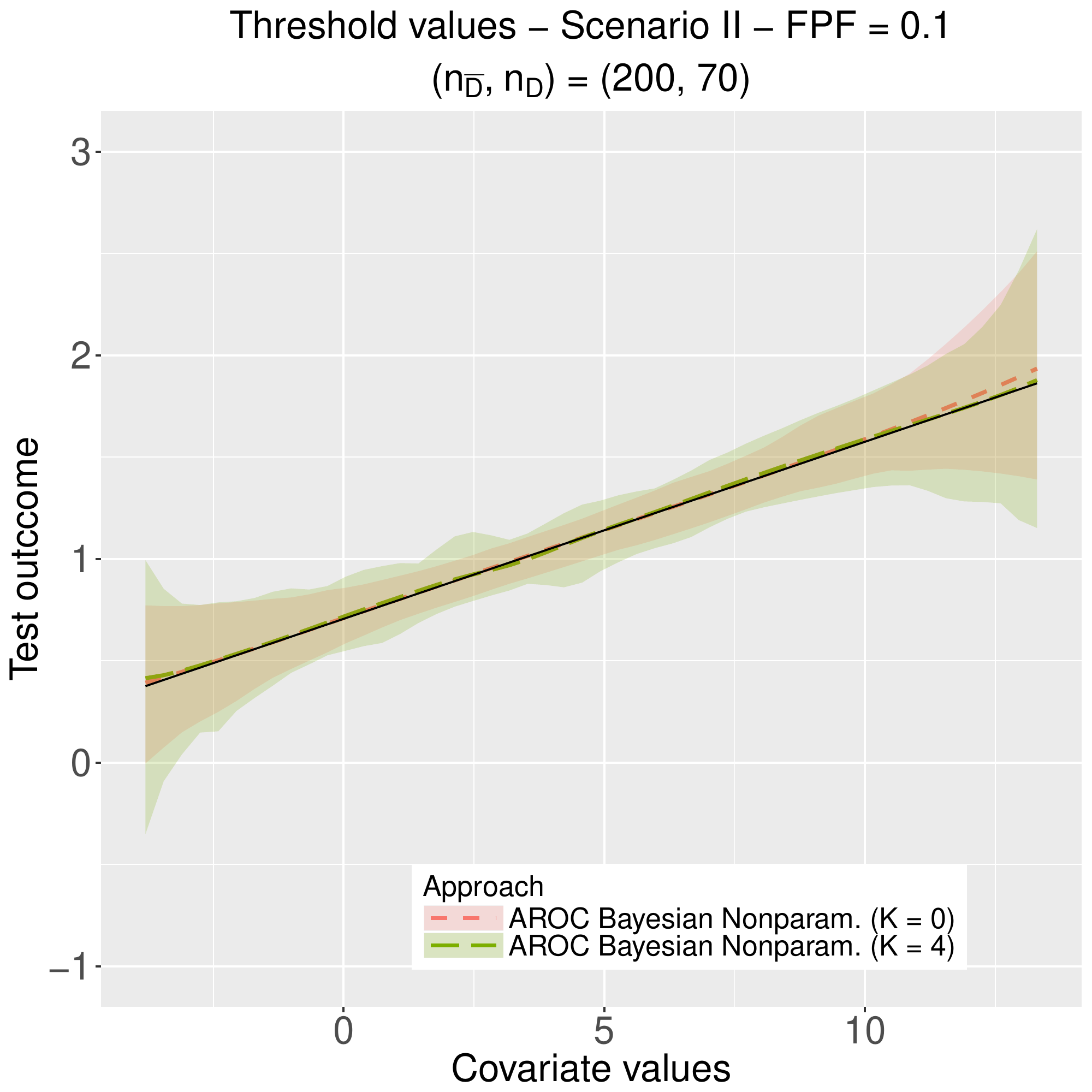}
		\includegraphics[height=4.5cm, page = 1]{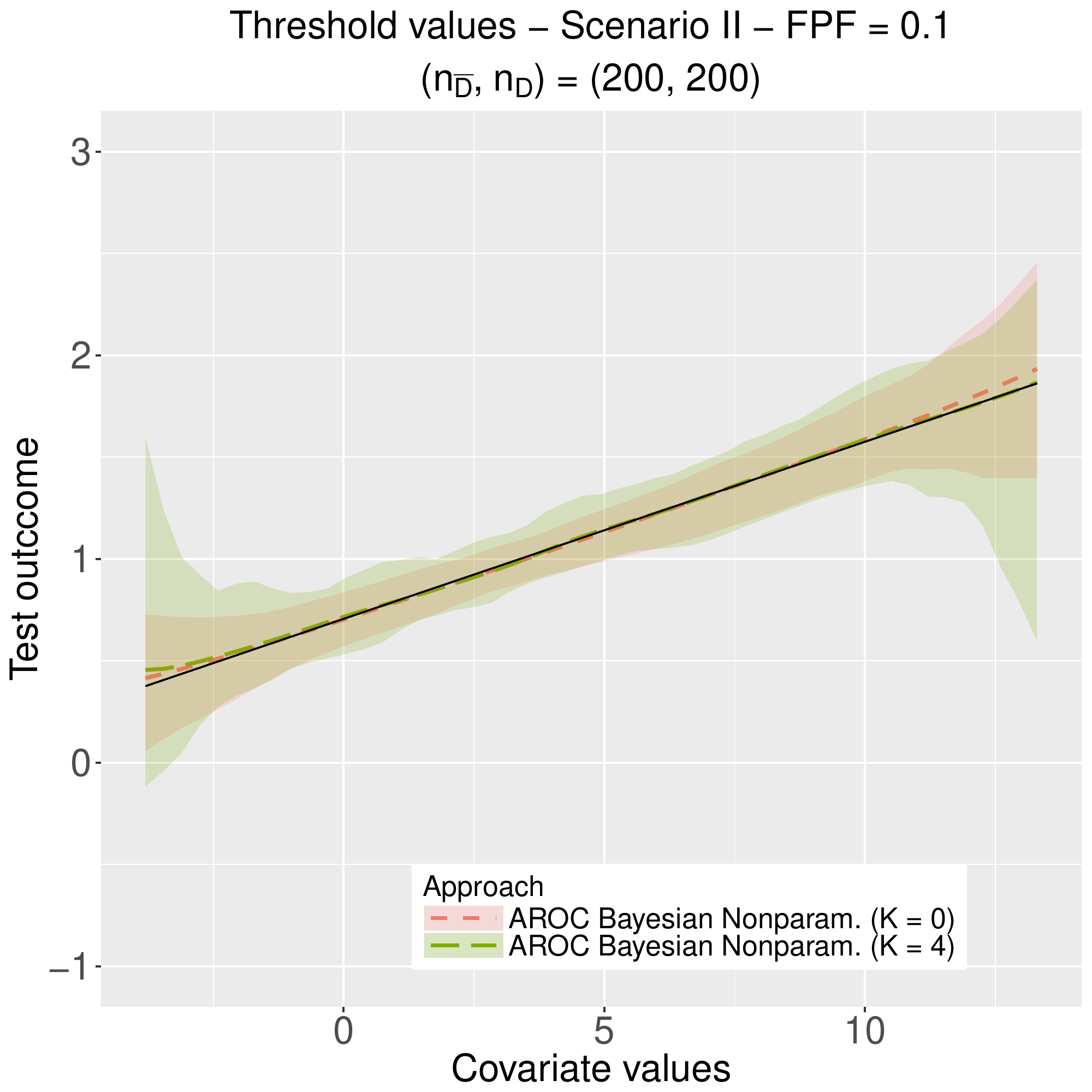}
		\includegraphics[height=4.5cm, page = 1]{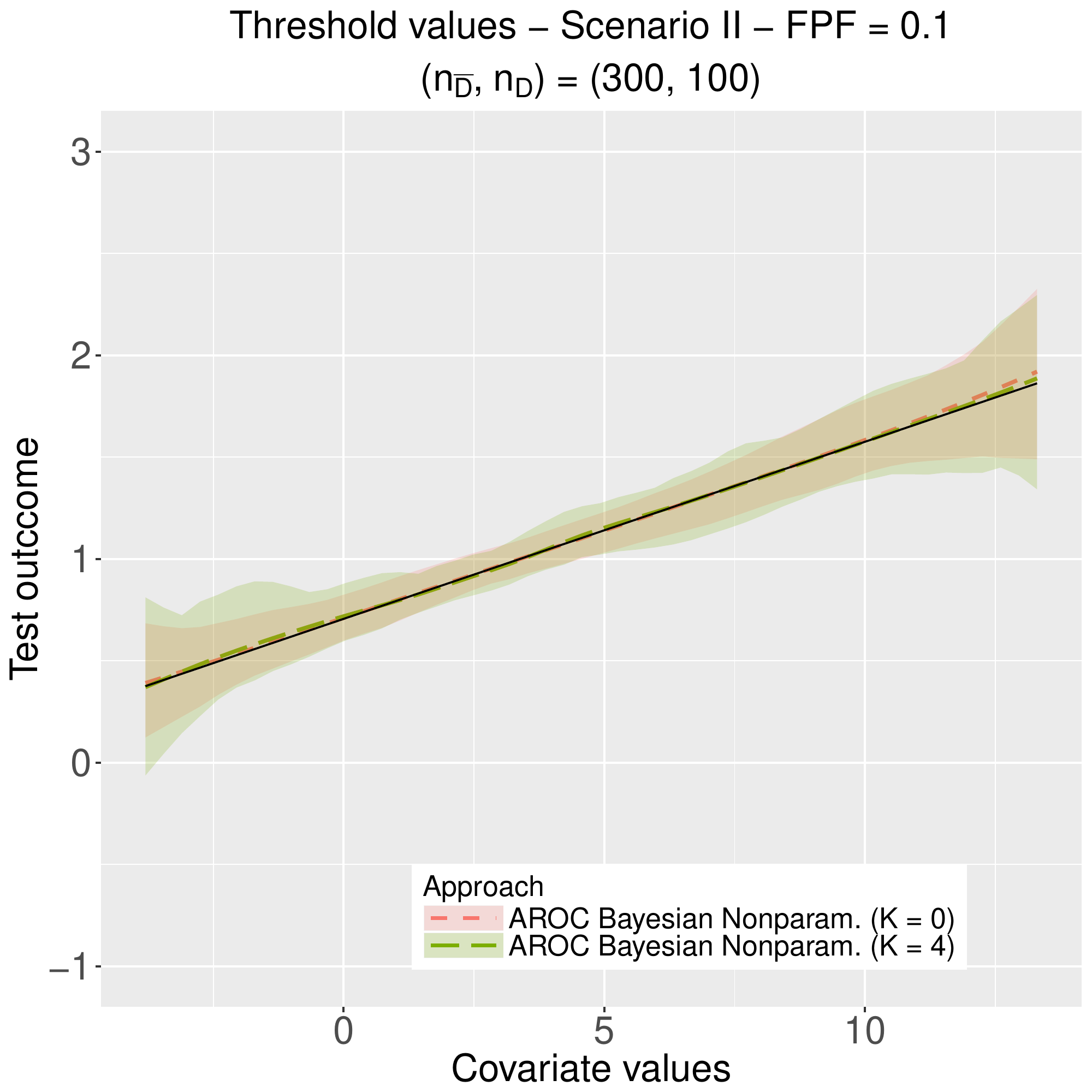}} \vspace{0.3cm}
		\subfigure[FPF = 0.3]{
		\includegraphics[height=4.4cm, page = 2]{sim_thresholds_II_50_50_ndx_1_5_bs.pdf}
		\includegraphics[height=4.4cm, page = 2]{sim_thresholds_II_200_70_ndx_1_5_bs.pdf}
		\includegraphics[height=4.4cm, page = 2]{sim_thresholds_II_200_200_ndx_1_5_bs.pdf}
		\includegraphics[height=4.4cm, page = 2]{sim_thresholds_II_300_100_ndx_1_5_bs.pdf}}
	\end{center}
		 \caption{Scenario II: true (solid black line) and average value of 100 simulated datasets (dashed lines) of the posterior mean of the thresholds used for defining a positive test result for the Bayesian nonparametric approach proposed in this paper. The shaded area are bands constructed using the pointwise $2.5\%$ and $97.5\%$ quantiles across simulations. The results are presented for no interior knots $(K = 0)$ and four interior knots $(K = 4)$ and for each sample size.}
		\label{thresholds_sim_ndx_1_II}
\end{figure}

\begin{figure}[H]
    \begin{center}
    	\subfigure[FPF = 0.1]{
		\includegraphics[height=4.5cm, page = 1]{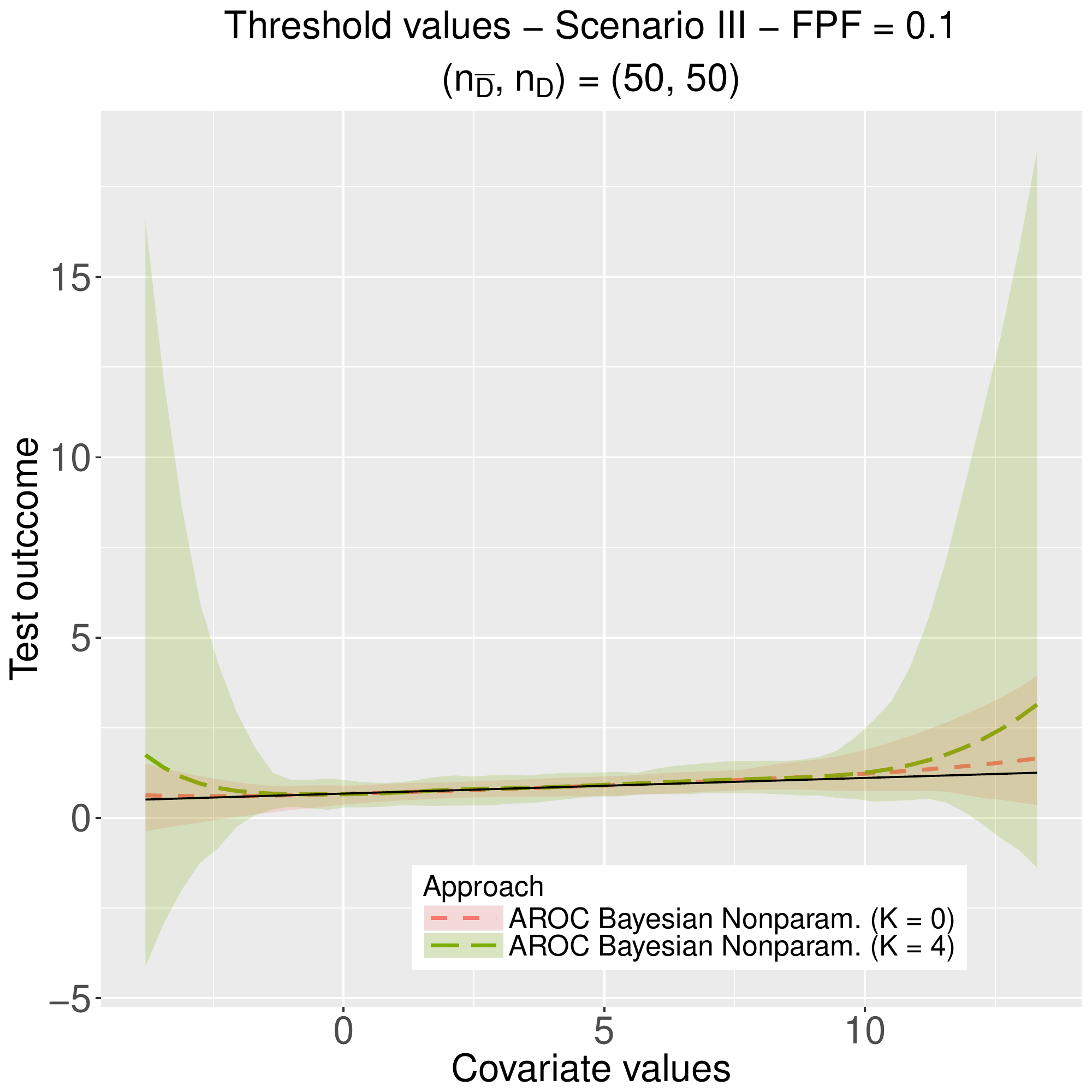}
		\includegraphics[height=4.5cm, page = 1]{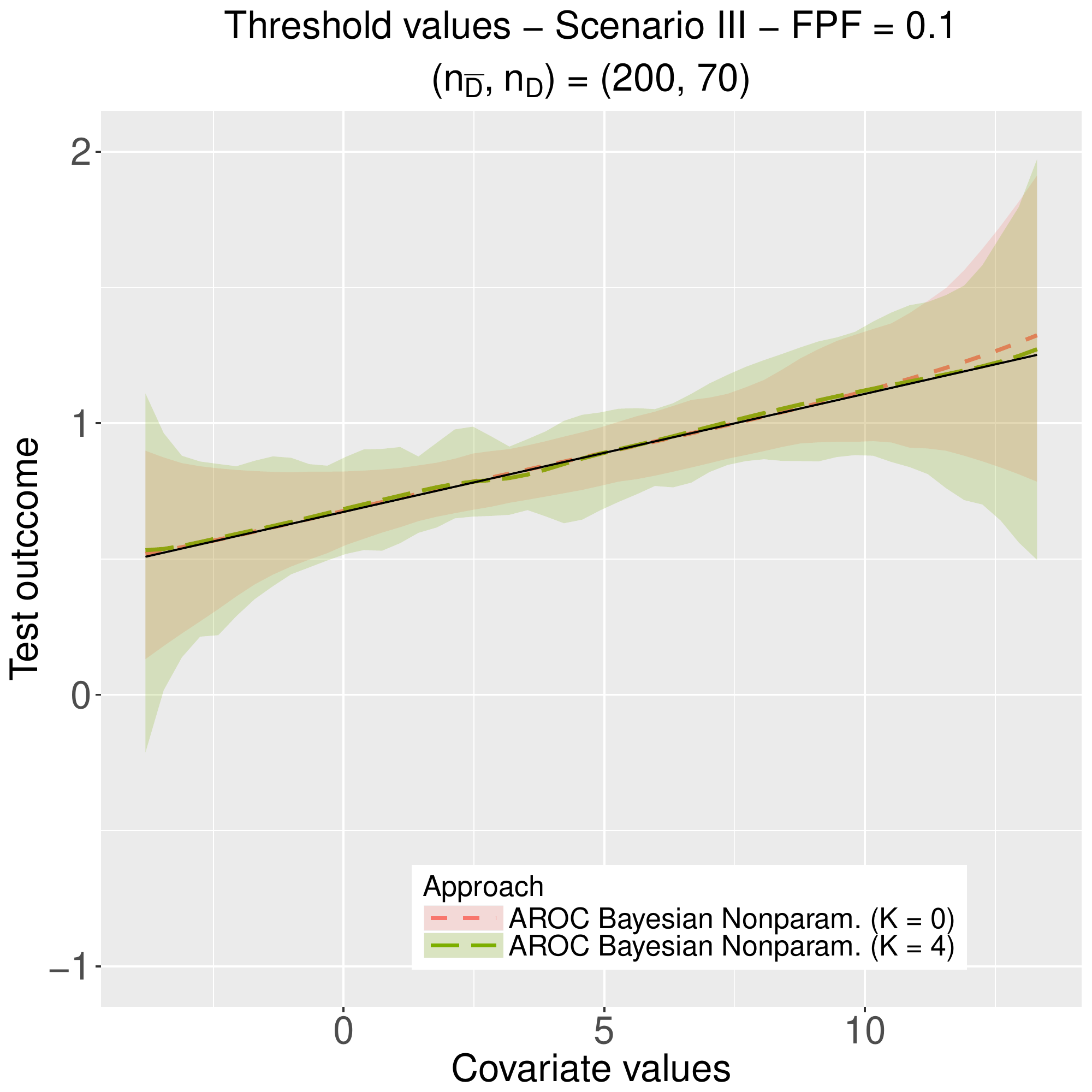}
		\includegraphics[height=4.5cm, page = 1]{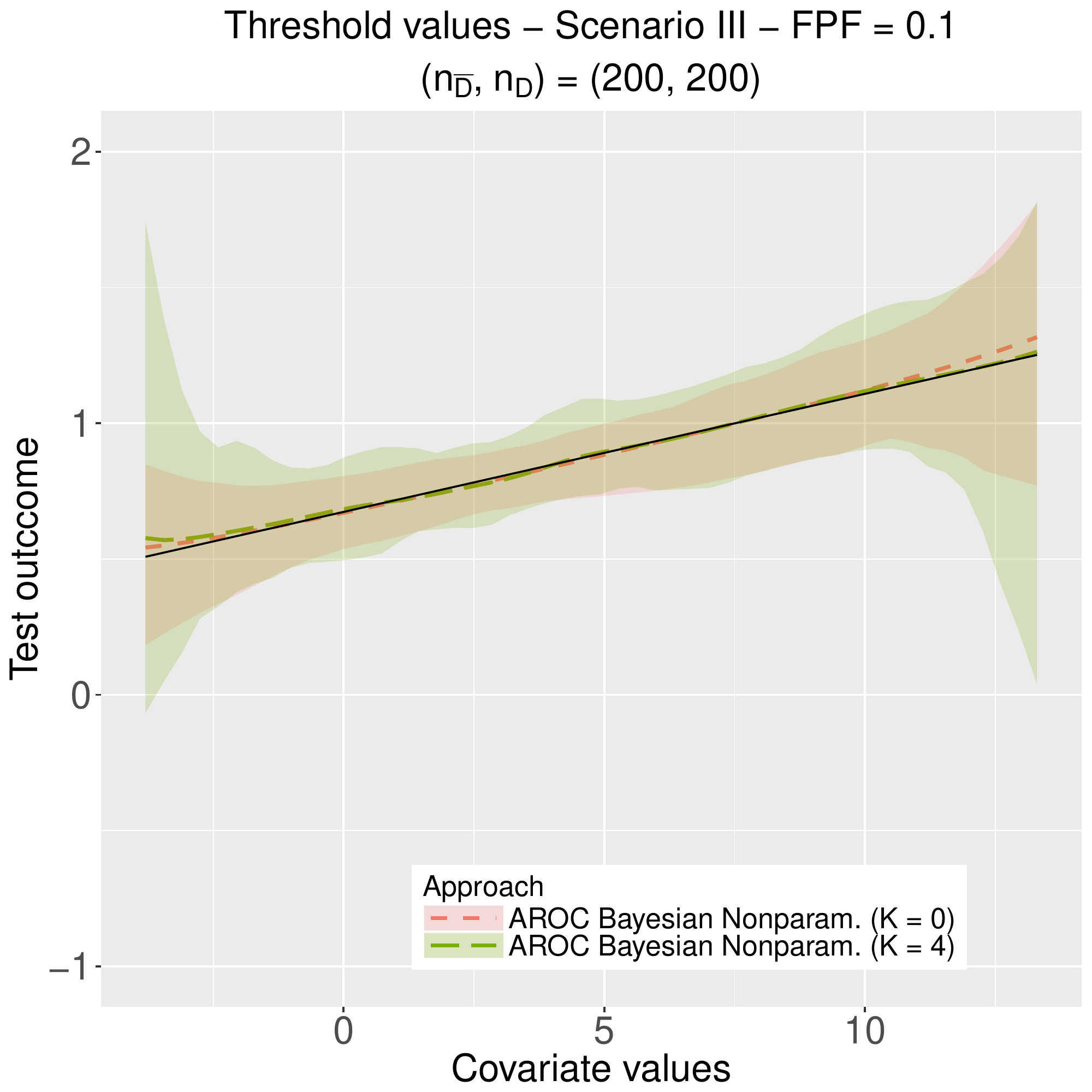}
		\includegraphics[height=4.5cm, page = 1]{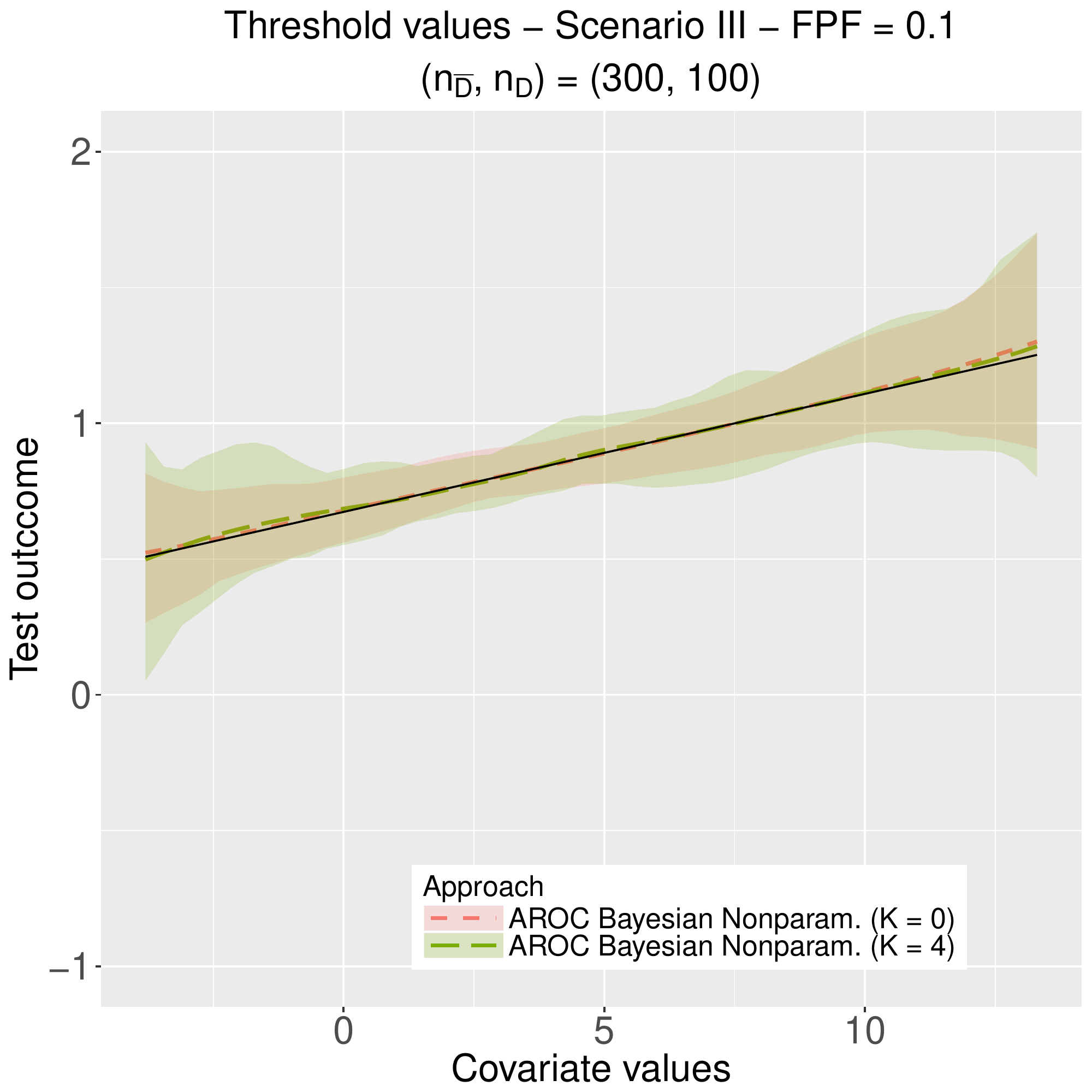}} \vspace{0.3cm}
		\subfigure[FPF = 0.3]{
		\includegraphics[height=4.4cm, page = 2]{sim_thresholds_III_50_50_ndx_1_5_bs.pdf}
		\includegraphics[height=4.4cm, page = 2]{sim_thresholds_III_200_70_ndx_1_5_bs.pdf}
		\includegraphics[height=4.4cm, page = 2]{sim_thresholds_III_200_200_ndx_1_5_bs.pdf}
		\includegraphics[height=4.4cm, page = 2]{sim_thresholds_III_300_100_ndx_1_5_bs.pdf}}
	\end{center}
		 \caption{Scenario III: true (solid black line) and average value of 100 simulated datasets (dashed lines) of the posterior mean of the thresholds used for defining a positive test result for the Bayesian nonparametric approach proposed in this paper. The shaded area are bands constructed using the pointwise $2.5\%$ and $97.5\%$ quantiles across simulations. The results are presented for no interior knots $(K = 0)$ and four interior knots $(K = 4)$ and for each sample size.}
		\label{thresholds_sim_ndx_1_III}
\end{figure}

\begin{figure}[H]
    \begin{center}
    	\subfigure[FPF = 0.1]{
		\includegraphics[height=4.5cm, page = 1]{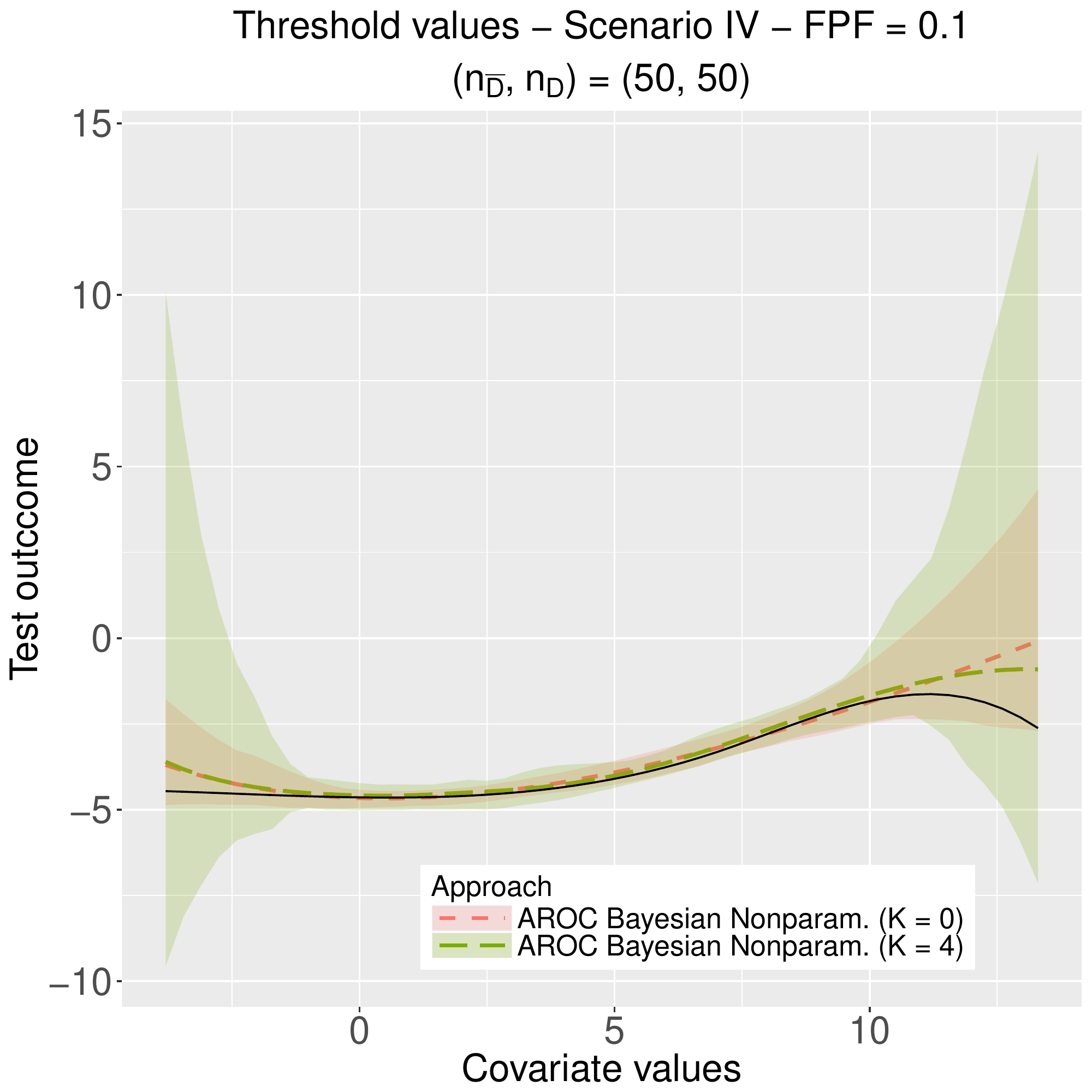}
		\includegraphics[height=4.5cm, page = 1]{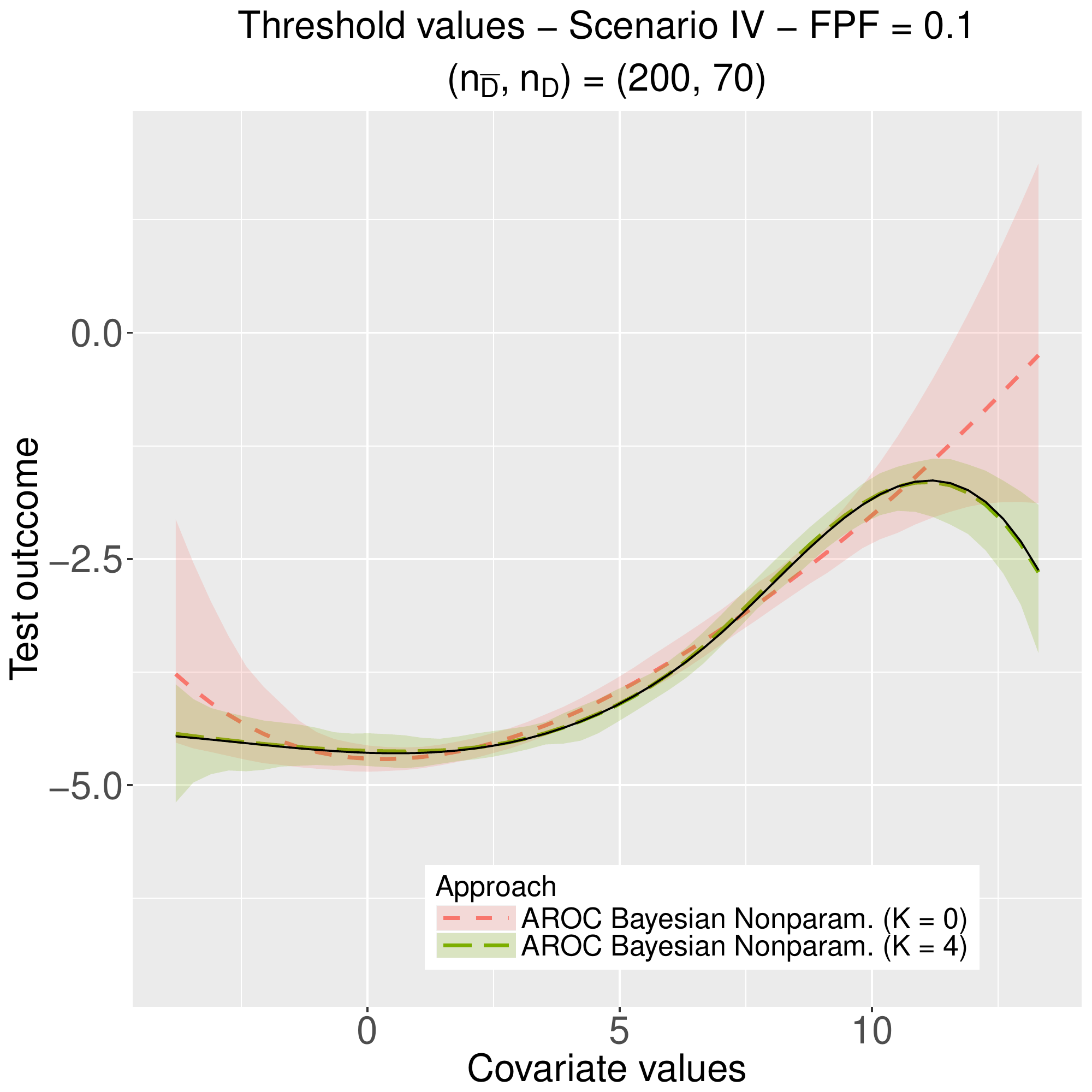}
		\includegraphics[height=4.5cm, page = 1]{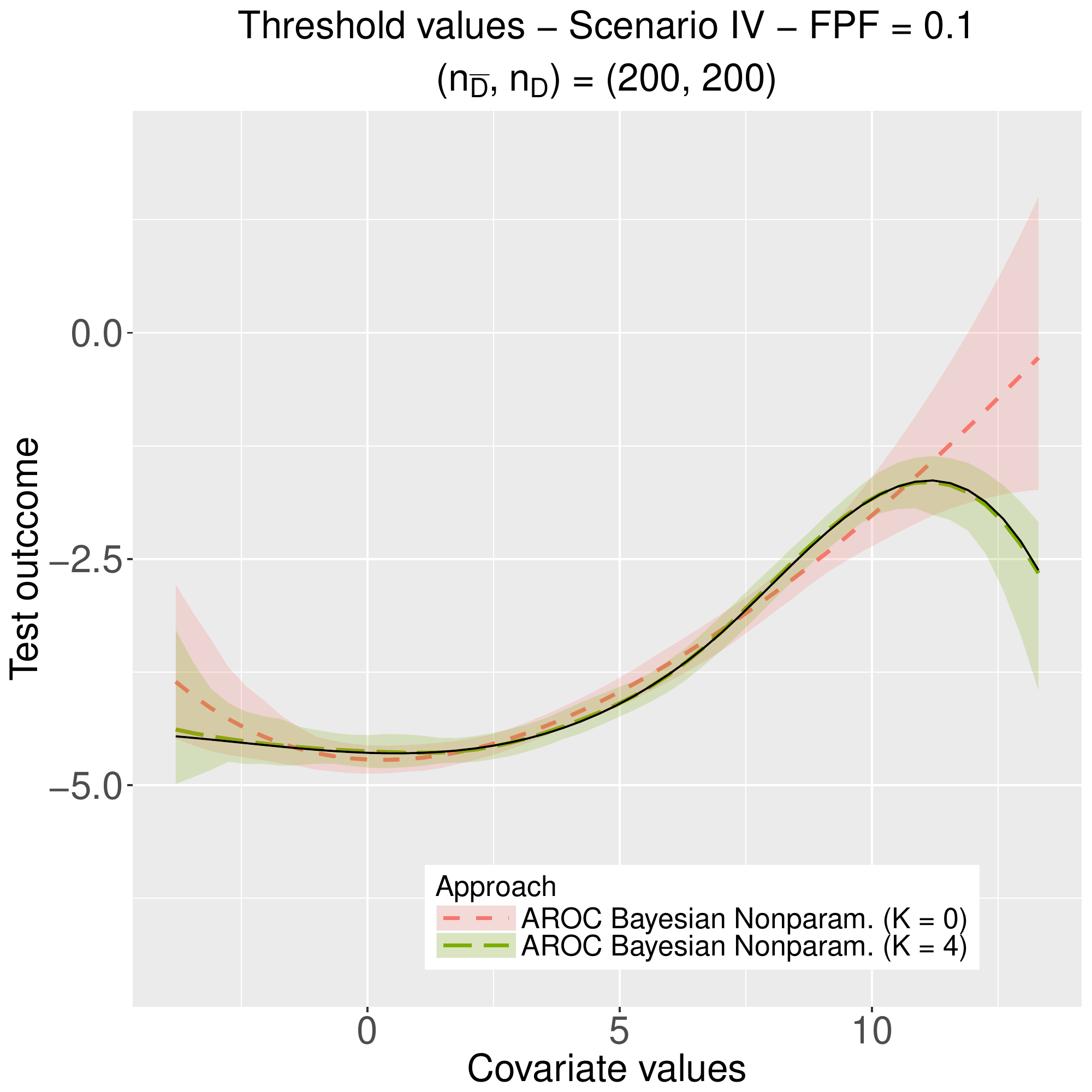}
		\includegraphics[height=4.5cm, page = 1]{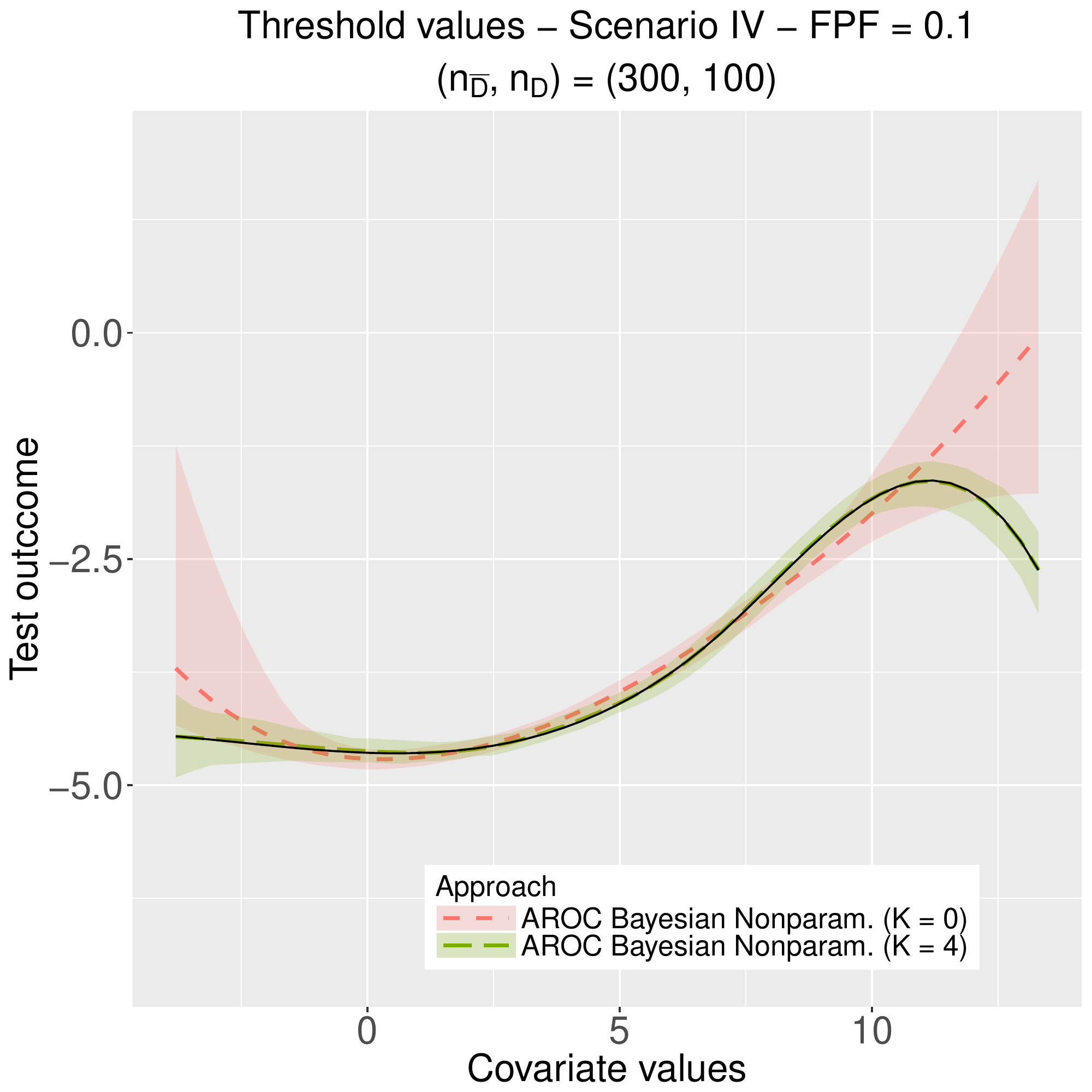}} \vspace{0.3cm}
		\subfigure[FPF = 0.3]{
		\includegraphics[height=4.4cm, page = 2]{sim_thresholds_IV_50_50_ndx_1_5_bs.pdf}
		\includegraphics[height=4.4cm, page = 2]{sim_thresholds_IV_200_70_ndx_1_5_bs.pdf}
		\includegraphics[height=4.4cm, page = 2]{sim_thresholds_IV_200_200_ndx_1_5_bs.pdf}
		\includegraphics[height=4.4cm, page = 2]{sim_thresholds_IV_300_100_ndx_1_5_bs.pdf}}
	\end{center}
		 \caption{Scenario IV: true (solid black line) and average value of 100 simulated datasets (dashed lines) of the posterior mean of the thresholds used for defining a positive test result for the Bayesian nonparametric approach proposed in this paper. The shaded area are bands constructed using the pointwise $2.5\%$ and $97.5\%$ quantiles across simulations. The results are presented for no interior knots $(K = 0)$ and four interior knots $(K = 4)$ and for each sample size.}
		\label{thresholds_sim_ndx_1_IV}
\end{figure}

\newpage

\section{\large{\textsf{SUPPORTING FIGURES AND TABLES FOR THE APPLICATION}}\label{supp_app}}
\begin{table}[H]
\centering
\begin{tabular}{lcc}
& \textbf{Diseased} & \textbf{Nondiseased}\\\hline
\textbf{Global sample} & $29.39\;(27.00, 32.60)$ & $25.14\;(22.44, 28.34)$\\
\hline
\textbf{Gender} & &\\
Female & $30.49\;(27.43, 34.57)$ & $24.39\;(21.78, 27.72)$\\
Male & $29.10\;(26.83, 31.79)$ & $25.89\;(23.58, 28.81)$\\\hline
\textbf{Age} & &\\
$\leq 29.6$ & $27.75\;(24.80, 32.09)$ & $22.96\;(21.00, 25.87)$\\
$(29.6,39.3]$ & $29.38\;(26.71, 32.84)$ & $24.78\;(22.48, 27.68)$\\
$(39.3,50.8]$ & $29.19\;(26.68, 32.69)$ & $25.99\;(23.73, 29.00)$\\
$> 50.8$  & $29.94\;(27.61, 32.58)$ & $28.09\;(25.59, 30.92)$\\
\hline
\end{tabular}
\vspace{0.4cm}
\caption{Median (interquartile range) of the body mass index (BMI) in diseased and nondiseased individuals, males and females, and for four gender strata based on quartiles.
\label{EndoDataDescriptive}}
\end{table}

\begin{table}[H]
\centering
\begin{tabular}{ccccccc}
\multirow{3}{*}{$K$}  & \multicolumn{2}{c}{\textbf{Global}} & \multicolumn{4}{c}{\textbf{Gender}}\\
& \multicolumn{2}{c}{\textbf{sample}} & \multicolumn{2}{c}{Female} & \multicolumn{2}{c}{Male} \\
& WAIC & LPML & WAIC & LPML & WAIC & LPML \\\hline
0 & 11736 & -5868 &  6904 & -3452 & 4836 & -2418 \\
1 & 11739 & -5870 &  6906 & -3454 & 4837 & -2419 \\
2 & 11744 & -5873 &  6908 & -3454 & 4838 & -2419 \\
3 & 11754 & -5877 &  6912 & -3456 & 4842 & -2421 \\
4 & 11762 & -5881 &  6916 & -3458 & 4845 & -2423 \\
\hline
\end{tabular}
\vspace{0.4cm}
\caption{WAIC and LPML for the Bayesian nonparametric approach for different number of inner knots and for the analysis using the global sample, and in females and males separately. \label{WAIC_LPML_knots_appl}}
\end{table}

\begin{table}[H]
\centering
\begin{tabular}{ccccccccc}
& $\alpha$ & $L$ & $\mathbf{m}_0$ & $\mathbf{S}_0$ & $\nu$ & $\Psi^{-1}$ & $a$ & $b$\\ \hline
Prior 1 & 0.1 & 30 & $\mathbf{0}_8$ & $100I_8$ & $10$ & $I_8$ & 2 & 0.5\\
Prior 2 & 10 & 30 & $\mathbf{0}_8$ & $100I_8$ & $10$ & $I_8$ & 2 & 0.5\\
Prior 3 & 1 & 10 & $\mathbf{0}_8$ & $100I_8$ & $10$ & $I_8$ & 2 & 3\\
\end{tabular}
\vspace{0.4cm}
\caption{Hyperparameter specification used in the sensitivity analysis.}\label{sens_analysis_app}
\label{sa}
\end{table}

\clearpage
\putbib[References_AROC]
\end{bibunit}
\end{document}